\documentclass[twoside,openany,12pt]{Latex/Classes/PhDthesisPSnPDF}
\usepackage{hyperref}
\usepackage{array}[=2016-10-06]
\hypersetup{
  colorlinks=true,
  linkcolor=blue,
  citecolor=blue,
  urlcolor=blue
}

\usepackage{xstring}
\usepackage[acronym,automake]{glossaries}
\makeglossaries

\newacronym{iot}{IoT}{Internet of Things}
\newacronym{iiot}{IIoT}{Industrial Internet of Things}
\newacronym{hiot}{H-IoT}{Healthcare-IoT}
\newacronym{iomt}{IoMT}{Internet of Medical Things}
\newacronym{ioht}{IoHT}{Internet of Healthcare Things}
\newacronym{g5}{5G}{fifth-generation}
\newacronym{lte}{LTE}{long-term evolution}
\newacronym{ddos}{DDoS}{Distributed Denial-of-Service}
\newacronym{sf}{SF}{Selective Forwarding}
\newacronym{mitm}{MITM}{Man-in-the-Middle}
\newacronym{lstm}{LSTM}{Long Short-Term Memory}
\newacronym{bilstm}{Bi-LSTM}{Bidirectional Long Short-Term Memory}
\newacronym{cnn}{CNN}{Convolutional Neural Network}
\newacronym{knn}{KNN}{K-Nearest Neighbors}
\newacronym{ml}{ML}{Machine Learning}
\newacronym{dl}{DL}{Deep Learning}
\newacronym{gdpr}{GDPR}{General Data Protection Regulation}
\newacronym{mqtt}{MQTT}{Message Queuing Telemetry Transport}
\newacronym{udp}{UDP}{User Datagram Protocol}
\newacronym{tcp}{TCP}{Transmission Control Protocol}
\newacronym{coap}{CoAP}{Constrained Application Protocol}
\newacronym{cia}{CIA}{Confidentiality, Integrity, and Availability}
\newacronym{nbiot}{NB-IoT}{Narrowband Internet of Things}
\newacronym{nr}{NR}{New Radio}
\newacronym{5gnr}{5G NR}{5G New Radio}
\newacronym{e2e}{E2E}{end-to-end}
\newacronym{hvac}{HVAC}{Heating, Ventilation, and Air Conditioning}
\newacronym{pii}{PII}{Personal Identifiable Information}
\newacronym{xss}{XSS}{Cross-Site Scripting}
\newacronym{rfid}{RFID}{Radio Frequency Identification}
\newacronym{rf}{RF}{Radio Frequency}
\newacronym{ble}{BLE}{Bluetooth Low Energy}
\newacronym{6lowpan}{6LoWPAN}{IPv6 over Low-Power Wireless Personal Area Networks}
\newacronym{nis2d}{NIS2D}{Network and Information Security Directive 2}
\newacronym{icmp}{ICMP}{Internet Control Message Protocol}
\newacronym{dnngan}{DNN-GANs}{Deep Neural Network-based Generative Adversarial Network}
\newacronym{ccn}{CCN}{Content-Centric Networking}
\newacronym{isp}{ISP}{Internet Service Provider}
\newacronym{ict}{ICT}{Information and Communication Technology}
\newacronym{url}{URL}{Uniform Resource Locator}
\newacronym{mac}{MAC}{Media Access Control}
\newacronym{ip}{IP}{Internet Protocol}
\newacronym{he}{HE}{High Efficiency}
\newacronym{vht}{VHT}{Very High Throughput}
\newacronym{eerp}{EERP}{Energy-Efficient Routing Protocol}
\newacronym{mumimo}{MU-MIMO}{Multi-User Multiple Input Multiple Output}
\newacronym{sql}{SQL}{Structured Query Language}
\newacronym{hse}{HSE}{Health Service Executive}
\newacronym{it}{IT}{Information Technology}
\newacronym{svm}{SVM}{Support Vector Machine}
\newacronym{ann}{ANN}{Artificial Neural Network}
\newacronym{ehealth}{e-Health}{electronic Health}
\newacronym{tcn}{TCN}{Temporal Convolutional Network}
\newacronym{restcn}{Res-TCN}{Residual Temporal Convolutional Network}
\newacronym{dt}{DT}{Decision Tree}
\newacronym{cnnbilstmgru}{CNN-BiLSTM-GRU}{Convolutional Neural Network-Bidirectional Long Short-Term Memory-Gated Recurrent Unit}
\newacronym{cnnbilstm}{CNN-BiLSTM}{Convolutional Neural Network-Bidirectional Long Short-Term Memory}
\newacronym{dnn}{DNN}{Deep Neural Network}
\newacronym{ids}{IDS}{Intrusion Detection System}
\newacronym{arp}{ARP}{Address Resolution Protocol}
\newacronym{smote}{SMOTE}{Synthetic Minority Oversampling Technique}
\newacronym{ue}{UE}{User Equipment}
\newacronym{wban}{WBAN}{Wireless Body Area Network}
\newacronym{bsn}{BSN}{Body Sensor Network}
\newacronym{wsn}{WSN}{Wireless Sensor Network}
\newacronym{sdn}{SDN}{Software-Defined Networking}
\newacronym{ehds}{EHDS}{European Health Data Space}
\newacronym{hiids}{HIIDS}{Hybrid Intelligent Intrusion Detection System}
\newacronym{ehr}{EHR}{Electronic Health Record}
\newacronym{fmmnn}{FMMNN}{Fuzzy Min-Max Neural Network}
\newacronym{hipaa}{HIPAA}{Health Insurance Portability and Accountability Act}
\newacronym{http}{HTTP}{Hypertext Transfer Protocol}
\newacronym{rpl}{RPL}{Routing Protocol for Low-Power and Lossy Networks}
\newacronym{jni}{JNI}{Java Native Interface}
\newacronym{ai}{AI}{Artificial Intelligence}
\newacronym{dos}{DoS}{Denial of Service}
\newacronym{rnn}{RNN}{Recurrent Neural Network}
\newacronym{gan}{GAN}{Generative Adversarial Network}

\newacronym{dscp}{DSCP}{Differentiated Services Code Point}
\newacronym{qos}{QoS}{Quality of Service}
\newacronym{tflite}{TFLite}{TensorFlow Lite}
\newacronym{gru}{GRU}{Gated Recurrent Unit}
\newacronym{ff}{FF}{Feedforward}
\newacronym{fp32}{FP32}{32-bit floating-point}
\newacronym{usb}{USB}{Universal Serial Bus}
\newacronym{arm}{ARM}{Advanced RISC Machines}
\newacronym{lr}{LR}{Logistic Regression}
\newacronym{nb}{NB}{Naïve Bayes}
\newacronym{el}{EL}{Ensemble Learning}
\newacronym{1dclstm}{1D-CLSTM}{One-Dimensional Convolution Long Short-Term Memory}
\newacronym{vae}{VAE}{Variational Autoencoder}
\newacronym{ece}{ECE}{Expected Calibration Error}
\newacronym{enisa}{ENISA}{European Union Agency for Cybersecurity}
\newacronym{shap}{SHAP}{Shapley Additive exPlanations}
\newacronym{lime}{LIME}{Local Interpretable Model-agnostic Explanations}
\newacronym{mscsl}{MSCSL}{Multi Step Convolutional Neural Network Stacked Long Short Term Memory Architecture}
\newacronym{ehms}{EHMS}{Enhanced Healthcare Monitoring System}
\newacronym{dhcp}{DHCP}{Dynamic Host Configuration Protocol}

\newacronym{b5g}{B5G}{Beyond Fifth-Generation}

\usepackage{textcomp} 
\usepackage{csquotes}         
\usepackage[
  backend=biber,
  style=ieee,
  sorting=none,
  backref=true 
]{biblatex}
\usepackage{url}
\usepackage{xurl}

\usepackage{alltt} 
\usepackage{needspace}
\usepackage{placeins} 
\usepackage{subcaption}
\usepackage{multirow}
\usepackage{xcolor}
\usepackage{colortbl}
\usepackage{booktabs, threeparttable, cellspace, tabularx}
\usepackage{caption}
\usepackage{longtable}
\usepackage{enumitem}
\usepackage{amssymb}
\usepackage{algorithm}
\usepackage{algpseudocode}
\usepackage{array}
\usepackage{graphicx,color}
\usepackage{float}
\usepackage{amsmath,amstext,amsgen,amsfonts} 
\usepackage{makecell}
\usepackage{adjustbox}
\usepackage{pdflscape}
\usepackage{multicol} 
\usepackage{pifont}
\usepackage{ragged2e} 
\newcolumntype{Y}{>{\RaggedRight\arraybackslash}X}
\newcolumntype{L}[1]{>{\RaggedRight\arraybackslash}p{#1}}
\newif\ifrevision
\revisionfalse 

\ifrevision
  \newcommand{\rev}[1]{\textcolor{blue}{#1}}
\else
  \newcommand{\rev}[1]{#1}
\fi

\newcommand{\urlwofont}[1]
{
\urlstyle{same}\url{#1}
}
\usepackage{mathtools}

\definecolor{c++red}{rgb}{0.6,0,0} 
\definecolor{c++green}{rgb}{0.25,0.5,0.35} 
\definecolor{c++purple}{rgb}{0.5,0,0.35} 
\usepackage{listings}
\makeatletter  
\def\url@leostyle{%
  \@ifundefined{selectfont}{\def\UrlFont{\sf}}{\def\UrlFont{\small\ttfamily}}}
\makeatother

\DeclareCaptionLabelSeparator*{bluecolon}{\textcolor{blue}{: }}
\ifpdf  
    \pdfcatalog { /PageMode (/UseOutlines)
                  /OpenAction (fitbh)  }
\fi

\title{Deep Learning for Cyber Threat Detection and Mitigation in Healthcare-IoT}

\usepackage{tikz}
\usepackage{pgfplots}
\pgfplotsset{compat=1.18}

\ifpdf
  \author{Mirza Akhi Khatun}
 \researchcentre{Data-Driven Computer Engineering (D\textsuperscript{2}ICE) Research Centre}
  \faculty{Department of Electronic and Computer Engineering}
  \department{Faculty of Science and Engineering}
  \university{University of Limerick}

  \crest{\includegraphics[scale=0.45]{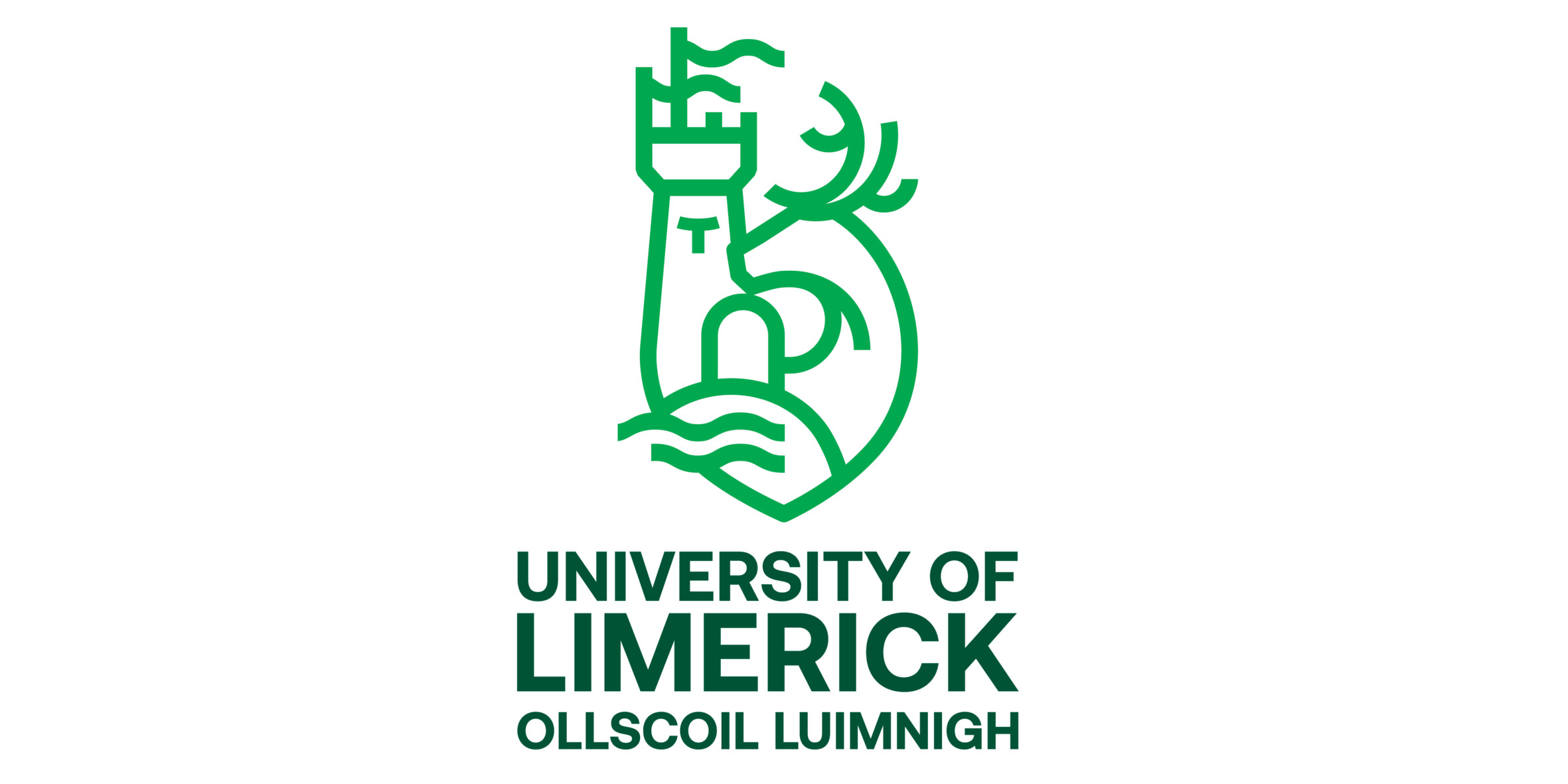}}

\degree{Philosophi\ae Doctor (PhD) 2026}
\degreedate{}

\begin{document}
\emergencystretch 3em
\makeatletter
\let\origdoublepage\cleardoublepage
\renewcommand{\cleardoublepage}{\clearpage} 
\makeatother

\maketitle

\renewcommand\baselinestretch{1.2}
\baselineskip=18pt plus1pt




\newpage

 \begin{tabbing}
1. Supervisor: \= Dr. Lubna Luxmi Dhirani \\
 \>  \= Department of Electronic and Computer Engineering \\
 \>  \= University \emph{of} Limerick \\
 \>  \= Ireland \\
\end{tabbing}

 \begin{tabbing}
2.  Supervisor: \= Professor Ciarán Eising    \\
 \>  \= Department of Electronic and Computer Engineering\\
 \>  \= University \emph{of} Limerick \\
 \>  \= Ireland \\
 \end{tabbing}

 \begin{tabbing}
1. External Examiner: \= Dr. Mona Jaber \\
 \>  \= School of Electronic Engineering and Computer Science \\
 \>  \= Queen Mary University \emph{of} London  \\
 \>  \=  UK \\
\end{tabbing}

\begin{tabbing}
2. Internal Examiner: \= Professor Thomas Newe \\
 \>  \= Department of Electronic and Computer Engineering\\
 \>  \= University \emph{of} Limerick \\
 \>  \= Ireland \\
\end{tabbing}

 \begin{tabbing}
1. Chair: \= Dr. Muzaffar Rao \\
 \>  \= Department of Electronic and Computer Engineering \\
 \>  \= University \emph{of} Limerick \\
 \>  \= Ireland \\
\end{tabbing}

Day of the defence: 2\textsuperscript{nd} April 2026

\vspace{10mm}
\hspace{60mm}Signature from head of PhD committee:

\makeatletter
\let\cleardoublepage\origdoublepage 
\makeatother


\frontmatter
\pagenumbering{roman}
\setcounter{page}{3}

\thispagestyle{empty}
\vspace*{6cm}
\begin{center}
\noindent\rule{0.3\textwidth}{0.4pt}\\[1em]
\textit{``Life always offers more than we expect, even beyond imagination.''}\\[1em]
--- \textbf{Mirza Akhi}\\[1em]
\noindent\rule{0.3\textwidth}{0.4pt}
\end{center}



\begin{declaration}        

I herewith declare that I have produced this thesis without the prohibited assistance of third parties and without making use of aids other than those specified; notions taken over directly or indirectly from other sources have been identified as such. This thesis has not previously been presented in identical or similar form to any other Irish or foreign examination board.

This doctoral thesis was conducted from 2022 to 2026 under the supervision of Dr. Lubna Luxmi Dhirani and Professor Ciarán Eising at the University of Limerick.

\vspace{10mm}
Mirza Akhi Khatun

Limerick, 2026

\end{declaration}


\begin{acknowledgements} 
I would like to express my sincere gratitude to my supervisors, Dr.~Lubna Luxmi Dhirani, and Professor Ciarán Eising for their invaluable guidance, generosity, and expertise throughout my PhD journey. I am especially grateful to Professor Ciarán~Eising for his generosity and prompt support whenever needed. Over the past three years, their insightful feedback has been instrumental in shaping my research and professional development. I would also like to express my gratitude to Professor James Gleeson, Funding Director of the Centre for Research Training (CRT), for his support in securing this PhD funding.

I am thankful to my colleagues from the \textit{D\textsuperscript{2}iCE} Research Centre and the \textit{CRT} for their kind collegiality. 

I am grateful to \textit{Tata Consultancy Services (TCS)} for my PhD internship, to Deepanand Saha Roy, Sreekanth Palagiri, and the team for their support and collaboration.

I sincerely appreciate everyone who offered kind words that motivated me throughout this non-linear path. I also acknowledge \textit{YouTube} for music such as \textit{Titanium~(I am bulletproof, nothing to lose)}, which helped me stay focused and resilient. During these years, the UL campus, where every lane has witnessed my happiness, struggles, and tears, will always hold a special place in my heart.

I am truly blessed to have a loving and supportive family whose belief and encouragement have been the foundation of my perseverance and success. This PhD stands on their unseen sacrifices and is as much theirs as mine.

This research is funded by the Science Foundation Ireland (SFI) CRT in Foundations of Data Science under Grant Number 18/CRT/6049.
\end{acknowledgements}




\begin{dedication} 

\vspace{31mm}

\begin{center}
\textit{Dedicated to my beloved Mom and Dad.}
\end{center}

\end{dedication}






\begingroup
  \thispagestyle{fancy}
    \fancyfoot{} 

  \markboth{\MakeUppercase{Publications}}{\MakeUppercase{Publications}}

\thispagestyle{fancy}
\fancyfoot{}
\markboth{\MakeUppercase{Publications}}{\MakeUppercase{Publications}}

\definecolor{Q1color}{RGB}{0,128,0}    
\definecolor{Q2color}{RGB}{255,247,0}    
\definecolor{Q3color}{RGB}{255,92,0}    
\definecolor{Q4color}{RGB}{205,28,24}    
\definecolor{confcolor}{RGB}{149, 117, 205} 
\definecolor{linkcolor}{RGB}{41, 128, 185} 
\definecolor{datasetcolor}{RGB}{155, 89, 182} 

\newcommand{\quartile}[1]{%
    \ifnum#1=1
        \textcolor{Q1color}{\textbf{[Q1]}}%
    \else\ifnum#1=2
        \textcolor{Q2color}{\textbf{[Q2]}}%
    \else\ifnum#1=3
        \textcolor{Q3color}{\textbf{[Q3]}}%
    \else\ifnum#1=4
        \textcolor{Q4color}{\textbf{[Q4]}}%
    \fi\fi\fi\fi
}

\newcommand{\confbadge}{%
    \textcolor{confcolor}{\textbf{[C]}}%
}

\newcommand{\datasetlink}[2]{%
    \StrBehind{#2}{doi.org/}[\shortdoi]%
    \par\vspace{0.2em}
    \hspace{0.5em}\textcolor{datasetcolor}{$\blacktriangleright$} \textbf{\small Dataset:} \emph{#1}, \textbf{DOI:} \href{#2}{\shortdoi}%
}


\vspace{1em}
\noindent The following publications are based on the research conducted by the author, who served as the primary contributor to this doctoral thesis.\footnote{Publications 5 and 6 are published under the author's previous name, \textbf{Mirza Akhi Khatun}. The author now publishes as \textbf{Mirza Akhi}.}

\vspace{1.5em}

\begin{enumerate}[leftmargin=1.2cm, label={\arabic*.}, itemsep=1.2em]

\item \textbf{Mirza Akhi}, Ciarán Eising, and Lubna Luxmi Dhirani, 2026. 
Securing IoT Using Lightweight TCN for Edge Deployment on Raspberry Pi 4. 
\textit{IEEE Open Journal of the Communications Society} \quartile{1}, PP(99):1--1, \textbf{DOI:} \href{https://doi.org/10.1109/OJCOMS.2025.3649498}{10.1109/OJCOMS.2025.3649498}


\item \textbf{Mirza Akhi}, Ciarán Eising, and Lubna Luxmi Dhirani, 2026. 
\textit{Residual Temporal CNNs for Emerging Cyber Threat Detection in Healthcare IoT.} 
\textit{Discover Internet of Things}$^{*}$ \quartile{2}, pp.~, \textbf{DOI:} \href{https://doi.org/10.1007/s43926-025-00271-w}{10.1007/s43926-025-00271-w}

\vspace{0.3em}
\datasetlink{H-IoT-Multiattack2025}{https://doi.org/10.5281/zenodo.17100208}

\item \textbf{Mirza Akhi}, Ciarán Eising, and Lubna Luxmi Dhirani, 2025. 
Datasets for Distributed Denial-of-Service Detection in Healthcare Internet of Things Environments. 
\textit{Data in Brief} \quartile{3}, vol. 63C, 112222, \textbf{DOI:} \href{https://doi.org/10.1016/j.dib.2025.112222}{10.1016/\allowbreak j.dib.2025.112222}

\vspace{0.3em}
\datasetlink{UL-ECE-MQTT-DDoS-H-IoT2024 \& UL-ECE-UDP-DDoS-H-IoT2024}{https://doi.org/10.5281/zenodo.15305814}

\item \textbf{Mirza Akhi}, Ciarán Eising, and Lubna Luxmi Dhirani, 2025. 
TCN-Based DDoS Detection and Mitigation in 5G Healthcare-IoT: A Frequency Monitoring and Dynamic Threshold Approach. 
\textit{IEEE Access} \quartile{1}, vol. 13, pp. 12709--12733, \textbf{DOI:} \href{https://doi.org/10.1109/ACCESS.2025.3531659}{10.1109/ACCESS.2025.3531659}

\item \textbf{Mirza Akhi Khatun}, Mangolika Bhattacharya, Ciarán Eising, and Lubna Luxmi Dhirani, 2024. 
Time Series Anomaly Detection with CNN for Environmental Sensors in Healthcare-IoT. 
\textit{IEEE 12th International Conference on Healthcare Informatics (ICHI)} \confbadge

\vspace{0.3em}
\datasetlink{WSN\_DDoS\_Attack\_H-IoT2023}{https://doi.org/10.5281/zenodo.15701881}

\item \textbf{Mirza Akhi Khatun}, Sanober Farheen Memon, Ciarán Eising, and Lubna Luxmi Dhirani, 2023. 
Machine Learning for Healthcare-IoT Security: A Review and Risk Mitigation. 
\textit{IEEE Access} \quartile{1}, vol. 11, pp. 145869--145896, \textbf{DOI:} \href{https://doi.org/10.1109/ACCESS.2023.3346320}{10.1109/ACCESS.2023.3346320}

\end{enumerate}



\noindent\fcolorbox{gray!40}{gray!5}{\begin{minipage}{0.97\textwidth}
\vspace{0.4em}
\centering
\textbf{\large Journal Rankings}
\vspace{0.5em}

\begin{tabular}{@{\hspace{1em}}c@{\hspace{0.8em}}l@{\hspace{2em}}c@{\hspace{0.8em}}l@{\hspace{1em}}}
\quartile{1} & Top 25\% journals & \quartile{2} & 25--50\% journals \\[0.3em]
\quartile{3} & 50--75\% journals & \quartile{4} & 75--100\% journals \\[0.3em]
\multicolumn{4}{c}{\confbadge~Peer-Reviewed Conference}
\end{tabular}
\\[0.2em]
\textit{\small $^{*}$Discover Internet of Things is a newly established journal (launched in 2021).}
\end{minipage}}

\vspace{0.8em}
\noindent\textbf{Note:} Parts of this research are also presented in a national conference and an international conference poster session; details are provided in Appendix~\ref{appendix:presentations}.

\endgroup


\begin{abstracts}     
 Cybersecurity is a fundamental requirement for protecting wearable devices used in healthcare Internet of Things (H-IoT) systems. As healthcare increasingly relies on resource-constrained, interconnected systems for medical monitoring, security failures can directly compromise patient safety and system reliability. Physiological data and network traffic, therefore, are frequent targets of cyber attacks in H-IoT environments. To address these risks, deep learning-based cybersecurity mechanisms for H-IoT often involve complex architectures with large parameter counts. The quality of existing datasets is also rarely considered, limiting their applicability in H-IoT environments. However, this research addresses these challenges by developing multiple realistic datasets and proposing lightweight deep learning models, namely the Temporal Convolutional Network (TCN) and Residual TCN (Res-TCN), for H-IoT. It includes two binary classification datasets for Distributed Denial of Service (DDoS) attacks and a multiclass dataset representing Selective Forwarding (SF), Man-in-the-Middle (MITM), and DDoS attacks. The datasets {\em UL-ECE-MQTT-DDoS-H-IoT2025} and {\em UL-ECE-UDP-DDoS-H-IoT2025} are generated in Cooja and ns-3 to capture transmission behaviours and protocol variations. The third dataset, {\em UL-ECE-MultiAttack-H-IoT2025}, integrates physiological and network features \rev{(e.g., total messages)} to represent multiple cyber threats in H-IoT. Building on this, the TCN model is designed to detect and mitigate DDoS attacks over the MQTT and UDP-based datasets. It incorporates a monitoring frequency-based detection mechanism and a dynamic threshold-based mitigation strategy. Extending this work, the Res-TCN is developed for detecting cyber threats in H-IoT systems. To enable edge deployment, the model is quantised and converted into TensorFlow Lite (TFLite) for real-time DDoS detection on Raspberry Pi 4, achieving low latency and power-efficient operation in H-IoT. Thus, this thesis establishes a DL-based cybersecurity defence mechanism encompassing realistic dataset generation, lightweight model design, and edge deployment for securing H-IoT systems.
\end{abstracts}
\nomenclature{Anomaly~Detection}{A technique in cybersecurity that identifies unusual patterns or behaviours in data that deviate from expected normal conditions, often indicating cyber attacks or malicious activities.}


\nomenclature{Mitigation}{A post-detection process that blocks or neutralises cyber attacks by applying defence mechanisms that generate a blacklist of malicious nodes and restore normal network operation.}

\nomenclature{H-IoT}{Integration of Internet of Things (IoT) technologies into digital healthcare, known as Healthcare-IoT (H-IoT), enabling remote monitoring and patient care.}

\nomenclature{DDoS}{Distributed Denial of Service (DDoS), a cyberattack that transmits high-volume traffic to a target (e.g., server) to disrupt service availability.}

\nomenclature{MITM}{Man-in-the-Middle (MITM), an attack where an adversary secretly intercepts and tampers with data while normal H-IoT nodes transmit health metrics, thereby manipulating the communication between legitimate devices.}  

\nomenclature{SF}{Selective Forwarding (SF), an attack where a malicious node selectively drops or forwards specific packets, disrupting normal data transmission in the network.}  

\nomenclature{Payload}{The portion of transmitted data that carries the actual information or message content (e.g., H-IoT sensor readings) within a data packet, excluding headers and control information.}

\nomenclature{MQTT}{Message Queuing Telemetry Transport (MQTT), a lightweight publish/subscribe protocol for IoT.}

\nomenclature{Synthetic\kern2.8ptData}{Artificial data that mimics real-world data. It is generated through network simulation using tools (e.g., Cooja) for both normal and malicious traffic to support ML/DL model training, testing, and evaluation in H-IoT settings.}  

\nomenclature{Inference\kern2.8ptTime}{The time or latency that a trained ML/DL model requires to process input data and generate the output, such as normal or malicious.}


\nomenclature{Quantisation}{The process of converting model parameters and computations from a higher-precision format (e.g., float32) to a lower-precision format (e.g., int8) in deep learning. This conversion reduces model size and power consumption, thereby enhancing inference efficiency on resource-constrained devices.}

\nomenclature{SMOTE}{Synthetic Minority Over-sampling Technique (SMOTE), a data augmentation method that generates synthetic samples for minority classes to balance imbalanced datasets.}

\nomenclature{Attack\kern2.8ptIntensity}{A measure representing how frequently malicious nodes attempt to send data packets during an attack. Higher intensity indicates more packet transmission attempts than normal nodes, leading to service disruption in H-IoT systems.}

\begingroup
  \renewcommand{\nomname}{Glossary of Key Terms}%
  \makeatletter
  \let\saved@makeschapterhead\@makeschapterhead
  \def\@makeschapterhead#1{}%
  \renewcommand{\nompreamble}{%
    \thispagestyle{fancy}%
    \markboth{\MakeUppercase{\nomname}}{\MakeUppercase{\nomname}}%
  }%
  \printnomenclature
  \let\@makeschapterhead\saved@makeschapterhead
  \makeatother
\endgroup


\setcounter{secnumdepth}{3} 
\setcounter{tocdepth}{3}    
\tableofcontents            


\listoftables  
\listoffigures	

\renewcommand*{\glstextformat}[1]{\textcolor{black}{#1}}
\printglossary[type=\acronymtype,title={List of Acronyms}]



\mainmatter

\renewcommand{\chaptername}{} 




\chapter{Introduction}
\label{chap:int}

\ifpdf
    \graphicspath{{1_introduction/figures/PNG/}{1_introduction/figures/PDF/}{1_introduction/figures/}}
\else
    \graphicspath{{1_introduction/figures/EPS/}{1_introduction/figures/}}
\fi


The integration of the \gls{iot}, including \gls{iiot}, and \gls{g5} networks, is transforming digital healthcare through high-speed, real-time communication and reliable connectivity~\cite{dhirani2021industrial}. This convergence integrates wearable sensors (e.g., body temperature, heart rate, and oxygen saturation) with edge devices, servers, and cloud platforms, ensuring seamless data exchange and efficient clinical management in \gls{g5}-enabled \gls{hiot} environments. The key advantage of this integration is its ability to support timely decision-making, reinforce precision healthcare, and extend specialist care to patients in remote or underserved areas.

However, the accelerated expansion of \gls{g5}-enabled \gls{hiot} systems also introduces significant challenges related to security, privacy, and reliability~\cite{goswami2024role}. The heterogeneous nature of connected \gls{hiot} devices, resource limitations, and continuous data transmission increase exposure to malicious activities. Specifically, emerging cyber threats such as \gls{ddos}, \gls{sf}, and \gls{mitm} attacks pose critical vulnerabilities and risks to the integrity and availability of precision healthcare data~\cite{madanian2024health}. \gls{hiot} systems also operate under low-latency and power-efficiency requirements, which constrain the deployment of complex deep learning models such as \gls{lstm}~\cite{akar2025l2d2}, \gls{bilstm}~\cite{zahid2025enhancing}, and hybrid \gls{cnn}–\gls{lstm}~\cite{altunay2023hybrid}. Particularly, landmark anomaly detection methods (e.g., \gls{lstm}), despite significant advancements, remain unsuitable for resource-constrained \gls{hiot} devices due to their high computational overhead~\cite{akar2025l2d2, aswad2023deep}.

Privacy and regulatory requirements (e.g., \gls{gdpr}) further restrict access to real-world data, creating challenges in obtaining sufficient labelled samples for \gls{ml}/\gls{dl} model training. In parallel, publicly available intrusion-detection datasets are typically released in aggregated traffic form, which limits the preservation of packet-level temporal characteristics such as inter-arrival timing and transmission rate variations~\cite{goldschmidt2025network}. Consequently, traffic dynamics caused by malicious activities are often underrepresented in such datasets~\cite{cordero2021generating}. These challenges highlight the need for realistic attack simulations, synthetic datasets, lightweight learning architectures, and practical deployment strategies to enable reliable cyber threat detection and mitigation in \gls{hiot} environments.

This thesis encompasses multiple aspects of \gls{hiot} research, spanning network topology design (e.g., star), simulation, security, and privacy~\cite{makina2024survey}. It also includes realistic experimentation with application and transport-layer protocols such as \rev{Message Queuing Telemetry Transport (MQTT), User Datagram Protocol (UDP)}, \gls{tcp}, and \gls{coap} under both normal and malicious conditions. Among these protocols, \gls{mqtt} plays a crucial role in \gls{iot} communications due to its lightweight publish–subscribe architecture and minimal bandwidth requirements. It enables reliable message exchange between multiple \gls{iot} devices and a central broker with minimal protocol overhead, such as a small header size and limited handshake or acknowledgement messages. Its efficiency in maintaining low latency and reliability, even in constrained networks, makes it well-suited for modelling the interaction patterns of \gls{hiot} devices. Hence, this research reflects the structural composition of \gls{hiot}, consisting of wearable and environmental sensors within the \gls{wban} and \gls{wsn} domains. These include physiological and ecological parameters such as body temperature, heart rate, room temperature, and humidity. In such interconnected systems, both \gls{wban} and \gls{wsn} nodes are vulnerable to adversarial manipulation, where malicious entities may alter sensor data or disrupt communication, affecting the reliability of healthcare monitoring~\cite{faris2023wireless, zhang2025survey}. The overall architecture facilitates seamless communication across edge, fog, and cloud infrastructures. However, during the simulation of physiological metrics (e.g., simulated heart rate of 80 bpm), \gls{gdpr}-related procedures are excluded, as the artificially generated data do not contain any subject-specific or patient information.

\begin{figure}[ht!]
  \centering
\includegraphics[width=\linewidth]{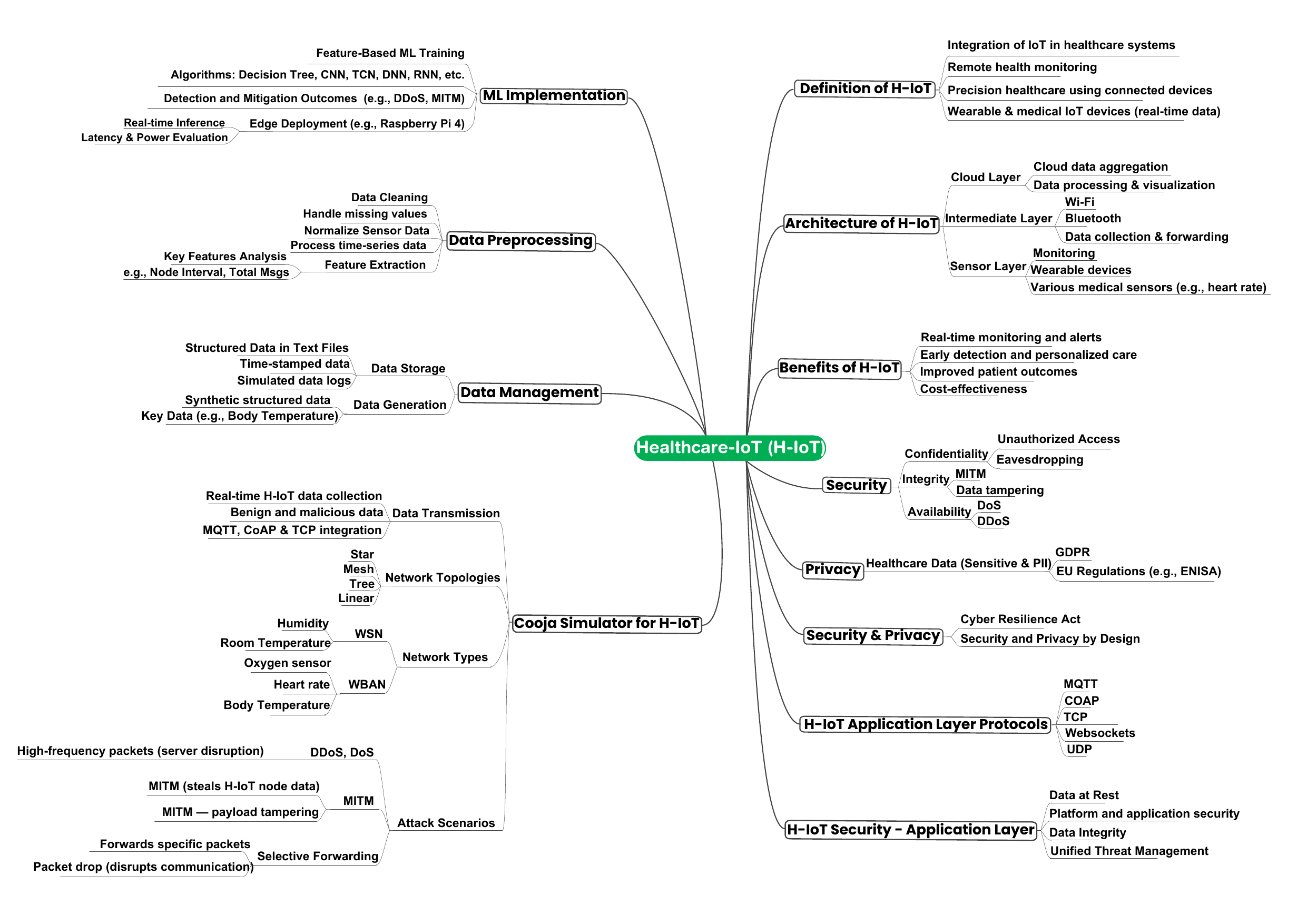}
\caption[Scope of the thesis in \gls{g5}-enabled \gls{hiot}]{Scope of the thesis in \gls{g5}-enabled \gls{hiot}. The diagram outlines the end-to-end \gls{hiot} ecosystem, covering simulation-driven data generation, data management and preprocessing, \gls{ml}/\gls{dl}-based attack detection and mitigation, edge deployment, and security and privacy aspects. Multiple application-layer protocols, including \gls{mqtt}, \gls{coap}, \gls{udp}, and \gls{tcp}, are experimentally investigated within a star network topology that reflects standard \gls{hiot} deployments. While experimental data are generated for all illustrated protocols, the published research focuses on \gls{mqtt} and \gls{udp}.}  
  \label{fig:h-iot-scope}
\end{figure}


The research scope and interconnections among components (e.g., data generation, threat detection, edge deployment) are depicted in Figure~\ref{fig:h-iot-scope}, illustrating how each element contributes to building a secure, privacy-preserving, and efficient \gls{hiot} ecosystem. The workflow integrates data preprocessing, feature extraction, and \gls{ml}/\gls{dl}-based model development for intelligent attack detection and mitigation. It includes the implementation and evaluation of shallow \gls{ml} models (e.g., \gls{dt}, Rule-Based DT) to compare with the proposed \gls{dl} architectures (e.g., \gls{cnn}, \gls{tcn}, and \gls{restcn}). Following the comparative evaluation, the proposed architecture (e.g., \gls{tcn}) is deployed on resource-constrained \gls{iot} devices such as the Raspberry Pi 4. The resulting performance, measured in terms of latency and power consumption, validates the feasibility of lightweight intelligent deployment in \gls{hiot} environments. In the final stage, the research focuses on multiclass classification of multiple attack types to provide a comprehensive evaluation of the proposed approach.



\rev{\section{Research Gaps}
Cyber-threat detection has been addressed using both traditional machine learning and deep learning approaches in \gls{hiot} environments. Traditional methods, such as \glspl{dt}~\cite{binbusayyis2022investigation} and \gls{knn}~\cite{wagan2023fuzzy}, offer low computational cost; however, they have limited capacity to capture time-dependent attack patterns. This limitation is evident in multiclass cyber threat detection scenarios, where rule-based \glspl{dt} can misclassify malicious samples as normal, as demonstrated in Chapter~\ref{chap:restcn}, Section~\ref{subsec:modelcomparison}. In contrast, deep learning models (e.g., \glspl{lstm}) improve detection performance by capturing complex patterns; however, they often introduce higher computational overhead and increased inference latency~\cite{akar2025l2d2, aswad2023deep}. As a result, prior approaches, including traditional machine learning and deep learning methods, do not consistently balance detection performance, inference latency, and model size for edge deployment in resource-constrained \gls{hiot} environments. Beyond model-level limitations, existing datasets often fail to capture transmission-frequency characteristics. This limits the representation of realistic traffic dynamics, resulting in degraded model performance, increased misclassification, and reduced generalisability. Therefore, a research gap exists in the design of cyber threat detection approaches that simultaneously achieve high accuracy, low latency, and resource efficiency for both binary and multiclass classification under realistic traffic dynamics in \gls{hiot} environments.} 

\section{Aims and Objectives} 
\label{sec:objective}
\rev{This research aims to address the challenges of detecting and mitigating cyber threats using deep learning-based approaches within \gls{hiot} environments. Deep learning is employed to learn complex, non-linear, and time-dependent cyber-attack patterns in network traffic. The study further considers the trade-off between detection performance, inference latency, and model size for edge deployment in resource-constrained \gls{hiot} environments.
To enable this, it analyses the data transmission behaviour of normal and malicious devices under realistic network conditions in \gls{hiot}.
Accordingly, it utilises emulated \gls{hiot} sensor data and simulated network traffic to capture device-to-device interactions within healthcare settings. In this context, a systematic simulation is adopted to generate \gls{hiot} sensor data with defined transmission patterns that differentiate normal and attack-related behaviour for cyber threat analysis.} The generated data are employed to train and evaluate the proposed detection and mitigation mechanism for cyber threats, including \acrlong{ddos}, \acrlong{sf}, and \acrlong{mitm} attacks in \gls{hiot} environments. In addition, this research investigates the design of lightweight deep learning architectures that maintain detection performance while remaining suitable for resource-constrained \gls{hiot} devices. The overarching research question is stated as follows:

\textbf{Main RQ: How can differences in data transmission behaviour between normal and malicious nodes support the development of realistic datasets and lightweight deep learning models for detecting and mitigating emerging cyber threats in \gls{hiot} environments?}

To address \gls{ddos} attack detection and mitigation, this research proposes a deep learning method, \gls{tcn}, that identifies \gls{ddos}-affected nodes and subsequently blocks the detected malicious nodes. Following this, the study extends toward real-world deployment to evaluate the feasibility of executing such deep learning models on resource-constrained devices. Accordingly, this research investigates the deployment of \gls{tcn} to assess latency and power efficiency on edge devices, such as the Raspberry Pi 4, in \gls{hiot} environments. Building on edge deployment, the research further examines the detection of multiple cyber attacks within a unified \gls{hiot} topology. In real-world scenarios, multiple threats may coincide; therefore, multiclass classification is a crucial step in anomaly detection. In line with the final research question, this dissertation proposes a deep learning model, \gls{restcn}, to classify emerging cyber threats such as \acrlong{sf}, \acrlong{mitm}, and \acrlongpl{ddos} in \gls{hiot}.

The aims and objectives of this dissertation are outlined through the following Research Questions:

\begin{itemize}

\item \textbf{RQ1:} \rev{What key challenges and research gaps are identified in existing literature on \gls{ml}/\gls{dl}-based cyber attack detection in \gls{hiot} systems, and how can a conceptual deep learning framework address these limitations?}

\item \textbf{RQ2:} How can realistic synthetic datasets be generated to represent \gls{ddos} attack behaviours in \gls{hiot} environments?

\item \textbf{RQ3:} Are deep learning techniques (e.g., \gls{tcn}) capable of detecting and mitigating evolving cyber threats (e.g., \gls{ddos}) across \gls{hiot} systems? 

\item \textbf{RQ4:} What approaches enable the edge deployment of deep learning models (e.g., \gls{tcn}) on resource-constrained devices (e.g., Raspberry Pi 4) in \gls{hiot}?

\item \textbf{RQ5:} How can deep learning models such as \gls{restcn} be utilised for multiclass classification to distinguish among multiple attack categories, including \acrlong{sf}, \acrlong{mitm}, and \acrlong{ddos} attacks, in \gls{hiot} environments?

\end{itemize}

\section{\rev{Dataset and Model Justification}}
\rev{This research constructs three synthetic datasets that incorporate both network traffic behaviour and physiological metrics, representing normal and malicious \gls{hiot} nodes. Synthetic datasets are generated due to the limited availability of publicly accessible datasets that capture realistic \gls{hiot} transmission behaviour and temporal characteristics. This research includes a detailed analysis of per-node communication behaviour, including the number of messages transmitted per node, total message volume, and the transmission patterns of emulated sensors (e.g., body temperature) to the server, aligned with practical \gls{hiot} operations. This analysis identifies consistent differences between normal and malicious nodes in transmission intervals and packet-generation frequency. These patterns are not captured at the node-level temporal granularity in commonly used datasets (e.g., CIC-IDS2017) \cite{sharafaldin2018toward, rosay2022network}. This facilitates the evaluation of detection models based on transmission dynamics rather than aggregated traffic patterns.}

\rev{The selection of \gls{tcn} is motivated by the need to capture temporal variations in node-level transmission behaviour while maintaining efficient inference for edge deployment. In this research, attack detection relies on identifying changes in transmission intervals and packet-generation frequency across nodes. This involves capturing temporal dependencies without introducing significant computational overhead. Unlike recurrent architectures such as \gls{lstm}, which process sequences sequentially, \gls{tcn} enables parallel processing, reducing inference latency. It captures temporal dependencies using dilated causal convolutions, which avoids the limitations of standard \gls{cnn} architectures in preserving temporal order~\cite{zhang2020multi}. Therefore, \gls{tcn} achieves a trade-off between detection performance and computational efficiency for edge deployment in resource-constrained \gls{hiot} environments.}

\rev{To support multiclass detection, the \gls{restcn} architecture is further justified by its ability to capture complex patterns from combined physiological and network features across multiple attack types, including \gls{ddos}, \gls{sf}, and \gls{mitm}. Residual connections facilitate stable training by improving gradient propagation.}

\section{Research Contribution}
This thesis primarily advances the field of cybersecurity, with a particular focus on \gls{hiot} systems and emerging cyber threat detection. It addresses how variations in data transmission behaviour between normal and malicious nodes support the development of realistic datasets, lightweight models, and edge-intelligent defence mechanisms in \gls{hiot}. The key contributions are as follows:

\begin{enumerate}
   \item \textbf{A review of \gls{ml}/\gls{dl}-based cybersecurity mechanisms in \gls{hiot}} – \rev{This review systematically analyses existing machine learning and deep learning techniques applied to \gls{hiot} security.
It identifies key challenges, limitations, and open issues that underpin the research question and support a conceptual framework.} It \rev{also} highlights the lack of realistic datasets, privacy constraints, and insufficient modelling of device-to-device interactions, which hinder the development of effective security solutions. Based on these insights, a conceptual convolutional neural network-based deep learning framework is introduced to address the identified gaps in securing \gls{hiot} systems. This conceptual framework motivates the subsequent experimental investigations presented in this thesis.

  This contribution is presented in Chapter~\ref{chap:review}, corresponds to \textbf{RQ1}, and is published in the following article: \textbf{Khatun, M.A.}, Memon, S.F., Eising, C. and Dhirani, L.L., 2023. Machine learning for healthcare-iot security: A review and risk mitigation. \textit{IEEE Access}, 11, pp.~145869-145896.


\item \textbf{Creation of realistic \gls{ddos} datasets for \gls{hiot} environments} – To overcome the dataset limitations identified in the review, two realistic \gls{ddos} attack datasets are developed that replicate the traffic behaviour of \gls{hiot} sensors under both normal and malicious conditions. The first dataset, generated using the Cooja simulator over \gls{mqtt}, includes three types of wearable sensors: body temperature, oxygen saturation, and heart rate, under both normal and malicious conditions. The second dataset, developed using the ns-3 simulator over the \gls{udp} protocol, captures both benign and malicious traffic in an \gls{hiot} environment. The data are preprocessed and labelled using a systematic pipeline to refine and organise the extracted \gls{hiot} communication features across both datasets.

This contribution is presented in Chapter~\ref{chap:dataset}, corresponds to \textbf{RQ2}, and is published in the following article: \textbf{Akhi, M.}, Eising, C. and Dhirani, L.L., 2025. Datasets for distributed denial-of-service detection in healthcare internet of things environments. \textit{Data in Brief}, p.~112222.

\item \textbf{Development of a \gls{tcn}-based \gls{ddos} detection and mitigation model for \gls{hiot}} – Using the previously developed datasets, this research trains and evaluates a \gls{tcn} model to detect and mitigate \gls{ddos} attacks in \gls{hiot} environments. The model uses monitoring frequency to identify \gls{ddos}-affected nodes and employs a dynamic threshold mechanism to block detected malicious nodes. This approach reliably distinguishes between normal and attack traffic, achieving high detection and mitigation performance across both \gls{mqtt} and \gls{udp}-based datasets.

This contribution is presented in Chapter~\ref{chap:tcn}, corresponds to \textbf{RQ3}, and is published in the following article: \textbf{Akhi, M.}, Eising, C. and Dhirani, L.L., 2025. \gls{tcn}-Based \gls{ddos} Detection and Mitigation in \gls{g5} Healthcare-\gls{iot}: A Frequency Monitoring and Dynamic Threshold Approach. \textit{IEEE Access}, 13, pp.~12709–12733.

\item \textbf{Edge deployment of the \gls{tcn} model on resource-constrained devices} – Extending the previous research, this study investigates the deployment of the trained \gls{tcn} model on a Raspberry Pi 4 to evaluate its feasibility in real-time \gls{hiot} applications. Using the previously developed datasets, the model’s inference latency and power consumption are analysed to assess its applicability for low-power healthcare devices. The findings demonstrate that the \gls{tcn} model operates efficiently on the Raspberry Pi 4, achieving low latency and reduced power consumption within \gls{hiot} constraints.

This contribution is presented in Chapter~\ref{chap:edge}, corresponds to \textbf{RQ4}, and is published in the following article: \textbf{Akhi, M.}, Eising, C. and Dhirani, L.L., 2026. Securing \gls{iot} using Lightweight \gls{tcn} for Edge Deployment on Raspberry~Pi~4 \textit{IEEE Open Journal of the Communications Society}, 7, pp.~442-460.

\item \textbf{Multiclass attack detection using Res-\gls{tcn} in \gls{hiot} environments} – Building on the earlier deployment research, this study focuses on distinguishing multiple cyber threats within a unified \gls{hiot} topology. To achieve this, the multiclass dataset is developed to represent diverse attack categories, including \gls{sf}, \gls{mitm}, and \gls{ddos}. A \gls{restcn} model is then employed to perform multiclass classification, accurately identifying each attack type based on its distinct communication characteristics. The results demonstrate the model’s strong generalisation across different attack categories while maintaining low inference latency suitable for real-time \gls{hiot} applications.

This contribution is presented in Chapter~\ref{chap:restcn}, corresponds to \textbf{RQ5}, and is published in the following article: \textbf{Akhi, M.}, Eising, C. and Dhirani, L.L., 2026. Residual temporal \glspl{cnn} for emerging cyber threat detection in healthcare \gls{iot}. \textit{Discover Internet of Things}.

\end{enumerate}

\section{Thesis Outline} 
This thesis is structured in an article-based format, as illustrated in Figure~\ref{fig:thesisoutline}. The remaining chapters are organised as follows:

\begin{figure}[ht!]
  \centering
  \includegraphics[width=\linewidth]{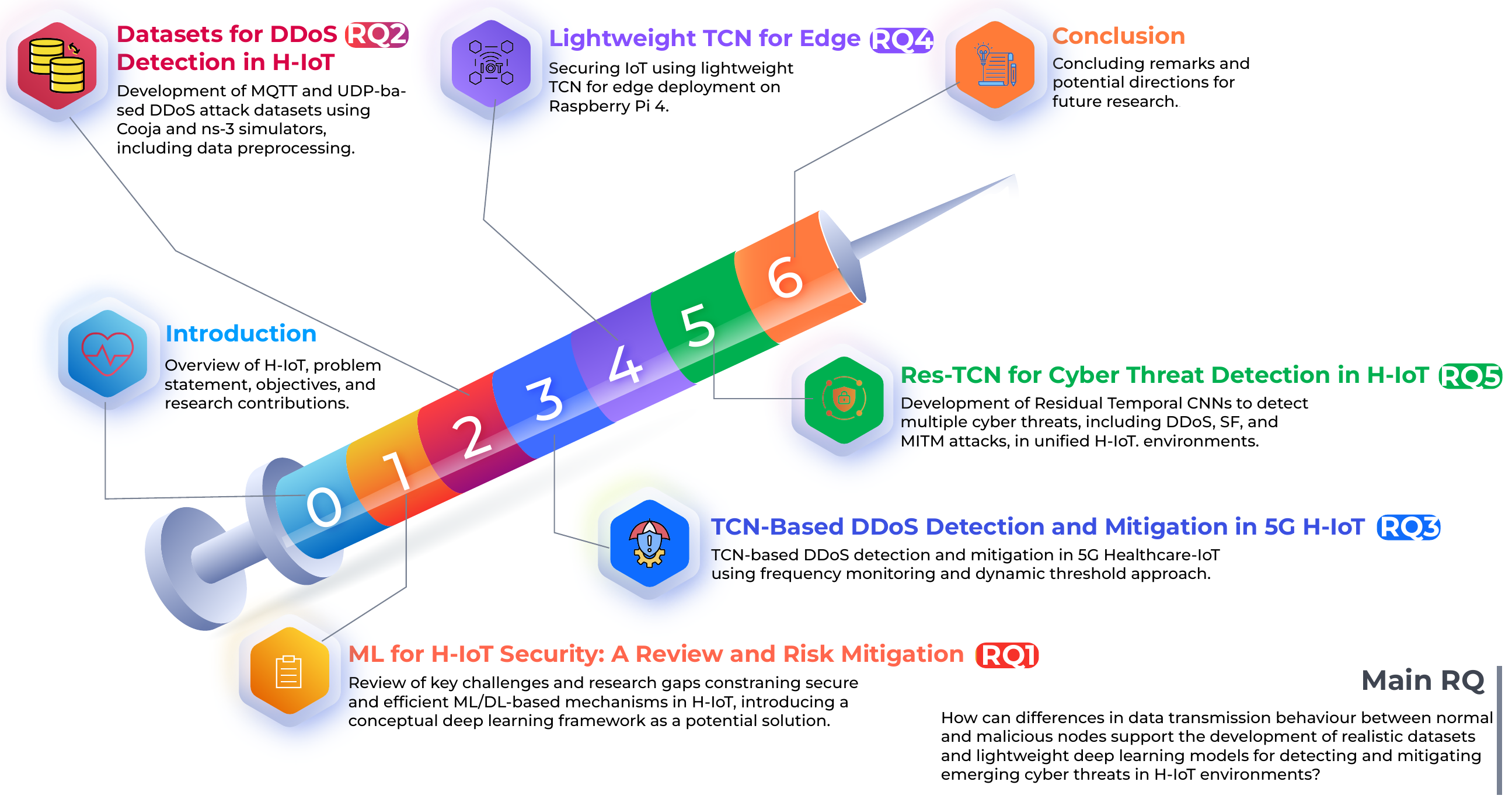}
  \caption[Thesis outline]{Thesis outline illustrating the article-based structure and core contributions aligned with the research questions across seven chapters.}
  \label{fig:thesisoutline}
\end{figure}

\begin{itemize}
    \item \emph{\textbf{Chapter Two}} is based on the journal publication~Khatun, M.A., Memon, S.F., Eising, C., and Dhirani, L.L., 2023. Machine learning for healthcare-\gls{iot} security: A review and risk mitigation. \textit{IEEE Access}, 11, pp.~145869--145896.

This chapter reviews the current landscape of \gls{hiot}, focusing on data security and privacy challenges associated with \gls{ml}/\gls{dl}-based solutions. It also highlights the scarcity of realistic and domain-specific datasets that constrain the development of effective detection mechanisms. In addition, it introduces a conceptual \gls{dl} framework (e.g., \gls{cnn}) for anomaly detection, which later informs a simplified \gls{cnn} implementation explored in early-stage research. The insights gained from these early studies lead to the development of realistic datasets and lightweight deep learning models in this thesis. Furthermore, this review outlines future research directions toward developing secure and efficient \gls{hiot} systems.

\item \emph{\textbf{Chapter Three}} is based on the journal publication~Akhi, M., Eising, C. and Dhirani, L.L., 2025. Datasets for distributed denial-of-service detection in healthcare internet of things environments. \textit{Data in Brief}, p.~112222.

This chapter addresses the dataset limitation discussed in Chapter Two by presenting two realistic \gls{ddos} datasets for \gls{hiot} environments. The first dataset is generated using the Cooja simulator over the \gls{mqtt} protocol with three types of wearable sensors: body temperature, oxygen saturation, and heart rate, under normal and attack scenarios. The second dataset is developed using the ns-3 simulator over the \gls{udp} protocol, containing benign and malicious traffic. It also systematically describes the preprocessing pipeline and the selected features, including four features for the \gls{mqtt} dataset and thirteen for the \gls{udp} dataset. Both datasets are designed to support evaluating \gls{ml}/\gls{dl}-based detection models.

\item \emph{\textbf{Chapter Four}} is based on the journal publication~Akhi, M., Eising, C. and Dhirani, L.L., 2025. \gls{tcn}-Based \gls{ddos} Detection and Mitigation in \gls{g5} Healthcare-\gls{iot}: A Frequency Monitoring and Dynamic Threshold Approach. \textit{IEEE Access}, 13, pp.~12709–12733.

This chapter utilises the datasets developed in Chapter Three over the \gls{mqtt} and \gls{udp} protocols in \gls{g5} \gls{hiot} systems. The \gls{tcn} model is trained using both datasets to perform \gls{ddos} detection and mitigation. The model uses monitoring frequency to accurately identify malicious nodes, while a dynamic threshold mechanism is applied during the mitigation phase to blacklist them. It achieves high prediction and mitigation accuracy across \gls{mqtt} and \gls{udp}-based datasets, outperforming conventional deep learning models such as BiLSTM and \gls{cnn}.

\item \emph{\textbf{Chapter Five}} is based on the journal publication~Akhi, M., Eising, C. and Dhirani, L.L., 2026. Securing \gls{iot} using Lightweight \gls{tcn} for Edge Deployment on Raspberry~Pi~4 \textit{IEEE Open Journal of the Communications Society}, 7, pp.~442-460.

This chapter introduces lightweight \gls{tcn}-based edge deployment for detecting \gls{ddos} attacks on Raspberry Pi~4 in \gls{hiot} environments. The previously developed \gls{mqtt} and \gls{udp}-based datasets are utilised to convert the model into TFLite format and optimise it through quantisation, reducing model size, inference latency, and power consumption. The study demonstrates the feasibility of deploying efficient, lightweight deep learning models on resource-constrained edge devices to secure \gls{hiot} systems.

\item \emph{\textbf{Chapter Six}} is based on the journal publication~Akhi, M., Eising, C. and Dhirani, L.L., 2026. Residual temporal \glspl{cnn} for emerging cyber threat detection in healthcare \gls{iot}. \textit{Discover Internet of Things}.

This chapter presents a \gls{restcn} model for classifying multiple concurrent cyber threats in \gls{hiot} systems. A realistic simulation is developed to model normal and attack scenarios involving \gls{sf}, \gls{mitm}, and \gls{ddos} behaviours. The resulting dataset is used to train the proposed Res-\gls{tcn} model, which combines residual and temporal learning components to enhance classification accuracy and stability. To address class imbalance and improve model generalisation, \gls{smote} is used during training. The study presents an efficient multiclass detection framework suitable for real-time, resource-constrained \gls{hiot} environments.

\item \emph{\textbf{Chapter Seven}} presents concluding remarks and highlights potential directions for future research.
\end{itemize}















\chapter{Machine Learning for 
Healthcare-IoT Security: A Review and Risk Mitigation} \label{chap:review} 


\ifpdf
    \graphicspath{{2_chapter2/figures/PNG/}{2_chapter2/figures/PDF/}{2_chapter2/figures/}}
\else
    \graphicspath{{2_chapter2/figures/EPS/}{2_chapter2/figures/}}
\fi



\vspace{5pt}

\noindent This literature review presented in this chapter has been previously published in:

\textbf{Khatun, M.A.}, Memon, S.F., Eising, C. and Dhirani, L.L., 2023. Machine learning for healthcare-IoT security: A review and risk mitigation. \textit{IEEE Access}, 11, pp.~145869--145896.

The content of the paper above has been presented as published in this chapter, except for correcting grammatical errors and making slight rewording so that the narrative is more appropriate for a thesis. A brief preamble is provided to explain its relevance to the thesis narrative.

\clearpage

\noindent\textbf{\underline{Preamble}}\\[0.5em]
In this chapter, \textbf{RQ1} is addressed:


\textbf{RQ1:} \rev{What key challenges and research gaps are identified in existing literature on \gls{ml}/\gls{dl}-based cyber attack detection in \gls{hiot} systems, and how can a conceptual deep learning framework address these limitations?} 

\vspace{5pt}

The primary objective of this chapter is to analyse security challenges and investigate \gls{ml} and \gls{dl}-based techniques for detecting, classifying, and mitigating cyber threats in \gls{hiot}.

Although \gls{ml} and \gls{dl} techniques have been applied in \gls{hiot} systems, existing methods remain inadequate for detecting and mitigating evolving cyber threats (e.g., \gls{ddos} attacks). Additionally, the limited availability of realistic and domain-specific \gls{hiot} datasets restricts the training and evaluation of \gls{ml}/\gls{dl} models. On the other hand, privacy regulations (e.g., \gls{gdpr}) restrict data sharing and access to medical records, creating additional challenges in developing and validating security solutions for precision healthcare systems.

According to \textbf{RQ1}, this chapter contributes by highlighting existing gaps and providing a comprehensive analysis of gls{ml} and \gls{dl} techniques for securing \gls{hiot} systems. It discusses key cybersecurity aspects, including big data, e-health, and cloud computing. The review examines critical use cases, including anomaly detection, intrusion detection, authentication, and access control. It also outlines realistic simulation tools (e.g., the Cooja simulator) and approaches for developing realistic datasets. It further discusses open challenges, practical considerations, and future research directions. Finally, this chapter introduces a conceptual framework derived from the analysis to illustrate how \gls{dl} architectures (e.g., \gls{cnn}) can support the development of secure and efficient \gls{hiot} ecosystems. This supports the overarching research question by framing \gls{dl}–based anomaly detection for securing \gls{hiot} systems.

\clearpage

\section{Abstract}
The \gls{hiot}, commonly known as Digital Healthcare, is a data-driven infrastructure that highly relies on smart sensing devices (i.e., blood pressure monitors, temperature sensors, etc.)~for faster response time, treatments, and diagnosis. However, with the evolving cyber threat landscape, \gls{iot} devices have become more vulnerable to the broader risk surface \rev{(e.g., synthetic data generation, adversarial manipulation, and automated attack generation enabled by generative \gls{ai}, and \gls{g5}-\gls{iot}, etc.)}, which, if exploited, may lead to data breaches, unauthorised access, a lack of command and control and potential harm. This chapter reviews the fundamentals of \gls{hiot}, its privacy and data security challenges, and the associated challenges with \gls{ml} and \gls{hiot} devices. The chapter further emphasises the importance of monitoring \gls{hiot} layers such as perception, network, cloud, and application. Detecting and responding to anomalies involves various cyber-attacks and protocols such as Wi-Fi 6, \gls{nbiot}, Bluetooth, ZigBee, LoRa, and \gls{5gnr}. A robust authentication mechanism based on \gls{ml} and \gls{dl} techniques is required to protect and mitigate \gls{hiot} devices from increasing cybersecurity vulnerabilities. Hence, in this chapter, security and privacy challenges and risk mitigation strategies for building resilience in \gls{hiot} are explored and reported.

\section{Introduction} \label{sec:introduction}
The \gls{iot} consists of interconnected physical devices exchanging data through sensors, software, and connectivity \cite{dhirani2021industrial},\cite{zardari2023iot}. The healthcare industry has undergone a significant transformation in recent years with advances in \gls{iot}, cloud, \gls{ai}, and \gls{ml}. According to several experts, the expanding horizon of \gls{iot} is expected to improve healthcare. \gls{iot} can revolutionise healthcare globally by providing affordable healthcare \cite{fi9040093}, remote health monitoring \cite{majumder2017wearable}, wellness management \cite{wickramasinghe2019delivering}, and virtual rehabilitation \cite{postolache2020remote}. Healthcare analytics can provide insight into disease and drug discovery while adding a new dimension \cite{gupta2021new}.

\begin{figure*}[t!]
    \centering
    \includegraphics[width=\textwidth]{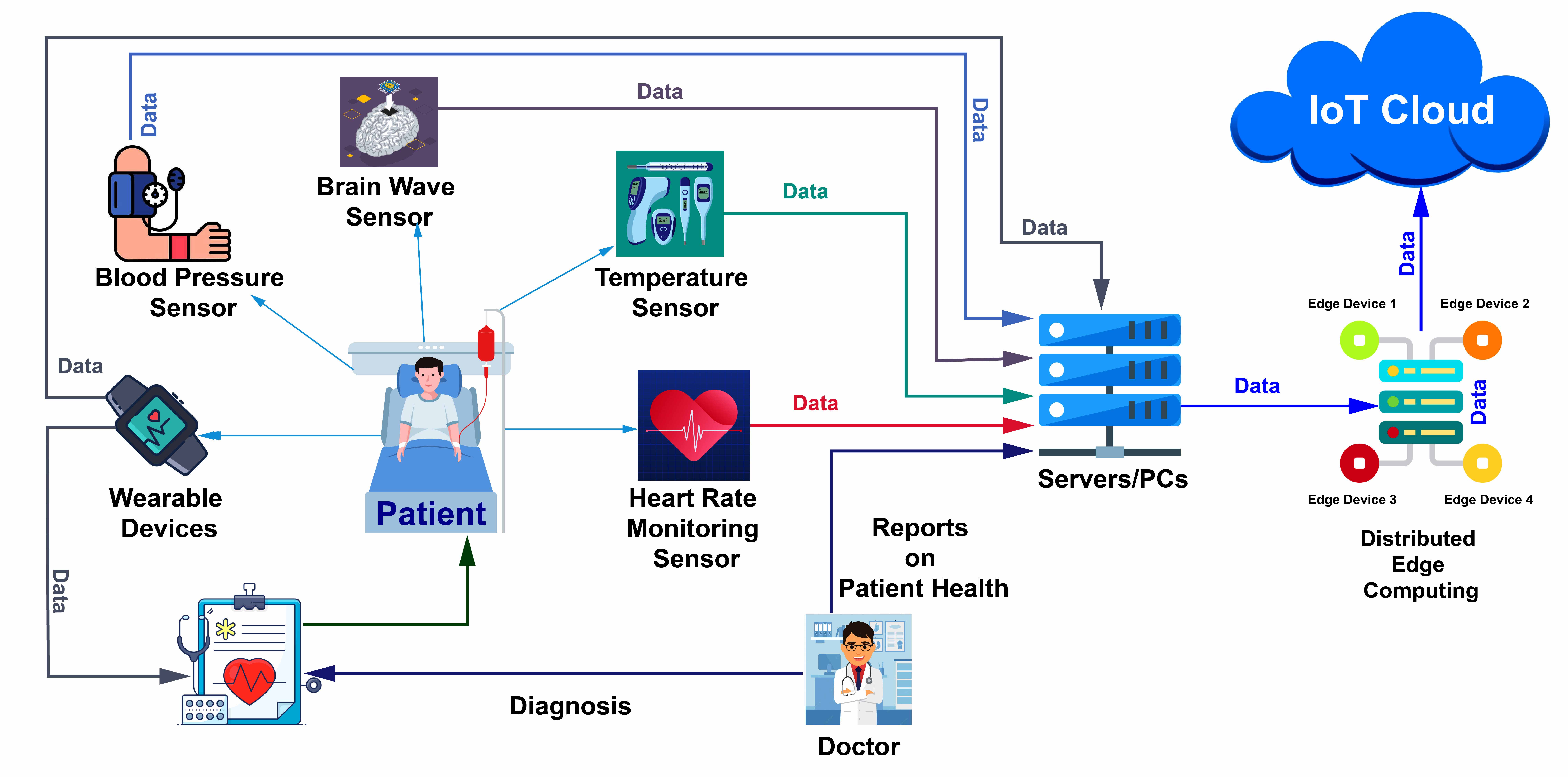}
    \caption[IoT-based healthcare architecture]{\gls{iot}-based healthcare architecture.}
    \label{fig:fig1}
\end{figure*}

The modern world requires more efficient and timely interventions to combat escalating health issues. While traditional healthcare systems are effective, these systems are often slow and inflexible \cite{jayashankara2021iot}. The COVID-19 pandemic has fueled the need for remote and precision healthcare, and such objectives could only be achieved using emerging technologies. Embedding an \gls{iot}-enabled architecture in a healthcare ecosystem may facilitate the collection and processing of real-time data from sensors (i.e., body sensors strategically placed on or within a patient's body, aiding real-time data collection) \cite{tavera2021wearable}. \rev{Different sensors are used for different applications, such as body motion, blood flow, and biomedical signals (e.g., ECG, EEG, and blood oxygen levels).} However, the ones used for healthcare applications are typically \gls{bsn}~\cite{gangwani2023ai}. In a \gls{bsn}, each sensor is connected within a group, forming a network; this configuration enhances data collection and efficiency by integrating \gls{iot} body sensors, such as brain waves, body temperature, and blood pressure sensors, as depicted in Figure \ref{fig:fig1}. Moreover, the diagram also illustrates how data is stored and transmitted to the cloud via edge nodes/layers for data processing and storage. \gls{iot} devices generate and transmit data with millisecond-level latency requirements and therefore require highly reliable, scalable, and low \gls{e2e} latency networks \cite{dhirani2021industrial}. Connecting the \gls{iot} devices with an edge-to-cloud environment is essential to facilitate the required/comprehensive health monitoring (security and reliability metrics).

As per recent forecasts, the number of connected \gls{iot} devices worldwide is expected to reach around 39 billion by 2030, reflecting sustained growth in sensor-driven systems across sectors such as healthcare, manufacturing, and smart infrastructure \cite{iotanalytics2026}. Because of its usability, efficiency, and applications,  these devices have the potential to further advance healthcare by reducing costs, improving patient well-being, and facilitating the efficient delivery of faster diagnostics, thereby improving medical services \cite{plageras2017algorithms}. Furthermore, \gls{iot} can provide the migration of patients from traditional healthcare management methods to new cloud-based systems, among other vital innovations \cite{usak2020health}.
Employing \gls{hiot}, remote monitoring of patients can save millions of lives and money, while other functions still have a crucial role across healthcare environments \cite{karunarathne2021security}. This type of remote monitoring of patients can assist in identifying health issues early on and making treatment plans more personalised. Similarly, \gls{iot} can benefit medical management, including medication and instrumental errors, and medication administration, allowing patients to take their medications at the right time and with the correct dosage \cite{hanson2022nursing}. Medication error prevention programs can improve patient outcomes and reduce healthcare costs. For example, \gls{hiot} could significantly improve health outcomes and healthcare costs for residents of care homes who frequently manage multiple medical conditions and take various medications \cite{ciampi2022intelligent}.

\rev{Healthcare-\gls{iot} frameworks utilise a variety of sensors, including diagnostic sensors (e.g., ECG, blood pressure), monitoring sensors (e.g., heart rate, SpO$_2$), and preventive sensors (e.g., fall detection, activity tracking), for healthcare applications~\cite{muhammad2020eeg}.} Over the past few years, healthcare researchers have primarily focused on finding ways to monitor patients remotely and transmit health reports in real-time to physicians. Researchers have identified several major challenges related to health surveillance systems. The challenges include data privacy, interoperability issues, data quality issues, and limitations associated with real-time analysis \cite{shrotriya2023apache},\cite{amin2020edge}. Patient monitoring concerns can be classified into two categories: static and dynamic monitoring. Smart hospitals use static monitoring systems to record patients' health status periodically. Moreover, as a static monitoring system, medical staff, including doctors and nurses, manually collect patients' vital signs during specific periods. The frequency of manual data collection within hospitals varies based on factors such as physicians' workload, working hours, patients' health conditions, and hospital leadership guidance \cite{shaik2023remote}. In contrast to static patient monitoring systems, dynamic patient monitoring systems can be used at home, work, or in the hospital. According to Figure \ref{fig:fig1}, \rev{healthcare systems consist of identifying (patient condition), locating (patient or device position), sensing (physiological data), and connecting (data transmission across networks).} \gls{hiot} components include emergency medical services, information technology, sensors, lab-on-a-chip technologies, wearable devices, connectivity devices, big data, and cloud computing \cite{hameed2020intelligent}.

Nowadays, smartphones enable rapid task completion by monitoring and collecting regular updates on patients' healthcare data \cite{al2022application}. For instance, an individual can promptly receive notifications when their heart rhythm changes when they wear a wristband linked to their smartphone \cite{mao2023health}. Furthermore, \gls{iot} allows the healthcare system to monitor and track community resources more efficiently and reliably \cite{subramaniyaswamy2019ontology}. However, \gls{hiot} has several risks, including privacy leakage during medical data processing and uploading \cite{he2023accelerated}. The risk of patient data disclosure may considerably discourage patients from sharing their medical information because confidentiality is at risk \cite{cha2018privacy}. Maintaining confidentiality and credibility in healthcare interactions is a fundamental factor for addressing these risks. Moreover, resolving such security concerns will facilitate seamless deployment and adoption of \gls{hiot}.

\rev{Over the years, several review papers have been published on securing \gls{hiot}~\cite{kayode2025enhancing, rehman2025internet, nabha2025internet, nithyavani2025comprehensive, bollineni2025iot, abbas2024federated, kaur2024unveiling, charfare2024iot, bala2024ensuring, singh2024comprehensive, wani2024security, chataut2023unleashing}.} There is, however, a noticeable gap in the existing review regarding the security of \gls{hiot}, often failing to address all necessary aspects holistically. Most of the discussions focus heavily on the technical aspects of \gls{hiot} devices, diving deep into their specs and applications and skipping over the essential issue of data security. During the discussion of \gls{ml}, the focus is on its potential application in healthcare instead of exploring how it can enhance security protocols, such as detecting unusual data patterns that may indicate a cybersecurity threat. Security issues related to remote patient monitoring in \gls{hiot} have not been discussed in existing reviews either. \rev{In \cite{8804790}, the authors analyse the \gls{iomt} security controls in a sustainable context but do not explicitly focus on application-specific scenarios such as remote patient monitoring or \gls{ai}-driven \gls{iomt} security frameworks. However, \gls{hiot} environments require the timely detection of abnormal data patterns. In such scenarios, \gls{ai}-based methods support real-time anomaly detection while operating within resource-constrained settings. In \cite{jagatheesaperumal2022holistic}, \gls{iot} and \gls{ai} are explored using a smart pill bottle case study. However, the study focuses on application design and does not compare security mechanisms such as anomaly detection, intrusion detection, or real-time monitoring approaches. In \cite{dhanvijay2019internet}, the authors review \gls{iomt} security and privacy challenges. However, the discussion remains conceptual and lacks practical validation through real-world deployment or experiments.}  Furthermore, although the discussion acknowledges security and privacy concerns, it does not provide in-depth solutions or strategies to address the identified challenges within \gls{iot} healthcare systems. In \cite{ketu2021internet}, a review of the \gls{ioht} covers its technologies, applications, and inherent challenges, particularly security and privacy concerns. Though it explores many issues, notably privacy and security, it modestly limits its explanations to practical solutions. In \cite{10201156}, the authors discuss the development of \gls{g5} technology within the \gls{hiot} space, including network slicing, security, and energy efficiency. Despite this, certain topics remain unresolved and require more exploration and innovation. These include data security, efficient resource management, and energy conservation. \rev{Healthcare services use large amounts of energy for \gls{hvac}~\cite{bawaneh2019energy}. \gls{hvac} systems can adjust settings using \gls{iot} sensors. This study excludes such energy management systems and focuses on H-IoT cybersecurity. These studies focus on network-level security, energy consumption, and system-level aspects, with limited attention to healthcare-specific threats (e.g., tampering with patient monitoring data or false emergency alerts), real-time detection, and healthcare-critical scenarios in \gls{hiot}.}

In \gls{hiot} research, secure data access presents a pivotal challenge, particularly regarding healthcare data regulations and strict policies that protect \gls{pii} and data confidentiality. Looking at the various ransomware attacks and data breaches \cite{databreaches2023} that healthcare sectors have suffered in the past, affecting millions of patients globally, demonstrates the urgency and need to mitigate these growing cyber risks. The COOJA simulator has reshaped this complex research area, allowing researchers to emulate \gls{hiot} device behaviours without interacting directly with sensitive data. Data collection from \gls{hiot} devices has not been explicitly discussed in existing surveys \cite{subhan2023ai}. In addition, there was a limitation on how malicious data could be collected for anomaly detection by \gls{ml}. This chapter reviews the security aspects of \gls{hiot} applications, pinpoints vulnerabilities, and uses \gls{ml} to suggest solutions, providing an enhanced understanding of the security issues associated with \gls{hiot} that incorporates both practical and theoretical considerations. In addition to highlighting the mitigation of \gls{hiot} security challenges, this review also illustrates the use of COOJA simulator tools, which allow a unique perspective to be gained that makes this survey distinctive from those previously conducted. 

This chapter aims to review current research on \gls{ai} and \gls{ml} techniques that can enhance \gls{hiot} security, build cyber resilience within the healthcare infrastructure, and enable it to detect, protect, and respond to novel cyber threats. The chapter also covers a broad spectrum of cybersecurity issues from the \gls{enisa} 2030 cyber threat landscape foresight \cite{ENISA2023} perspective, including complexities associated with \gls{iot} layers. The chapter highlights the following key contributions:

\begin{itemize}
    \item Reviews key aspects of cybersecurity, big data, e-health, and cloud computing in \gls{hiot}.
    
    \item Discusses \gls{ml} techniques such as anomaly detection, device classification, and critical use cases for security enhancement, such as intrusion detection, authentication, and access control.
    
    \item Highlights the challenges and potential solutions related to \gls{hiot} security, along with future research directions.
\end{itemize}

\begin{figure*}[!htbp]
    \centering
   \includegraphics[width=\textwidth]{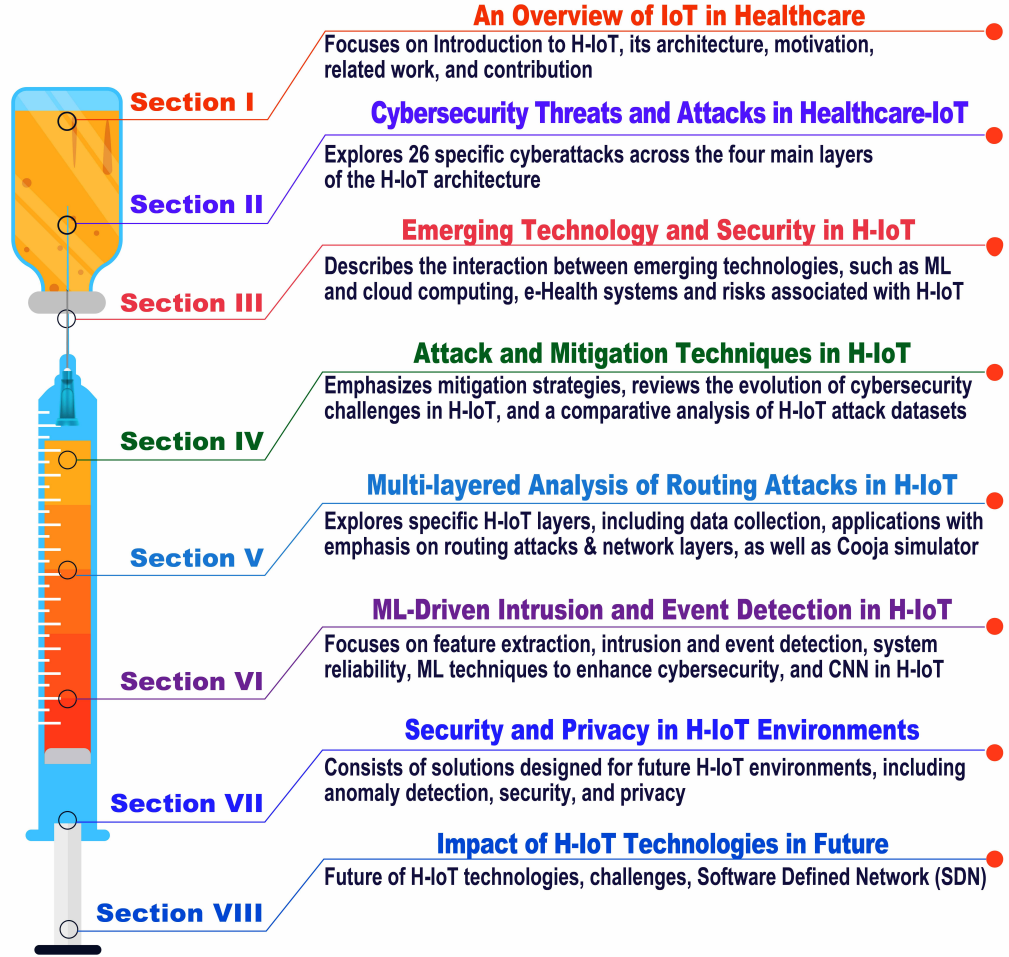}
   \caption[Structure and outline]{Structure and outline of the chapter.}
   \label{fig:fig2}
\end{figure*}
\FloatBarrier
Figure \ref{fig:fig2} presents the structure of this chapter as follows: Section \ref{sec:cybersecurity2} explains 26 different types of cyber attacks carried out across the four main layers of the \gls{hiot} architecture, Section \ref{sec:emerging3} describes the interaction between emerging technologies, such as \gls{ml} and cloud computing, e-Health systems and also discusses the risks associated to \gls{hiot}, Section \ref{sec:attacknmitigation4} emphasizes risk mitigation strategies, reviews the evolution of cybersecurity challenges in healthcare over the last five years, and provides a comparative analysis of \gls{hiot} attack data-sets, Section \ref{sec:multilayer5} explores specific \gls{hiot} layers, including data collection, applications with an emphasis on routing attacks, and network layers, as well as COOJA simulator, Section \ref{sec:ml-driven6} focuses on feature extraction, intrusion and event detection, system reliability, and \gls{ml} techniques to enhance cybersecurity, and convolutional neural network in the \gls{hiot}, Section \ref{sec:security7} consists of solutions designed for future \gls{hiot} environments, including anomaly detection, security, and privacy, section \ref{sec:impact8} outlines the future of \gls{hiot} technologies, software defined networks, challenges, and mitigation strategies and finally, Section \ref{sec:conclude9} concludes the chapter, exploring future directions in \gls{hiot} security.

\section{Cybersecurity Threats and Attacks in \gls{hiot}} \label{sec:cybersecurity2}
\rev{Healthcare-\gls{iot} is distinct because it integrates life-critical data, resource-constrained devices, and real-time patient-monitoring requirements, posing unique cybersecurity challenges.}

\begin{figure}[t!]
    \centering
    \includegraphics[width=0.7\textwidth]{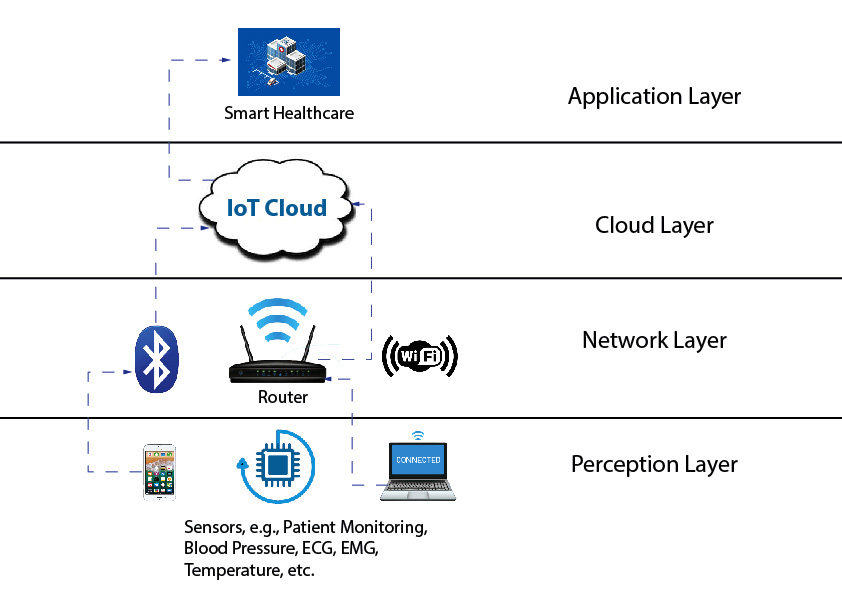}
    \caption[The layers of \gls{hiot}]{The layers of \gls{hiot}.}
    \label{fig:fig3}
\end{figure}

Building on this discussion, this section examines cybersecurity threats and attacks across different layers of the \gls{hiot}. The \gls{hiot} system is typically categorised into four layers. These layers include (I) the perception layer, (II) the network layer, (III) the cloud or processing layer, and (IV) the application layer \cite{chaurasia2023comprehensive}. Cybersecurity threats and attacks emerge at these layers, causing significant challenges (i.e., routing attacks, impersonating, tampering, or data transit attacks at the perception layer, \gls{dos}, \gls{ddos} or \gls{mitm} attacks at the network layer, Cloud-malware injection or Brute-force attacks at the Cloud-\gls{iot} Layer and \gls{sql} injection or \gls{xss} attacks at the application layer) \cite{yavuz2018deep},\cite{puthal2016threats},\cite{abdulqadir2021study},\cite{chordiya2018man},\cite{alkhathami2022detection},\cite{bhushan2017man},\cite{malik2023security}. This section discusses 26 different cyber attacks and demonstrates their specific mechanisms and impact on the infrastructure. Figure \ref{fig:fig3} illustrates the details of healthcare \gls{iot} layers. It also provides a detailed explanation and insights into the most prevalent threats and attacks at different \gls{hiot} layers.

\subsection{The Perception Layer}
\begin{figure}[ht!]
    \centering
    \includegraphics[width=0.7\textwidth]{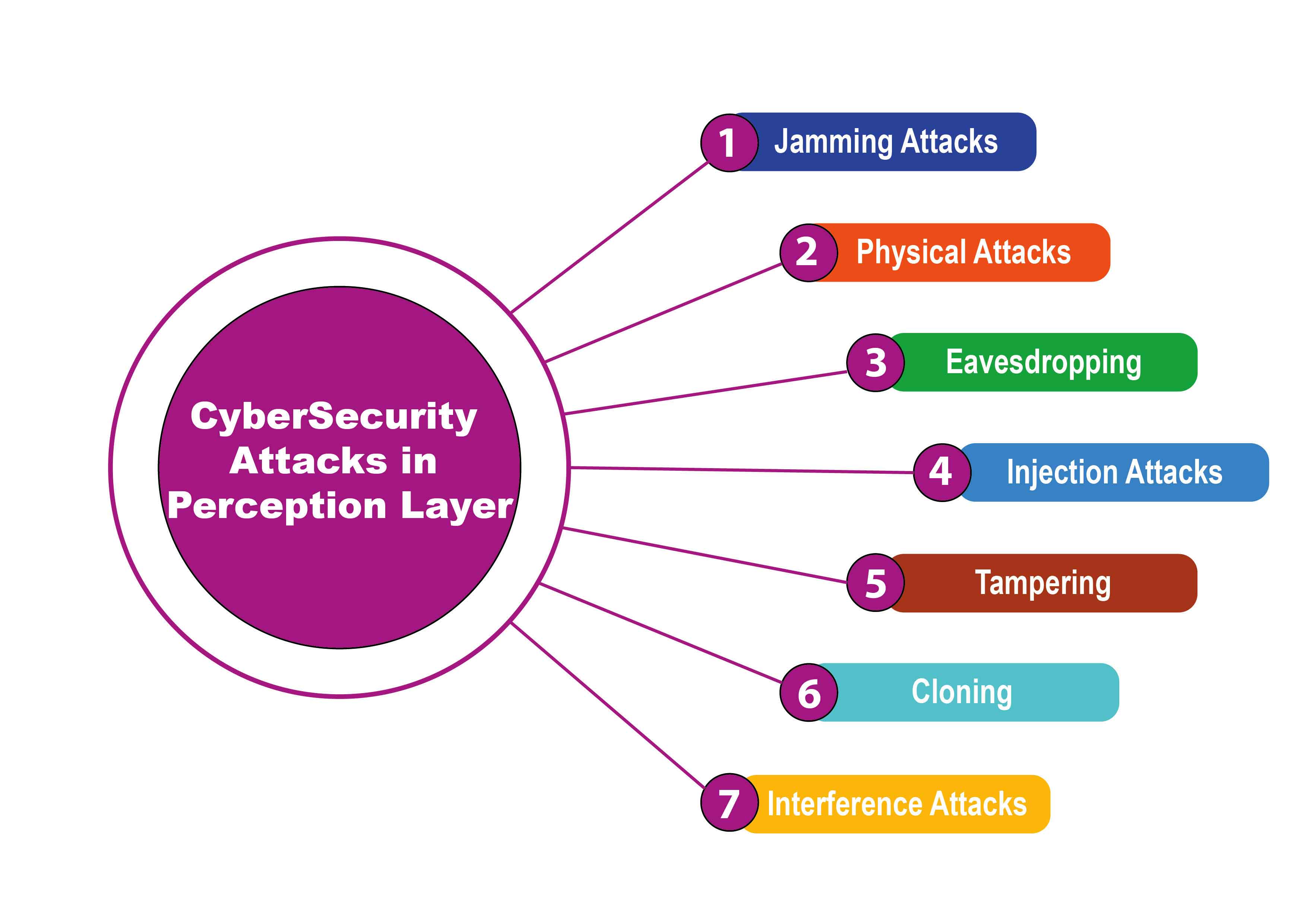}
    \caption[Cybersecurity attacks in perception layer]{Cybersecurity attacks in perception layer.}
    \label{fig:fig4}
\end{figure}

 This layer can also be termed the ``physical layer,'' or the ``sensor layer,'' as it refers to the physical devices \cite{sarker2023internet}, \cite{mahmoud2015internet} that serve the purpose of sensing and collecting essential information about patients, such as their medical history. \gls{iot} healthcare frameworks enable the interconnection of key stakeholders such as doctors, nurses, technicians, pharmacists, and medical devices as part of the perception layer. A specialist can monitor the data that the perception layer collects in real time over the Internet. There are multiple communication protocols (i.e., \gls{rfid}, \gls{ble}, \glspl{wsn}, Zigbee, and \gls{6lowpan} used for collecting and transmitting data from \gls{iot} nodes \cite{askar2022architecture},\cite{albalawi2022assessing},\cite{bekkali2022systematic},\cite{zaman2022thinking}. This layer is susceptible to multiple attacks affecting the infrastructure in various ways, as depicted in Figure \ref{fig:fig4}. 

Here, each threat in this layer is summarised briefly: 
\begin{itemize}
    \item \textbf{Physical Attacks:} These attacks focus on the \gls{iot} architecture's perception layer or physical layer, consisting of hardware devices. To attack this layer, the malicious actor must remain closer to the network infrastructure or gain unauthorised physical access to execute an attack \cite{mohindru2021security}. These exploitations may facilitate the threat actor to tamper with implanted medical devices (i.e., Wi-Fi-enabled pacemakers and insulin pumps).
    \item \textbf{Eavesdropping:} \gls{iot} devices are vulnerable to eavesdropping if such attacks are executed successfully, which gives the malicious actor direct access to retrieve confidential information exchanged between the devices or \gls{rfid} tags and readers \cite{khader2021survey}, \cite{hu2017cooperative},\cite{luong2016data}. Eavesdropping attacks may have a variety of malicious intents, such as taking intellectual property, biometrics, and genome data for personal gain or using personally identifiable information/genetic data for espionage \cite{cvitic2015classification}. Such attacks have been carried out in the past \cite{dimaggio2022art}.
    \item \textbf{Jamming:} The majority of wireless devices communicate with each other via \gls{rf} signals, which stronger signals can obstruct. In this scenario, a malicious actor may interrupt and block communications between sensors and receivers, leading to an absolute downtime and lack of availability \cite{roohi2019ddos},\cite{chen2018internet}. Jamming stands out as one of the most common attacks carried out at perception layer attack, alongside physical tampering, false data injection, and eavesdropping \cite{adil2020energy},\cite{khan2020improving},\cite{qasem2021multi},\cite{almaiah2020new},\cite{al2020improved},\cite{almaiah2022lightweight},\cite{almaiah2022novel},\cite{almudaires2021data}. The impact of jamming attacks in \gls{hiot} can be catastrophic as it could disrupt on-going surgeries, medical diagnostics, access to online-systems. In short, it can lead the healthcare facility to an absolute downtime, affecting human lives. These types of cyber-attacks have an escalated impact on an infrastructure's operational and financial aspects and the ethical/social dimensions.    
    \item \textbf{\gls{rfid} Cloning:} These attacks have emerged as a major security concern at the \gls{iot}'s perception layer, as they camouflage themselves in different forms, such as RFID cloning and tag cloning. \gls{rfid} cloning involves replicating and deceiving readers with duplicate tags that simulate the original to gain unauthorised access to information\cite{alotaibi2022security},\cite{chahid2017internet}. The concept of tag cloning, however, encompasses duplicating various tag-based identification systems beyond \gls{rfid} \cite{abdullah2019cybersecurity}. These two forms of hacking endanger the system's integrity and increase risks related to bio-hacking \cite{HelpNetSecurity2023}.
    \item \textbf{Injection Attacks:} These attacks can occur on various \gls{iot} layers; however, at the perception layer, it is carried out by injecting malicious code and modifying the firmware of an \gls{iot} device. \gls{iot} infrastructures face the greatest challenge when a single compromised/malicious node can spread across and infect the entire network, causing complete operational disruptions \cite{varadharajan2018study}. As per the European Cyber Resilience Act, digital and critical infrastructures must have the ability to detect, protect, respond, mitigate cyber threats, and maintain operational resilience simultaneously \cite{CyberResilienceAct2022},\cite{meagher2023cyber}.
    \item \textbf{Interference:} An intruder can disrupt network communication by interrupting network traffic and constantly broadcasting radio waves \cite{sinha2021impact}, for spreading false information. As per the \gls{nis2d}, the healthcare infrastructures fall under Annexe I (Essential Sectors) and must comply with the high common level of cybersecurity across the European jurisdiction \cite{EuropeanCommissionNIS2023}. These types of attacks can potentially spread misinformation and catalyse panic situations.
    \item \textbf{Tampering:} Attackers who manipulate nodes' memory to alter their functionality are called node tamperers. Furthermore, attackers may physically tamper with a device by turning it on and off, restarting it, stealing its key code, and manipulating the data \cite{puthal2016threats}, affecting the data integrity of the environment. Due to the misuse of generative \gls{ai}, adversaries have been orchestrating and executing such attacks more frequently. As per the European \gls{gdpr} \cite{ProtonGDPR2023}, if any sensitive, \gls{pii} or healthcare data is breached in terms of integrity (data altered, amended, and tampered with), it is considered a data breach. An individual can suffer catastrophic consequences if their personal information is misused.
    
\end{itemize}

Defences against perception-layer cyber-attacks in \gls{iot} are discussed in \cite{nasralla2020defenses}. However, the approaches may not effectively mitigate the adversarial and escalating cyber threats due to generative \gls{ai}. 

\subsection{The Network Layer}
Data packets are received and processed through this layer by the perception layer. Afterwards, the layer transmits the made-trust data to the cloud layer immediately above it. The cloud layer's role is to store and share/build trust mechanisms between the smart devices \cite{sugadev2022implementation},\cite{alihamidi2019proposed}. The network layer can be enabled with different wireless networks (Wi-Fi 6, \gls{g5}, Bluetooth, \gls{nbiot}, and \gls{lte}) \cite{ahmed2019survey}. 
As part of the IEEE wireless communication standards, such as 802.11a, 802.11g, 802.11n (Wi-Fi 4), 802.11ac (Wi-Fi 5), 802.11ax (Wi-Fi 6), and 802.11p (for vehicular communications), orthogonal frequency division multiplexing (OFDM) is employed for the effective transmission of data across multiple frequency bands \cite{hagras2023physical}. In the case of \gls{iot} devices, high-speed Wi-Fi has recently been recognised as an attractive option due to its compatibility with existing infrastructure \cite{zhu2023enhancing}. The IEEE 802.11ax, a sixth-generation protocol popularly known as \gls{he}, reports a 30\% improvement in throughput over the older IEEE 802.11ac protocol known as \gls{vht} \cite{chotalia2023performance}. A \gls{g5}-enabled \gls{eerp} that uses Wi-Fi 6 makes it easier for medical devices to communicate with one another, while \gls{mumimo} allows parallel communications \cite{sefati2023ultra}. A data-centric protocol prioritises the transmission of critical health information, and \gls{ccn} improves the efficiency of the distribution of healthcare information between \gls{iot} devices \cite{anjum2023towards}.

The vulnerability to \gls{ddos} attacks has increased dramatically with the advent of \gls{g5}. A distributed network architecture with multiple remote edge computing sites is essential for achieving low latency, which is critical for vehicle-to-vehicle communication. Additionally, the dense network of connected devices, open-source applications, and additional access points contribute to increasing ``attack surfaces''. Furthermore, higher wireless speeds dramatically increase the impact of \gls{ddos} attacks on the network since each device can transmit an order of magnitude higher attack throughput to the network (at MECs and the core). As mentioned, this effect is amplified by the plurality of \gls{iot} devices, which is expected in \gls{g5}, due to the vulnerability of \gls{iot} devices to cyber attacks \cite{valadares2023security}. Figure \ref{fig:fig5} illustrates seven attacks that are most encountered at this layer. The following discussion highlights key attacks in the network layer of the \gls{iot}:


\begin{figure}[ht!]
    \centering
    \includegraphics[width=0.7\textwidth]{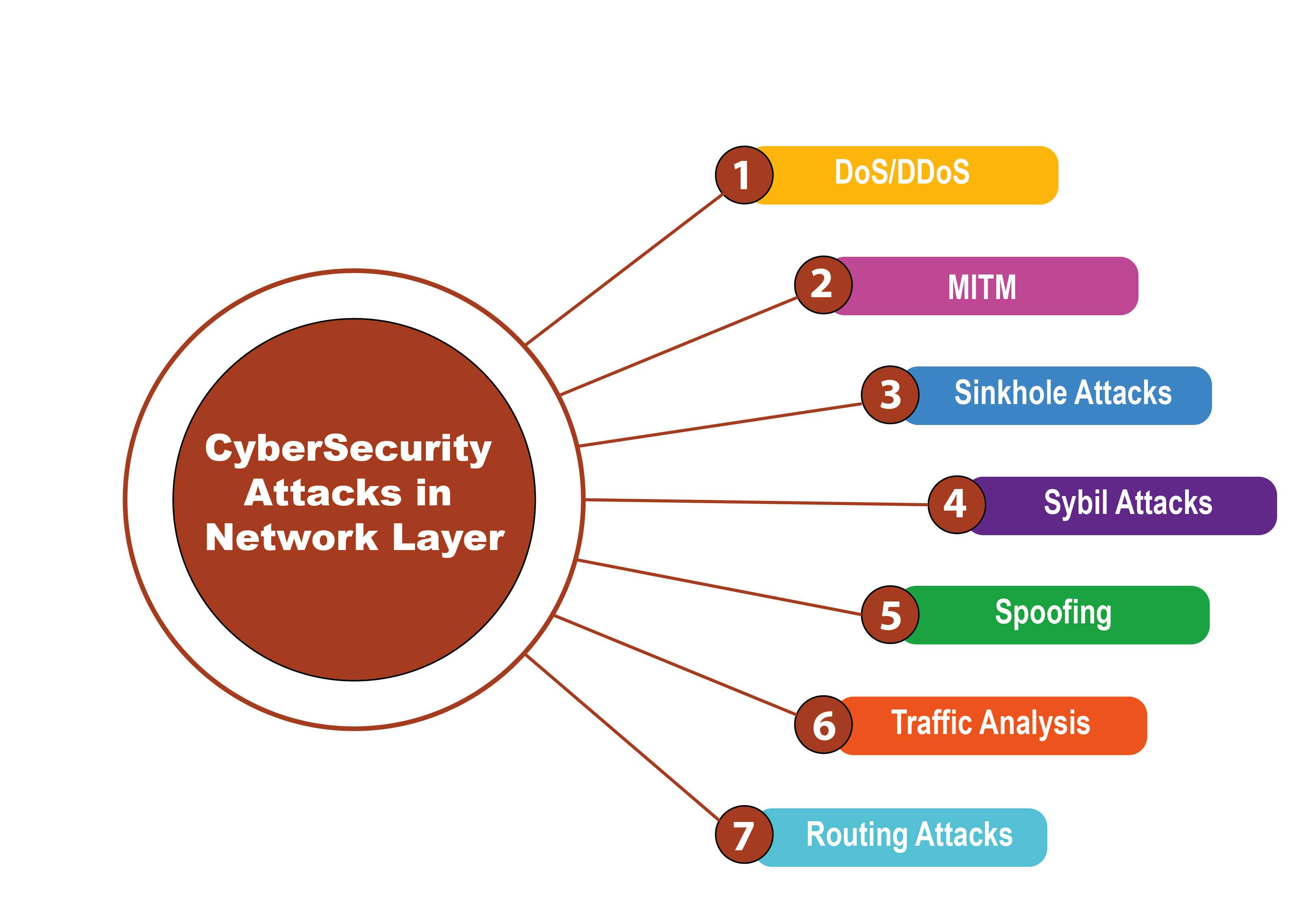}
    \caption[Cybersecurity attacks in the network layer]{Cybersecurity attacks in the network layer.}
    \label{fig:fig5}
\end{figure}

\begin{itemize}
    \item \textbf{\gls{dos}/\gls{ddos} Attacks:} In smart ecosystems, denial of service and distributed denial of service attacks are prevalent \cite{sinha2021multi}. A hacker attempts to consume legitimate network resources or bandwidth during a \gls{dos} attack. This attack is referred to as a \gls{ddos} when it originates from multiple compromised nodes \cite{sonar2014survey} and falls into different categories, such as traffic/fragmentation attack, bandwidth attack, and application attack~\cite{9798301}. Wireless and wired networks are critical to \gls{ddos} attacks, which target the network layer \cite{9096818}. Flooding the \gls{icmp} and \gls{udp} with excessive data leading to delays, resource consumption, potential system crashes, and ultimately disrupting services are common methods for executing these attacks \cite{dvzaferovic2019and},\cite{porambage2014pauthkey}. These attacks would lead to absolute downtime and operational disruption.

    \item \textbf{Routing Attacks:} Throughout the Internet of Things, IPv6 is widely used, especially in wireless sensor networks. IPv6-centric \glspl{wsn} are particularly vulnerable to routing attacks. Furthermore, \gls{wsn} sensors are frequently constrained by memory limitations, narrow bandwidths, and energy consumption \cite{yavuz2018deep}. These attacks are generally carried out at the \gls{isp} level, and \gls{ict} service providers must be \gls{nis2d} compliant \cite{CyberRiskGmbH2022}. The routing attacks are discussed in detail in Section \ref{sec:multilayer5C}.
    \item \textbf{Traffic Analysis:} Network traffic is analysed for detecting and monitoring malicious behaviour and anomalies. Depending on the network traffic, routers manage data packets. Each router is equipped with a certain amount of capacity. Clogging or unusually high traffic may indicate a traffic analysis or even a denial-of-service attack \cite{ganesh2020improved}. Malicious actors sometimes observe the traffic to guess passwords by analysing the packets sent during each keystroke and assessing the duration between them. By using such a pattern, they can reconstruct the users' passwords.
    \item \textbf{Spoofing Attacks:} Several types of spoofing are possible, such as email, \gls{url}, and frame spoofing. However, \gls{mac} or \gls{ip} address spoofing is most widespread \cite{hijazi2019address}. \gls{iot} uses the \gls{mac} address to authenticate wireless networks at the data link layer. Spoofing attacks occur when malicious entities imitate legitimate users' \gls{mac} addresses to gain network access illegally, affecting the data confidentiality and integrity of the environment \cite{madani2020mac}. Lack of data confidentiality means unauthorised or malicious actors may know the contents of data, whereas lack of integrity means the data may have been amended, altered, tampered with, or discarded. As per \gls{gdpr}, a lack of security (\gls{cia}) metrics and controls is considered a \gls{gdpr} breach.    
    \item \textbf{Sybil Attacks:} \gls{iot} devices are vulnerable to the Sybil attack \cite{murali2019lightweight},\cite{makhdoom2018anatomy}. In this attack, legitimate nodes are impersonated by malware to redirect traffic towards malicious ones. An individual node can appear in multiple locations simultaneously or impersonate several others \cite{avila2022analytical}. During this attack, a malicious node claims to have multiple identities. Due to their ability to control the flow of information within a network, these types of attacks affect data integrity and resource allocation.
    \item \textbf{Sinkhole Attacks:} In sinkhole attacks, the attacker redirects or discards traffic, preventing the base station from receiving full data transmission. This attack spreads misleading routing details from one node to another, resulting in energy drain and reduced network durability. Energy depletion is particularly damaging to wireless sensor networks \cite{al2023passive}. These types of attacks affect the integrity, availability, and reliability of data in a network.
    \item \textbf{\gls{mitm} Attacks:}
    Man-in-the-middle attacks involve intercepting communications between two devices by an attacker \cite{chordiya2018man}. The main target of \gls{mitm} attacks is the confidential details of users \cite{bhushan2017man}. Malicious actors exploit existing or new vulnerabilities in \gls{iot} systems to carry out this attack. An example would be a temperature reading of a sensor or a malfunctioning of a device enabling hackers to steal sensitive information \cite{deep2022survey}.

\end{itemize}

\subsection{The Cloud Layer}
The cloud allows for convenient and secure backup and preservation of confidential health information, and sharing between authorised parties (such as doctors, insurance providers, medical staff, and pharmacies) is convenient. Similarly, independent and public healthcare providers can maintain confidential data online and share it with trusted colleagues to improve the quality of treatment. These protocols include \gls{tcp}, \gls{udp} \cite{hong2018cloud}, and protocols for data analysis, prediction, and \gls{ml} \cite{hooshmand2022network}. Figure \ref{fig:fig6} depicts several attacks related to cloud layers. These specific vulnerabilities are described as follows:

\begin{figure}[ht!]
    \centering
    \includegraphics[width=0.7\textwidth]{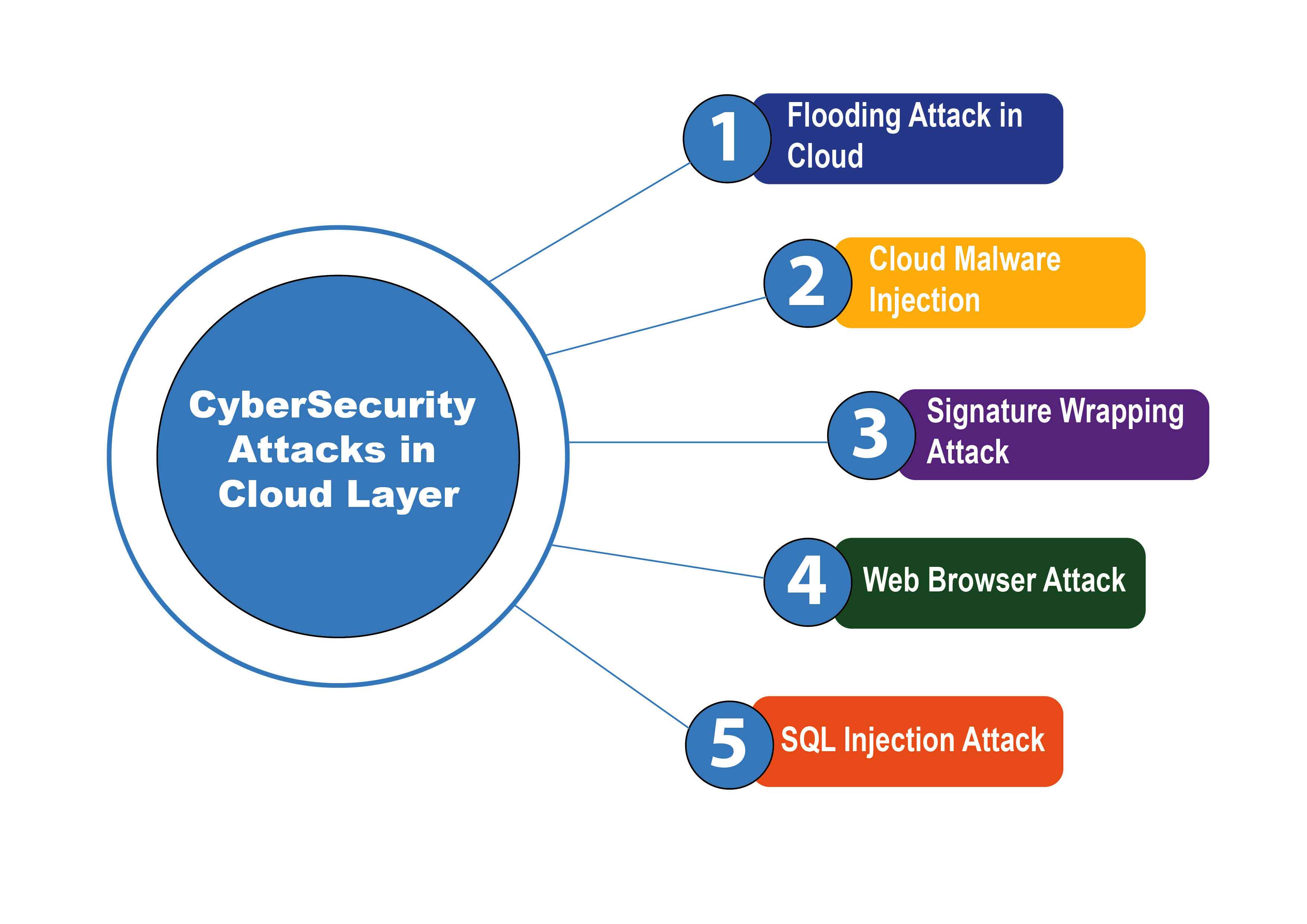}
    \caption[Cybersecurity attacks in the cloud layer]{Cybersecurity attacks in the cloud layer.}
    \label{fig:fig6}
\end{figure}

\begin{itemize}
    \item \textbf{Flooding Attacks:} Flooding attacks are a type of \gls{ddos} attack that disrupts services by overloading the servers with enormous traffic from compromised computers. During these attacks, excessive messages are sent to servers, which causes legitimate users to be denied access to the Internet \cite{kumar2022distributed}. In the cloud environment, attackers often rely on sophisticated methods to exploit its inherent scalability and flexibility \cite{wani2019analysis}. 
    
    \item \textbf{Web Browser Attacks:} Throughout the modern digital era, web browsers have become integral tools that allow users to access various online services and connect to the vast internet. Nowadays, browsers are a crucial component of almost every computer. Nevertheless, browsers are not immune to vulnerabilities \cite{baako2020integrated,singh2019taxonomy}. As a result of these vulnerabilities, attackers often gain access to a victim's computer, steal \gls{pii}, corrupt files, or use the hacked machine to attack others, potentially becoming a part of a botnet \cite{ibrahim2022review}. 
    
    \item \textbf{Signature Wrapping Attacks:} Cloud infrastructure attacks may provide attackers with root-level access to systems without targeting the cloud environment directly \cite{hu2021seapp}. The authentication systems of cloud interfaces are vulnerable to advanced cross-site scripting methods and signature wrapping \cite{kaur2022cross}. In addition to manipulating virtual machine operations and password resets, signature-wrapping attacks are also known as XML signature-wrapping attacks \cite{krishnamoorthy2021experimental}. In contrast, attackers can steal login credentials by exploiting Cross-Site Scripting (XSS) vulnerability \cite{jannat3proposed}.
    
    \item \textbf{Cloud Malware Injection Attacks:} Attackers perform malware injections by inserting malicious code or services into a network, appearing as if these components were legitimate components already present \cite{malik2023security}. As a result of this attack, users are often deceived into downloading software or opening malicious emails. This type of attack is sometimes driven by downloading or metadata spoofing. An attacker may trick users into downloading harmful software by creating a deceptive environment. In the aftermath of such an attack, users may have limited access to their systems, while attackers may be able to execute malicious activities in the cloud and gain unauthorised control \cite{geetha2020cloud}.
    
    \item \textbf{\gls{sql} Injection Attacks:} Cloud computing often exposes web services to various security threats due to their inherent internet accessibility. \gls{sql} injection is one of the most widespread risks \cite{alkhathami2022detection}. By manipulating \gls{sql} queries, \gls{sql} injection attacks can achieve malicious objectives. Consequently, sensitive data may be modified, unauthorised and confidential details may be accessed, or the server may crash due to such manipulation \cite{ponnumani2022various}.
    
\end{itemize}

\subsection{Application Layer}
Within the \gls{iot} architecture, the application layer resides at the top level. This layer provides the user interface, which includes services such as healthcare, smart homes, and connected cars \cite{abbasi2022security},\cite{padmanaban2023role}. It recognises spam and filters out the malicious content \cite{kumar2023machine}. Figure \ref{fig:fig7} illustrates \gls{iot} applications' primary security attacks. Application layer vulnerabilities include the following:

\begin{figure}[ht!]
    \centering
    \includegraphics[width=0.7\textwidth]{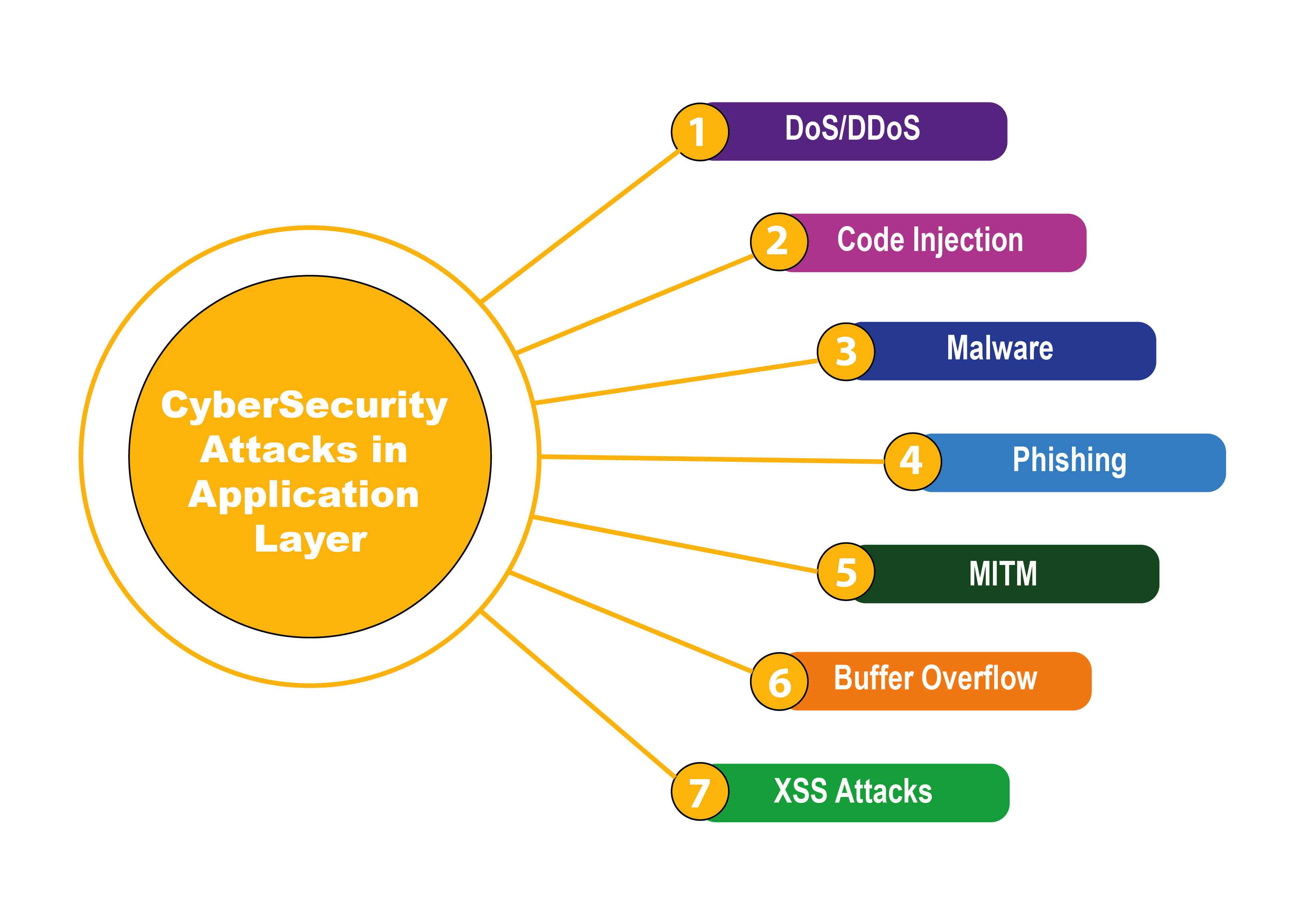}
    \caption[Cybersecurity attacks in the application layer]{Cybersecurity attacks in the application layer.}
    \label{fig:fig7}
\end{figure}

\begin{itemize}
    \item \textbf{\gls{dos}/\gls{ddos} Attacks:} \gls{ddos} attacks are carried out by a group of compromised computers operating from multiple locations that flood a particular target with traffic. In this attack, the primary goal is to overwhelm and render unavailable a server, website, or online service \cite{gaurav2022comprehensive},\cite{ahuja2023improved}. In a \gls{ddos} attack, \gls{iot} devices such as smart appliances, healthcare monitors, and industrial sensors can be rendered useless \cite{shaikh2018internet}. In addition to affecting their immediate functionality, disruptions can jeopardise critical operations they oversee, like monitoring patient health or maintaining optimal building conditions \cite{jia2019adopting}.
    \item \textbf{Phishing Attacks:} Phishers can use compromised devices such as smartphones, appliances, and smart cars to conduct phishing attacks at the application layer. Through this method, attackers impersonate legitimate devices and send messages that appear to be authentic. A malicious attacker can then manipulate or obtain credentials to authenticate or identify, which can be used for criminal purposes \cite{shifa2016multimedia}. For example, the Irish \gls{hse} Conti Cyber attack \cite{ncsc2021}, which was initially a phishing attack, soon escalated to a ransomware attack after an employee accidentally clicked on the malicious email, disrupting the healthcare facility operations and \gls{it} outages across the country for days. However, it took four months for the Irish HSE to recover from the aftermath completely. This is why it is mandatory to build cyber resilience within such infrastructures.    
    \item \textbf{Buffer Overflow Attacks:} Buffers hold data while it is transferred from one location to another. In buffer overflows, data exceeds the buffer's storage capacity. As a result, the application may crash, memory access errors may occur, and results may be incorrect. Memory overwrite vulnerabilities can be exploited by attackers. It can affect execution paths, leak confidential information, and corrupt files \cite{akhtar2022systemic}. Legacy systems in healthcare environments are most vulnerable to these attacks as they have limited memory \cite{dhirani2021industrial}.
    \item \textbf{Malware:} A malware attack attempts to use a malicious program to commit an offence with \gls{iot} applications, and recently, many malicious programs have been released to attack \gls{iot} devices, including rootkits, spyware, and adware \cite{khatun2019malicious},\cite{9941250}. In \cite{dhirani2021industrial, dhirani2023ethical}, the authors outline different types of malware and their impact on national security (i.e., Red October), social (transportation, communications, energy, and water sectors), financial (BaFin) \cite{comfort2023german} and economic domains (manufacturing, fintech) affecting human lives \cite{raj2023critical}.
    
   \item \textbf{\gls{xss} Attacks:}
    In cross-site scripting attacks, harmful code is embedded into an authentic and trusted website \cite{nagaty2023iot}. An attacker can alter the content of an application through this potentially dangerous intrusion \cite{gupta2017cross}. Due to the inability of the targeted browser to distinguish between genuine and malicious code, the infected code is executed. Consequently, this malicious code can access cookies, session identifiers, or other confidential information. In addition, the attacker can control the device directly, sending users to malicious websites or causing direct damage to the device.

    \item \textbf{Unauthorised Scripts Attacks:} These attacks occur when unwanted or malicious scripts are executed without the user's consent. In contrast to XSS attacks, which target the user's browser, unauthorised scripts can run anywhere in the system or application. Data breaches, system malfunctions, or other vulnerabilities may result from this \cite{5207654}.
    
    \item \textbf{Code Injection Attacks:}
    Code injection attacks that use \gls{sql} injection break the data-code isolation rule by inserting malicious \gls{sql} code into input fields \cite{abdulqadir2021study}. Attackers can embed these malicious \gls{sql} commands in web forms, URLs, or page requests. Without proper filters, a web application may process these harmful commands incorrectly. As a result, unwanted access to the database can occur \cite{lu2023semantic}.
 
\end{itemize}

\section{Emerging Technology and Security in \gls{hiot}} \label{sec:emerging3}
Technological advancements, including the \gls{iot}, \gls{ml}, and cloud computing, are gradually reshaping healthcare. This development creates innovative healthcare solutions but also challenges, particularly when it comes to data security. \rev{\gls{ai} and \gls{ml} are applied to analyse IoT data patterns and detect cybersecurity threats, such as intrusion and malware attacks, in \gls{hiot} environments \cite{zhang2022explainable}.} This section explores the conjunction of \gls{ml} and cloud technologies, investigating emerging trends and attendant security considerations within e-health applications to establish a secure and innovative framework for \gls{hiot} implementations. It further explores the connections between \gls{ml} and cloud technologies within e-health, navigating through emerging trends and assessing related security concerns. \gls{dl} and \gls{ml} are increasingly integral to mitigating cyber threats; however, how these technologies impact e-health is a crucial issue \cite{mathews2019explainable}. It aims to provide a secure and innovative foundation for \gls{hiot} implementations.

\subsection{\acrlong{ml} with Cloud Computing}
Researchers all over the world have applied \gls{ml} to a variety of applications and domains \cite{islam2022systematic}. \rev{\gls{ml} has gained significant attention among \gls{hiot} researchers \cite{kumar2023healthcare}.} In the context of \gls{hiot}, \gls{ml} is beneficial for remote monitoring and real-time treatment of diseases \cite{amin2020edge},\cite{ram2019machine}. \gls{ml} algorithms such as \glspl{svm}, \glspl{dt}, Random Forests, and \glspl{ann} can analyse huge volumes of medical data collected by healthcare-related smart devices, including vital signs and medical histories \cite{vanin2022study}. In this process, \gls{ml} techniques are applied to analyse massive datasets to find patterns and generate insights that may assist clinical decisions, improve patient outcomes, and reduce healthcare expenditures. Cloud computing, however, provides computing power and storage resources for \gls{hiot} to support \gls{ml} based on big data collected from \gls{iot} devices \cite{dhirani2020hybrid}. \gls{ml} algorithms utilise the cloud to process and analyse data and provide a secure and scalable computing environment \cite{sharma2022modelling}.  
Forum et al. \cite{desai2022healthcloud} developed a Health-Cloud platform using \gls{ml} and cloud computing to track patients suffering from heart-related diseases. Moreover, the researchers built a live data analysis iOS App using Google Cloud Firebase.  

Abdelaziz et al. \cite{abdelaziz2019machine} investigate the use of \gls{iot} and cloud computing in healthcare to predict chronic kidney illness in a city of the future. \gls{iot} devices transmit chronic kidney disease (CKD) data to cloud storage, improving prediction accuracy. The hybrid model, which combines linear regression and neural networks, predicts CKD with 97.8\% accuracy. Cloud \gls{iot} offers tremendous potential in healthcare services, benefiting patients and smart city stakeholders.

\subsection{\gls{ehealth} Systems}
\begin{figure}[ht!]
    \centering
    \includegraphics[width=0.7\textwidth]{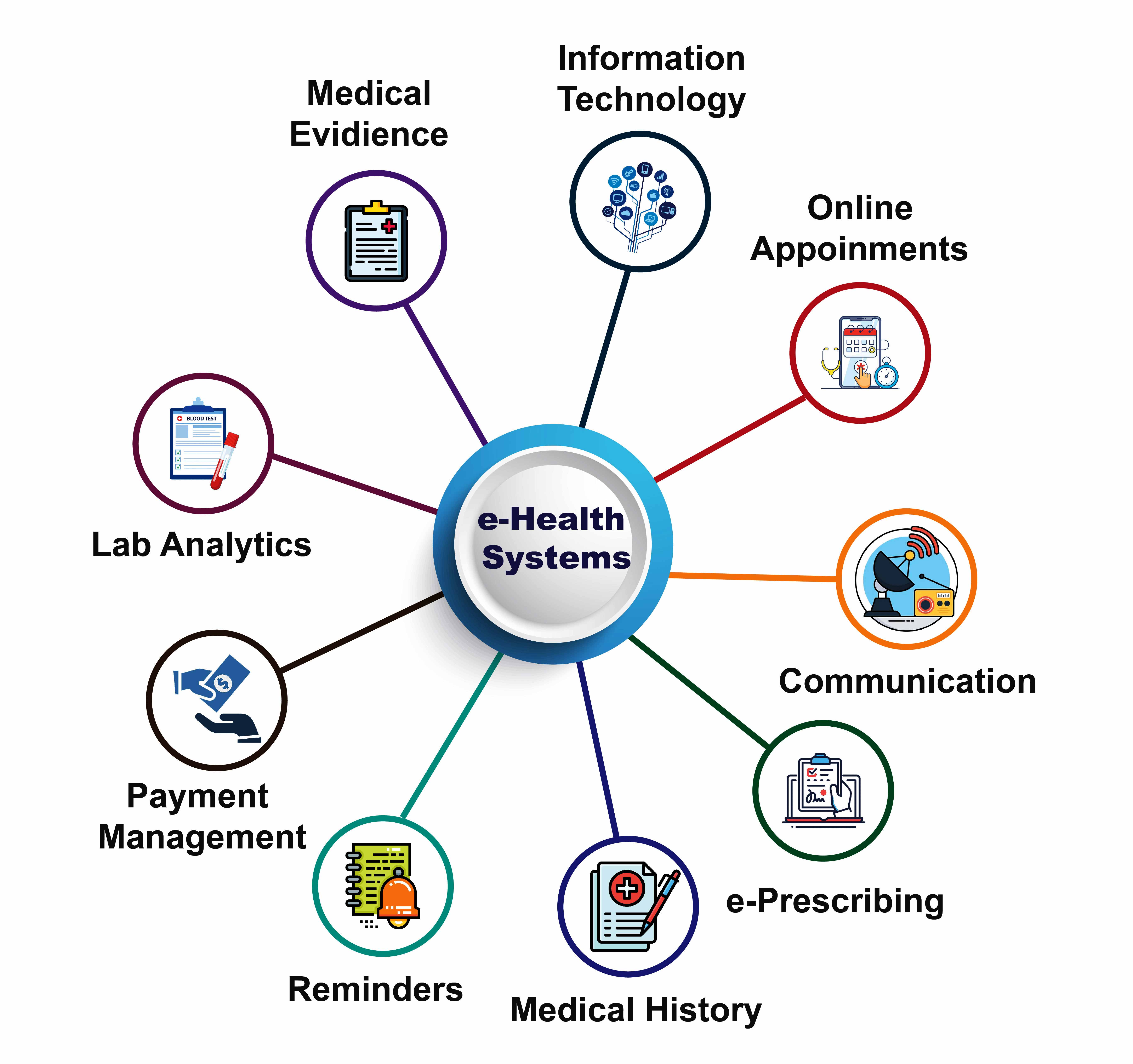}
    \caption[e-Health systems]{e-Health systems.}
    \label{fig:fig8}
\end{figure}

The exchange of health records between doctors and patients is now largely electronic, enabling efficient data transfer and communication. As illustrated in Figure \ref{fig:fig8}, e-health application systems can solve several problems associated with traditional healthcare systems, including online appointments, medical evidence, information technology, communication, e-prescribing, medical history, reminders, payment management, and lab analysis. e-healthcare services can enhance patient data and health information management \cite{butpheng2020security}. As a result, patients can access their medical records online, receive remote consultations, and use mobile health applications to manage their health, which increases accessibility and empowers patients. 

Maksimović et al. \cite{maksimovic2017internet} describe the gradual adoption of e-health platforms primarily due to infrastructure and political restrictions. Despite these challenges, e-health and \gls{iot} convergence are progressing. However, there are hurdles like consistency, security, and interoperability in \gls{iot} integration. Privacy concerns and regulations further complicate the adoption of large-scale technologies. The chapter highlights the influence of \gls{iot} in e-health and outlines these crucial challenges, emphasising the importance of overcoming barriers to implementation.
Zhang et al. \cite{9566565} explore how \gls{g5} technology can revolutionise e-health. While \gls{g5} promises reliable access to e-health, current efforts remain insufficient due to limited practical deployment, integration challenges, and unresolved technical constraints. This chapter also discusses technological aspects, practical use cases, research trends, and challenges in advancing \gls{g5} e-health.

\subsection{Big Data in \gls{hiot}}
\gls{iot} networks generate enormous amounts of data at every moment. Therefore, manipulating so many datasets requires considerable technical expertise. \gls{iot} architecture and \gls{ml} techniques boost big data capacity, and advanced \gls{dl} models are essential \cite{hossain2019applying}. Smart healthcare systems have dynamically transformed mathematical modelling with data collection using big data analytics. Besides collecting and storing data, big data is also about empowering machines to think and act like humans to simplify complex tasks \cite{hewson2015internet},\cite{asri2022iot}. Wearable sensors continuously collect data, such as sleep patterns, exercise levels, walking distances, heart rates, etc. The latest \gls{iot} sensors can also monitor heart rate, blood sugar levels, and pulse. A major benefit of big data for the medical industry is that it could reduce the costs of accommodations, travel time, and transportation delays. In addition, big data analytics have improved healthcare facilities and enabled patients to recover at home \cite{solfa2023big}. \gls{hiot} devices can also decrease medical expenses by reducing personnel and transport costs with big data \cite{arji2023identifying},\cite{8603570}. 

\rev{Asri} et al. \cite{asri2023toward} developed a real-time disease prediction platform that enhances patient care and reduces healthcare costs with a big data platform. It helps users make better decisions in real-time, improving health and safety. Heart attacks, obesity, and miscarriages can be detected using this approach. Long-term treatment and hospitalisation costs can be reduced by early detection and accurate prediction of health problems.  

Sasubilli et al. \cite{9120906} utilise \gls{ml} and big data to analyse the use of electronic health record applications in neuro- and cardiac-related fields. Integrating \gls{ml} approaches and big data frameworks in healthcare architecture enhances data management while improving patient care. Advanced technologies enable more precise diagnostics, personalised treatment plans, and better outcomes, elevating patient care standards.

\subsection{Risk Categories in \gls{hiot}}
\begin{figure}[ht!]
    \centering
    \includegraphics[width=0.9\textwidth]{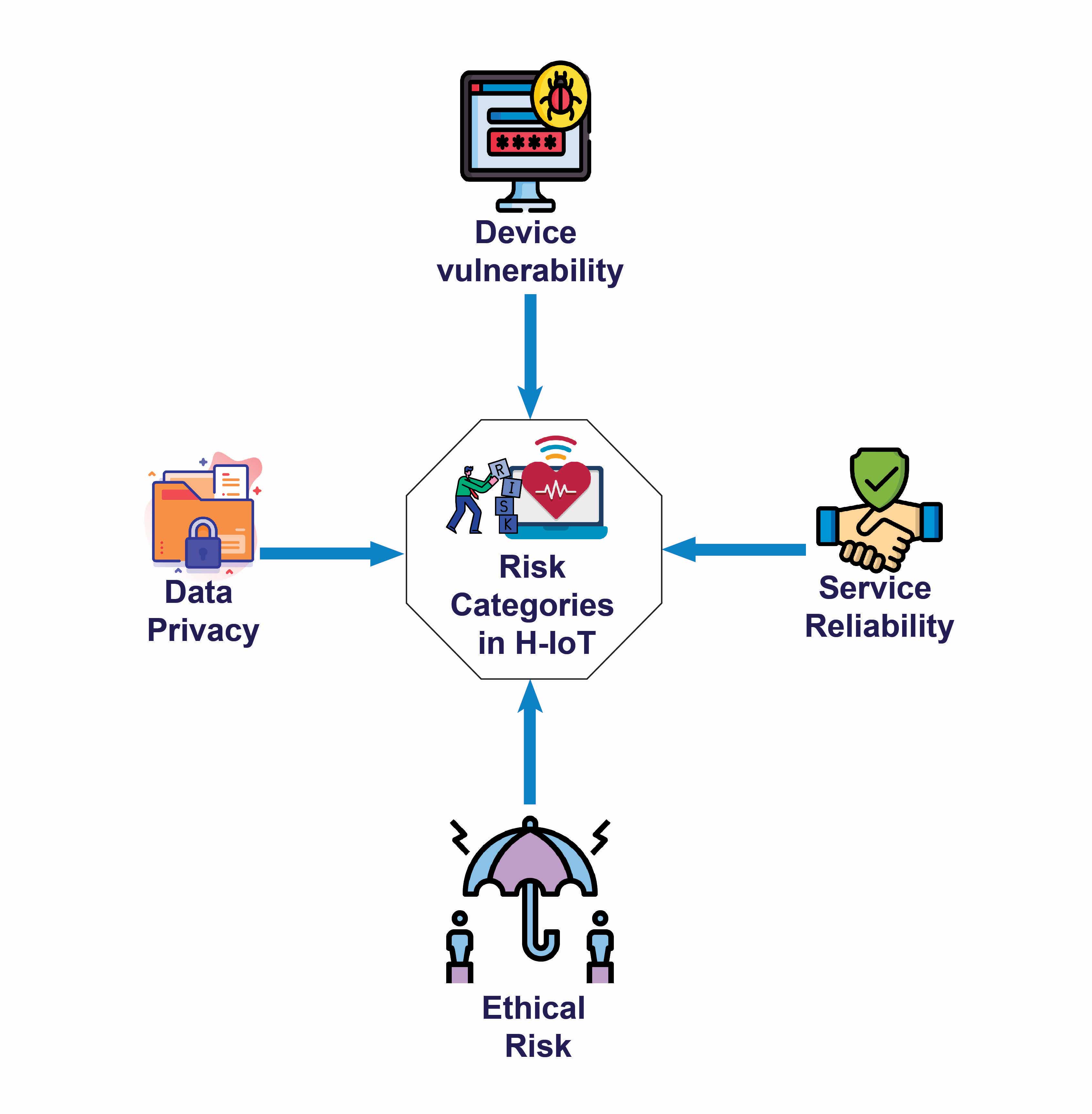}
    \caption[Risk categories in H-IoT]{Risk categories in \gls{hiot}.}
    \label{fig:fig9}
\end{figure}

\gls{hiot} brings multiple risks associated with medical devices integrated with advanced technologies. Data privacy, device vulnerability, service reliability, and ethical issues are all included in these risks. Both security and user trust depend on addressing these issues. As depicted in Figure \ref{fig:fig9}, these risk categories apply to the \gls{hiot} industry. 

The following risk categories and associated impacts within the context of \gls{hiot} are discussed:

\begin{itemize}
    \item \textbf{Data Privacy Risk:} During the COVID-19 pandemic, several countries reduced their data privacy guidelines as part of their emergency response \cite{do2020covid}. As per GDPR, implementing data security, governance, risk, and control metrics has been mandatory since the regulation was enacted in 2018; the regulation also enabled and improved the cybersecurity posture of organisations working across Europe in comparison to other countries/jurisdictions, and this is why Europe suffered fewer data breaches in comparison to the rest of the world. In the context of digitally enabled precision healthcare 5.0, securing systems and enforcing strong data controls are essential to mitigate risks associated with emerging technologies \cite{suganthi2020end}; this requires end-to-end data security across data-in-use, data-in-transit, and data-at-rest. To mitigate the risk of data breaches, sensitive information should be encrypted, anonymised, and pseudonymized.  As a result, even if adversaries gain unauthorised access, the compromised data remains largely unusable for malicious purposes \cite{dhirani2023ethical}.
    
    \item \textbf{Device Vulnerability Risk:} \gls{iot} has many security vulnerabilities, especially in the \gls{hiot} space, because these devices would be manufactured by different suppliers who would not have considered security by design and privacy-by-design metrics, leading to unencrypted network services, inadequate password protocols, and user interface credentials. These issues have been identified as serious security lapses in 70\% of \gls{hiot} devices \cite{gelenbe2023real}. In addition, 90\% of these devices collect personal data \cite{williams2016always}, illustrating the complexity and diversity of this issue. Healthcare sectors worldwide are particularly vulnerable to cyber threats due to their diverse operating environments and intricate regulatory frameworks. However, \gls{enisa} has enforced stronger regulations to combat supply chain, digital services, digital market, and third-party risks \cite{dhirani2023ethical}.

   \item \textbf{Service Reliability Risk:} In \gls{hiot}, service reliability risks refer to service interruptions or failures, which can cause critical delays or data loss. This is a major concern for emergency response and ongoing health monitoring. \gls{hiot} services require effective interference mitigation and optimisation of \gls{g5} networks. To fully benefit from \gls{g5} and \gls{iot}, these steps will enhance reliability and efficiency \cite{pons2023utilization}. With \gls{g5} and beyond \gls{g5} (\gls{b5g}) deployment, service interruptions, data loss, and seamless communications between multiple devices will be prevented in \gls{hiot} or the \gls{iomt} \cite{10144626}. The ability to transmit real-time, accurate patient data to medical personnel through \gls{g5} and \gls{b5g} is especially crucial in medical emergencies \cite{rai2023paradigms}.
    
    \item \textbf{Ethical Risk:} A potential ethical risk associated with \gls{iot} devices is that of unethical actions \cite{vermanen2022ethical}. There is an example from the automotive industry that serves as a cautionary tale, even though it isn't directly related to \gls{hiot}. In violation of the Clean Air Act, Volkswagen developed and installed software to manipulate diesel emissions tests \cite{kandasamy2020iot}. Reputational and financial damages resulted from this breach of ethics. When it comes to \gls{hiot}, where patients' health and well-being are at risk, cyber ethical breaches such as harm to privacy, harm to property, and misuse of technical resources \cite{dhirani2023ethical} can be detrimental \cite{karale2021challenges}.
    
\end{itemize}


\section{Attack and Mitigation Techniques in \gls{hiot}} \label{sec:attacknmitigation4}
With cutting-edge technologies in the healthcare industry, the security and privacy risks have drastically increased \cite{dhirani2023ethical}. As per \cite{Phillips2023}, in 2022, more than 200 cyber attacks were carried out across healthcare organisations in the world daily by malicious actors, impacting millions of people worldwide. The number of cyber attacks on \gls{hiot} increased by an alarming 74\% in 2023 \cite{McKeon2023}. Demystifying cyber threats and attack scenarios is crucial to preventing these escalating risks. Therefore, this section explores the cybersecurity landscape of \gls{hiot}. Table~\ref{tab:CybersecurityH-IoT} surveys cybersecurity challenges within the healthcare sector over the last five years. Furthermore, it discusses attack datasets for detecting and mitigating anomalies in \gls{hiot} environments and threats, potential impacts, and mitigation strategies.

\begin{table}[htbp]
\footnotesize
\centering
\setlength{\tabcolsep}{3pt}
\renewcommand{\arraystretch}{1.2}
\caption[Cybersecurity challenges in \gls{hiot}]{Cybersecurity challenges in \gls{hiot}.}
\label{tab:CybersecurityH-IoT}
\begin{tabular}{@{}%
  >{\RaggedRight\arraybackslash}p{3.5cm}  
  >{\RaggedRight\arraybackslash}p{3.5cm}  
  >{\RaggedRight\arraybackslash}p{5cm}    
@{}}
\toprule
\textbf{Key Publications} & \textbf{Area} & \textbf{Security and Privacy Issues} \\ 
\hline
Garcia-Perez~et~al. \cite{garcia2023resilience} (2023) & \gls{hiot} & The paper highlights the lack of cyber risk awareness, management, and security risks related to emerging technologies. \\
\hline
Cartwright~et~al. \cite{cartwright2023elephant} (2023) & \gls{iomt}, COVID-19 & Security gaps in protected health information (PHI), \gls{iomt} challenges, COVID-19 amplification, and funding shortages. \\
\hline
Javaid~et~al. \cite{javaid2023towards} (2023) &
\gls{hiot} & Data security risks identified in the paper include ransomware attacks, unauthorised access to patient information, device vulnerabilities, and healthcare data source complexity.\\ 
\hline
Kumar~et~al. \cite{kumar2023healthcare} (2023)
& 
\gls{hiot} & \gls{hiot} devices face various security threats, scalability, and latency challenges. \\ 
\hline  
Sharma et al. \cite{sharma2023anomaly} (2023) & \gls{dl}, Cyber-attacks, \glspl{ids}, Healthcare  & A \gls{dl}-based \gls{ids} detecting cyber attacks achieves 84\% accuracy, increasing to 91\% with Generative Adversarial Networks (GANs) balanced data. \\ 
\hline
\end{tabular}
\end{table}

\begin{table}[htbp]
\ContinuedFloat
\footnotesize
\centering
\setlength{\tabcolsep}{3pt}
\renewcommand{\arraystretch}{1.2}
\caption[]{Cybersecurity challenges in \gls{hiot} (continued).}
\begin{tabular}{@{}%
  >{\RaggedRight\arraybackslash}p{3.5cm}  
  >{\RaggedRight\arraybackslash}p{3.5cm}  
  >{\RaggedRight\arraybackslash}p{5cm}    
@{}}
\toprule
\textbf{Key Publications} & \textbf{Area} & \textbf{Security and Privacy Issues} \\ \hline

Khatkar et al. \cite{khatkar2023parametric} (2023) & \gls{dl}, Intrusion Detection, Healthcare & For intrusion detection in healthcare, evaluating a \gls{dl}-based \gls{lstm} algorithm against a \gls{dt}, \gls{svm}, \gls{knn}. \\ 
\hline
Ksibi et al. \cite{10182654} (2023) & \gls{iomt}, Cybersecurity, \gls{ml}, Risk Assessment & Analyse the security risks in the \gls{iomt} using \gls{ml} algorithms for anomaly detection and introduce a new model of risk assessment. \\ 
\hline
Bouabida et al. \cite{bouabida2022telehealth} (2022) & Pandemic-driven, patient data risks & A significant concern raised by the COVID-19 crisis is the security and confidentiality of patient data. \\
\hline
Wang et al. \cite{9020632} (2019) & Big Data, \gls{hiot} & An Analysis of Big Data Analytics in healthcare is presented, including its advantages and challenges. \\ 
\hline
Coventry et al. \cite{coventry2018cybersecurity} (2018) & Cybersecurity in Healthcare & Due to cybercrime risks, healthcare technology needs a holistic security program. \\ \hline
\end{tabular}
\vspace{.1cm}
\end{table}
\FloatBarrier

\subsection{Mitigation Techniques}
\begin{figure}[htbp!]
    \centering
    \includegraphics[width=0.9\textwidth]{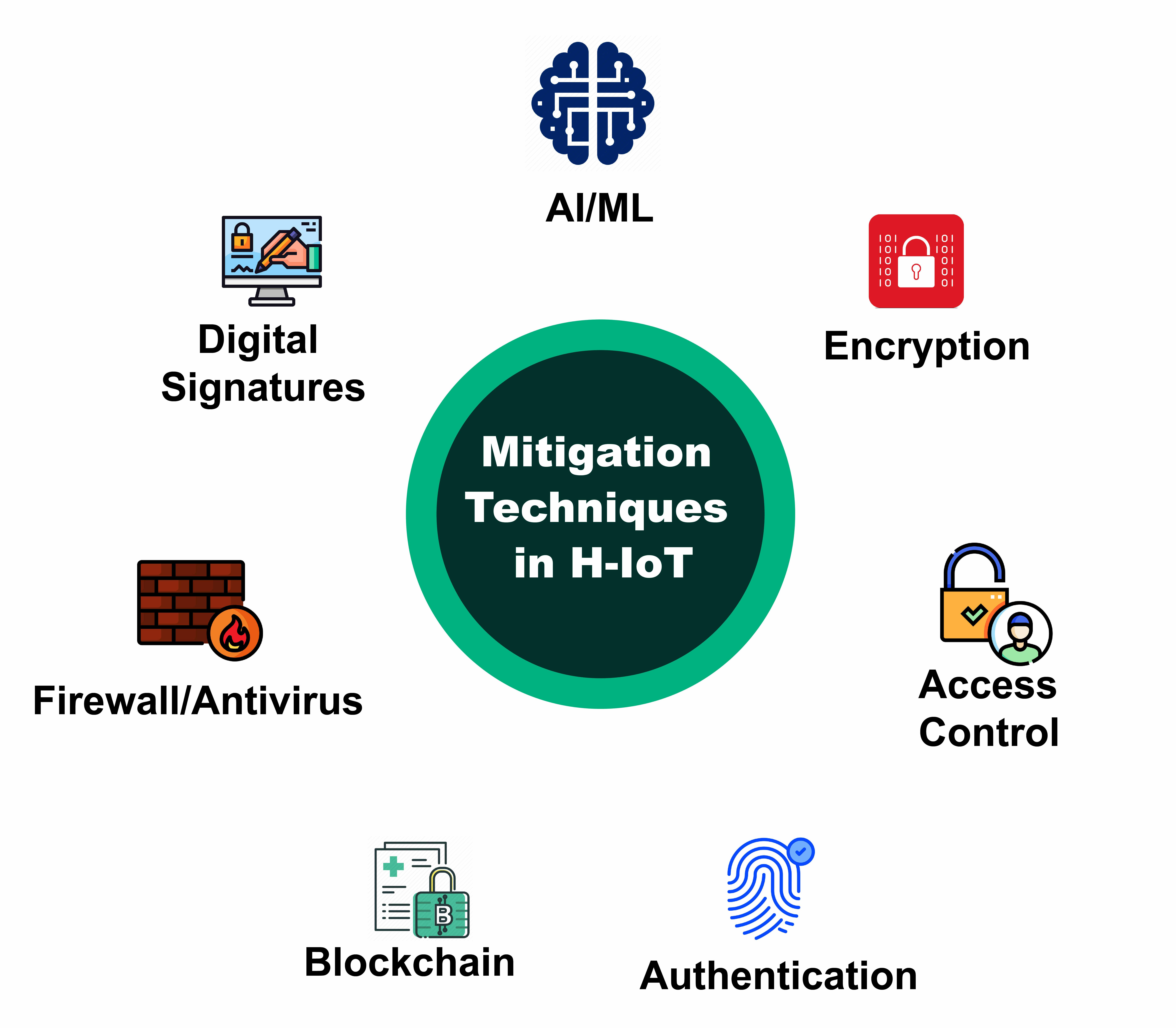}
    \caption[Mitigation techniques in H-IoT]{Mitigation techniques in \gls{hiot}.}
    \label{fig:fig10}
\end{figure}

This section presents findings on mitigation strategies employed in prior research to mitigate cybersecurity threats associated with \gls{iot} in healthcare. \rev{Different types of threats across perception, network, and application layers require corresponding mitigation techniques tailored to each layer.} Figure~\ref{fig:fig10} illustrates the mitigation techniques for access control, encryption, authentication, \gls{ai}/\gls{ml}, blockchain, digital signatures, and firewalls/antivirus. 

Techniques such as differential privacy and federated learning are also used for privacy-preserving \gls{hiot} systems, although they are not shown in Figure~\ref{fig:fig10}.

\gls{hiot} solutions also integrate modern \gls{ai} and \gls{dl} technologies. For instance, Al-Garadi et al. \cite{al2020survey} described that \gls{ml} models have been incorporated into security solutions to identify vulnerabilities and attack surfaces. Qiu et al. \cite{8656914} explore the integration of healthcare and smart cities, enhancing health practices in countries worldwide. Blockchain technology addresses this integration's security concerns, ensuring the confidentiality of patient data. The authors discussed the benefits and drawbacks of each model used in the research at different \gls{iot} layers. 


\subsection{Comparative Review of \gls{hiot} Attack Datasets}
Datasets are crucial in identifying and mitigating anomalies in the ever-expanding world of the \gls{iot} in healthcare, industry, and general applications. \gls{hiot} attack datasets are reviewed in this section, including simulations and real-world data from \gls{hiot} sensors. In the constantly evolving area of \gls{hiot}, these datasets serve as essential tools for research \cite{raeesi2021iot}. Similarly, \gls{ml} applications require high-quality healthcare data, and exact labelling improves their accuracy \cite{kempa2020healthcare},\cite{haq2020intelligent},\cite{vaishya2020artificial}. Table \ref{tab:AnalysisDataset} provides a detailed overview of datasets, attacks, mitigation techniques, benefits, and limitations.


\begin{table}[!htbp]
\centering
    \setlength{\tabcolsep}{3pt}
\renewcommand{\arraystretch}{1.2}
\caption[Comparative analysis of techniques]{Comparative analysis of techniques and targeted attacks in published research.}
\label{tab:AnalysisDataset}
\resizebox{\textwidth}{!}{%
\begin{tabular}{@{}%
p{2.4cm}  
p{3.4cm}  
p{3.5cm}  
p{4.4cm}  
p{4.3cm}  
p{4.1cm}  
@{}}
\toprule

\textbf{Reference} & \textbf{Techniques Used} & \textbf{\shortstack{Targeted\\Attacks}} & \textbf{Key Benefits} & \textbf{Limitations} & \textbf{Dataset Employed} \\
\hline
Wagan et al. \cite{wagan2023fuzzy} (2023) & Dynamic Fuzzy C-Means clustering, Customised Bi-\gls{lstm} technique. & Botnet-based \gls{ddos}, Zero-day network attacks. & 
- Separates attacks from regular \gls{iomt} data.\par
- Securely handles medical data.\par
- Analyses attack patterns in \gls{iomt}.\par
&
- Inequity in the model.\par
- Model decisions are not comprehensible.\par
- Possibility of corruption of data.\par 
& 
WUSTL-EHMS-2020 is a heart disease dataset with 36 attributes and 18940 instances developed with the \gls{ehms}.
\\ \hline
Thulasi et al. \cite{thulasi2023lso} (2023) & \gls{mscsl}, 
light spectrum optimiser (LSO) for optimising \gls{mscsl}’s hyperparameters. & Port scans, Brute force, \gls{dos} attacks. & 
- Data adaptability improved.\par
- Reduced computational complexity.\par
- High precision and accuracy.\par
&
- Privacy issues are not addressed. \par
- Not optimised for multi-cloud/fog-based dynamic. environments with multiple devices.\par 
& 
Supervisory Control and Data Acquisition International Electrotechnical Commission (SCADA IEC 60870-5-104) datasets.
\\ \hline
Gupta et al. \cite{gupta2022tree} (2022) & Tree classifier-based network intrusion detection model. & \gls{mitm} Attack \par (Data Alteration and Data Spoofing). & 
- Reduced input data dimension effectively.\par
- Accelerates anomaly detection.\par
&
- The dataset only consists of data alterations and data spoofing attacks.\par 
- Tested on a restricted network.\par 
- Difficulties in formulating \gls{iomt} security design.\par

& 
A combined network flow and biometrics dataset by \cite{hady2020intrusion}
\\ \hline

\end{tabular}
}
\end{table}


\begin{table}[!htbp]
\ContinuedFloat
    \centering
    \caption[]{Comparative analysis of techniques and targeted attacks in published research (continued).}
\resizebox{\textwidth}{!}{%
\begin{tabular}{@{}%
p{2.4cm}  
p{3.4cm}  
p{3.5cm}  
p{4.4cm}  
p{4.3cm}  
p{4.1cm}  
@{}}
\toprule
\textbf{Reference} & \textbf{Techniques Used} & \textbf{\shortstack{Targeted\\Attacks}} & \textbf{Key Benefits} & \textbf{Limitations} & \textbf{Dataset Employed} \\
\hline
Ullah et al. \cite{ullah2025hybrid} (2025) &
\gls{cnn}-\gls{lstm} hybrid model with SHAP-based explainability in \gls{iot} environments. &
\gls{ddos} and other \gls{iot} cyber-attacks. &
- High detection accuracy.\par
- Captures spatial and temporal features.\par
- Improved interpretability using SHAP.\par
&
- Limited to specific attack types.\par
- Higher computational cost.\par
- Depends on the SDN framework.\par
&
CICDDoS2019 and IoT Healthcare Security datasets.
\\ \hline
Kumar et al. \cite{kumar2023augmenting} (2023) &
\gls{ml}-based framework using Random Forest for \gls{iot} healthcare security. &
\gls{iot} botnet and \gls{ddos} attacks. &
- Early botnet detection.\par
- Improves \gls{iot} security.\par
- Handles imbalanced data.\par
&
- Limited attack scope.\par
- Does not incorporate \gls{dl} techniques.\par
- Lacks real-time validation.\par
&
IoT botnet dataset.

\\ \hline
Khan et al. \cite{khan2024fl} (2024) &
Federated learning-based framework using \gls{ml} classifiers for RPL-based \gls{iot} security. &
Selective forwarding and RPL routing attacks. &
- Detects \gls{rpl}-based routing attacks.\par
- Lightweight and scalable.\par
- Enhances data privacy.\par
&
- Binary classification only.\par
- Limited attack diversity.\par
- Depends on RPL protocol.\par
&
IoT Routing Attack Dataset (IRAD).

\\
\hline

RM et al. \cite{rm2020effective} (2020) & \gls{ml} algorithms (Supervised \& unsupervised), \gls{dnn}. & Replay, \gls{mitm}, Impersonation, Privileged-insider, Remote hijacking, Password guessing, \gls{dos} attacks, Malware attacks. & 
- Classification \& detection (network/host).\par
- Detects anomalies reliably.\par
- DNN compared to other \gls{ml} algorithms.
&
- Unpredictable attacks.\par
- Scalability issues.\par
- Network behaviour changes rapidly.\par
- Evolving attacks quickly.
& 
Kaggle benchmark dataset for intrusion detection. 
\\
 \hline
\end{tabular}
}
\end{table}
\FloatBarrier

\section{Multi-layered Analysis of Routing Attacks in \gls{hiot}} \label{sec:multilayer5} 
This section focuses on routing attacks in the \gls{hiot} domain, reviewing key components sequentially. The first step is to discuss the vital process of collecting health-related data in the layer for medical data collection. After collecting raw medical data, the medical application layer transforms it into meaningful healthcare services. Next, it discusses the role and vulnerabilities of the routing and network layers in the context of routing attacks. The next step is emphasising that the medical application layer is crucial to translating data into actionable insights. In addition, it describes the role of the COOJA simulator in collecting data and generating datasets in the \gls{hiot} environment.

\subsection{Medical Data Collection Layer}
\begin{figure}[!htbp]
    \centering
    \includegraphics[width=0.7\textwidth]{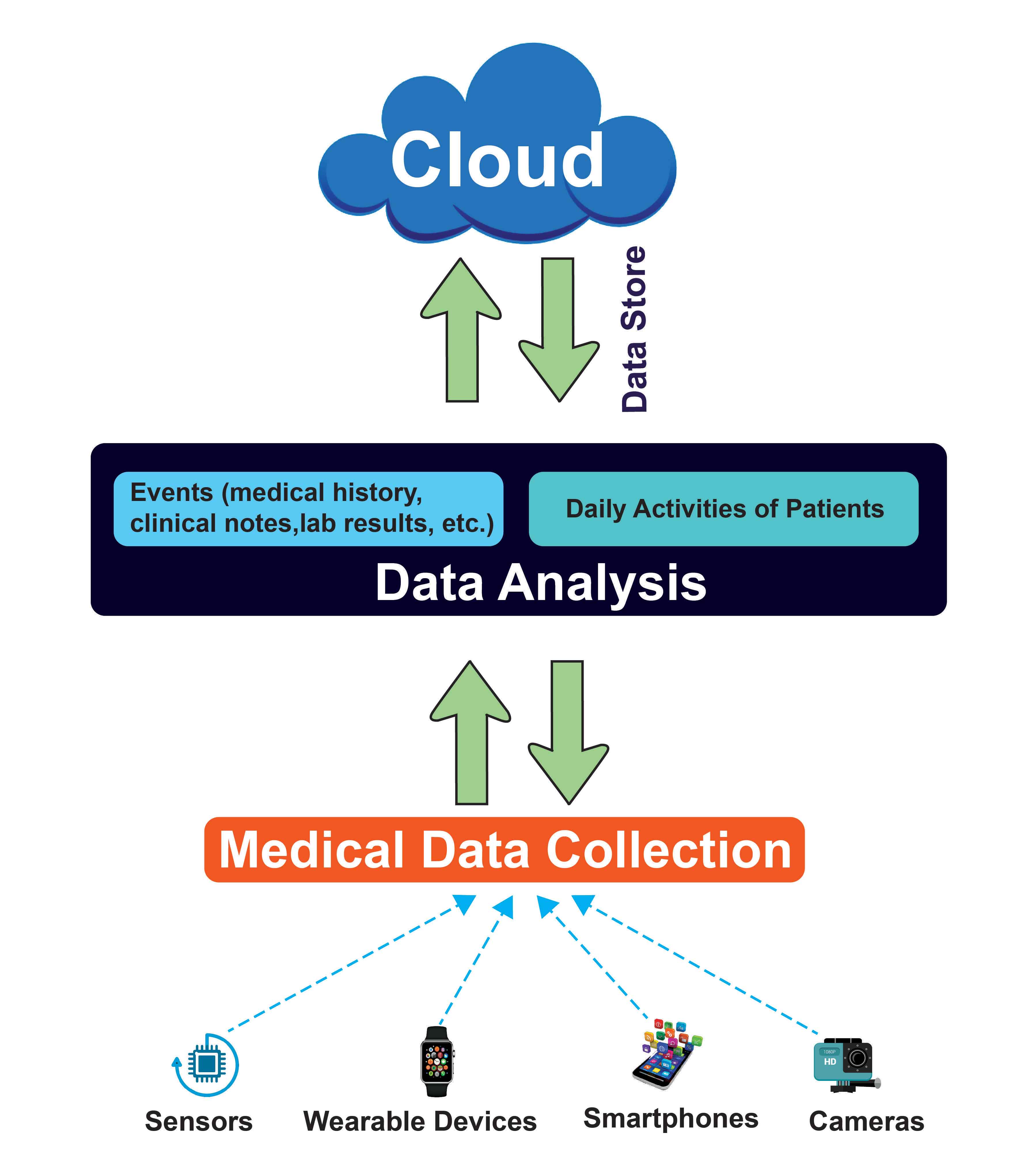}
    \caption[Data collection layers in healthcare]{Data collection layers in healthcare.}
    \label{fig:fig11}
\end{figure}

Healthcare data is collected by smart healthcare sensors, such as devices for tracking brain activity and blood pressure. A doctor can track the progress of an elderly patient by using \gls{iot} devices that record every detail and activity. Staff members and nurses who care for elderly patients receive information. \gls{hiot} integrates healthcare devices at home for disabilities. Therefore, the doctor may follow and monitor patients' cases at times. Despite this, the data generated and collected are susceptible to various threats. As a result of the attacks, hackers can gain complete command and control (C\&C), disrupting the devices \cite{dhirani2021industrial}. These attacks can drain sensors' power, causing them to fail or disrupt the entire \gls{iot} network in the healthcare industry. Medical data collection includes collecting, analysing, and monitoring data, as shown in Figure \ref{fig:fig11}. 

\rev{Edge computing and federated learning are also important components of systems for distributed processing and privacy-preserving data analysis in modern \gls{hiot}, although they are not illustrated in Figure~\ref{fig:fig11}.}

As a starting point, several types of devices are
used for collecting data within the medical data collection layer. These devices include sensors, cell phones, wearable devices, and cameras, among other things. The second phase involves data analysis, which analyses every event and activity of an elderly patient. A secure server then processes and stores the data. Monitoring healthcare data remotely and automatically recording it includes a patient's medical history and status \cite{kamel2020mitigating}. 

\subsection{Medical Application Layer}
The traditional application layer in \gls{iot} environments handles data processing, storage, and user interfaces. 
On the other hand, the medical application layer is uniquely designed for handling healthcare-related applications, data, and services, ensuring it follows the highest regulations following the EU \gls{gdpr} or \gls{hipaa}~\cite{ProtonGDPR2023},\cite{terry2017existential} while processing United States healthcare data. This layer presents unique challenges and complexities. For example, \glspl{ehr} for patient data, remote consultation platforms, real-time monitoring systems, medical resources, medical care, and diagnosis personnel. Moreover, the medical application layer processes, organises, and maintains medical records. This layer ensures that patients' data is authentic, secure, private, and reliable when transmitted over the communication system \cite{bansal2020application}. Numerous application layer protocols exist, including \gls{mqtt}, \gls{http}, WebSockets, RESTful, Secure-\gls{mqtt}, and \gls{coap}~\cite{9365708}. Choosing the appropriate protocol for \gls{hiot} depends on its application \cite{mittelstadt2017designing}.

\subsection{Routing Attack and Network Layer} \label{sec:multilayer5C}  
The network layer defines paths for the routing of data over the network. It has been demonstrated that low-power wireless networks with multiple hops are susceptible to a wide range of attacks, with routing attacks emerging as one of the most significant threats in the \gls{hiot} environment \cite{al2022systematic}. There are many routing attacks, including selective forwarding attacks and replay attacks \cite{saminathan2023blockchain}. In the selective forwarding attack, control packets are deliberately forwarded within the \gls{rpl} while data packets are dropped. In conjunction with sinkhole attacks, this strategy can disrupt established routing paths and lead to severe consequences for the network~\cite{patel2020rpl}. A replay attack entails capturing and re-transmitting packets captured from nearby nodes by an unauthorised node or attacker~\cite{tomic2017survey}. The purpose of these attacks is to manipulate or obstruct the transmission of data packets to impair the integrity of the network~\cite{butt2019iot}. Furthermore, rank and wormhole attacks are susceptible to \gls{iot} routing protocols due to their lightweight nature and limited computational resources. Using these attacks, \gls{iot} infrastructures can be devastated by attacks targeting control messages and resources~\cite{zahra2022rank}.

Routing and network layers receive data sent by medical data collection layers. These layers use protocols such as Wi-Fi 6, \gls{nbiot}, Bluetooth, IPv6 Over Low-Power Wireless Personal Area Networks (6LoWPAN), RPL, WiMAX, ZigBee, Sigfox, LoRa, and \gls{g5} \gls{nr} to transmit data to the medical application layer \cite{tariq2023critical},\cite{haghrah2023survey},\cite{ghazal2021positioning}. This setup requires the analysis of routing attacks, potentially compromising the confidentiality and accessibility of medical data. To ensure reliable and secure data transmission, protocol configurations designed for power-efficient and unstable networks should be carefully reviewed for vulnerability to various routing attacks \cite{peivandizadeh2023compatible}.
In this context, \gls{ml} techniques are explored for detecting cyber attacks in \gls{hiot} environments.

As highlighted by \cite{9738218}, \gls{hiot} networks are vulnerable to routing attacks, such as sinkholes and wormholes. The study proposes multiple detection approaches for identifying such attacks. Due to the limited resources of \gls{hiot} devices, traditional intrusion detection systems are often impractical. The proposed method efficiently detects and handles network attacks using \gls{ml} and \gls{dl} while considering device limitations.

According to \cite{anand2021efficient}, especially in e-health applications, protecting sensitive patient data from routing attacks and security threats is pivotal in \gls{g5}-\gls{iot}. Cloud-based e-health data is at risk of various cyber-threats (i.e., ransomware, loss of privacy, and digital identity fraud) and attacks mentioned in Section \ref{sec:cybersecurity2}. \gls{cnn}-DMA is a \gls{dl} model that detects malware attacks using a \gls{cnn}. The e-health apps enabled with \gls{g5}-\gls{iot} and \gls{dl} models, such as \gls{cnn}-DMA, are useful for ensuring data safety and system security against cyber-attacks. The \gls{dl} models can be trained directly from original data, such as images and text \cite{ahsan2022cybersecurity}. Therefore, raw data does not need to be preprocessed before being used for training. Moreover, \gls{dl} algorithms enable the seamless execution of security protocols without consuming significant computing resources, which is one of the major benefits of \gls{g5} cybersecurity \cite{mahalle2023data}.

In light of the delicate nature of healthcare data, enhancing security within \gls{hiot} network layers is imperative. The examples show that leveraging \gls{ml} and \gls{dl} can neutralise and mitigate these risks. The COOJA simulator further serves as a useful tool for analysing these attacks in controlled \gls{hiot} network environments.

\subsection{COOJA Simulator}
The COOJA simulator runs on Contiki OS, a portable operating system with limited resources designed specifically for devices such as sensor nodes \cite{8276353}. It is built upon an event-driven kernel while also supporting multi-threading. It supports a complete \gls{tcp}/IP stack through IP and programming protothreads. Simulating sensor nodes based on their real characteristics is the main advantage of the COOJA simulator. The process executes program code from ContikiOS and TinyOS using \gls{jni}. A Java Virtual Machine and C programming (a programming language commonly used in firmware sensor nodes) are interconnected by the \gls{jni}. As a result, COOJA can provide accurate simulations of sensor nodes or devices across multiple platforms, including \gls{hiot}, closely replicating real-world functionality. Moreover, it is crucial to note that the primary objectives of COOJA are extensibility and plug-in support. In contrast to the interface, the plug-in allows users to interact with the simulator. Plug-ins, for example, enable users to control simulation speeds or observe network traffic between nodes \cite{shahra2020comparative}. COOJA also has several inherent tools, including the ``radio logger,'' which records all packets dispatched by nodes in the simulation and associates them with a universal timestamp \cite{finne2021multi}. As a versatile simulation tool, COOJA allows for research across diverse environments by generating and collecting data to thoroughly assess \glspl{wban} in the \gls{hiot} domain. Using emulated \gls{hiot} devices, the COOJA simulation can even construct attack scenarios to uncover cybersecurity vulnerabilities within H-\gls{iot} networks.

Hence, \gls{ml} and \gls{dl} techniques have been explored to prevent routing attacks on \gls{hiot} networks. By reviewing examples, it becomes apparent that artificial intelligence models significantly defend against routing attacks in \gls{g5}-\gls{iot} e-health. COOJA has also been demonstrated to collect data from \gls{hiot} devices.

\section{\gls{ml}-Driven Intrusion and Event Detection in \gls{hiot}} \label{sec:ml-driven6}
To ensure data security within the \gls{hiot}, it is necessary to explore a variety of strategic mechanisms. The section endeavours to illustrate the pivotal role of \gls{ml} in establishing and reinforcing robust security protocols in healthcare, generalised \gls{ml} techniques, and exploring potential cybersecurity applications. Following this, the discussion discusses how feature extraction is essential to identifying threats accurately and mitigating risks. Afterwards, intrusion and event detection systems designed for healthcare environments will be offered, emphasising mechanisms for identifying potential breaches. A discourse on the reliability of \gls{hiot} systems is required to achieve consistent and reliable risk mitigation. The last topic of this section is the potential and challenges associated with \glspl{cnn} in the context of \gls{iot} security.

\subsection{\gls{ml} Techniques in \gls{hiot} Cybersecurity} 
\begin{figure}[ht!]
    \centering
    \includegraphics[width=0.9\textwidth]{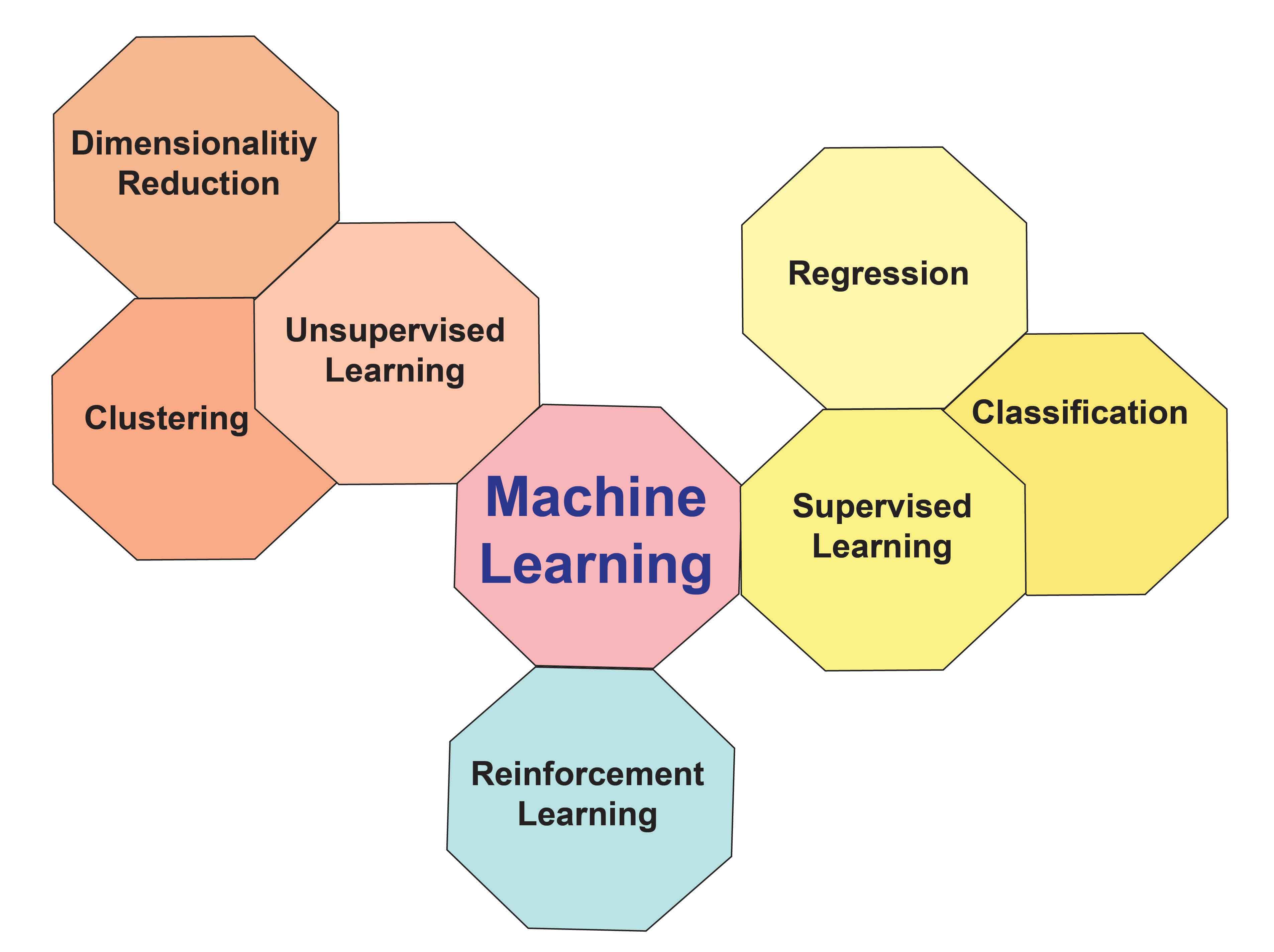}
    \caption[\gls{ml} techniques for H-IoT cybersecurity]{\gls{ml} techniques for \gls{hiot} cybersecurity.}
    \label{fig:fig12}
\end{figure}

The \gls{ml} method involves using data and algorithms to create models replicating human learning processes. \gls{ml} algorithms can be refined and optimised by improving an algorithm's loss function \cite{wang2020comprehensive}. As depicted in Figure \ref{fig:fig12}, \gls{ml} is traditionally classified into three categories: supervised, unsupervised, and reinforcement learning \cite{sarker2021machine},\cite{olowononi2020resilient}. 

\begin{figure}[htbp!]
    \centering
    \includegraphics[width=0.9\textwidth]{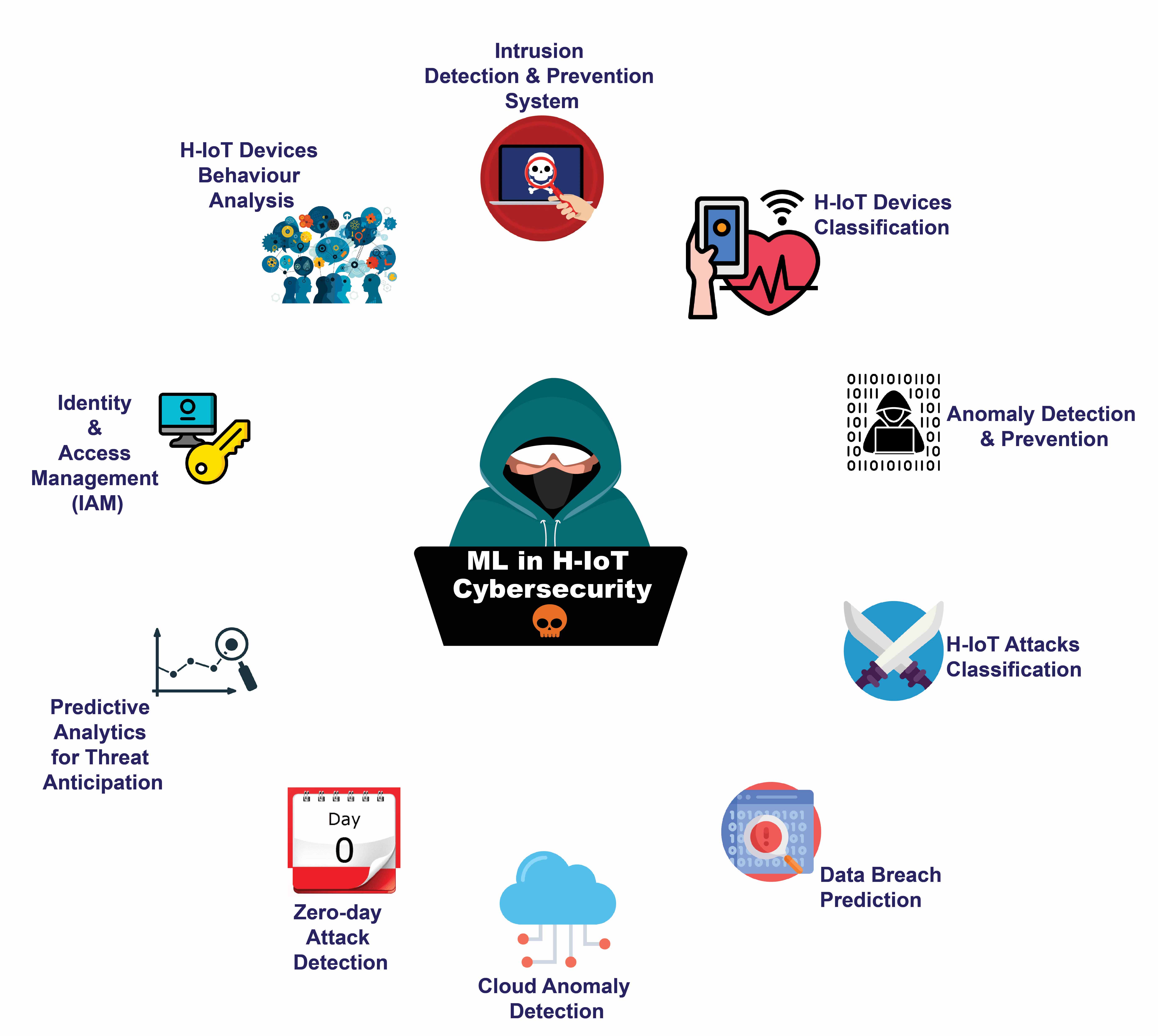}
    \caption[\gls{ml} use cases for H-IoT cybersecurity]{\gls{ml} use cases for \gls{hiot} cybersecurity.}
    \label{fig:fig13}
\end{figure}

Over the past several years, \gls{ml} techniques have been successfully applied to a diverse spectrum of cybersecurity challenges \cite{neupane2022explainable}. In the specialised domain of \gls{hiot} cybersecurity, these techniques have emerged as key assets, significantly enhancing the security level of healthcare devices and systems \cite{mazhar2023analysis}. As illustrated in Figure \ref{fig:fig13}, several potential use cases are encompassed, including intrusion detection and prevention system, classification of \gls{hiot} devices, anomaly detection and prevention, \gls{hiot} attacks classification, zero-day attack detection, predictive analytics for threat anticipation, identity, and access management (IAM), data breach prediction, Cloud anomaly detection, as well as \gls{hiot} devices behavior analysis. According to Table \ref{tab:AnalysisDataset}, defenders can identify, evaluate, and mitigate possible attacks more precisely with \gls{ml}, as outlined in Section \ref{sec:attacknmitigation4}. Due to this, \gls{ml} algorithms can automate various tasks, including identifying vulnerabilities, deceiving, and disrupting attacks \cite{sarker2022machine}. This section discusses several potential use cases for \gls{ml} in \gls{hiot} cybersecurity.

\begin{itemize} [label=–]
    \item \textbf{Classifying Anomalies in Cybersecurity:} In classifying threats, \gls{ml} offers a swift and efficient method for processing huge amounts of data. \gls{ml} analyzes behavior and continuously changes parameters to find anomalies that could indicate attacks \cite{sarker2021cyberlearning}. \gls{ml} allows for intelligent security services learning that learn to detect or predict cyber attacks and anomalies.

    \item \textbf{Predicting and Responding to Data Breaches and Cyber Attacks in Real-time:} With \gls{ml}, large amounts of data can be analysed from multiple sources to predict cyber threats \cite{alahakoon2020self}. A \gls{ml} system can quickly build defensive patches in response to an attack after identifying a cyber threat without human intervention \cite{kumar2022applications}.

    \item \textbf{Automating processes:} In the \gls{hiot} cybersecurity sector, \gls{ml} is becoming increasingly crucial as time-consuming tasks such as vulnerability assessments, malware analysis, network log scrutiny, and intelligence assessments continue to multiply \cite{ravi2022attention}. Automating security workflows enables organisations to respond quickly to threats and incidents, a feat that cannot be achieved through manual efforts alone \cite{singla2019deep}. Furthermore, automation offers sustainability by enhancing efficiency and scalability features and reducing unnecessary costs.

    \item \textbf{Authentication and Access Control:} Authentication technology is essential for medical devices since it validates a user's credentials with those stored in authorised user systems or authentication servers, enabling access to multiple systems simultaneously \cite{roobini2022cyber}. \gls{ml} determines when to request multi-factor authentication with intelligent authentication.

\end{itemize}

\subsection{Feature Extraction}
\gls{hiot} feature extraction involves selecting and transforming relevant data attributes. Moreover, it is necessary to determine the most crucial features before embarking on any \gls{ml} design process \cite{qiao2023hybrid}. Specific attributes can reduce computing costs and improve storage efficiency \cite{sharma2022feature}. For feature selection, the following methods are used:

\begin{itemize}
    \item \textbf{Filter Methods:} This method computes attribute relationships by comparing their pairwise correlations. Attributes are selected based on statistical method scores and predefined threshold values. There are several popular methods for feature extraction, such as information gain and Chi-Square tests, as well as correlation analysis \cite{rohith2022sentiment}. 

    \item \textbf{Wrapper Methods:} An \gls{ml} model trains on selected features. The model's accuracy determines whether to include or exclude attributes depending on their reliability within the subset. Eliminating backward and selecting forwards are the most common methods \cite{ul2022improving}. 
    \item \textbf{Embedded Methods:} This method combines the benefits of both filters and wrappers (i.e., extracts the best features while maintaining the computational cost). These methods include Least Absolute Shrinkage and Selection Operator (LASSO) regularisation, Random Forest \cite{doreswamy2020feature}. 
\end{itemize}
In addition, this stage calculates parameter weights for unique features based on the training set using a \gls{dl} algorithm. Each \gls{dl} algorithm has several invisible layers embedded within the inputs and outputs, allowing for the extraction of detailed features \cite{he2019intrusion}.

\subsection{Intrusion Detection System in \gls{hiot}}
Intrusion detection systems for \gls{hiot} are described in this section \cite{si2023survey}. An intrusion typically targets a network or device's integrity, availability, or confidentiality by impairing its security. Due to the sensitive nature of healthcare data, \gls{hiot} is a lucrative target for external or internal attackers. Monitoring and analysing device and network activities is automated by \gls{ids} solutions, which are available as hardware or software solutions. \gls{ids} consists of three components: information source, analysis, and response. Information sources feed data to the analysis component, which, upon identifying an attack, triggers a response - either passively, such as notifications, or actively, such as disabling communication.

In \cite{taouali2023intelligent}, the authors propose a \gls{g5}-driven healthcare landscape where the Internet of Medical Things enables remote patient monitoring. Nevertheless, ensuring the security of data remains a challenge. A new lightweight Intrusion Detection System for \gls{iomt}, utilising kernel techniques for feature selection and a kernel extreme learning machine for decision-making, is introduced. This \gls{ids} detected anomalies with 99.9\% accuracy on the dataset WUSTL-EHMS-2020.

In \cite{thamilarasu2020intrusion}, the authors used mobile agents to detect intrusions in a medical environment. Furthermore, a simulation of hospital network topology was conducted to simulate \gls{iomt} experiments, which included the identification of network-level intrusions using \gls{ml} and regression algorithms and the implementation of various digital imaging and communications in medicine (DICOM) protocols by connected devices such as ultrasound scanners and MRI machines, utilising wireless body area networks. These networks are composed of wearable and implantable devices that transmit physiological data continuously, allowing patients to be monitored, diagnosed, and treated continuously. In total, 72 independent simulations were conducted across 216 network types, resulting in an overall best- and worst-case detection accuracy of 99.9\% and 92.91\%, respectively.

In \cite{kumar2021ensemble}, the proposed system integrates ensemble learning and cloud-based architecture to detect cyber attacks. In this ensemble setup, \glspl{dt}, \glspl{nb}, and Random Forest are used as individual learners at the initial level. In the following level, XGBoost uses the classification results to differentiate between normal and attack cases. ToN-\gls{iot} is a large-scale \gls{iot} cybersecurity dataset that provides diverse telemetry and network traffic for model evaluation. Experimental results indicate that the proposed framework can achieve a detection rate of 99.98\%, a precision of 96.98\%, and a reduction in false alarms.

By demonstrating significant accuracy and reduced false alarms in cybersecurity, these examples show that \gls{ml} enhances intrusion detection systems in \gls{hiot}.

\subsection{Event Detection System}
A \gls{hiot} system uses event detection to identify and capture significant occurrences and incidents. An event detection algorithm analyses sensor data, network traffic, and device interactions to detect patterns, anomalies, and predefined triggers that indicate critical events, such as emergencies, device malfunctions, and abnormal patient conditions. Several \gls{ml} techniques investigate classification, including supervised, unsupervised, and reinforcement learning. A supervised learning algorithm employs a labelled dataset for training that is used to determine the relationship between inputs and outputs. Based on the correlation between the input samples, unsupervised learning algorithms can classify the provided data into clusters. The third category of algorithms is reinforcement learning and online learning. These algorithms utilise current knowledge and explore the environment to classify the data \cite{abdellatif2019edge},\cite{alsheikh2014machine}. 

\subsection{Reliability in \gls{hiot} Systems}
Data exchange between e-health systems must be reliable and efficient in \gls{hiot} for enabling cyber resilience, security metrics (\gls{cia} Triad), and controls within the ecosystem \cite{dhirani2021industrial},\cite{dhirani2023ethical},\cite{ukil2016iot}:  

\begin{itemize}
    \item \textbf{Confidentiality:} It is illegal for unauthorised individuals to access medical information. 
   \item \textbf{Integrity:} Data integrity ensures that information remains accurate and unchanged during transit or storage, preventing adversarial modification.

    \item \textbf{Non-repudiation:} Data transmission and reception are inevitable. 
    \item \textbf{Data Freshness:} There is no way to re-generate old data.
    \item \textbf{Resilience to Attacks:} The system should adapt to failed nodes, and there should be no single point of failure. 

    \item \textbf{Data Authentication:} It is necessary to authenticate the addresses and object information retrieved to prevent spoofing.
    \item \textbf{Access Control:} Providing access control to data is a requirement for information providers. 
    \item \textbf{Client Privacy:} A lookup system should only be able to infer its purpose from the information provider.
    \item \textbf{Fault Tolerance and Self-healing:} An \gls{iot} device or battery energy failure should not disrupt health service continuity. 
\end{itemize}

\subsection{Convolutional Neural Networks in \gls{hiot}}
\begin{figure}[ht!]
    \centering
    \includegraphics[width=0.9\textwidth]{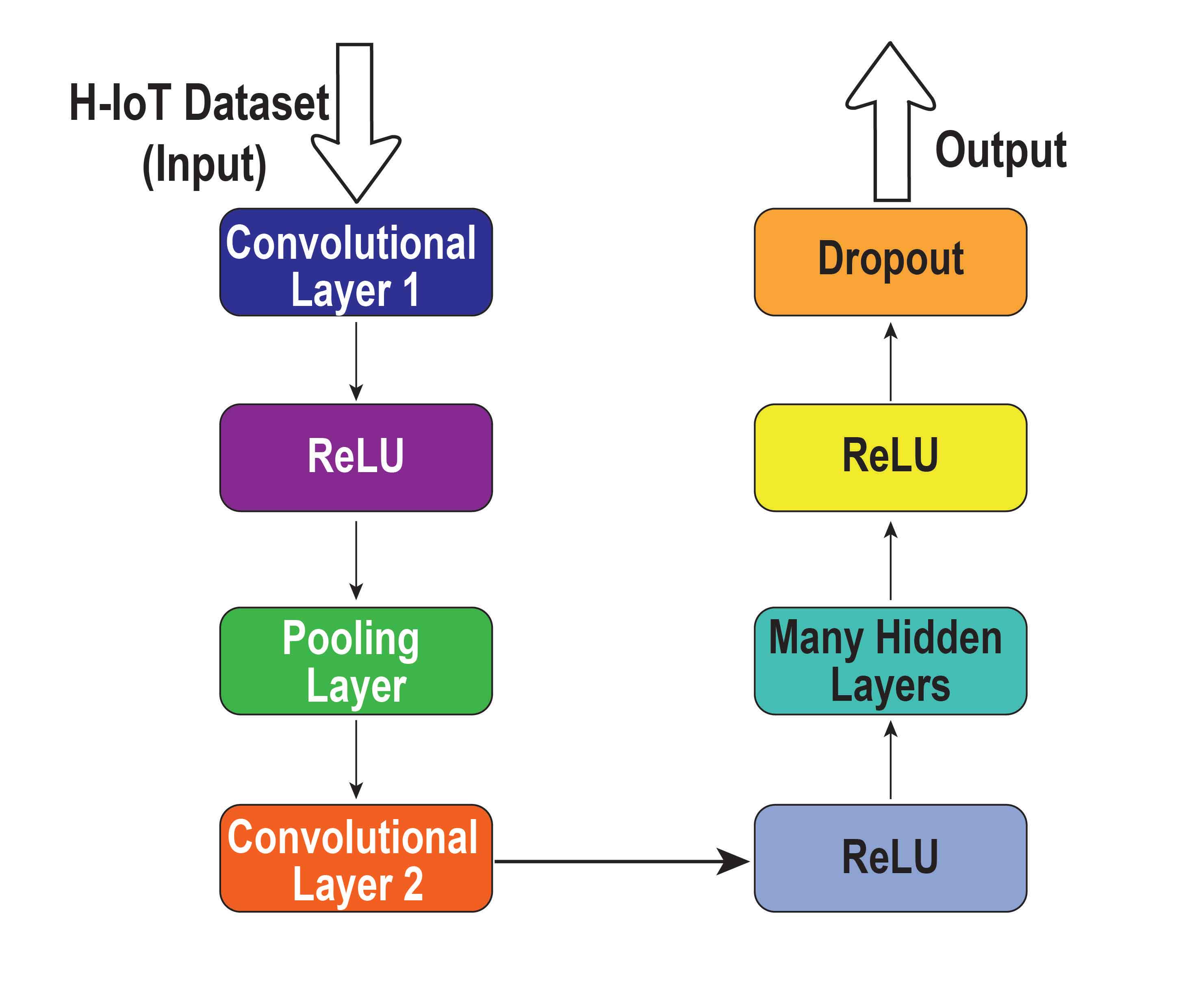}
   \caption[\rev{Neural network illustration}]
{\rev{Conceptual illustration of a CNN-based framework for anomaly detection in \gls{hiot} systems.}}
\label{fig:fig14}
\end{figure}

Using Convolutional Neural Networks has been demonstrated to be significantly more effective than previous methods in identifying threats, surpassing techniques like \glspl{dt}, \glspl{svm}, and \glspl{knn} \cite{deliu2017extracting}. 
In classification problems, \glspl{cnn} are used as feed-forward networks \cite{mehra2022correction}.
Essentially, Convolutional Neural Networks are characterised by two processes. First, convolution involves transforming inputs into outputs using filters or kernels. In particular, \glspl{cnn} have streamlined image recognition to be self-contained, removing the need for external image processing software. Furthermore, it can recognise outcomes efficiently and adapt seamlessly to changing identification criteria. Moreover, their ability to leverage preexisting networks enhances system versatility and applicability \cite{mohammed2022intrusion}. Neurons are arranged in layers in a three-dimensional configuration, spanning width, height, and depth. As shown in Figure \ref{fig:fig14}, which illustrates a \gls{cnn} architecture for analysing \gls{hiot} data.

As part of the architecture of Convolutional Neural Networks, the initial layer, called Layer 1, is composed of a significant number of neurons to effectively process input data \cite{gil2019review}. There are three fundamental parameters: dataset size, quality, and type. Multiple receptive layers can be used to process input layers effectively. This configuration creates overlapping input regions, allowing high-resolution outputs from the original input. For detecting features, \glspl{cnn} employ operations such as pooling and convolution. Using the fully connected layer as a classifier follows extracting features. Typically, the architecture of \gls{cnn}, convolutional, and pooling layers is merged, though it is not mandatory.

\textbf{Rectified Linear Unit (ReLU):} In neural networks, ReLUs follow convolution modules. The primary function of ReLU is to activate specific neurons within the network. The ReLU keeps the values of active neurons while converting inactive neurons to \textit{0} using a simple threshold \cite{datta2020survey},\cite{andrearczyk2018convolutional}. The activation information retained by this segment of the neural network remains significant for future inferences made by other network segments.

An activation function like the ReLU is essential for introducing nonlinear operations into inherently linear neural networks. Adding nonlinearity to the network increases its ability to fit data and accurately capture intricate patterns and relationships.

The ReLU activation function does not alter the dimensions of the input or output data. A mathematical expression for this function can be found in equation \eqref{eq:relu}
\begin{equation} 
    y = \max\{0, x\}
    \label{eq:relu}
\end{equation}

\textbf{Pooling Layer:} Pooling layers contribute significantly to \gls{cnn} performance by strategically inserting them between \gls{cnn} layers. By reducing the number of network parameters and downsampling the data, this integration achieves two objectives: increasing computational efficiency and reducing computation time. Furthermore, the downsampling process, which is more precise than average pooling, preserves texture information by minimising the differences in evaluation values resulting from the convolutional layers \cite{sze2017efficient},\cite{khalil2022designing}. Max pooling is essential because it detects and retains the maximum value within each receptive field. Consequently, it preserves essential features while reducing data dimensions, and in each window, it displays the maximum value \cite{nirthika2022pooling}. 

\textbf{Fully Connected Layer:} The Fully Connected Layer follows the pooling and convolution layers. Based on the features extracted by the previous layers, this layer classifies the input data. The activation functions of the neurons in this layer play a significant role in processing the information from the preceding layer. Fully connected neurons allow the network to recognise complex patterns and relations. According to \cite{anand2021efficient}, \gls{cnn} \gls{dl} classifiers can be used to identify malware instances in healthcare datasets. This research demonstrates the networks' capability to analyse intricate data and provide valuable insight into several domains vital to human health and security.

\textbf{Softmax:} The Softmax activation functions are commonly assigned to the neural networks' final layer to perform multi-class classification. Data can be transformed using the Softmax function into a probability distribution from \textit{0} to \textit{1} with a sum of \textit{1}. As a result of the function, data classification is easier since it emphasises differences between input values. According to Sharma et al., \cite{sharma2023efficient}, a \gls{cnn}-Bidirectional \gls{lstm} architecture is proposed for using \gls{dl} in smart healthcare networks for intrusion detection. The model detects \gls{ddos} attacks using softmax activation functions by classifying traffic flows as benign or malicious. Faruqui et al. \cite{faruqui2023safetymed} propose SafetyMed, a specialised Intrusion Detection System for securing the Internet of Medical Things. \gls{iomt}-specific intrusions are identified with SafetyMed using \gls{cnn} and \gls{lstm}. In the output layer, Softmax activation functions convert raw scores into probabilities ranging from 0 to 1. This enhances the accuracy of the decision-making process. \gls{iomt} vulnerabilities are effectively addressed by SafetyMed, which has an average detection accuracy of 97.63\%.    

This review encompasses a wide range of \gls{ml} techniques and applications, with particular emphasis on \gls{cnn}-based mechanisms for protecting \gls{hiot} devices.

\section{Security and Privacy in \gls{hiot} Environments}\label{sec:security7}
This section emphasises the importance of effective safeguards and rigorous protocols to protect sensitive data and preserve user confidentiality in \gls{hiot} environments. In-depth analysis explores key challenges, effective strategies, and emerging trends. \gls{iot} is being transformed into a network of interconnected devices, where security and privacy are becoming crucial challenges. Therefore, the \gls{iot} network is built around anomaly detection techniques. Anomaly detection algorithms detect abnormal patterns immediately in the \gls{iot}, preventing serious problems. However, healthcare and the \gls{iot} combine to raise the stakes even higher. As a result, \gls{ml} could help address these challenges. Additionally, a \gls{ml}-based architecture for anomaly detection is described in detail in the section.

\subsection{Anomaly Detection}
\begin{figure}[ht!]
    \centering
    \includegraphics[width=0.9\linewidth]{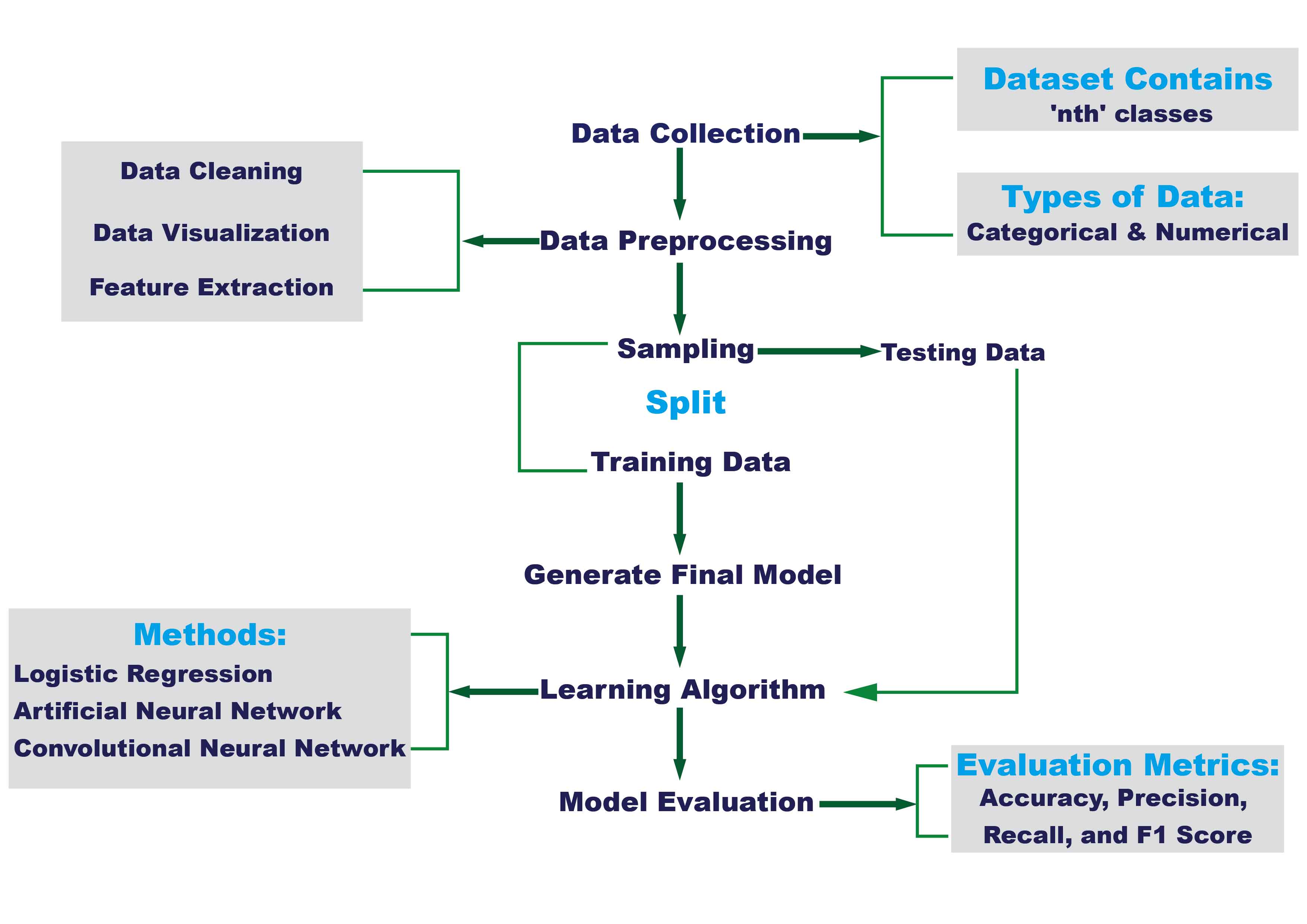}
    \caption[ML-based anomaly detection architecture in H-IoT]{\gls{ml}-based anomaly detection architecture in \gls{hiot}.}
    \label{fig:fig15}
\end{figure}

The term anomaly refers to data segments that do not follow the expected pattern or features of the rest of the data. Anomaly detection is crucial since anomalous activities contain critical information and relevance \cite{hasan2019attack}. Anomaly detection involves several independent processes in \gls{hiot}, as illustrated in Figure \ref{fig:fig15}. Data collection is the first step in this architecture. Following the dataset's creation, it should be analysed thoroughly to determine the data types.  

It is also crucial to preprocess the data before analysing it \cite{khatun2019malicious}. The preprocessing of data includes filtering, visualising, and extracting features. These steps convert the data into feature vectors. It is then necessary to split the feature vectors into training and test sets at a 80:20 ratio. The learning algorithm develops an optimal final model by utilising the training set. It is possible to use a variety of classifiers to experiment with the most accurate accuracy~\cite{patgiri2019empirical}.

With \gls{g5} networks facilitating rapid, high-volume data exchanges, the \gls{iomt} has gained momentum in the healthcare sector. In \cite{wang2022anomaly}, the focus is on detecting anomalies using \gls{dl} to enhance security within \gls{g5} contexts in \gls{iomt}. By utilising multi-model autoencoders for feature extraction, the complexity of traffic feature information has been significantly reduced. A new algorithm for detecting multi-featured sequence anomalies, optimised to cope with the high data volume of \gls{g5} networks, is also presented. It has been demonstrated that \gls{dl} combined with \gls{g5} technology can identify and mitigate anomalies efficiently, thereby increasing \gls{iomt} systems' resilience and reliability.

\subsection{Best practices for enabling data privacy, security and building risk mitigation strategies in smart environments} \label{sec:security7B} 
Digital healthcare environments \cite{chaudhary2022taxonomy} are becoming more prevalent despite numerous security and privacy concerns, and they are susceptible to a unique set of challenges (i.e., securing the massive amount of healthcare data harvested every second from thousands of connected \gls{iot} devices). Healthcare procedures and decisions are made based on this data; in situations where this data is compromised or breached, this may result in a wrong diagnosis/treatment affecting human life. Securing smart environments requires implementing defense-in-depth strategies and multiple layers of security across the ecosystem such as: (i) Standards and data governance risk and control (DGRC) policies for access control, data protection, encryption, remote access (ii) Monitoring and response mechanisms for Security Information and Event Management (SIEM), identity and access management, log analysis, cyber forensics, (iii) Perimeter controls for intrusion detection, prevention and next generation firewalls, (iv) Securing the network using VPN services, web-filtering, securing remote services and site-to-site connections (v) Hardening end-point security by enabling Zero Trust (ZT) mechanisms, patching the systems with latest updates to mitigate vulnerabilities, using an up-to-date antivirus tool to detect and protect the end-point for new vulnerabilities, (vi) the human element is the weakest link and its essential to build cyber awareness within the workplace environment and train them so that they are cyber prepared, (vii) enabling application security using vulnerability scanners and making sure that only inventoried tools are used, (viii) last but not the least, to ensure data security mechanisms are in place (i.e. end-to-end secured data, data classification and have a data loss prevention/backup strategy) \cite{dhirani2021industrial}. However, the impact and frequency of cyber attacks have escalated due to generative \gls{ai}. Since the attacks are constantly evolving and new threats are coming to the surface, there is a possibility that the novel threats would go undetected. This is why there is an essential need for a versatile machine-learning solution to neutralise and mitigate these threats.

\subsection{Mitigating \gls{hiot} Security and Privacy Threats using \gls{ml}}
\begin{figure}[ht!]
    \centering
    \includegraphics[width=0.9\linewidth]{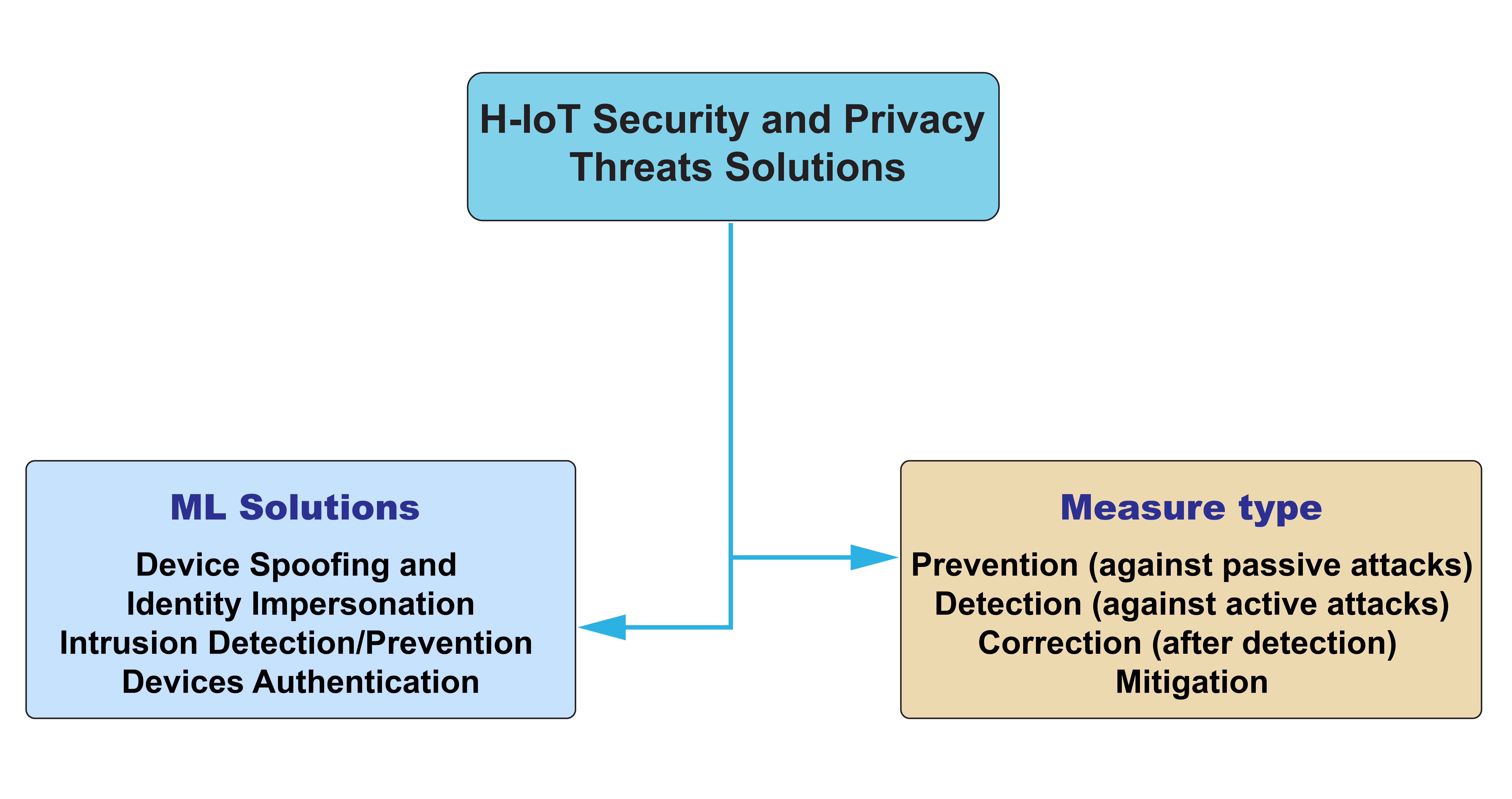}
    \caption[H-IoT security and privacy threat solution]{\gls{hiot} security and privacy threat solution.}
    \label{fig:fig16}
\end{figure}

The security and privacy of sensitive health information have become a worldwide concern. In healthcare security and privacy, \gls{ml}-based solutions are essential among cryptographic, password-based, and digital signature-based solutions. As shown in Figure \ref{fig:fig16}, \gls{ml} models can detect device spoofing and identity impersonation, perform intrusion detection and prevention, and authenticate devices. A \gls{ml} model can prevent anomalies after detection. As a result, \gls{ml} models perform a significant role in \gls{hiot} \cite{rajendran2022prescreening}. 

In healthcare systems, Hau et al. applied \gls{ml} techniques to increase security and privacy \cite{hau2019exploiting}. Begli et al. implemented \gls{svm}-based intrusion detection to monitor remote healthcare \cite{begli2019layered}. Sengan et al. developed an adaptive, \gls{ml}-driven routing system for healthcare data security \cite{sengan2022secured}. 

\section{Impact of \gls{hiot} Technologies in the Future} \label{sec:impact8}
Healthcare and smart cities benefit greatly from the \gls{iot}. This technology will significantly change everyday life and urban environments in the coming years. This section discusses the potential impact of \gls{hiot} on smart cities, including relevant challenges and innovative solutions. \gls{qos} is integral to this discussion, which entails ensuring the reliability of services. Furthermore, \gls{hiot} networks can be made more flexible, scalable, resource-efficient, and secure with the help of \glspl{sdn}.  

\subsection{Future of \gls{hiot} Technologies}
Generally, people expect \gls{hiot} systems to integrate with their living environments. The future of infrastructure lies in smart cities. As a result, affiliating the \gls{hiot} system within modern city architecture facilitates widespread deployment of the \gls{hiot} \cite{8993839}. Software-defined networking plays a significant role in smart cities and \gls{hiot} systems. Using \gls{sdn} in \gls{hiot}, cities can gain more efficient management, more secure provision of services, effective resource allocation, and overall better performance.

\subsection{Challenges}
Despite the growing popularity of \gls{iomt} applications in healthcare, several challenges restrict their implementation \cite{sharma2022software}. Key challenges associated with \gls{hiot} development are examined.

\textbf{Heterogeneous Networks:} Developing \gls{hiot} requires devices to interact and interoperate across diverse networks and operating systems. Among the technologies used in communication are Bluetooth, Infrared Data Association (IrDA), UWB, Zigbee, WLANs, Wi-Fi 6, and \gls{g5}. Batteries drain out faster when communicating over long distances \cite{ajmani20235g,broell2023iot}. It is essential that these devices, regardless of their heterogeneity in implementation and communication, can identify and communicate with each other.

\textbf{\gls{qos}:} \gls{hiot} is characterised by low latency and low power consumption. There is a risk to a patient's health when dealing with issues related to the Internet of Things. Consequently, \gls{hiot} signals should be transmitted and processed with the least possible delay. High-bandwidth networking resources can achieve this \cite{gowda2022industrial}. A \gls{g5} network, with its ability to ensure \gls{qos} through rapid data transmission, minimal latency, and extensive connectivity, presents itself as an ideal solution for \gls{hiot} applications \cite{sutradhar2023dynamic}. \gls{qos} refers to the ability of a system to maintain a basic level of performance and reliability, ensuring that data is transmitted and received accurately and promptly.

\textbf{Privacy and Security:} In \gls{hiot}, the privacy and security of patients and their data are paramount, as breaches may disrupt real-time patient monitoring, manipulate actuator-driven medical devices, delay emergency responses, or corrupt physiological data streams used for clinical decision-making. In many cases, \gls{hiot} uses heterogeneous networks to communicate and store data, exposing it to attack over long distances \cite{adil2023covid}. Thus, the security and privacy of data require anomaly detection, mitigation, robust encryption, and authentication techniques, as mentioned in Section \ref{sec:security7B}. However, in designing and implementing the defence-in-depth techniques, it is necessary to consider the resource constraints of the devices involved in \gls{hiot}, and the developed solutions should be lightweight and energy-efficient. 

\begin{figure}[ht!]
    \centering
    \includegraphics[width=0.9\textwidth]{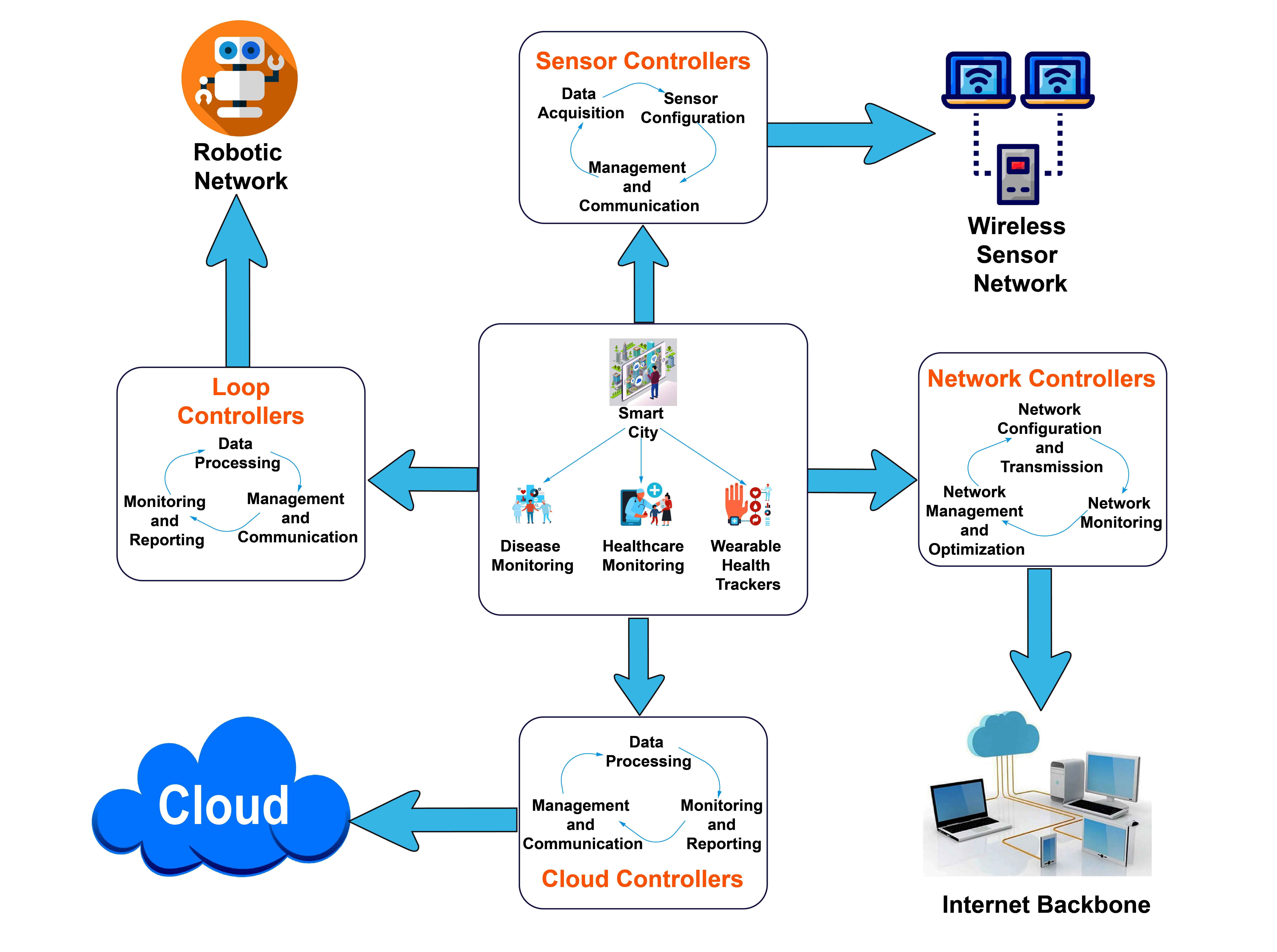}
    \caption[Software defined network-based H-IoT]{Software defined network-based \gls{hiot}.}
    \label{fig:fig17}
\end{figure}

\subsection{Software-Defined Network in \gls{hiot}}
An emerging paradigm in computer networking is software-defined networking. Data planes, control planes, and application planes are the three tiers of \gls{sdn}. Data and control planes are separated to improve network performance. In contrast, traditional architectures combine data and control planes onto one device, while \gls{sdn} separates them to simplify network management \cite{krishnamoorthy2023role,sallabi2018managing}. \glspl{sdn} forward packets according to predefined policies \cite{abawajy2017federated}. 

\rev{Although this chapter focuses on anomaly detection, \gls{sdn}-based architectures are included to indicate their potential role in enabling automated mitigation in H-IoT environments. In this context, such architectures consist of sensor controllers, cloud controllers, network controllers, and closed-loop controllers, which provide four types of services: data acquisition, network configuration and transmission, processing, and actionable feedback, as illustrated in Figure \ref{fig:fig17}.}

As an example, Google uses an \gls{sdn} to manage its wide-area network traffic \cite{taheri2023deep}. 
The network infrastructure increasingly relies on \glspl{sdn}, and protecting a network from attack is crucial to ensure its integrity and availability \cite{elubeyd2023hybrid}. In \cite{9776486}, a novel software-defined network architecture is presented, incorporating a new controller and load-balancing algorithms based on \gls{ml} for the \gls{hiot}. A support vector machine algorithm, which achieved 95.1\% accuracy through extensive simulations, validates the architecture's effectiveness. In preparation for broader \gls{g5} integration, this approach represents an important advance in \gls{hiot}. Adding \gls{dl} to \glspl{sdn} makes \gls{hiot} services, such as telemedicine, more reliable and secure \cite{lakshminarayanan2023health}. A software-defined network system illustrates the communication process between medical services and nano-networks inside individuals' bodies  \cite{ahmed2023development}. A certain set of software rules controls this method, which makes healthcare more efficient.

\section{\rev{Review and Limitations of Cyber Threat Detection for \gls{hiot}}}
\rev{This research reviews studies that employ dataset-driven intrusion detection approaches in \gls{hiot} environments, with an emphasis on \gls{ddos} attacks, given their impact and prevalence. Datasets such as CICIoT2023 \cite{CICIoT2023} and healthcare-specific datasets such as CICIoMT2024 \cite{dadkhah2024ciciomt2024} are commonly used in existing studies, with a primary focus on \gls{ddos} attack scenarios. These studies integrate deep learning-based detection mechanisms to improve classification performance \cite{ullah2025hybrid, yang2025ddos}. However, existing research predominantly concentrates on \gls{ddos} attacks \cite{ullah2025hybrid, yang2025ddos}, with limited consideration of other threats such as \gls{mitm} and \gls{sf} attacks. In addition, existing studies rely on benchmark datasets such as CIC-IDS2017 \cite{cicids2017} and BoT-\gls{iot} \cite{botiot2023}, which do not adequately represent healthcare-specific attack behaviours, such as \gls{mitm} and \gls{sf}, thereby limiting their applicability to \gls{hiot} environments \cite{abdallah2025advancing}. This limitation indicates the need for datasets and detection methods capable of classifying multiple attack types in a multiclass setting.}

\section{Conclusion} \label{sec:conclude9}
This chapter systematically dives into digital healthcare's fundamental elements and practical applications. It discusses the essential technologies and methodologies for optimising these systems within Healthcare 5.0. The research focuses on various cyber threats, vulnerabilities, and cyber attacks carried out at the different \gls{iot} layers and explores the security and privacy of \gls{hiot} through deep neural networks and \gls{ai}. A demonstration of a cloud-based solution and big data analytics using \gls{iot} and software-defined networks is provided and illustrates the advantages of a precise, efficient, and secure healthcare ecosystem. This thesis also presents insights into the security challenges related to wireless and communication technologies (i.e., \gls{g5}) in \gls{hiot} and how \gls{ml} algorithms can enable the digital healthcare environment to bridge those security gaps.


\chapter{Datasets for Distributed Denial-of-Service Detection in Healthcare Internet of Things Environments} \label{chap:dataset} 

\ifpdf
    \graphicspath{{3_chapter3/figures/PNG/}{3_chapter3/figures/PDF/}{3_chapter3/figures/}}
\else
    \graphicspath{{3_chapter3/figures/EPS/}{3_chapter3/figures/}}
\fi


\vspace{5pt}

\noindent The datasets presented in this chapter have been previously published in:

\textbf{Akhi, M.}, Eising, C. and Dhirani, L.L., 2025. Datasets for distributed denial-of-service detection in healthcare internet of things environments.~\textit{Data in Brief}, p.112222.

The content of the paper above has been presented as published in this chapter, except for correcting grammatical errors and making slight rewording so that the narrative is more appropriate for a thesis. A brief preamble is provided to explain its relevance to the thesis narrative.

\rev{\noindent\textbf{Thesis Structure and Publication Context:}
Parts of this thesis, including Chapters~\ref{chap:dataset}, \ref{chap:tcn}, and \ref{chap:edge}, are based on related published research that share a common scope and methodological foundation. Therefore, some overlap in background, simulation setup, and methodological descriptions is included to ensure completeness.}

\clearpage

\noindent\textbf{\underline{Preamble}}\\[0.5em]
In this chapter, \textbf{RQ2} is addressed:

 \textbf{RQ2:} How can realistic synthetic datasets be generated to represent \gls{ddos} attack behaviours in \gls{hiot} environments? 

The preceding chapter reviews security and privacy challenges in \gls{hiot} systems and outlines a wide range of cyber threats, including \gls{ddos} attacks. It also introduces simulation tools such as Cooja for emulating \gls{hiot} devices and generating data. Building on this foundation, this chapter implements \gls{ddos} attack behaviours in \gls{hiot} environments using two simulators, Cooja and ns-3. It captures network traffic data and health metrics generated by emulated wearable \gls{hiot} devices and focuses on dataset generation for subsequent analysis.

The main objective of this dataset chapter is to generate realistic synthetic datasets that simulate \gls{ddos} behaviours in \gls{hiot} environments. The datasets capture communication patterns, message frequencies, and transmission intervals of normal and malicious nodes from three \gls{wban} sensor types, including body temperature, oxygen saturation, and heart rate, simulated in two separate tools, Cooja and ns-3.
In the Cooja simulator, 14 wearable \gls{hiot} sensor nodes emulate the three sensor types, along with one \gls{mqtt} broker acting as the server and five \gls{ddos}-attack nodes generating malicious traffic. The resulting dataset,~\emph{UL-ECE-MQTT-DDoS-H-IoT2025}, includes four extracted communication features suitable for training \gls{ml}/\gls{dl}-based detection models.
In the ns-3 simulation, a \gls{g5}-enabled \gls{hiot} topology is modelled over the \gls{udp} protocol, consisting of one server node, multiple eNodeBs, 14 normal client nodes, and 5 \gls{ddos}-attack nodes. To emulate adaptive attack dynamics, three additional malicious nodes are introduced at 10-second intervals, beginning 20 s into a 120-second simulation, to increase the attack frequency and intensity over time.

In line with \textbf{RQ2}, this study focuses on developing detailed traffic datasets that represent both normal and malicious behaviours in \gls{hiot} environments. In this context, this chapter constructs datasets that distinguish between normal and malicious nodes’ data transmission behaviour, addressing the overarching research question. During preprocessing, the \gls{udp}-based \gls{ddos} dataset is analysed at 5-second monitoring intervals to enable fine-grained observation of both normal and \gls{ddos}-affected activity. This configuration also facilitates the detection of newly introduced malicious nodes with comparatively lower message volumes. The resulting dataset,~\emph{UL-ECE-UDP-DDoS-H-IoT2025}, serves as a foundational resource for developing and evaluating \gls{ml}/\gls{dl}-based detection and mitigation models in \gls{hiot} environments.


\section{Abstract}
The growing number of \gls{iot} devices in healthcare settings raises critical concerns about security, particularly in defending against \gls{ddos} attacks. These attacks can cause operational downtime in \gls{iot} environments. To mitigate \gls{ddos}-based attacks, advanced defence-in-depth strategies and well-labelled datasets are required for \gls{hiot}, \gls{iot}, and other distributed computing contexts. This chapter presents two labelled datasets,~\emph{UL-ECE-MQTT-DDoS-H-IoT2025} and \emph{UL-ECE-UDP-DDoS-H-IoT2025}, generated by simulating realistic traffic patterns under both normal and \gls{ddos} conditions using the Cooja and ns-3 simulators. In Cooja, the raw dataset records healthcare-specific \gls{mqtt} traffic (e.g., simulated oxygen level of 100\%) randomly generated by emulated \gls{hiot} sensors. It also includes message counts and network metadata that enable detailed analysis across both application and network layers. In ns-3, the raw data comprises \gls{g5}-enabled \gls{hiot} network traces from all nodes, capturing timestamps, payload size, and the header details of the \gls{udp}. Existing benchmark datasets mainly consist of generic network traffic attributes, including packet IDs, protocol types, and timestamps. In contrast, the proposed datasets address this gap by incorporating \gls{hiot}-specific communication parameters that closely resemble real-world conditions, such as node-level message counts and monitoring frequencies. This inclusion provides a realistic representation of communication patterns for security and performance research in \gls{hiot}. The datasets enable detailed analysis of key features for detecting \gls{ddos} threats, including \gls{udp} flood variants extending beyond the \gls{hiot} domain. This characteristic makes them directly usable for developing, testing, and comparing \gls{ml} and \gls{dl} models across diverse \gls{iot} security contexts. The \gls{mqtt}-based dataset is derived from a 5-hour simulation run using the Cooja simulator, which emulates wearable sensors such as body temperature, heart rate, and oxygen saturation. In this setup, normal \gls{hiot} nodes transmit data to the server at 60-second intervals, while \gls{ddos}-affected nodes publish data at 20-second intervals to simulate a higher transmission frequency. The \gls{udp}-based dataset is derived from a 120-second simulation conducted using the ns-3 simulator, which simulates a \gls{g5}-enabled \gls{hiot} environment. In this scenario, normal and malicious nodes transmit data at 124 kbps and 248 kbps, respectively. Both datasets are processed from raw simulation logs converted into structured CSV files using Python scripts. The CSV files contain features such as timestamp, payload size, message frequency, and node-level communication statistics. The \emph{UL-ECE-MQTT-DDoS-H-IoT2025} and \emph{UL-ECE-UDP-DDoS-H-IoT2025} datasets contain approximately 20,080 and 99,887 records, respectively. The primary objective of creating these datasets is to enhance security in \gls{hiot} ecosystems by enabling robust detection of advanced cyber threats. In line with this objective, the datasets support the development of \gls{ml}/\gls{dl}-based cybersecurity mechanisms. In addition, this resource forms a foundation for future research, motivating the creation of new datasets for emerging attack scenarios.

This research summarises the key characteristics and data acquisition details of the \gls{ddos} \gls{hiot} datasets, as outlined in Table~\ref{tab:specifications}.

\begin{table}[ht!]
\centering
\small
\caption{Specifications Table.}
\label{tab:specifications}
\setlength{\tabcolsep}{3pt}
\begin{tabular}{p{4.8cm}p{9cm}}
\toprule
\textbf{Subject} & Computer Sciences \\
\textbf{Specific subject area} & \acrlong{hiot} focuses on cybersecurity, \gls{ddos} detection using \gls{ml} methods \\

\textbf{Type of data} & Structured tabular data (Processed and Analyzed) \\

\textbf{Data collection} & Data is collected using two simulators, Cooja for \gls{mqtt} and ns-3 for \gls{udp}. The \gls{mqtt}-based dataset includes 14 normal and 5 \gls{ddos}-affected \gls{hiot} nodes emulating wearable sensors (e.g., heart rate, body temperature) over Contiki OS 3.0. The \gls{udp}-based dataset simulates a \gls{g5} \gls{hiot} environment with 14 clients, 5 \gls{ddos}-affected nodes, and 3 additional \gls{ddos} attackers introduced at 20 s with 10 s intervals. Simulations run in Oracle VirtualBox using Windows 11 and Ubuntu 24.10 on a system with an Intel Core i5-12500H CPU, 16 GB RAM, and a NVIDIA GeForce RTX GPU. \\

\textbf{Data source~location} & University of Limerick, Limerick, Ireland \newline
Latitude: 52.6736° N, Longitude: 8.5721° W \\

\textbf{Data accessibility} & \textbf{Repository name:} Zenodo \newline 
\textbf{Data identification number (DOI):} 10.5281/zenodo.15305814 \newline 
\textbf{Direct URL to data:} 
{\small \url{https://zenodo.org/record/15305814}} \\

\textbf{Related research article} & M. Akhi, C. Eising, L.L. Dhirani, \gls{tcn}-based DDoS detection and mitigation in \gls{g5} healthcare-\gls{iot}: A frequency monitoring and dynamic threshold approach, IEEE Access. 13 (2025) 12709--12733. 
{\small \url{https://doi.org/10.1109/ACCESS.2025.3531659}} \\
\bottomrule
\end{tabular}
\end{table}

\FloatBarrier

\section{Value Of The Data}
\begin{itemize}
    \item Privacy regulations and ethical concerns limit the ability to generate malicious data from physical \gls{hiot} devices (e.g., body temperature and oxygen saturation) \cite{dhirani2023ethical}. As an alternative, synthetic data generated through simulation provides a practical way to continue cybersecurity research in \gls{hiot} environments. It enables the creation of realistic \gls{ddos} attack scenarios where parameters like payload size, data traffic interval, and protocol type are systematically adjusted, capturing both regular and malicious traffic without compromising data security or violating privacy regulations (e.g., \gls{gdpr}).
    \item Synthetic data provides a cost-effective solution that reduces reliance on both expensive physical \gls{hiot} device-based records and the need for access to sensitive healthcare information. However, its simulated nature might not accurately capture the unpredictability of real-world settings, including device-to-device interactions, malfunctions, traffic across multiple protocols, and dynamic sensor behaviours (e.g., fluctuating readings). This limitation should be considered when interpreting results for practical deployment. Nonetheless, the datasets remain a scalable, openly shareable, and manageable resource applicable across cyberattack scenarios in \gls{hiot} contexts. 
    \item Researchers, cybersecurity analysts, and data scientists working on anomaly detection and \gls{ddos} mitigation in \gls{hiot} networks may utilise these datasets to develop, test, and validate \gls{ml} and \gls{dl} models across both \gls{mqtt} and \gls{udp} protocols in broader \gls{iot} ecosystems. \rev{This enables applicability beyond healthcare-specific scenarios, as the datasets capture protocol-driven communication and \gls{ddos} attack patterns that are not limited to healthcare sensor data.} \rev{Accordingly, the novelty is in protocol-level implementation of \gls{mqtt} and \gls{udp}, enabling the generation of normal traffic and \gls{ddos} attack patterns within the network.} Therefore, educators and students may leverage these datasets for academic projects (e.g., Master’s theses) to analyse network security and \gls{iot} attacks using \gls{ai}/\gls{ml} techniques.

    \item The \emph{UL-ECE-MQTT-DDoS-H-IoT2025} and \emph{UL-ECE-UDP-DDoS-H-IoT2025} datasets enable assessment of detection accuracy by comparing the proposed models with existing anomaly detection techniques. The datasets also provide a basis for evaluating model generalisation by comparing detection performance against other publicly available \gls{ddos} datasets.
    \item Analysis of variations in traffic parameters (e.g., packet interval, payload size, attack duration) and their impact on key features, such as total messages, node-specific message counts, and transmission frequency in \gls{hiot} networks, can provide deeper insights into network behaviour. The datasets also provide ample and valuable data for evaluating intrusion detection systems in network security, contributing to ongoing efforts to address emerging cybersecurity threats. Prior research has demonstrated the importance of such features in improving \gls{ml}/\gls{dl}-based detection performance~\cite{khatun2019malicious}.

    \item In the broader \gls{iot} environment, these open-access \gls{ddos} attack datasets serve as a valuable resource, filling a critical gap by enabling protocol-specific analysis with controlled configurations (e.g., traffic interval). 
\end{itemize}

\section{Background}
Creating realistic attack scenarios on \gls{hiot} devices is challenging in real-world settings due to privacy, ethical, and safety concerns, as well as limited access to physical infrastructure for testing vulnerabilities. However, continued research in this domain requires analysing how \gls{iot} nodes interact under normal and attack conditions. In such cases, simulated or synthetic health-related data are essential for analysing device-to-device interactions and protocol-specific vulnerabilities. Moreover, synthetic data generation facilitates the development of \gls{ml}/\gls{dl} models in domains where access to real-world data is limited or restricted. To address this gap, this research constructs two datasets over the \gls{mqtt} and \gls{udp} protocols using the Cooja and ns-3 simulators in \gls{hiot} environments. The \emph{UL-ECE-MQTT-DDoS-H-IoT2025} and \emph{UL-ECE-UDP-DDoS-H-IoT2025} datasets support the detection and mitigation approach presented in Chapter~\ref{chap:tcn}, which employs a Temporal Convolutional Network (TCN) model. In comparison, this chapter documents the simulation setup and dataset construction process. It describes the experimental configuration and provides detailed information on the generated raw and processed datasets. It also includes the Python scripts used for data preprocessing to support reproducibility and transparent evaluation within this thesis.

\section{Data Description}
This section describes the \emph{UL-ECE-MQTT-DDoS-H-IoT2025} and \emph{UL-ECE-UDP-DDoS-H-IoT2025} datasets of the UL-ECE-DDoS-\gls{hiot}-Datasets2025. It details the folder structure, data files, and corresponding preprocessing scripts located under the \gls{mqtt} and \gls{udp} directories. Each dataset consists of three components: raw simulation logs, processed CSV files containing labelled traffic records, and Python scripts for preprocessing these records. It describes the key features of the \emph{UL-ECE-MQTT-DDoS-H-IoT2025} dataset and presents a sample table to illustrate its processed data format. It highlights the 13 key features (e.g., ids, total messages, frequency, and mean frequency) of the \emph{UL-ECE-UDP-DDoS-H-IoT2025} dataset and presents a sample table of these parameters. It also explains the 4 features excluded from the \gls{tcn} model’s training phase. Finally, it compares with existing datasets (e.g., CIC-IDS2017). 

\subsection{Dataset Details for \gls{mqtt}} 
This section details the \gls{mqtt} folder of the UL-ECE-DDoS-H-IoT-Datasets2025, including the processed CSV file (\emph{UL-ECE-MQTT-DDoS-H-IoT2025}), raw communication logs, and the Python script used for preprocessing. 
The file \emph{UL-ECE-MQTT-DDoS-H-IoT2025}.csv contains labelled \gls{mqtt} traffic data generated using the Cooja simulator in a \gls{hiot} context. This CSV file includes both normal and \gls{ddos}-affected traffic records, where each row represents data from a single \gls{hiot} node and contains 4 feature columns (e.g., ids, total messages, total messages each node, frequency). The corresponding raw traffic is stored in \emph{UL-ECE-MQTT-DDoS-H-IoT2025}-raw.txt, located in the same \gls{mqtt} folder, which contains the original \gls{mqtt} communication logs captured during the simulation. 
Additionally, the \gls{mqtt} folder contains the file (\path{data_preprocessing_ddos_mqtt_cooja.py}), which is the Python script used to preprocess the raw \gls{mqtt} traffic logs. It generates a labelled CSV file containing extracted features for \gls{ml} operations. All files described above are archived on Zenodo~\cite{akhi2025ulece}. 

\subsection{Data Preprocessing and Feature Extraction for \gls{mqtt}}
This research generates raw \gls{mqtt} communication logs in the Cooja simulator and preprocesses the obtained raw data using the Python script named \path{data_preprocessing_ddos_mqtt_cooja.py} to produce the labelled dataset. This preprocessing pipeline parses the simulation output, cleans the data, and extracts key features. The resulting dataset comprises 4 features: 

\textbf{Node IDs:} Numeric identifiers assigned by the Cooja simulator to each simulated sensor node, used to distinguish individual \gls{hiot} devices or sensors in the network. 

\textbf{Total Messages:} This feature tracks the cumulative number of messages recorded at each observation during data transmission in the \gls{hiot} network. The total message count $T$ is computed using the following formula:

\begin{equation}
T = \sum_{i=1}^{N} M_i
\tag{i}
\end{equation}
where $M_i$ represents a message indicator at the $i^{\text{th}}$ observation. Further details of this metric and its derivation are provided in Chapter~\ref{chap:tcn}.

\textbf{Total Messages Each Node:} This feature represents the total number of messages transmitted by each individual node in the \gls{hiot} network. It reflects the communication activity of a node by summing up all messages it transmits throughout the data transmission period. The total messages sent by node $i$ are calculated using the following expression:

\begin{equation}
total\_messages\_each\_node_i = \sum_{j=1}^{N} M_{node_i,j}
\tag{ii}
\end{equation}

where $M_{node_i,j}$ denotes whether node $i$ transmitted a message at the $j^{\text{th}}$ logged transmission.

A detailed explanation of the feature derivation and thresholding logic is provided in Chapter~\ref{chap:tcn}.

\textbf{Message Frequency Proportion (frequency):} This feature represents the proportion of messages sent by each node as a percentage of the total number of messages in the network. It reflects the relative activity level of each node and is computed using the following equation:

\begin{equation}
frequency~(\%) = \left( \frac{M_{node_i}}{M_{total}} \right) \times 100
\tag{iii}
\end{equation}

where $M_{node_i}$ is the total number of messages transmitted by node $i$ and $M_{total}$ is the cumulative number of messages across all nodes. This feature plays a crucial role in identifying \gls{ddos}-affected nodes, as an abnormally high frequency value may indicate anomalous or malicious behaviour. A threshold-based mechanism is applied to flag nodes that exceed a frequency of 5\%, as illustrated in the sample dataset provided in Table~\ref{tab:table2}. A detailed discussion of the thresholding method and feature behaviour is provided in Chapter~\ref{chap:tcn}.

\begin{table}[htbp]
\centering
\caption[Example entries from the \emph{UL-ECE-MQTT-DDoS-H-IoT2025}]{Example entries from the \emph{UL-ECE-MQTT-DDoS-H-IoT2025} dataset highlighting key traffic features.}
\label{tab:table2}
\begin{tabular}{llll}
\toprule
\textbf{ids} & \textbf{total\_message} & \textbf{total\_message\_each\_node} & \textbf{frequency} \\
\midrule 
11 & 2476 & 90 & 3.634894991922456 \\

18 & 2477 & 252 & 10.173597093257973 \\

8 & 2478 & 89 & 3.591606133979015 \\

18 & 2479 & 253 & 10.205728116175877 \\

16 & 2480 & 247 & 9.959677419354838 \\

20 & 2481 & 269 & 10.842402257154374 \\

14 & 2482 & 93 & 3.746978243352135 \\

15 & 2483 & 68 & 2.738622633910592 \\

10 & 2484 & 144 & 5.797101449275362 \\

8 & 2485 & 90 & 3.621730382293762 \\
\bottomrule 
\end{tabular}

\medskip
\begin{tablenotes}
    \small
    \item \textbf{Note:} The \textbf{ids} mentioned in the first column of the table correspond to the \textbf{node} column in the dataset, which represents unique identifiers for each node in the network.
\end{tablenotes}
\end{table}

\subsection{Dataset Details for \gls*{udp}}
This section outlines the contents of the \gls{udp} directory within the UL-ECE-DDoS-\gls{hiot}-Datasets2025~\cite{akhi2025ulece}, which includes the processed CSV file and the Python script used during data preprocessing.  
The file \emph{UL-ECE-UDP-DDoS-H-IoT2025}.csv contains \gls{udp}-based network traffic data labelled as either normal or \gls{ddos} attack-affected in a \gls{hiot} context. The raw simulation log file corresponding to this dataset is generated using the ns-3 simulator. The directory also includes the Python script (data\_preprocessing\_ddos\_udp\_ns\_3.py), which transforms the raw simulation output into the structured CSV format.  
The README.md file in the root directory provides an overview of the \gls{mqtt} and \gls{udp} datasets, detailing their structure, including files, and usage instructions. 

\begin{table}[!htbp]
\centering
\caption[Representative sample entries]{Representative sample entries from the \emph{UL-ECE-UDP-DDoS-H-IoT2025} dataset.}
\label{tab:table3}
\renewcommand{\arraystretch}{1.4}
\resizebox{\textwidth}{!}{%
\begin{tabular}{lllllllllllll}
\toprule
\rotatebox{90}{\textbf{ids}} & 
\rotatebox{90}{\textbf{time}} & 
\rotatebox{90}{\textbf{protocol}} & 
\rotatebox{90}{\textbf{source\_ip}} & 
\rotatebox{90}{\textbf{destination\_ip}} & 
\rotatebox{90}{\textbf{payload\_size}} & 
\rotatebox{90}{\textbf{total\_messages}} & 
\rotatebox{90}{\textbf{total\_messages\_same\_node}} & 
\rotatebox{90}{\textbf{frequency}} & 
\rotatebox{90}{\textbf{mean\_frequency}} & 
\rotatebox{90}{\textbf{monitoring\_total\_messages}} & 
\rotatebox{90}{\textbf{monitoring\_total\_messages\_same\_node}} & 
\rotatebox{90}{\textbf{monitoring\_frequency}} \\
\midrule
15 & 108.6924 & 0 & 3 & 0 & 512 & 89981 & 3291 & 3.6574 & 5.4325 & 4400 & 152 & 3.4545 \\

16 & 108.6924 & 0 & 4 & 0 & 512 & 89982 & 3291 & 3.6573 & 5.4325 & 4400 & 152 & 3.4545 \\

17 & 108.6924 & 0 & 5 & 0 & 512 & 89983 & 3291 & 3.6573 & 5.4325 & 4400 & 152 & 3.4545 \\

18 & 108.6924 & 0 & 6 & 0 & 512 & 89984 & 3291 & 3.6573 & 5.4325 & 4400 & 152 & 3.4545 \\

19 & 108.6924 & 0 & 7 & 0 & 512 & 89985 & 6582 & 7.3145 & 5.4325 & 4400 & 303 & 6.8863 \\

20 & 108.6924 & 0 & 8 & 0 & 512 & 89986 & 6582 & 7.3144 & 5.4325 & 4400 & 303 & 6.8863 \\

21 & 108.6924 & 0 & 9 & 0 & 512 & 89987 & 6582 & 7.3143 & 5.4325 & 4400 & 303 & 6.8863 \\

22 & 108.6924 & 0 & 10 & 0 & 512 & 89988 & 6582 & 7.3143 & 5.4326 & 4400 & 303 & 6.8863 \\

23 & 108.6924 & 0 & 11 & 0 & 512 & 89989 & 6582 & 7.3142 & 5.4326 & 4400 & 303 & 6.8863 \\

25 & 108.6994 & 0 & 13 & 0 & 512 & 89990 & 4766 & 5.2961 & 5.4326 & 4400 & 303 & 6.8863 \\
\bottomrule
\end{tabular}%
}
\medskip
\begin{tablenotes}
    \small
    \item \textbf{Note:} The \textbf{ids} column corresponds to the \textbf{node\_id} field in the \textbf{\emph{UL-ECE-UDP-DDoS-H-IoT2025}.csv} dataset.
\end{tablenotes}
\end{table}

\subsection{Data Preprocessing and Feature Extraction for \gls{udp}} 
Raw \gls{udp} communication logs from the ns-3 simulator are processed to build the labelled dataset. The preprocessing pipeline (data\_preprocessing\_ddos\_udp\_ns\_3.py) applies a rule-based parser and a 5-second sliding-window aggregation method. It performs data cleaning and extracts features specific to \gls{udp} traffic. Within this pipeline, timestamps are extracted to calculate elapsed times, and per-node message counts are continuously tracked. Categorical fields (protocol, source \gls{ip}, destination \gls{ip}) are encoded as numeric codes, and the corresponding original string values (e.g., source\_ip\_des = 7.0.0.13) are also retained in separate columns for reference. The pipeline computes both running message frequencies and the running mean and aggregates monitoring totals from the last 5 s to determine the monitoring frequency. If the 5-second monitoring frequency is greater than or equal to the running mean frequency, the labelling rule assigns a binary outcome label, malicious = 1 and normal = 0. The resulting CSV file \emph{UL-ECE-UDP-DDoS-H-IoT2025}.csv contains key time-based parameters (e.g., timestamp, time elapsed) alongside node-specific parameters such as node ID, protocol type, \gls{ip} addresses, payload size, and message frequency. 
The \emph{UL-ECE-UDP-DDoS-H-IoT2025} dataset contains 17 features. A subset of 13 features, listed in Table~\ref{tab:table3}, is selected and used in Chapter~\ref{chap:tcn} to achieve optimal performance with the \gls{tcn} model, where their definitions and associated equations are discussed in detail. In this chapter, all 17 features are defined as part of the \emph{UL-ECE-UDP-DDoS-H-IoT2025} dataset and are summarised in Table~\ref{tab:table4}.

\begin{table}[!htbp]
\centering
\small
\caption[The 13 selected features]{The 13 selected features from the \emph{UL-ECE-UDP-DDoS-H-IoT2025} dataset are presented without underscores, alongside definitions relevant to \gls{ddos} detection.}
\label{tab:table4}
\begingroup
\setlength{\tabcolsep}{3pt}
\begin{tabular}{p{5.5cm}p{8.5cm}}
\toprule
\textbf{Feature Name} & \textbf{Definition} \\
\midrule
Node IDs & Unique identifiers are assigned to each node in the network topology. \\

Time Elapsed & The time elapsed indicates the difference between the current timestamp and the initial timestamp. \\

Protocol & Numerical code representing the communication protocol (0 for \gls{udp} in this dataset). \\

Source \gls{ip} Address & Unique \gls{ip} address of each node that transmits packets to the destination server. \\

Destination \gls{ip} Address & Node IDs transmit data to the same destination node (e.g., server at 7.0.0.2, labelled as 0 for all \gls{hiot} nodes in the dataset). \\

Payload Size & Size of each packet in bytes (e.g., 512). \\

Total Messages & Cumulative count of messages sent by the node. \\

Total Messages from Same Node & Total number of messages transmitted by a specific source node. \\

Frequency & Percentage of messages from the current node ID compared to the total messages processed until the current log line. \\

Mean Frequency & Average percentage of messages from all nodes up to the current log line. \\

Monitoring Total Messages & It is calculated by summing the messages monitored from each \gls{hiot} node in the network. \\

Monitoring Total Messages from Same Node & Rate of messages from a single \gls{hiot} node, calculated using total messages sent, simulation time, and monitoring interval. \\

Monitoring Frequency & The proportion of messages from the same \gls{hiot} node within the last 5 s compared with all messages in that period. This message-based proportion helps detect newly appearing nodes, which may initially have a lower message frequency than established ones. \\
\bottomrule
\end{tabular}
\endgroup
\end{table}

Moreover, while Table~\ref{tab:table3} presents the numerical values of the selected features, Table~\ref{tab:table4} lists these features in plain text along with their definitions. These concise definitions clarify the meaning and role of each feature in detecting \gls{ddos} attacks across \gls{hiot}, general \gls{iot}, and intrusion scenarios.

The remaining 4 features (timestamp, protocol\_des, source\_ip\_des, and destination\_ip\_des) store the original string values of timestamps, protocol types, and \gls{ip} addresses to support querying and manual review of the generated dataset, as shown in Table~\ref{tab:table5}. The ids column, one of the 13 selected features retained for model training, is also included in Table~\ref{tab:table5}. It is presented primarily to give context for the excluded features and to illustrate their values for specific \gls{hiot} nodes. These 4 features are retained for reference purposes in the \emph{UL-ECE-UDP-DDoS-H-IoT2025} dataset; however, they are excluded from the \gls{tcn} model’s training phase to improve generalisation and maintain optimal accuracy, as discussed in Chapter~\ref{chap:tcn}. The specific roles of these features and the rationale for their exclusion are described below.

\begin{table}[ht!]
\centering
\caption[Representative sample entries of the 4 excluded features]{Representative sample entries of the 4 excluded features (timestamp, protocol\_des, source\_ip\_des, and destination\_ip\_des) from the \emph{UL-ECE-UDP-DDoS-H-IoT2025} dataset.}
\label{tab:table5}
\begin{tabular}{lllll}
\toprule
\textbf{ids} & \textbf{timestamp} & \textbf{protocol\_des} & \textbf{source\_ip\_des} & \textbf{destination\_ip\_des} \\
\midrule
15 & 108.709 & ns3::Ipv4Header & 7.0.0.13 & 7.0.0.2 \\

16 & 108.709 & ns3::Ipv4Header & 7.0.0.14 & 7.0.0.2 \\

17 & 108.709 & ns3::Ipv4Header & 7.0.0.15 & 7.0.0.2 \\

18 & 108.709 & ns3::Ipv4Header & 7.0.0.16 & 7.0.0.2 \\

19 & 108.709 & ns3::Ipv4Header & 7.0.0.17 & 7.0.0.2 \\

20 & 108.709 & ns3::Ipv4Header & 7.0.0.18 & 7.0.0.2 \\

21 & 108.709 & ns3::Ipv4Header & 7.0.0.19 & 7.0.0.2 \\

22 & 108.709 & ns3::Ipv4Header & 7.0.0.20 & 7.0.0.2 \\

23 & 108.709 & ns3::Ipv4Header & 7.0.0.21 & 7.0.0.2 \\

25 & 108.716 & ns3::Ipv4Header & 7.0.0.23 & 7.0.0.2 \\
\bottomrule
\end{tabular}
\begin{tablenotes}
    \small
    \item \textbf{Note:} The \textbf{ids} column is among the 13 selected features retained for model training and is not excluded.
\end{tablenotes}
\end{table}

\textbf{Timestamp:} This feature represents the simulation time when an \gls{hiot} node sends a data packet to the server. While each record has a unique timestamp, the values primarily reflect the simulation period rather than meaningful attack-related patterns. As a result, the model may learn time-dependent correlations instead of genuine network traffic characteristics (e.g., packet size, message frequency). This introduces noise and bias without predictive value, thereby degrading generalisation. Including this feature during \gls{tcn} model training reduces accuracy; therefore, it is excluded from the final feature set.

\textbf{Protocol\_des:} This feature contains the descriptive string ``ns3::Ipv4Header" for the IPv4 header (e.g., the network layer packet structure, such as version (IPv4 vs. IPv6), source and destination \gls{ip} address, etc.) in every record. Since this value is constant across all \gls{hiot} node entries, it provides no variation for the model to learn from. Moreover, \gls{ml} models cannot directly utilise raw string values without encoding. Any protocol differences (e.g., \gls{udp} vs. \gls{tcp}, \gls{mqtt}, or other protocols) are already captured by the numeric protocol feature (e.g., 0 for \gls{udp}). This allows detection if an external \gls{iot} node uses a different protocol. For this reason, protocol\_des is omitted from the final set of features used for training.

\textbf{Source\_ip\_des:} This feature represents the dotted-decimal \gls{ip} address of the source node (e.g., 7.0.0.13), which is already numerically encoded in the source\_ip feature (e.g., 1, 2, 3). For example, taking the first row of Tables \ref{tab:table3} and \ref{tab:table5} for \gls{hiot} node 15, if the log line shows /NodeList/15/ 7.0.0.13 $>$ 7.0.0.2, the source \gls{ip} is 7.0.0.13, which belongs to group 3 as shown in the tables. These groups are created in the same way for all \gls{hiot} nodes. Hence, source\_ip\_des is excluded from the features used for model training to avoid redundancy.

\textbf{Destination\_ip\_des:} This feature contains the dotted-decimal \gls{ip} address of the destination node, which is constant in this dataset (7.0.0.2, the server \gls{ip}) for all records. Since it offers no variation for the model to learn from, and its numeric form is already represented in the destination\_ip feature (e.g., 0 for all \gls{hiot} nodes), destination\_ip\_des is not included in the features used for model training.

\subsection{Comparison with Existing Datasets}
\gls{ddos} attacks pose a significant threat within the cybersecurity landscape, prompting researchers to develop datasets that capture both normal and attack behaviours. Over the past ten years, several publicly available datasets have been used to evaluate \gls{ddos} detection techniques with advanced \gls{ml}/\gls{dl} models in \gls{iot} and \gls{hiot} environments. These include broader network intrusion detection scenarios such as BoT-\gls{iot}, TON\_IoT, and CICIoT2023 \cite{alex2023comprehensive}. Table~\ref{tab:table5} provides a comparative overview of datasets, including CIC-IDS2017~\cite{rosay2022network}, BoT-\gls{iot}~\cite{koroniotis2019towards}, and WSN-DDoSAttack-\gls{hiot}2023 \cite{khatun2024time}, which are frequently used in \gls{ml}/\gls{dl}-based \gls{ddos} detection research. It also presents the detection accuracy achieved using the \gls{tcn} model on the proposed \emph{UL-ECE-MQTT-DDoS-H-IoT2025} and \emph{UL-ECE-UDP-DDoS-H-IoT2025} datasets. The comparison highlights key aspects, including \gls{iot}/\gls{hiot} focus, protocol, attack type, class type, model type, and accuracy. 

While CIC-IDS2017~\cite{cicids2017} offers general \gls{ddos} coverage, it is primarily designed for intrusion detection based on fundamental network traffic attributes such as timestamps, source and destination IPs, protocols, and attack types. However, it lacks a specific focus on \gls{iot}-based scenarios. The dataset is built primarily around protocols like HTTP and HTTPS and includes email-related traffic alongside typical enterprise activities such as web browsing, FTP, and SSH. As a result, it may not adequately capture the complexity of modern \gls{iot} and \gls{hiot} systems or reflect the evolving nature of contemporary \gls{ddos} attacks targeting such systems.  

The BoT-\gls{iot} dataset is developed in a simulated \gls{iot} environment at the Cyber Range Lab of UNSW Canberra~\cite{botiot2023}. It contains both benign and botnet traffic, including \gls{dos} and \gls{ddos} attacks. While it offers a broad range of attack types relevant to \gls{iot}, the dataset is fully synthetic and known to be imbalanced ~\cite{ibrahimi2022improving}, with certain classes having very few samples. This imbalance poses challenges for training reliable \gls{ml} models. Moreover, the dataset lacks detailed protocol-level features, which limits its usefulness for analysing protocol-specific \gls{ddos} behaviours in resource-constrained \gls{iot} deployments.  

The WSN\_DDoS\_Attack\_\gls{hiot}2023~\cite{akhi2025ulwsn} dataset is developed using \gls{wsn}, focusing on environmental parameters such as humidity and room temperature within \gls{hiot} scenarios. It captures both benign and \gls{ddos} traffic over the \gls{udp} protocol using the Cooja simulator, representing communication behaviours among sensor nodes under attack conditions. However, a key limitation of this dataset is the absence of direct healthcare sensor data, such as body temperature, heart rate, or other vital signs. Including this simulated data would be representative of actual \gls{hiot} device communications and enhance the dataset’s relevance. 

As outlined in Table~\ref{tab:table6}, benchmark datasets like CIC-IDS2017 and BoT-\gls{iot} have been widely used to evaluate traditional \gls{ml} models such as \gls{dt} and Random Forest, often yielding high accuracy. The WSN\_DDoS\_Attack\_\gls{hiot}2023 dataset also enables the evaluation of \gls{dl} models, including \gls{cnn}, which are reported to achieve high detection accuracy~\cite{khatun2024time, akhi2025ulwsn}. In this comparison, the \gls{tcn} model demonstrates improved accuracy over the existing datasets, except for \gls{dt}, which achieves 100\% on one dataset (e.g., BoT-\gls{iot} 2018). These findings underscore the importance of creating protocol-specific datasets tailored for modern \gls{iot} and \gls{hiot} \gls{ddos} detection research. 

\begin{table}[ht!]
\centering
\caption[Comparison of existing public datasets]{Comparison of existing public datasets with the proposed \emph{UL-ECE-MQTT-DDoS-H-IoT2025} and \emph{UL-ECE-UDP-DDoS-H-IoT2025} datasets, evaluated using the \gls{tcn} model.}
\label{tab:table6}
\renewcommand{\arraystretch}{1.1}
\setlength{\tabcolsep}{3pt}
\resizebox{\textwidth}{!}{%
\begin{tabular}{llclllcl}
\toprule
\textbf{Ref.} & \textbf{Dataset} & \textbf{\shortstack{\gls{iot}/\gls{hiot}\\Focus}} & \textbf{Protocol} & \textbf{\shortstack{Attack\\Type}}
 & \textbf{\shortstack{Class\\Type}}
& \textbf{Model} & \textbf{\shortstack{Acc.\\(\%)}} \\
\midrule
\cite{rosay2022network} (2017) & CIC-IDS2017 & \ding{53} & HTTP/HTTPS & \gls{ddos} & Binary & RF & 99.31 \\

\cite{botiot2023} (2018) & BoT-\gls{iot} 2018 & \checkmark & N/A & \gls{dos}/ \gls{ddos} & Binary & DT & 100 \\

\cite{khatun2024time} (2024) & WSN\_DDoS\_Attack\_\gls{hiot}2023 & \checkmark & \gls{udp} & \gls{ddos} & Binary & \gls{cnn} & 92 \\

Chapter~\ref{chap:tcn} (2025) & \emph{UL-ECE-MQTT-DDoS-H-IoT2025} & \checkmark & \gls{mqtt} & \gls{ddos} & Binary & \gls{tcn} & 99.98 \\

Chapter~\ref{chap:tcn} (2025) & \emph{UL-ECE-UDP-DDoS-H-IoT2025} & \checkmark & \gls{udp} & \gls{ddos} & Binary & \gls{tcn} & 99.99 \\
\bottomrule
\end{tabular}
}
\end{table}

\section{Experimental Design, Materials and Methods}
This section presents the data generated through simulation-based experiments using the Cooja and ns-3 simulators. The Cooja simulator generates \gls{mqtt}-based traffic, while the ns-3 simulator generates \gls{udp}-based traffic. Both are configured in a \gls{hiot} environment with normal and \gls{ddos}-affected nodes. This section also describes simulation design, node configurations, communication protocols, traffic generation behaviour, and the tools used to capture and preprocess raw data. 

\begin{figure}[!htbp]
    \centering
    \includegraphics[width=0.7\textwidth]{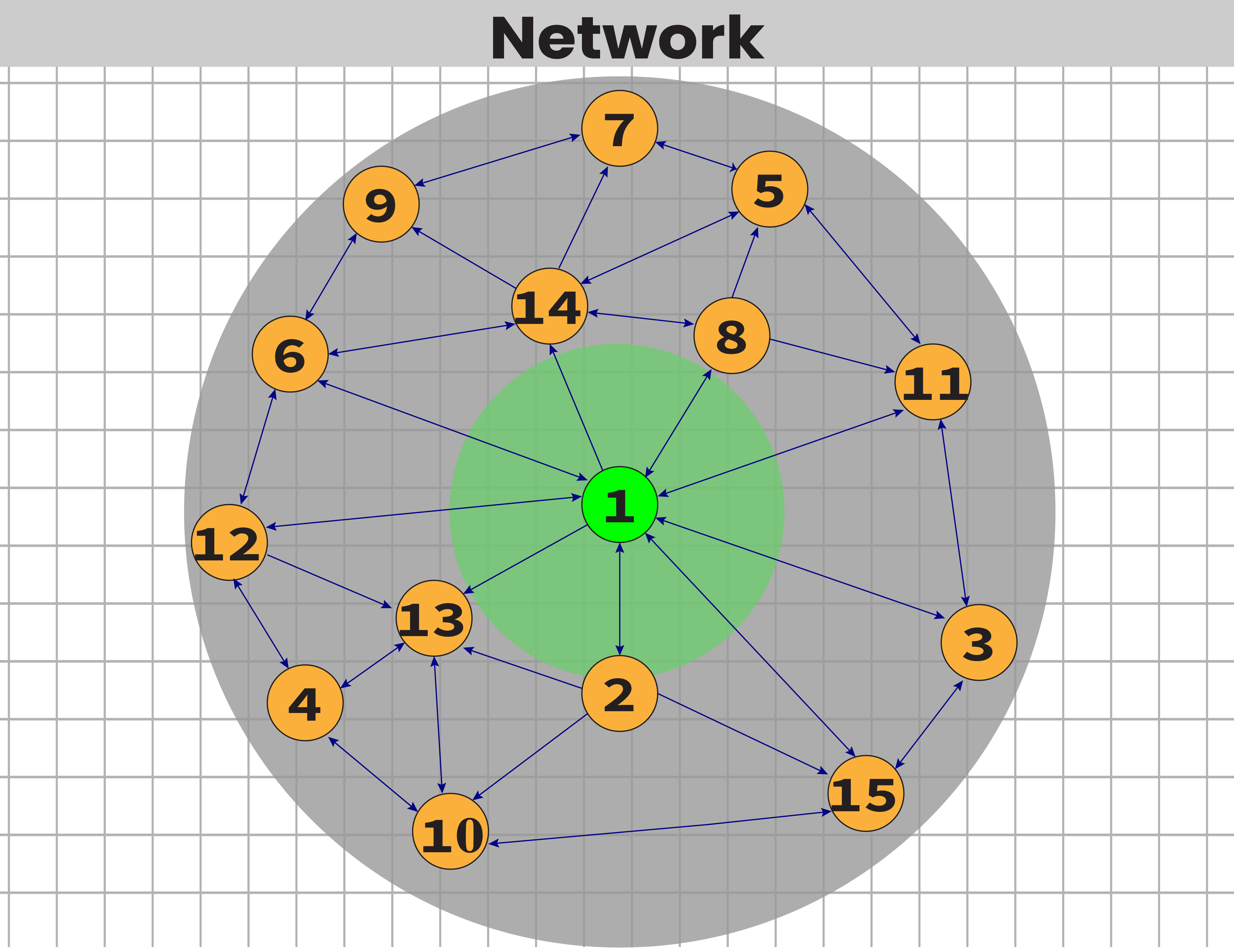}
 \caption[The baseline network topology over the MQTT]{The baseline network topology over the \gls{mqtt} protocol using the Cooja simulator in an \gls{hiot} environment. The green node represents the \gls{mqtt} broker/server, and the yellow nodes denote normal \gls{hiot} sensor nodes. This topology diagram is derived from the configuration presented in Chapter~\ref{chap:tcn}.}
    \label{fig:withoutattack}
\end{figure}

\begin{figure}[!htbp]
    \centering
    \includegraphics[width=0.7\textwidth]{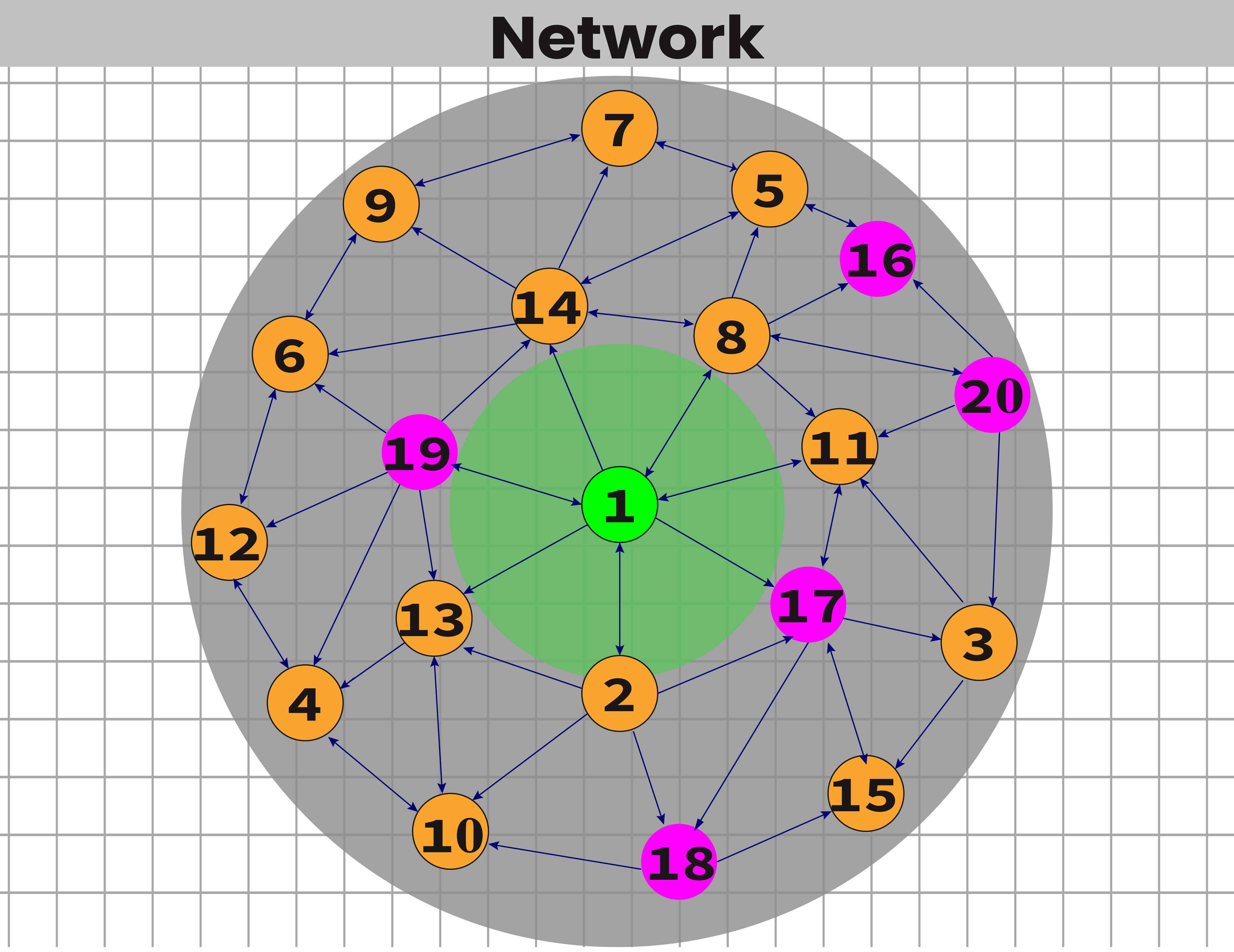}
 \caption[This network topology shows normal and \gls{ddos}-affected nodes]{This network topology shows normal and \gls{ddos}-affected nodes over the \gls{mqtt} protocol in the \gls{hiot} environment. The purple nodes indicate \gls{ddos} attack-affected nodes. This figure is adapted from the configuration described in Chapter~\ref{chap:tcn}.}
    \label{fig:withattack}
\end{figure}

\subsection{Simulation Setup for \gls{mqtt}}
The \gls{mqtt}-based \gls{ddos} attack simulation in the \gls{hiot} environment follows the configuration outlined in Chapter~\ref{chap:tcn}. In brief, the baseline topology consists of 14 emulated \gls{hiot} nodes representing sensors such as body temperature, oxygen saturation, and heart rate, along with one \gls{mqtt} broker acting as the server, as depicted in Figure~\ref{fig:withoutattack}. The malicious topology includes five \gls{ddos} attack-affected \gls{hiot} nodes, as illustrated in Figure~\ref{fig:withattack}. In the figures, normal nodes are shown in yellow, the \gls{mqtt} broker (server) is shown in green, and \gls{ddos} attack-affected nodes are shown in purple. The simulation is conducted using the Cooja simulator and the Contiki-NG operating system. Normal nodes transmit health-related \gls{mqtt} messages at regular intervals (e.g., one packet every 60 s), while malicious nodes generate traffic at shorter intervals (e.g., one packet every 20 s) to emulate \gls{ddos} behaviour. The \gls{ddos}-affected nodes transmit data more frequently, every 20 s, to simulate aggressive traffic conditions. Such high-frequency transmissions can degrade server responsiveness. This behaviour may also limit the network’s ability to handle legitimate traffic, making it a common technique to disrupt normal operations.

\subsection{Data Generation Using Cooja Simulator}
The synthetic raw data, including regular and malicious traffic, is generated over the \gls{mqtt} protocol using the Cooja simulator in an \gls{hiot} environment. It is collected from a simulation of 14 \gls{mqtt}-enabled sensor nodes emulating three types of healthcare monitoring devices (body temperature, heart rate, and oxygen saturation). Each node transmits data (e.g., a body-temperature reading of 98.6 °F) to the server at selected intervals, with different transmission rates assigned to normal and \gls{ddos}-affected nodes to reflect distinct communication behaviours. The simulation logs include timestamps, node identifiers, and published \gls{mqtt} messages, which are captured from the node output console and stored as unprocessed raw text. The simulation runs for 5 h, producing raw data that is saved in a plain text file for further processing.

\subsection{Simulation Setup for \gls{udp}}
This research simulates a \gls{ddos} attack over the \gls{udp} protocol in a \gls{g5}-enabled \gls{hiot} environment using ns-3, an open-source discrete-event network simulator. A detailed configuration is provided in Chapter~\ref{chap:tcn}. The simulation is based on a hierarchical network topology within a \gls{g5}-enabled \gls{hiot} environment, which organises client \gls{hiot} devices, a central server, and multiple \gls{lte} base stations (eNodeBs), with communication established via \gls{udp}.

\begin{figure}[ht!]
    \centering
    \includegraphics[width=0.8\textwidth]{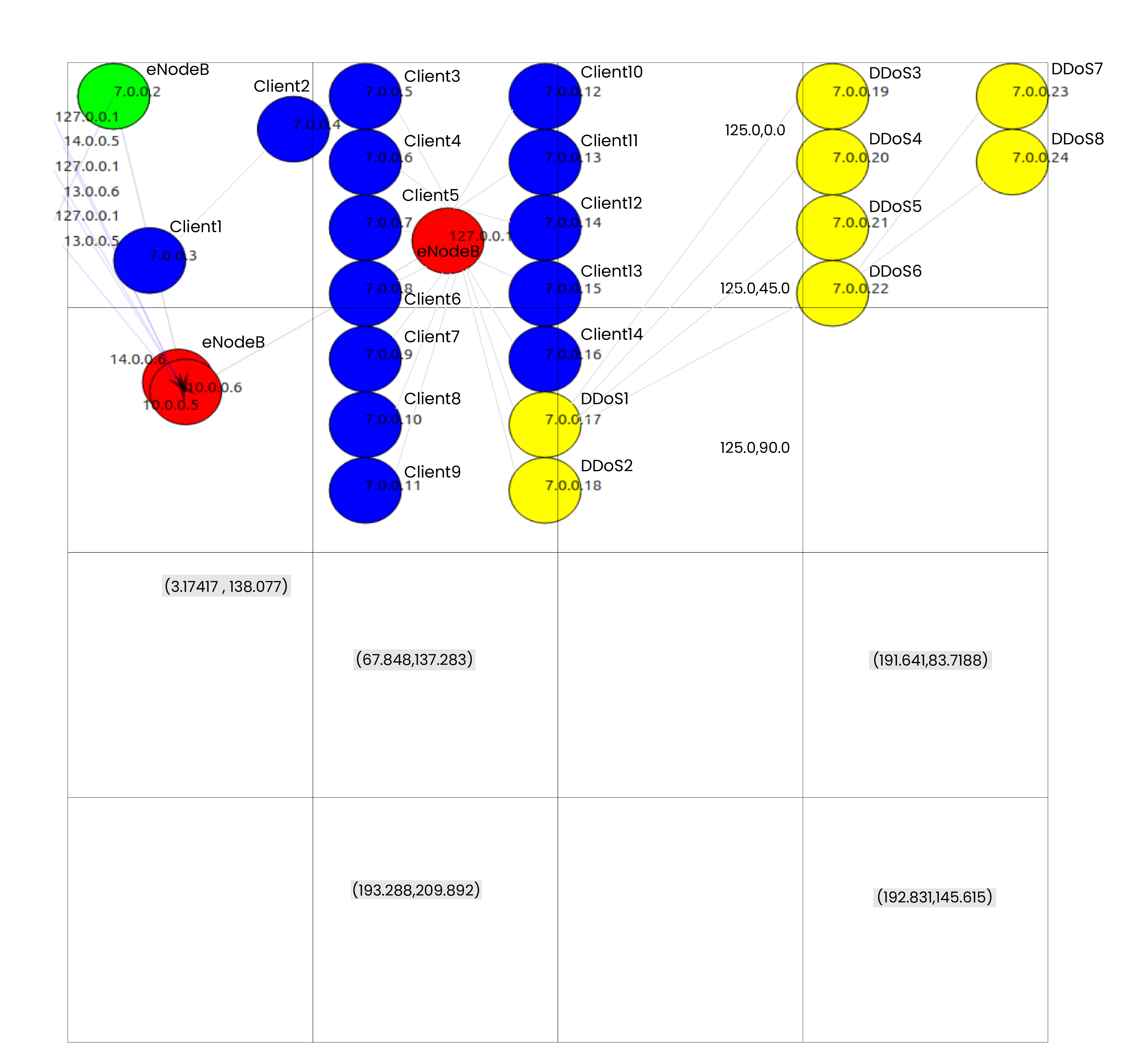}
 \caption[Simulated DDoS attack topology over the \gls{udp} protocol using ns-3]{Simulated \gls{ddos} attack topology over the \gls{udp} protocol using ns-3 in an \gls{hiot} environment. The green node denotes the server, red nodes represent eNodeBs, blue nodes indicate regular \gls{hiot} clients (Client: 1–14), and yellow nodes represent \gls{ddos} attackers (\gls{ddos}: 1–5). Additional attackers (\gls{ddos}: 6–8) are introduced within the same topology to intensify the attack. This figure is derived from the configuration presented in Chapter~\ref{chap:tcn}.}
    \label{fig:ns3simulation}
\end{figure}

The \gls{hiot} setting includes multiple node types, such as regular client nodes (5–18) generating real-time health-related traffic over the \gls{udp} protocol. These nodes are assigned to \gls{ip} addresses from 7.0.0.3 to 7.0.0.16 and transmit data to the central server (target UE) at \gls{ip} address 7.0.0.2. \gls{ddos} attack-affected nodes (19–23), with \gls{ip} addresses from 7.0.0.17 to 7.0.0.21, are configured to send high-volume traffic to overwhelm the server, as illustrated in Figure~\ref{fig:ns3simulation}. 

Additional \gls{ddos} attack-affected \gls{hiot} nodes (24–26, referred to as \gls{ddos}-6 to \gls{ddos}-8) are dynamically introduced during the simulation to escalate the attack intensity. These nodes, assigned \gls{ip} addresses from 7.0.0.22 to 7.0.0.24, are activated at 10-second intervals starting 20 s into the simulation. This setup reflects a realistic scenario where new malicious nodes emerge within the same network topology over time. As the number of attacking nodes increases, the central server faces progressively higher volumes of malicious traffic, challenging its capacity to maintain service availability. Evaluating the server's performance under these conditions is essential for assessing its resilience in a \gls{g5}-enabled \gls{hiot} environment. Additionally, this scenario facilitates the evaluation of the \gls{dl} model’s ability to detect newly introduced malicious actors within the same \gls{hiot} topology. The simulation spans 120 s and generates over one million data records, capturing benign and malicious traffic activities.

\subsection{Data Generation Using ns-3 Simulator}
The simulation records detailed communication activity from each \gls{hiot} node to generate raw data, including timestamps, transmitted and received packets, \gls{ip} and \gls{udp} header information, packet sizes, and protocol types. Each entry also captures source and destination \gls{ip} addresses, port numbers, and payload length. The simulation output includes control-plane messages such as GtpcCreateSessionRequestMessage~\cite{mohammadkhan2020cleang}, which represent healthcare device session establishment procedures under a \gls{g5}-enabled \gls{hiot} environment. These recorded elements serve as metadata and reflect realistic traffic patterns across various \gls{hiot} nodes, supporting protocol-level inspection and \gls{ml}–based anomaly detection.

\section{Limitations}
The data is not collected from physical \gls{hiot} devices; rather, it is generated through simulations using Cooja and ns-3 simulators. Hence, it may not fully reflect the variability or complexity found in real-world healthcare settings. The attack scenarios and traffic behaviours are predefined based on simulation design, which may not encompass the full range of variability in real-world \gls{ddos} events. \gls{ml}/\gls{dl} models trained on these datasets may require additional hyperparameter tuning (e.g., adjusting learning rate, batch size, or dropout) or further validation with real-world \gls{hiot} data (e.g., testing on actual healthcare network traffic) to enhance their generalizability and robustness in practical deployments. \rev{Moreover, the dataset is generated with a fixed number of nodes and predefined transmission behaviour, which may not fully reflect varying node densities and communication patterns in real-world \gls{iot} environments.}

\section{\rev{Clarification and Comparative Summary}}
\rev{This section clarifies the use of reported accuracy values and the placement of \gls{tcn}-based results alongside values reported in the literature in Table~\ref{tab:table6}.}

\rev{The accuracy of the \gls{tcn} model shown in Table~\ref{tab:table6} is derived from the evaluation presented in Chapter~\ref{chap:tcn}, which was published prior to this dataset-focused chapter. As the datasets are associated with that study and published separately, including these accuracy values demonstrates the applicability of the proposed datasets for \gls{dl}-based \gls{ddos} detection.}

\rev{The results in Table~\ref{tab:table6} include reported accuracy values from existing studies alongside the \gls{tcn}-based evaluation of the proposed datasets. The accuracy values for existing datasets (e.g., BoT-\gls{iot}) are taken from their respective studies, which use models such as \gls{dt}~\cite{botiot2023} and \gls{cnn}~\cite{khatun2024time}, whereas the proposed datasets are evaluated using the \gls{tcn} model. The inclusion of these values demonstrates that the proposed datasets can be effectively used with \gls{dl} models and achieve high detection accuracy.}

\rev{However, this research compares the proposed datasets with an existing DDoS dataset developed from environmental sensor data, including humidity and room temperature~\cite{khatun2024time}. The WSN\_DDoS\_Attack\_\gls{hiot}2023 dataset comprises 14 nodes, including 1 server node and 3 malicious nodes, and operates over the \gls{udp} protocol~\cite{akhi2025ulwsn}. While it captures attack behaviour in a wireless sensor network setting, it differs from the proposed \gls{wban}-based datasets in the number of nodes, the type of emulated sensor data, and the dataset size, as summarised in Table~\ref{tab:dataset_comparison_summary}.}

\begin{table}[htbp]
\centering
\small
\caption{Comparative summary of datasets.}
\setlength{\tabcolsep}{5pt}
\resizebox{\textwidth}{!}{
\begin{tabular}{l lccc}
\toprule
\textbf{Ref.} & \textbf{Dataset} & \textbf{Protocol} & \textbf{Attack Type} & \textbf{Total Samples} \\
\midrule
\cite{akhi2025ulwsn} (2025) & WSN\_DDoS\_Attack\_\gls{hiot}2023 & \gls{udp} & \gls{ddos} & 15,988 \\
Chapter \ref{chap:dataset} & \emph{UL-ECE-MQTT-DDoS-H-IoT2025} & \gls{mqtt} & \gls{ddos} & 20,080 \\
Chapter \ref{chap:dataset} & \emph{UL-ECE-UDP-DDoS-H-IoT2025} & \gls{udp} & \gls{ddos} & 99,887 \\
\bottomrule
\end{tabular}}
\label{tab:dataset_comparison_summary}
\end{table}

\section{Conclusion}
This chapter presents two synthetic \gls{ddos} datasets,~\emph{UL-ECE-MQTT-DDoS-H-IoT2025} and \emph{UL-ECE-UDP-DDoS-H-IoT2025}, that enable realistic traffic analysis in \gls{hiot} environments. The datasets are generated using Cooja and ns-3 simulators to represent application-layer (\gls{mqtt}) and transport-layer (\gls{udp}) behaviours under both normal and attack conditions. Each dataset includes communication-level parameters such as message frequency, transmission intervals, and node activity, providing a realistic representation of \gls{ddos} patterns in heterogeneous \gls{hiot} networks. These datasets address a key research gap by providing domain-specific resources for evaluating data-driven cybersecurity mechanisms in healthcare-oriented \gls{iot} systems.			

\chapter{\gls*{tcn}-Based \gls*{ddos} Detection and Mitigation in \gls*{g5} \gls*{hiot}: A Frequency Monitoring and Dynamic Threshold Approach} \label{chap:tcn} 


\ifpdf
    \graphicspath{{4_chapter4/figures/PNG/}{4_chapter4/figures/PDF/}{4_chapter4/figures/}}
\else
    \graphicspath{{4_chapter4/figures/EPS/}{4_chapter4/figures/}}
\fi


\noindent This research presented in this chapter has been previously published in:

\textbf{Akhi, M.}, Eising, C. and Dhirani, L.L., 2025.~\textit{\gls{tcn}-Based DDoS detection and mitigation in 5G Healthcare-IoT: A frequency monitoring and dynamic threshold approach.}~\textit{IEEE Access}, 13, pp.~12709-12733.

The content of the paper above has been presented as published in this chapter, except for correcting grammatical errors and making slight rewording so that the narrative is more appropriate for a thesis. A short preamble is provided to explain its relevance to the narrative of the thesis.

\rev{\noindent\textbf{Thesis Structure and Publication Context:}
Parts of this thesis, including Chapters~\ref{chap:dataset}, \ref{chap:tcn}, and \ref{chap:edge}, are based on related published research that share a common scope and methodological foundation. Therefore, some overlap in background, simulation setup, and methodological descriptions is included to ensure completeness.}

\clearpage

\noindent\textbf{\underline{Preamble}}\\[0.5em]
In this chapter, \textbf{RQ3} is addressed:

\textbf{RQ3:} Are \gls{dl} techniques (e.g., \gls{tcn}) capable of detecting and mitigating evolving cyber threats (e.g., \gls{ddos}) across \gls{hiot} systems? 

\vspace{5pt}

The preceding chapter addresses \textbf{RQ2} by detailing the simulation design and the generation of realistic synthetic datasets that represent \gls{ddos} attack behaviours in \gls{hiot} environments. It explains how protocol-specific communication patterns are modelled and used to construct the~{\em UL-ECE-MQTT-DDoS-H-IoT2025} and {\em UL-ECE-UDP-DDoS-H-IoT2025} datasets. Building on this foundation, the current chapter addresses \textbf{RQ3} by applying a \gls{tcn} model to detect and mitigate \gls{ddos} attacks using these datasets. The chapter, therefore, presents the detection and mitigation performance of the \gls{tcn} model against \gls{mqtt} and \gls{udp}-protocol–specific \gls{ddos} attacks in \gls{hiot}. 

This chapter contributes to the main research question by focusing on the detection and mitigation of \gls{ddos} attacks using a \gls{tcn}-based \gls{dl} model in \gls{hiot} environments. It examines how the \gls{tcn} model detects \gls{ddos}-affected nodes and enforces mitigation by isolating them from subsequent network communication.

The main objective of this study is to employ a \gls{tcn} model to detect and mitigate \gls{ddos} attacks within \gls{hiot} environments. The model is trained and evaluated on the datasets~{\em UL-ECE-MQTT-DDoS-H-IoT2025} and {\em UL-ECE-UDP-DDoS-H-IoT2025}, introduced in Chapter~\ref{chap:dataset}. These datasets encompass diverse communication parameters, including total message count, transmission frequency, and monitoring frequency, across both \gls{mqtt} and \gls{udp} protocols, providing a robust foundation for cross-protocol analysis. 

In the {\em UL-ECE-MQTT-DDoS-H-IoT2025} dataset, the \gls{tcn} model utilises four extracted communication features, among which message frequency plays a crucial role in identifying abnormal node behaviour. A node is considered potentially malicious or \gls{ddos}-compromised if its message transmission frequency exceeds a threshold (e.g., frequency $>$ 5), indicating unusually high traffic intensity compared to normal nodes. In contrast, the \gls{tcn} method in the {\em UL-ECE-UDP-DDoS-H-IoT2025} dataset employs a monitoring frequency mechanism that calculates the percentage of the total message count for each \gls{hiot} node over the past five seconds, enabling the identification of malicious nodes. This five-second monitoring interval also allows the model to detect additional \gls{ddos}-affected nodes that are introduced every 10 seconds, beginning 20 seconds into the 120-second simulation on ns-3, as discussed in Section~\ref{subsubsec:ddosns-3}. It also incorporates a dynamic threshold-based mitigation strategy that automatically blacklists malicious nodes in real-time, while minimising false positives. 

This research, in line with \textbf{RQ3}, investigates how the \gls{tcn} model leverages critical features to detect and mitigate \gls{ddos} attacks in \gls{hiot} environments. The detection and mitigation results provide valuable insight into how the \gls{tcn} model effectively captures temporal dependencies in \gls{hiot} traffic, thereby enabling efficient performance over the \gls{mqtt} and \gls{udp}-based \gls{hiot} datasets. Temporal dependencies refer to the sequential correlations between communication patterns over time. Moreover, this model outperforms traditional architectures (e.g., \gls{cnn}, \gls{bilstm}) due to its capability to learn and utilise these time-dependent patterns for both detection and mitigation tasks. Additionally, the model’s lightweight architecture, with fewer parameters and low latency, makes it suitable for edge deployment on resource-constrained \gls{hiot} devices.  

\section{Abstract}
The \gls{iot} revolutionises precision healthcare by enhancing patient care and reducing costs. However, this technology poses challenges in securing \gls{hiot} devices, as low-latency packet transmission increases vulnerability to attacks (e.g., \gls{ddos}) in \gls{g5} networks. This research develops a monitoring frequency-based detection and dynamic threshold mitigation method using \glspl{tcn} in \gls{g5} \gls{hiot} environments. A monitoring frequency-based detection method calculates each \gls{hiot} node's message count over the past five seconds as a percentage of total traffic. A dynamic threshold strategy also provides adaptive security by averaging detected malicious behaviour messages across all nodes to determine the base threshold. This approach enhances classification accuracy, effectively mitigates actual \gls{ddos} attack nodes, and blacklists malicious nodes without false positives in \gls{hiot} environments.
Moreover, this research creates \gls{ddos} attack models and monitoring parameters using two simulators (Cooja and ns-3) on \gls{hiot} devices. Data is collected over the \gls{mqtt} and \gls{udp} in a realistic \gls{g5}-based \gls{hiot} environment, creating two datasets with malicious and benign data. The proposed \gls{tcn}-based \gls{ddos} prediction and mitigation method achieves 99.98\% and 95\% accuracy on the \gls{mqtt} dataset. This model also attains a slightly higher prediction accuracy of 99.99\% and mitigation of 80\% on the \gls{udp} dataset. This research evaluates the proposed model against prior methods, including \gls{bilstm} and \glspl{cnn}, demonstrating improved accuracy. Thus, the proposed \gls{tcn} model is a versatile and resilient security solution for \gls{hiot} environments. 

\section{Introduction} 
The future of a smart and sustainable healthcare ecosystem depends on the security of infrastructures enabled by the \gls{iiot} \cite{dhirani2021industrial}. The \gls{iot} consists of physical devices connected by sensors, software, and networks \cite{dhirani2021industrial,article}. In recent years, implementing \gls{iot} has enabled the vision to achieve precision healthcare. The deployment of wearable \gls{hiot} devices has enabled real-time surveillance and monitoring of essential health parameters (e.g., body temperature, oxygen saturation, etc). \glspl{wban} powered by \gls{mqtt} are employed to enhance the security of these \gls{iot} devices \cite{taher2022certificateless}. However, the magnitude and scope of security risks, e.g., \gls{dos}/\gls{ddos}, associated with \gls{iot} devices have massively increased \cite{Statista2022IoTVulnerabilities}. The growing number of cyber risks requires urgent advancements in \gls{hiot} security. Malicious and threat actors, now equipped with generative \gls{ai} capabilities, exploit internet-connected devices in sophisticated ways that traditional security measures cannot adequately address \cite{Alqahtani2024, article}. Such attacks can completely disrupt essential healthcare services \cite{hamza2022hyperparameter}. Hence, an \gls{ai}-based defence method is necessary to identify, assess, and neutralise these novel threats. This approach may consider any deviation from an expected behaviour, indicating a potential security risk or anomaly \cite{palaniappan2024dynamic}. Various \gls{dl} techniques, including \glspl{rnn} \cite{yousuf2022ddos}, \glspl{cnn} \cite{ibrahim2023anomaly, khatun2024time}, and \glspl{gan} \cite{alabsi2023conditional}, have been employed in the prevention of \gls{ddos} attacks and anomaly detection in \gls{iot} networks. 
With healthcare's growing reliance on \gls{iot} devices, addressing the challenges of high latency, mobility, and resource constraints is critical in \gls{hiot} environments \cite{hassen2020home}. \rev{This research considers an \gls{hiot} \gls{ddos} detection and mitigation method, as existing \gls{dl}-based approaches have limitations in handling latency, mobility, and resource constraints, particularly in real-time scenarios. This limitation arises because conventional \gls{dl} models, such as \glspl{lstm} and \glspl{cnn}, often fail to capture long-range temporal dependencies and to provide low-latency inference in resource-constrained environments \cite{ghosh2025novel}.}

\begin{figure}[ht!]
    \centering
    \includegraphics[width=\textwidth]{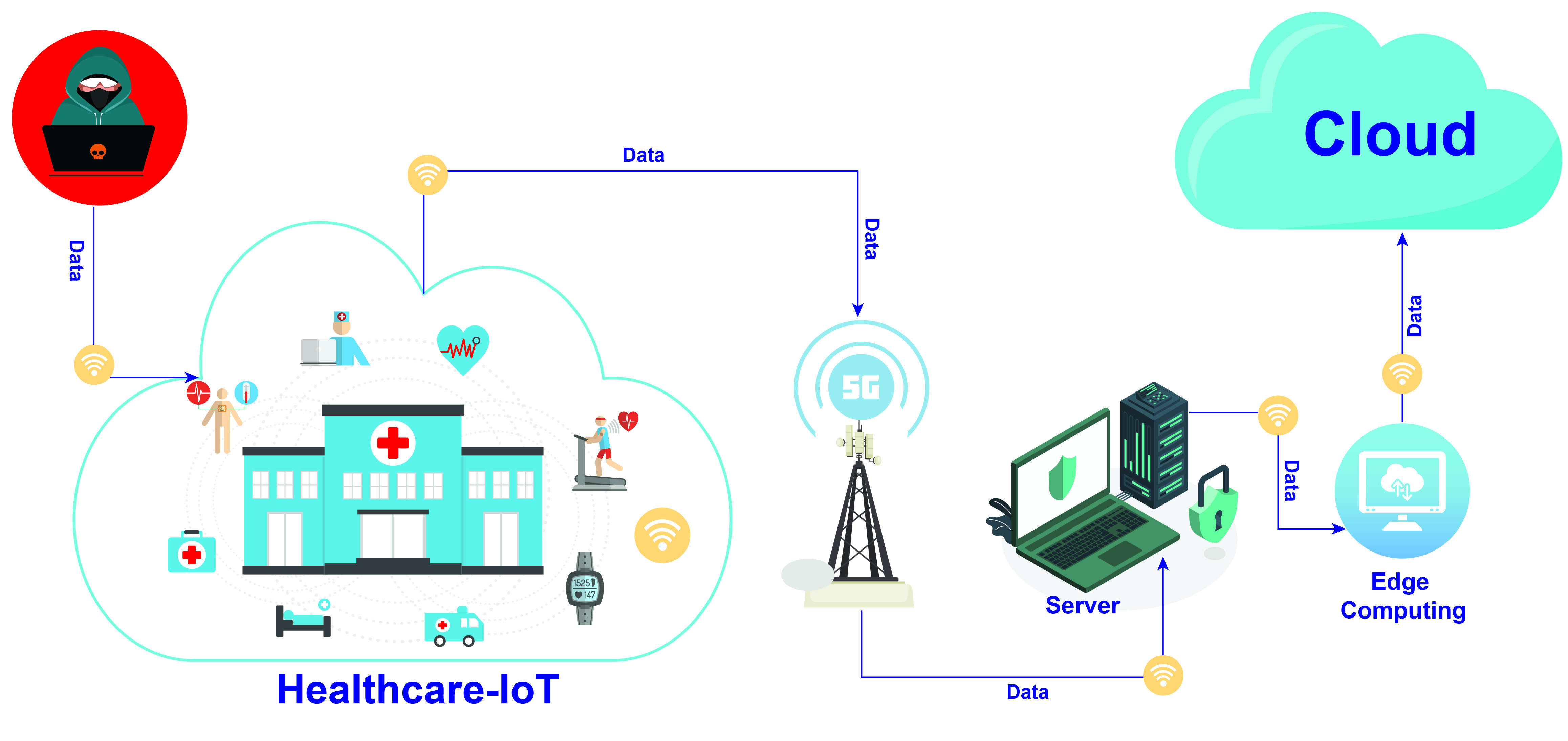}
    \caption[Security risks with 5G, Edge, and Cloud integration in H-IoT]{Security risks with \gls{g5}, Edge, and Cloud integration in \gls{hiot}.}
    \label{fig:Flagship}
\end{figure}

The emergence of \gls{g5} networks significantly enhances \gls{hiot} data transmission speeds and volumes \cite{putro2024future}. \gls{iot} devices powered with \gls{g5} may facilitate healthcare services with high speed and lower latency \cite{ahad2020technologies}; however, it can also enable more powerful and potentially damaging cyber attacks (e.g., \gls{ddos}, proximity service intrusions, etc.) \cite{triplett2024cybersecurity} \cite{khraisat2021critical}. Integrating \gls{g5}, edge computing, and cloud in these paradigms broadens the cyber threat landscape, introducing advanced and emerging security risks posed by malicious or harmful actors, as depicted in Figure~\ref{fig:Flagship}. Malicious actors are involved in monitoring or sniffing using sophisticated tools, like Wireshark, to monitor and analyse a device's traffic, capturing details such as data packet size, payload length, and protocol \cite{anthi2024investigating}, \cite{s23031151}. Attackers also employ this technique for harvesting sensitive user information within a network, such as usernames and passwords \cite{anthi2024investigating, sharma2023exploratory, bastille2016mousejack}, typically as a preliminary step before launching a \gls{ddos} attack. 

Health monitoring systems increasingly rely on cloud and edge computing for transmitting sensitive healthcare data across multiple \gls{hiot} devices and networks \cite{singh2024integration}. The increasing number of data transmission devices and networks significantly expands the attack surface for cybercriminals, making it easier for them to harvest and exploit vulnerabilities \cite{shah2024iot}. As a result, managing and securing this \gls{hiot} infrastructure has become more complex \cite{zhang2019towards}. It is, therefore, necessary to develop robust security mechanisms that address vulnerabilities to cyber-attacks and data breaches among \gls{hiot} devices at every layer, from edge devices to cloud infrastructure.

To address these complexities, this research uses the Cooja simulator to create a realistic \gls{ddos} attack simulation in \gls{hiot} environments, using emulated \gls{wban} \gls{hiot} sensor data, such as body temperature, oxygen saturation, etc. In addition, the ns-3 simulator is used to create a \gls{ddos} attack model within a \gls{g5}-based \gls{hiot} environment, focusing on real-time traffic and network performance over the \gls{udp} protocol. In this instance, simulated data, commonly known as synthetic data, is artificially generated to mimic real-world \gls{hiot} sensor data closely. This research generates two datasets from simulated data, reflecting both \gls{ddos} attacks and normal behaviours of \gls{hiot} sensors. The first dataset focuses on the \gls{mqtt} protocol and contains malicious and non-malicious \gls{hiot} \gls{wban} data from a simulated \gls{ddos} attack on the Cooja simulator. The second dataset relies on the \gls{udp} protocol and includes malicious and benign data collected from a realistic \gls{g5}-based \gls{hiot} environment with a simulated \gls{ddos} attack on the ns-3 simulator. This research uses two simulators, ns-3 and Cooja, to capture diverse \gls{hiot} network environments. The ns-3 allows us to carefully control the flow and types of data in the network and set specific rules for how data is sent and received, making the simulation more realistic \cite{venkateswaran2024ieee}.
In contrast, Cooja addresses \gls{iot}-specific challenges. For example, the ns-3 simulator does not allow users to manually select and configure individual \gls{hiot} nodes in a graphical environment. This dual approach ensures a comprehensive simulation of \gls{ddos} attack scenarios across general and \gls{iot} networks.

\rev{Hence, this research employs \glspl{tcn} to address real-time challenges in detecting and mitigating \gls{ddos} attacks within \gls{hiot} environments, due to their ability to effectively model temporal dependencies in network traffic. While the focus of this research is on \gls{hiot}, the proposed TCN-based approach generalises to broader \gls{iot} networks, as discussed in Section~\ref{sec:evaluation}, with supporting results on non-\gls{hiot} datasets such as BoT-\gls{iot} and CIC-IDS2017, as shown in Table~\ref{tab:comparedataset}.} The \glspl{tcn}' outputs offer advanced \gls{iot} security solutions, improving cybersecurity and supporting precision healthcare by addressing vulnerabilities within \gls{g5} networks. This is achieved through real-time attack prediction, message frequency proportion, monitoring frequency, and dynamic thresholding. Specifically, this research proposes security solutions that detect unusual traffic patterns by calculating each \gls{hiot} node's message frequency as a percentage of total network activity. It continuously monitors node behaviour over past intervals (e.g., every 5 seconds) to identify anomalies in traffic patterns, including the appearance of new nodes with lower message volumes compared to established ones.

In the mitigation phase, a dynamic threshold-based strategy is applied. This strategy calculates a base threshold based on the average count of malicious message activity among nodes affected by the \gls{ddos} attack. This threshold automatically captures real-time changes in malicious behaviour within \gls{hiot} environments. A blacklist is created for nodes that exceed the dynamic threshold, flagging those nodes as compromised by the \gls{ddos} attack. During the mitigation phase, a filtering approach improves prevention accuracy. If the initial base threshold set by the \gls{tcn} model flags non-malicious nodes, it can be adjusted to reduce incorrect detections or false positives. This adjustment also makes the blacklist more selective, minimising the risk of mistakenly flagging non-malicious nodes as malicious. The adjusted threshold is dynamically calculated to capture only accurate malicious activities. This adaptability ensures the mitigation strategy remains effective in real-world \gls{g5} \gls{hiot} environments, where precise detection is essential under fluctuating network conditions. This filtering strategy is also crucial for maintaining the integrity and availability of data from \gls{g5} \gls{hiot} devices.

The \gls{mqtt} protocol uses a publisher/subscriber model to facilitate communication between \gls{iot} devices in a lightweight approach \cite{hintaw2023mqtt}. Due to its architecture, this protocol is inherently vulnerable to \gls{dos} attacks that compromise the reliability and efficiency of \gls{iot} systems. Specifically, \gls{mqtt}'s centralised broker is a single point of failure, and the lack of payload encryption and authentication allows attackers to overwhelm it with malicious traffic easily \cite{lakshminarayana2024securing}. Even high-priority messages can be flooded into the broker by exploiting various QoS levels. Consequently, \gls{ddos} attacks increase the risk of severe patient safety and privacy breaches on \gls{hiot} devices. One of the biggest challenges is to develop security measures for \gls{mqtt} that can counteract \gls{ddos} attacks without compromising the battery-powered \gls{iot} device. With \gls{g5}-\gls{iot}, risks related to unauthorised access are amplified, as these devices lack sophisticated security functions, such as real-time monitoring, frequency-based detection and dynamic thresholding mitigation mechanisms. Hence, developing sustainable security solutions (e.g., a dynamic threshold mechanism) is imperative for securing \gls{g5}-enabled \gls{hiot} devices.

\gls{udp} is a connectionless protocol that prioritises speed over reliability \cite{gupta2024ns2}, \cite{tagliaro2024large}. Due to the lack of mechanisms such as acknowledgement and retransmission, data transfer is faster but less dependent. \gls{udp}'s simplicity and lack of flow control make it vulnerable to attackers. It cannot differentiate between authorised and unauthorised traffic, which enables malicious exploitation. This vulnerability highlights the need for effective security mechanisms, such as monitoring frequency, regularly checking network activity at specified intervals (e.g., 5 seconds) to protect healthcare data against potential attacks (e.g., \gls{ddos}) in a \gls{g5}-based \gls{hiot} environment.

As \gls{ddos} attacks increase in \gls{hiot} environments, exploring existing detection and mitigation methods is essential. The following section reviews prior research, highlights its limitations, and explains how this research addresses those gaps.

The key contributions of this research are outlined as follows:

\begin{itemize}
    \item This research develops a \gls{ddos} attack model using the Cooja simulator to test \gls{mqtt} vulnerabilities in \gls{hiot} environments, emulating realistic \gls{wban} sensor data such as oxygen saturation, body temperature, and heart rate. Hence, the synthetic dataset, ``{\em UL-ECE-MQTT-DDoS-H-IoT2025},'' is generated explicitly for \gls{hiot} security.

\item This research develops an additional \gls{ddos} attack model for the \gls{udp} protocol using the ns-3 simulator within a \gls{g5}-enabled \gls{hiot} environment. The attack model adopts a two-dimensional grid layout in ns-3 to control distances among clients, \gls{ddos} attack-affected nodes, and the server, enabling proximity-based network performance analysis. Traffic data generated by the simulation is preprocessed at 5-second monitoring intervals for detailed analysis of normal and \gls{ddos}-affected patterns. This preprocessing strategy also facilitates the detection of new nodes with lower message volumes than existing ones. The resulting dataset, {\em UL-ECE-UDP-DDoS-H-IoT2025}, is constructed using critical features.

    \item This research proposes a \gls{tcn}-based prediction model that uses message frequency proportions to predict \gls{ddos} attacks on the {\em UL-ECE-MQTT-DDoS-H-IoT2025} dataset, achieving a high prediction accuracy of 99.98\%. To the best of current knowledge, this is the first study to propose a monitoring frequency-driven \gls{tcn} prediction model for detecting \gls{ddos} attacks in \gls{g5}-based \gls{hiot} environments, representing the core novelty of this research. The proposed model achieves a prediction accuracy of 99.99\% on the {\em UL-ECE-UDP-DDoS-H-IoT2025} dataset. 
    
    \item Based on the proposed \gls{tcn} prediction model, a dynamic threshold-based strategy provides an adaptive security solution for detecting and mitigating \gls{ddos} attack-affected nodes in \gls{hiot} environments. It automatically calculates the average, which serves as the base threshold, by analysing malicious activity across all detected nodes. A novel aspect of this approach is its automatic mitigation and blacklist creation capabilities, achieving detection and mitigation accuracy of 96.94\% and 95\% on the {\em UL-ECE-MQTT-DDoS-H-IoT2025} dataset. On the second dataset, {\em UL-ECE-UDP-DDoS-H-IoT2025}, this approach achieves detection and mitigation accuracies of 75\% and 80\%, respectively. Thus, the \gls{tcn} model automatically identifies and blacklists compromised \gls{hiot} nodes in both proposed datasets, which enhances security by isolating unauthorised traffic.
 
\end{itemize}

This chapter is organised as follows. Section~\ref{sec:bg} discusses related work in anomaly detection systems for \gls{hiot} environments, Section~\ref{sec:mqttdataset} outlines the structured data collection process using the Cooja simulator, encompassing \gls{ddos} attack modeling over the \gls{mqtt} protocol and data preprocessing to create a dataset that supports reliable anomaly detection in \gls{hiot} environments, Section~\ref{sec:udpdataset} explains the data collection process using the ns-3 simulator, where \gls{ddos} attack modeling is conducted over the \gls{udp} protocol in a \gls{g5}-based \gls{hiot} environment, followed by data preprocessing to create another dataset for \gls{ddos} detection and mitigation, Section~\ref{sec:tcnmethod} details implementing \glspl{tcn} to detect \gls{ddos} attacks and dynamically mitigate malicious nodes in \gls{hiot} sensor data, with a focus on efficiency and precision, Section~\ref{sec:evaluation} outlines the performance evaluation of the proposed \gls{tcn} model, comparisons with previous models and datasets, a discussion of the mitigation strategy, and a comparison of the two proposed datasets across \gls{mqtt} and \gls{udp} protocols, Section~\ref{sec:conclu} concludes the chapter by indicating future research directions.

\section{Related Work} \label{sec:bg}
This section reviews existing \gls{hiot} anomaly detection methods, including different attack types (e.g., \gls{dos}/\gls{ddos}), datasets employed, gaps, and limitations. This review also highlights the novelty and distinguishing aspects of the present research compared to previous studies. Moreover, Table~\ref{tab:litreview} summarises a survey of published methods in \gls{hiot} over the last five years, offering additional insights.

\begin{table}[!htbp]
\centering
\setlength{\tabcolsep}{8pt}
\renewcommand{\arraystretch}{1.3}
\caption[Survey of classifiers, attack types, datasets]{Survey of classifiers, attack types, datasets, limitations, and benefits in \gls{hiot}.}
\label{tab:litreview}
\resizebox{\textwidth}{!}{
\begin{tabular}{@{}%
p{1.8cm}  
p{2.1cm}  
p{1.9cm}  
p{3.4cm}  
p{3.4cm}  
p{2.4cm}  
@{}}
\hline
\textbf{Ref.} & \textbf{Techniques} & \textbf{Attacks} & \textbf{Benefits} & \textbf{Limitations} & \textbf{Dataset} \\
\hline
Khatun et al. \cite{khatun2024time} (2024) & \gls{cnn} & 
\gls{ddos}

&
Accurate Anomaly Detection with high accuracy.\par
A new dataset of \gls{wsn} \gls{ddos} attacks is created.

& 
The \gls{wsn} \gls{ddos} dataset is specific to environmental \gls{iot} data and may not generalise sufficiently to other \gls{iot} sensors.

&
Dataset Created by Authors \cite{khatun2024time}. \\ 

\hline

 Altulaihan et al. \cite{altulaihan2024anomaly} (2024)
& \gls{dt}, \gls{knn} & \gls{dos} & Enhanced security of \gls{iot} networks. \par

 & 
Computational efficiency varies; \gls{dt} performs faster in training and testing, affecting real-time efficiency.  \par
 & IoTID2020 \cite{ullah2020scheme}.
\\ \hline

Wagan et al. \cite{wagan2023fuzzy} (2023) &  Customized \gls{bilstm}  & Man-in-the-middle attacks & 
Securely handles medical data.\par
Analyses attack patterns in \gls{iomt}.\par
&
No \gls{ddos} attack data in the dataset\par
Model decisions are not comprehensible.\par
Possibility of corruption of data.\par 
& 
Heart disease dataset with 36 attributes and 18940 instances developed with \gls{ehms}~\cite{hady2020intrusion}.
\\ \hline
Kumar et al. \cite{kumar2023novel} (2023) & \gls{cnn}, \gls{bilstm}, and \gls{gru}  & Botnet & 

High accuracy with the hybrid \gls{dl} model.
 &  
There is no mitigation strategy for \gls{ddos} and botnet attacks in \gls{iomt}. \par
Old dataset & CIC-IDS2017 \\

\hline

Chaganti et al. \cite{chaganti2022particle} (2022) & \gls{cnn}-\gls{lstm} & Network traffic & Addresses \gls{iomt} security and privacy issues.\par
Combines network traffic and patient sensing data for detection.
 & 
 Does not evaluate critical \gls{ddos} attacks.
 & Dataset combines network traffic and patient biometrics using the ARGUS tool \cite{argus_openargus}. \\ \hline
\end{tabular}}
\end{table}

 \begin{table}[ht!]
 \ContinuedFloat
\centering
\setlength{\tabcolsep}{6pt}
\renewcommand{\arraystretch}{1.3}
\caption[]{Survey of classifiers, attack types, datasets, limitations, and benefits in \gls{hiot} (continued).}
\resizebox{\textwidth}{!}{
\begin{tabular}{@{}%
p{1.8cm}  
p{2.1cm}  
p{1.9cm}  
p{3.4cm}  
p{3.4cm}  
p{2.4cm}  
@{}}
\hline
\textbf{Ref.} & \textbf{Techniques} & \textbf{Attacks} & \textbf{Benefits} & \textbf{Limitations} & \textbf{Dataset} \\
\hline
Binbusayyis et al. \cite{binbusayyis2022investigation} (2022) & \gls{dt} & \gls{dos} & Uses BoT-\gls{iot} dataset for comprehensive testing. \par & Achieves 100\% with the traditional \gls{ml} algorithm. \par

Notably, it lacks coverage of modern \gls{iot}-specific protocols such as \gls{mqtt}. & BoT-\gls{iot} 2018 \cite{koroniotis2019towards} \\ \hline

Faruqui et al. \cite{faruqui2023safetymed} (2022) & Conv-\gls{lstm} & \gls{dos}/\gls{ddos} & Protects \gls{iomt} devices from cyber intrusions. \par
Achieves high accuracy. \par
 & 
Complex architecture.\par
High costs. \par
Experimental setups that lack mirror real-world scenarios
Old datasets. & CIC-IDS2017 \\\hline

 Rosay et al. \cite{rosay2022network} (2022) & \gls{knn}, RF & \gls{dos}/\gls{ddos} & Enhanced feature extraction with the new tool, LycoSTand. \par
Achieves high accuracy. \par
 & 
Uses old datasets.\par
Uses traditional \gls{ml} algorithms.  \par
 & CIC-IDS2017 \cite{sharafaldin2018toward}

\\ \hline
\end{tabular}}
\end{table}

A prior study by Khatun et al. \cite{khatun2024time} investigates the use of \glspl{cnn} for anomaly detection in time-series data in healthcare-\gls{iot} environments. The study simulates \gls{ddos} attacks using the Cooja simulator and focuses on environmental sensing data, including temperature and humidity. The proposed \gls{cnn}-based approach achieves 92\% accuracy in attack detection. The authors also introduce the WSN\_DDoS\_Attack\_H-IoT2023 dataset, which is tailored to environmental \gls{iot} scenarios and may have limited generalisability to heterogeneous healthcare sensing devices.

Altulaihan et al. \cite{altulaihan2024anomaly} propose an \gls{ids} for enhancing \gls{iot} security against \gls{dos} attacks. The approach focuses on anomaly detection using supervised \gls{ml}, particularly leveraging \glspl{dt} and Random Forest classifiers. The system trains on the IoTID20 dataset with feature selection \cite{ullah2020scheme}. While DT provides superior efficiency in training and testing times, the study highlights the challenge of balancing performance and computational efficiency in \gls{iot} environments. This consideration is underscored by previous research.

Chaganti et al. \cite{chaganti2022particle} developed a DNN-based system for detecting network attacks in the \gls{iomt}. Employing a \gls{dnn} within a comprehensive framework that includes Particle Swarm Optimisation for feature selection, the study utilised the ``Network Traffic and Patient Biometric Combined Dataset" to achieve a notable 96\% accuracy rate in network intrusion detection. As mentioned by the authors \cite{chaganti2022particle}, this high accuracy outperforms algorithms like \gls{cnn} and \gls{lstm}, effectively integrating network traffic with patient sensing data to secure \gls{iomt} systems. However, the study did not cover \gls{ddos} attack detection, a critical and persistent attack on \gls{iomt}. Furthermore, \gls{iomt} devices are not well-secured, making them easy targets for attacks like \gls{ddos}, which illustrates the need for advanced detection and prevention methods \cite{fagan2020foundational}. In contrast to earlier approaches, the proposed \gls{dl} approach, \gls{tcn}, uses a dynamic strategy to detect and mitigate \gls{ddos} attacks by blacklisting malicious nodes. As a result, it is a more effective solution for securing \gls{hiot} networks than previous methods.

Wagan et al. \cite{wagan2023fuzzy} develop the Duo-Secure \gls{iomt} framework, marking a notable advance in \gls{iomt} security. A \gls{bilstm} technique processes multi-modal sensory data while identifying attack patterns within \gls{iomt} networks. This approach effectively combines general network-flow metrics with patient biometrics collected from real healthcare monitoring systems, where devices and sensors continuously measure vital signs such as heart rate, blood pressure, and oxygen saturation, using the WUSTL-\gls{ehms}-2020 dataset \cite{hady2020intrusion}. \rev{The study also utilises a ``heart attack'' label generated by a fuzzy clustering algorithm rather than an actual medical diagnosis, indicating that the model is trained to predict clustering-based labels rather than clinically confirmed outcomes; therefore, the reported medical prediction performance does not represent true clinical prediction accuracy. 
From a network perspective, the authors utilise data from physical \gls{iomt} sensors; however, the underlying communication protocol used in the dataset is not explicitly defined or analysed, making it difficult to assess model performance under specific communication protocols in IoT environments.} \rev{Existing studies also rely on deep learning models for cyber threat detection; however, their evaluations are often conducted on datasets that do not explicitly account for communication-layer characteristics, limiting their effectiveness in capturing \gls{iot} network-level attack behaviours.}

\rev{To address these limitations, two datasets are generated using synthetic data that incorporate both \gls{mqtt} and \gls{udp} protocols, simulating realistic \gls{hiot} sensor behaviour.} The first dataset, {\em UL-ECE-MQTT-DDoS-H-IoT2025}, is created using the Cooja simulator, emulating \gls{wban} sensors with \gls{mqtt} protocol interactions. The second dataset, {\em UL-ECE-UDP-DDoS-H-IoT2025}, focuses on \gls{udp} protocol interactions to enable further analysis and assess the impact of \gls{ddos} attacks in \gls{hiot} environments.

Using simulated or synthetic data to train \gls{ml} and \gls{dl} models is a standard practice \cite{liu2024best}. Researchers widely use synthetic datasets, such as image data and text manipulation\cite{agrawal2024review, khatun2024time}, particularly attack data, as creating an attack model within physical \gls{hiot} devices is challenging. In this research, the simulated data includes physiological metrics such as body temperature, oxygen saturation, heart rate, time, and data per node to reflect real-world health data. This analysis enhances the understanding of \gls{mqtt} protocols and \gls{hiot} sensor activities. It ensures the dataset captures critical parameters for detecting and mitigating \gls{ddos} attacks in \gls{hiot} environments. Additionally, creating a dataset significantly contributes to cybersecurity because it helps improve attack detection, supports research, and enhances the training of security systems using \gls{ml}/\gls{dl} algorithms.

\rev{In \cite{faruqui2023safetymed}, the SafetyMed system, which employs Conv-\gls{lstm} for \gls{iomt} security, is characterised by a complex architecture and associated computational cost that may hinder deployment in resource-constrained \gls{iomt} environments.} \rev{Furthermore, the system is evaluated using the CIC-IDS2017 dataset~\cite{sharafaldin2018toward, rosay2022network}, which is designed for general intrusion detection rather than healthcare-specific \gls{iomt} environments.} \rev{This dataset lacks specific features, such as \gls{iomt} device data, \gls{g5}-based transmission characteristics, and the latest protocols (e.g., \gls{mqtt}) necessary for effectively representing \gls{g5}-based \gls{iomt} environments \cite{faruqui2023safetymed, elhanashi2022machine}.} \rev{This mismatch reduces the suitability of such datasets for evaluating security mechanisms in healthcare-specific \gls{iomt} environments.} \rev{Therefore, this research addresses this gap by generating protocol-specific sensor data and creating datasets to secure \gls{g5}-based \gls{hiot} environments.} \rev{In \cite{rosay2022network}, some traditional \gls{ml} algorithms, such as the \gls{knn} and Random Forest, are implemented using the CIC-IDS2017 dataset and achieve high accuracy; however, these results are obtained on a general-purpose dataset, which may not be representative of realistic \gls{iomt} environments.}


In \cite{kumar2023novel}, Kumar et al. propose an \gls{iomt} security model using \gls{cnn}, Bi-\gls{lstm}, and \gls{gru}, excelling in botnet attack detection. However, this research highlights several challenges. One major challenge is the diverse device configurations and network conditions that make real-life applications difficult in \gls{iomt} environments. The rapidly changing \gls{hiot} landscape underscores the need for adaptable and reliable security solutions. Additionally, resilient solutions are needed to defend against multiple simultaneous malware attacks in \gls{iot} and \gls{iomt} scenarios. Scalability is another critical challenge, particularly for protecting \gls{iot}-enabled cloud servers that process Electronic Health Records (EHRs) from various \gls{wban} devices. Therefore, despite the effectiveness of this approach in detecting botnet attacks, it lacks a mitigation strategy, leaving a gap in comprehensive security measures for critical attacks such as \gls{ddos} and botnet attacks in \gls{iomt} environments.

Binbusayyis et al. \cite{binbusayyis2022investigation} address significant vulnerabilities (e.g., \gls{dos}/\gls{ddos}) within \gls{iomt} by investigating various \gls{ml} algorithms, such as \gls{knn}, \gls{nb}, and \glspl{svm}, for intrusion detection. Based on the BoT-\gls{iot} dataset \cite{koroniotis2019towards}, the authors propose a DT model with 100\% accuracy. The dataset includes \gls{dos}-\gls{ddos}, scan, and theft attacks.
The research comprehensively addresses traditional attack vectors on the \gls{http} and \gls{tcp}. However, it notably lacks coverage of modern \gls{iot}-specific protocols such as \gls{mqtt}, which is increasingly popular in \gls{iomt} applications. Due to its lightweight nature, \gls{mqtt} is vulnerable to distinct security vulnerabilities that distinguish it from traditional web protocols \cite{hintaw2023mqtt,kant2021analysis}. Thus, this research gap presents a significant limitation. 

While prior studies have effectively addressed various \gls{iomt} attack detection methods, this proposed method focuses on \gls{tcn}-based \gls{ddos} attack detection and dynamic threshold-based mitigation. Unlike existing models, this approach achieves high prediction and detection accuracy and introduces an effective real-time mitigation strategy, significantly enhancing the security and performance of \gls{hiot} systems.

\section{Dataset Details for \gls*{mqtt}} \label{sec:mqttdataset}
This section describes the data collection experiment conducted using the \gls{iot} simulator Cooja. This research systematically develops \gls{ddos} attack models and collects data representing malicious and benign activities over the \gls{mqtt} protocol in an \gls{hiot} environment. Following data collection, preprocessing steps, including feature extraction, prepare the dataset for analysis, with a tabular summary of samples from the proposed {\em UL-ECE-MQTT-\gls{ddos}-H-IoT2025} dataset included for reference.

\subsection{Experimental Setup with Cooja Simulator}
The Cooja simulator in the experiment runs on Contiki OS, which supports the \gls{mqtt} protocol. Before hardware deployments, such as servers and motes, the simulation setup emulates MSP430-based SkyMote devices to compile code, evaluate, and perform system testing. This setup allows radio transmissions to be visualised. During the simulation stage, the system operates with the following technical specifications: 

\begin{itemize}
    \item CPU: 12th Gen Intel(R) Core(TM) i5-12500H   2.50 GHz
    \item Random Access Memory: 16 GB
    \item Operating System: Windows 11
    \item GPU: NVIDIA GeForce RTX
    \item Emulated Environment: Oracle VM VirtualBox 7.0 Platform
    \item Simulation Spotlight: Contiki 3.0 OS  
\end{itemize}

\begin{table*}[!htbp]
\caption{Simulation setup and parameters on Cooja.}
\label{tab:simulation}
\small
\adjustbox{max width=\textwidth}{
\begin{tabular}{l|l}
\hline
\multicolumn{1}{c|}{\textbf{Category}} & \multicolumn{1}{c}{\textbf{Value}}         \\ \hline
Platform                               & Cooja Contiki 3.0                            \\
Number of nodes                        & 14 nodes, 1 \gls{mqtt} Broker                           \\
Topology                               & star                                         \\
Area                                   & 400 m X 400 m                                \\
Sending rate (normal nodes)                          & 1 packet/60 Seconds                          \\
Sending rate (malicious nodes)                          & 1 packet/20 Seconds                          \\
Simulation run time                    & 5 hours                                       \\
Radio Medium Model                     & Unit Disk Graph for Distance Loss \\
Mote Type                              & SkyMote                                     \\
Duty Cycle                             & ContikiMAC                                   \\
Range of Nodes                         & Rx and Tx: 50 m, Interference: 100 m         \\
Physical Layer                         & IEEE 802.15.4                                \\
Protocol                             & \gls{mqtt} Broker \\ 
MAC Layer                              & ContikiMAC                                   \\
MAC Driver                             & CSMA/CA                                      \\
Channel Selection                      & By Default, 26                               \\
Network Layer                          & ContikiRPL                                   \\
Trafﬁc model                           & Constant Bit Rate                            \\ \hline
\end{tabular}}
\end{table*}

 \begin{figure}[ht!]
    \centering
\includegraphics[width=0.9\textwidth]{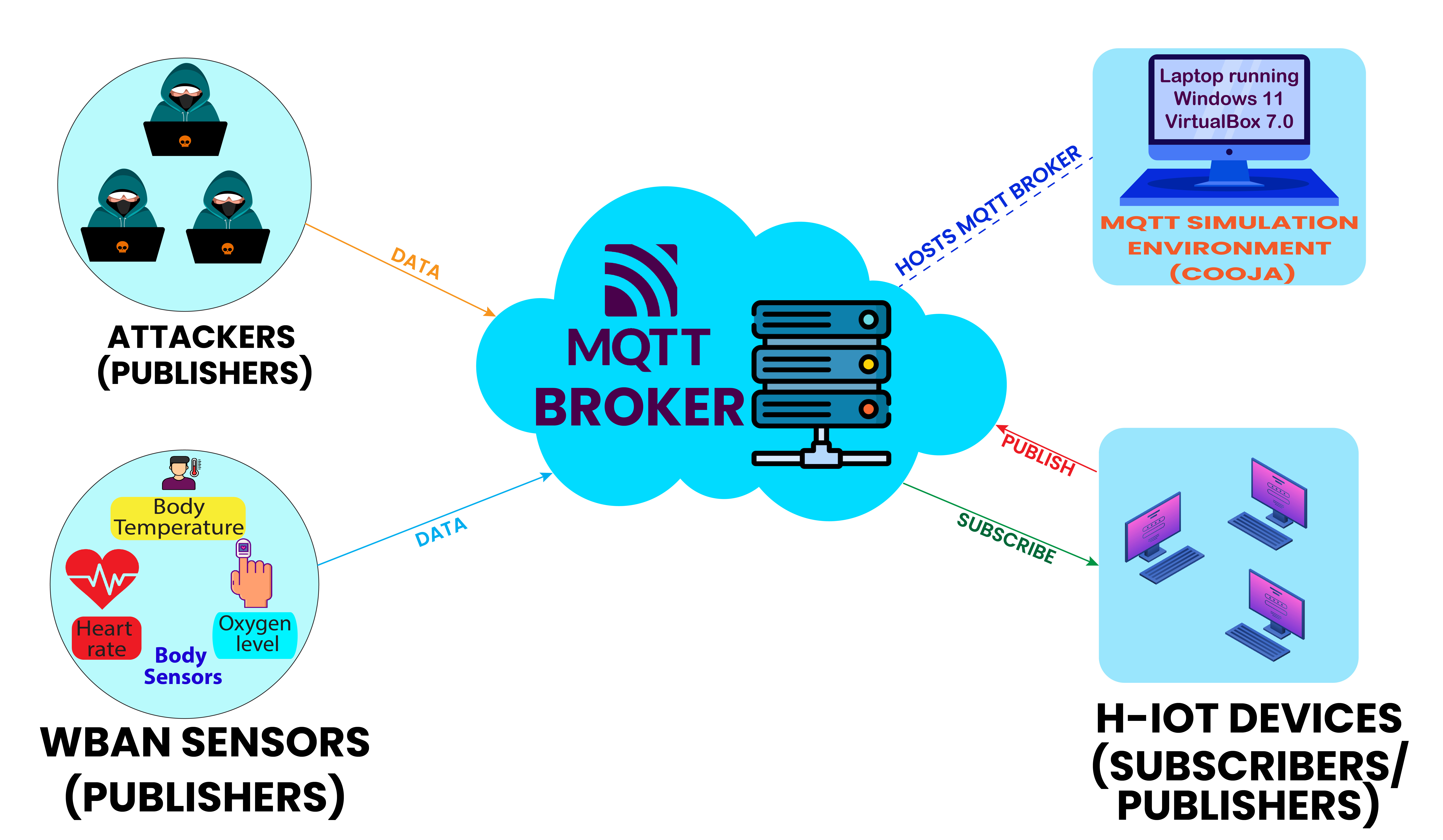}
    \caption[The illustration shows data capturing in a simulated H-IoT]{The illustration shows data capturing in a simulated \gls{hiot} environment, where \gls{wban} sensors (e.g., body temperature) and attackers publish data to the \gls{mqtt} broker, which forwards it to \gls{hiot} devices for monitoring, processing, and application-layer tasks.} 
    \label{fig:datacapture}
\end{figure}

\subsubsection{\gls{ddos} Attack Simulation Setup Over \gls*{mqtt} with Cooja}
The Cooja simulator simulates the network setup and assesses topology metrics for \gls{mqtt}-based communication in an \gls{hiot} environment. Table~\ref{tab:simulation} outlines the simulation setup details, including the selection of emulated nodes used to depict the network behaviour of \gls{hiot} devices. As depicted in Figure~\ref{fig:datacapture}, publishers in the \gls{mqtt} system, such as \gls{wban} sensors, publish data (e.g., body temperature) to the broker using a ``PUBLISH'' request, specifying the topic name (e.g., body temperature). Upon receiving this data, the broker forwards it to all subscribed devices. These \gls{hiot} devices (subscribers/publishers) can subscribe to specific topics by sending the broker a ``SUBSCRIBE'' request \cite{alotaibi2024secure}. For instance, they might subscribe to the `body temperature' topic to access data for visualisation and detailed analysis in the application layer. These devices can also act as publishers by communicating processed information or triggering alerts, allowing real-time monitoring and decision-making in \gls{hiot} systems.

\subsubsection{Topology} 
The network topology, organised in a star configuration, illustrates unique scenarios in \gls{mqtt}-enabled \gls{wban} \gls{hiot} sensors, with some operating normally and others experiencing malicious activity. Therefore, the regular and adversarial nodes are strategically distributed in both scenarios to simulate realistic and adaptable \gls{hiot} environments. The simulations are conducted over 5 hours and repeated several times to ensure the scalability and reliability of the agile network topology. An agile network design can adapt to fluctuations in data packet rates between normal and malicious nodes, providing accurate data transmission during \gls{ddos} attacks \cite{rafiq2024delay}.

\begin{figure}[ht!]
    \centering
    \includegraphics[width=0.9\textwidth]{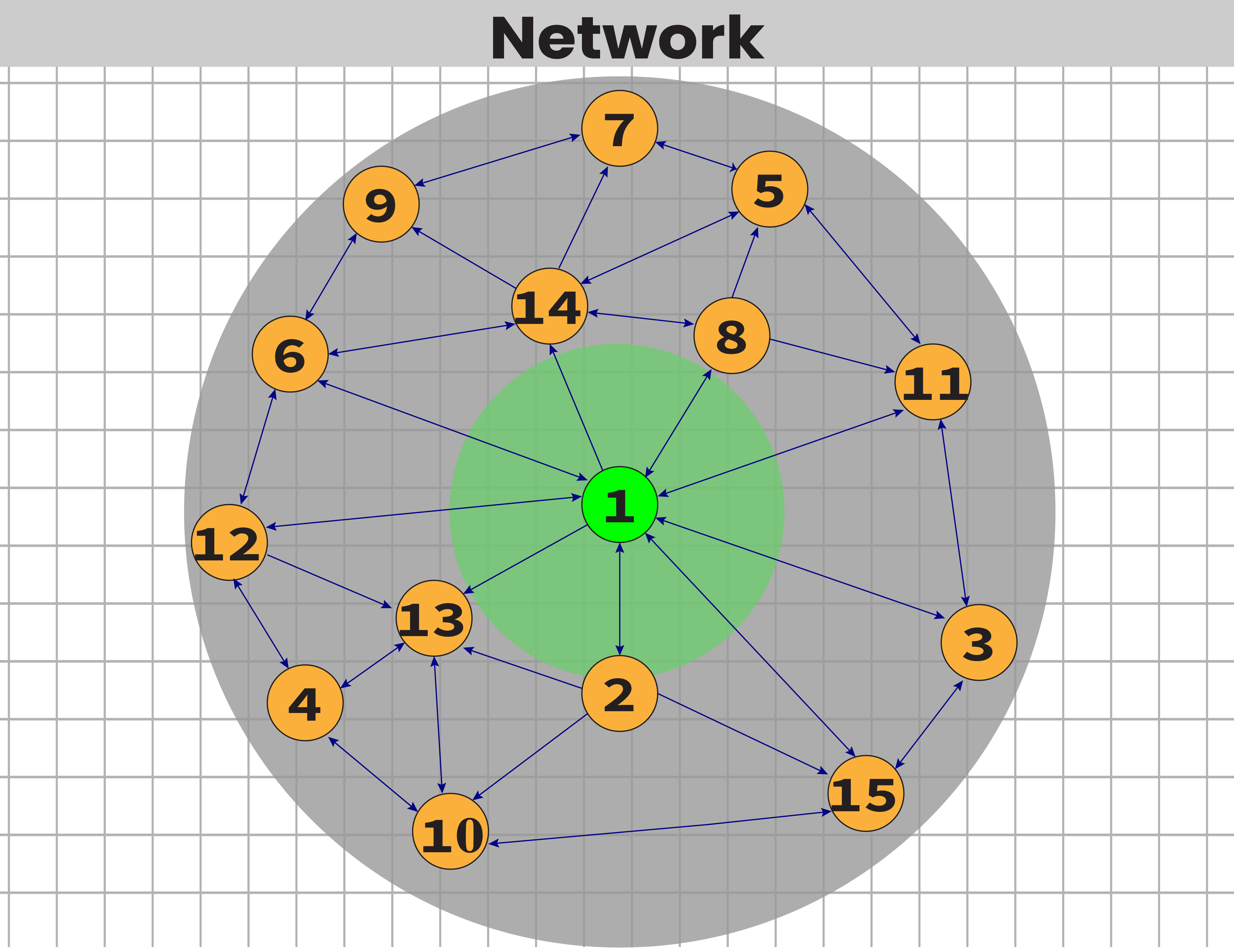}
 \caption[The baseline network topology in a simulated H-IoT]{The baseline network topology in a simulated \gls{hiot} environment over the \gls{mqtt} protocol using the Cooja simulator, with the green node representing the broker/server and the yellow nodes denoting normal \gls{hiot} sensors.}
    \label{fig:setup1}
\end{figure}

 \subsubsection{Scenario 1} Using the Cooja simulator in the baseline \gls{mqtt} configuration, 14 motes are strategically distributed to simulate a typical \gls{hiot} environment. It includes medical sensors that monitor key physiological parameters such as body temperature, oxygen saturation levels, and heart rate. Node ID 1, representing the \gls{mqtt} broker, a message-routing hub, is highlighted with a green circle, as depicted in Figure~\ref{fig:setup1}. This broker coordinates communication among sensor nodes (IDs 2 to 15), each positioned at specific locations where medical devices monitor patient vital signs, such as heart rate, oxygen saturation, and body temperature, thereby contributing to precision healthcare, as depicted in Figure~\ref{fig:setup1}. Additionally, normal nodes transmit data at intervals of 60 seconds, reflecting typical communication behaviour.

\subsubsection{Scenario 2} 
\begin{figure}[hbt!]
    \centering
    \includegraphics[width=0.9\textwidth]{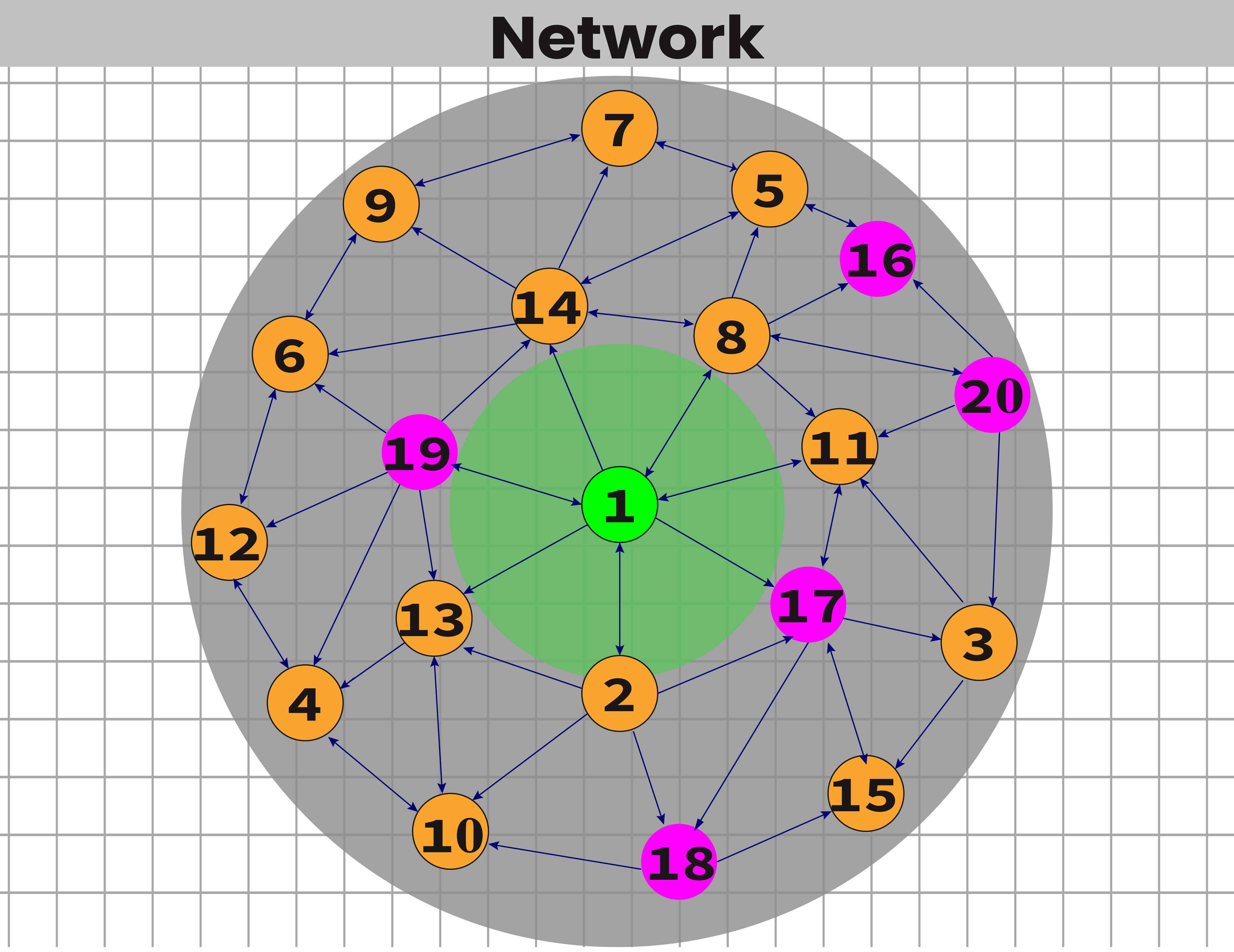}
    \caption[DDoS attack on the network topology in a simulated H-IoT]{\gls{ddos} attack on the network topology in a simulated \gls{hiot} environment over the \gls{mqtt} protocol using the Cooja simulator. The green node represents the broker/server, the yellow nodes denote normal \gls{hiot} sensors, and the purple nodes represent \gls{ddos} attackers.}
    \label{fig:setup2}
\end{figure}

In a simulated \gls{hiot} environment, a \gls{ddos} attack scenario is presented, targeting an \gls{mqtt} broker system. The attack involves a group of malicious motes, specifically nodes with IDs from 16 to 20, highlighted as purple motes in the visualisation, as depicted in Figure~\ref{fig:setup2}. These nodes rapidly send data to the \gls{mqtt} broker (Node ID 1), consuming significant radio transmission energy to overload the server. The malicious nodes transmit data at shorter intervals, every 20 seconds, to generate aggressive traffic patterns. These aggressive traffic patterns indicate malicious nodes send data much more frequently than normal, potentially exceeding the network’s capacity. This frequent data transmission can cause slower server response times and, in many cases, prevent legitimate devices from accessing the network. Thus, it is a common attack strategy to disrupt regular network activity.

\subsection{Data Acquisition with Cooja simulator}
This research uses the Cooja simulator to collect data through simulations~\cite{khatun2024time}. Three \gls{mqtt}-enabled \gls{wban} \gls{hiot} sensors are emulated to monitor body temperature, heart rate, and oxygen levels. This simulation generates raw data representing patient health metrics, such as a body temperature of 98.6°F, a heart rate of 72 beats per minute, and an oxygen saturation level of 98\%. Additionally, it provides timestamps and node IDs to capture information for each node's activity. It also generates data at randomised intervals to mimic real-world variations, enabling a thorough analysis of each sensor's values in healthcare monitoring.

\subsection{Data Preprocessing of \gls*{mqtt} Dataset}
The first step is establishing a baseline profile for generating datasets from \gls{hiot} sensors’ raw data packets. The baseline profile consists of three features, including \textbf{Time}, \textbf{Mote}, and \textbf{Message}. Four new features are subsequently extracted based on these baseline features, as outlined in Table~\ref{tab:DDoSdataset}.

\begin{table}[!htbp]
\centering
\caption[Sample from the \textbf{{\em UL-ECE-MQTT-DDoS-H-IoT2025}}]{Sample from the \textbf{{\em UL-ECE-MQTT-DDoS-H-IoT2025}} dataset.}
\renewcommand{\arraystretch}{1.2} 
\small
\begin{tabular}{lccc}
\toprule
\textbf{ids} & \textbf{total\_messages} & \textbf{total\_messages\_each\_node} & \textbf{frequency} \\ \hline
\midrule
18 & 10006  & 1008 & 10.073956626624025 \\ \hline
3  & 10007  & 350  & 3.49755171380034   \\ \hline
18 & 10008  & 1009 & 10.08193445243805  \\ \hline
16 & 10009  & 926  & 9.251673493855531  \\ \hline
14 & 10010  & 355  & 3.546453464535465  \\ \hline
17 & 10011  & 967  & 9.659374687843373  \\ \hline
3  & 10012  & 351  & 3.5057930483419897 \\ \hline
7  & 10013  & 337  & 3.3656246879057226 \\ \hline
12 & 10014  & 324  & 3.2354703415218693 \\ \hline
17 & 10015  & 968  & 9.665501747378931  \\ \hline
20 & 10016  & 986  & 9.844249201277954  \\ \hline
18 & 10017  & 1010 & 10.082859139462913 \\ \hline
4  & 10018  & 344  & 3.433819125573967  \\ \hline
20 & 10019  & 987  & 9.851282563130052  \\ \hline
20 & 10020  & 988  & 9.860279441117765  \\ \hline
10 & 10021  & 577  & 5.75790839237601   \\ \hline
18 & 10022  & 1011 & 10.087806824985034 \\ \hline
3  & 10023  & 352  & 3.5119225787070438 \\ \hline
20 & 10024  & 989  & 9.86632083000798   \\ \hline
16 & 10025  & 927  & 9.246882793017457  \\ 
\bottomrule
\end{tabular}
\vspace{-4pt}
\begin{tablenotes}
    \item \textbf{Note:} \textbf{ids:} Node IDs, \textbf{total\_messages:} Total Messages, \textbf{total\_messages\_each\_node:} Total Messages Each Node, \textbf{frequency:} Frequency. Each feature listed in the table corresponds to a column in the dataset, with dataset column names presented in monospace font (e.g., \texttt{total\_messages}) to represent the features.
\end{tablenotes}
\label{tab:DDoSdataset}
\end{table}

\subsubsection{Feature Calculation}
Four features, such as node IDs, total messages, average total messages, and total messages for each node, are extracted from raw data descriptions, such as packet traces and log data, through pattern matching \cite{khatun2019malicious}, \cite{khatun2024time}. 

\subsubsection{Node IDs}
In the Cooja simulation, IDs serve as unique identifiers for nodes and servers, facilitating efficient tracking and management of client-server interactions.
    
\subsubsection{Total Messages}  
The ``Total Messages'' metric counts the number of messages sent between nodes (e.g., 20 nodes) in a network. When this count increases, it indicates that an additional message has been sent. Thus, it is possible to determine how effectively the network communicates.
    \begin{itemize}
        \item Let \( M_i \) denote the presence of a message at \( i \)-th interval, with a value of 1 if a message is present and 0 otherwise.
        \item Let \( N \) denote the total number of data records.
    \end{itemize}
In this case, the total number of messages \( T \) up to the \( N \)-th observation is defined in Equation \ref{equation:i}:
    \begin{equation}
        T = \sum_{i=1}^{N} M_i
        \label{equation:i}
    \end{equation}
The equation sums each message indicator \( M_i \) together to determine the total number of messages.

\subsubsection{Total Messages Each Node} 
The ``Total Messages Each Node'' metric quantifies the number of messages each node in the network has sent. This metric reflects the activity level of each node by counting all the messages it sends. For each node \( i \), the total number of messages sent is calculated as:
    \begin{equation}
        total\_messages\_each\_node_i = \sum_{j=1}^{N} M_{\text{node}_i,j}
        \label{equation:ii}
    \end{equation}
where \( M_{\text{node}_i,j} \) counts the messages sent by node \( i \) at the \( j \)-th time period. The sum of these counts from 1 to \( N \) provides the total messages sent by node \( i \).

\subsubsection{Message Frequency Proportion (frequency)}
This feature displays each node's number of messages as a percentage of the total number of messages in the network. The equation calculates each node's share of the total volume of messages. Let \( M_{\text{node}_i} \) be the total messages from each node, and \( M_{\text{total}} \) be the total number of messages. The frequency percentage can be calculated using Equation \ref{equation:freq}, where \(\text{Frequency (\%)}\) represents the percentage of total messages sent by a specific node:
    \begin{equation}
        \text{frequency (\%)} = \left( \frac{M_{\text{node}_i}}{M_{\text{total}}} \right) \times 100
        \label{equation:freq}
    \end{equation}
Therefore, in this detection mechanism, the message frequency of each node is compared against a predefined threshold. The threshold value is set at five frequency values. It is interpreted as a potential attack or an anomaly if the frequency of messages from a node exceeds this threshold (e.g.,\textit{ frequency \textgreater{} 5}). In this case, it may indicate that a node is acting abnormally or is infected. For instance, as detailed in the sample dataset in Table~\ref{tab:DDoSdataset}, the ``frequency'' of messages each node sends is highlighted as a crucial metric. 

\subsubsection{Handling Class Imbalance with \gls{smote}} \label{subsubsec:smote}
The \gls{smote} addresses the imbalance between ``Malicious'' and ``Normal'' classes in the training data. \gls{smote} generates synthetic samples for the minority class, ``Normal,'' as illustrated in Table~\ref{tab:mqttclasssample}. This process involves selecting random samples from the minority class and creating new instances by calculating their differences, as shown in Equation \ref{eq:smote}. Consequently, the number of ``Normal'' samples increases to match the number of ``Malicious'' samples, resulting in a balanced training set \cite{khalaf2024detection}. This step is critical for reducing bias and enhancing the model's ability to detect both classes effectively. \rev{To address class imbalance, \gls{smote} is selected over common strategies such as random oversampling, undersampling, and class weighting (i.e., assigning higher loss weights to minority-class samples during training)~\cite{gao2026comprehensive}. For instance, random oversampling repeats existing minority-class samples, which may increase the likelihood of overfitting \cite{alkhawaldeh2023challenges}. In contrast, undersampling reduces class imbalance by removing samples from the majority class, which may lead to the loss of important majority-class patterns, such as normal network traffic behaviour \cite{altalhan2025imbalanced}. Compared to more complex variants such as Adaptive Synthetic Sampling (ADASYN), which may introduce noisy samples near decision boundaries between normal and malicious traffic \cite{alkhawaldeh2023challenges}, \gls{smote} provides a simpler approach for generating synthetic minority samples to balance the minority class distribution relative to the majority class in numerical network traffic data. Therefore, \gls{smote} is applied only to the training data after the train–test split to avoid bias and ensure fair evaluation. Specifically, the synthetic sample generation process in \gls{smote} is mathematically defined as follows:}
\begin{equation}
\mathbf{x}_{\text{new}} = \mathbf{x}_i + \delta \times (\mathbf{x}_j - \mathbf{x}_i)
\label{eq:smote}
\end{equation}
Where:
\begin{itemize}
    \item $\mathbf{x}_i$ is a randomly selected minority class sample.
    \item $\mathbf{x}_j$ is one of the k-nearest neighbors of $\mathbf{x}_i$ (also from the minority class).
    \item $\delta$ is a random number between 0 and 1.
    \item $\mathbf{x}_{\text{new}}$ is the newly generated synthetic sample.
\end{itemize}

\section{Dataset Details for \gls*{udp}} \label{sec:udpdataset}
This section describes the data collection experiment using the ns-3 network simulator, focusing on \gls{ddos} attack scenarios over the \gls{udp} protocol. \gls{ddos} attack models are developed to capture both malicious and benign activities in a \gls{g5}-based \gls{hiot} environment. Following data collection, preprocessing steps, such as feature extraction, are conducted to prepare the dataset for training the proposed \gls{tcn} model. A tabular overview of samples from the created dataset, {\em UL-ECE-UDP-DDoS-H-IoT2025}, is provided for reference in Table~\ref{tab:NS3DDoSdataset}. 

\subsection{Experimental Setup with ns-3 Simulator}
The ns-3 simulator, an open-source tool for discrete-event network simulation, is used in this experimental setup \cite{larionov2024methodology}. This simulation involves creating a model of \gls{ddos} attacks over the \gls{udp} protocol within a \gls{g5}-based \gls{hiot} environment. It runs on a Windows 11 host system, with Ubuntu 24.10 installed through Oracle VM VirtualBox 7.0. The simulation utilises ns-3 version 3.39, a simulator with modules that support many protocols. These include \gls{tcp}, \gls{udp}, IPv4, IPv6, routing protocols (e.g., OLSR), and wireless protocols (e.g., \gls{lte} and \gls{g5} \gls{nr}) \cite{larionov2024methodology}. The experimental parameters and system specifications used for the simulation are detailed in Table~\ref{tab:ns3simulationsetup}.

\begin{table}[!htbp]
\caption[Simulation setup and parameters]{Simulation setup and parameters on ns-3.}
\label{tab:ns3simulationsetup}
\resizebox{\columnwidth}{!}{%
\begin{tabular}{@{}ll@{}}
\toprule
\multicolumn{1}{c}{\textbf{Simulation System Setup}} & \multicolumn{1}{c}{\textbf{Value}}                            \\ \midrule
Operating System                                     & Ubuntu 24.10 on Oracle VM VirtualBox 7.0 \\
CPU Specifications       & 12th Gen Intel Core i5-12500H @ 2.50 GHz   \\
GPU                      & NVIDIA GeForce RTX                         \\
RAM Specifications       & 16 GB                                      \\
ns-3 Version             & 3.39                                                    \\ \midrule
\multicolumn{2}{c}{\textbf{Simulation Experimental Parameters}}       \\ \midrule
Network Environment      & \gls{g5}                                         \\
Protocol                & \gls{udp} 
                    \\
Topology                 & Star  
                    \\ 
Data Rate (Normal nodes) & 124 kb/s                                   \\
Data Rate (\gls{ddos} attack-affected nodes)   & 248 kb/s                                   \\
Number of Clients        & 14                                         \\
Number of \gls{ddos} attack-affected nodes     & 5                                          \\
Extra \gls{ddos} attack-affected nodes         & 3                                          \\
Extra \gls{ddos} Attack Start Time    & 20 s                                       \\
Extra \gls{ddos} Attack Interval      & 10 s                                       \\
Payload Size             & 512 bytes                                  \\
Max Simulation Time      & 120 s \\ \bottomrule
\end{tabular}%
}
\end{table}

\subsubsection{\gls*{ddos} Attack Simulation Setup Over \gls*{udp} with ns-3} \label{subsubsec:ddosns-3}
\begin{figure*}[ht!]
    \centering
    \includegraphics[width=\textwidth]{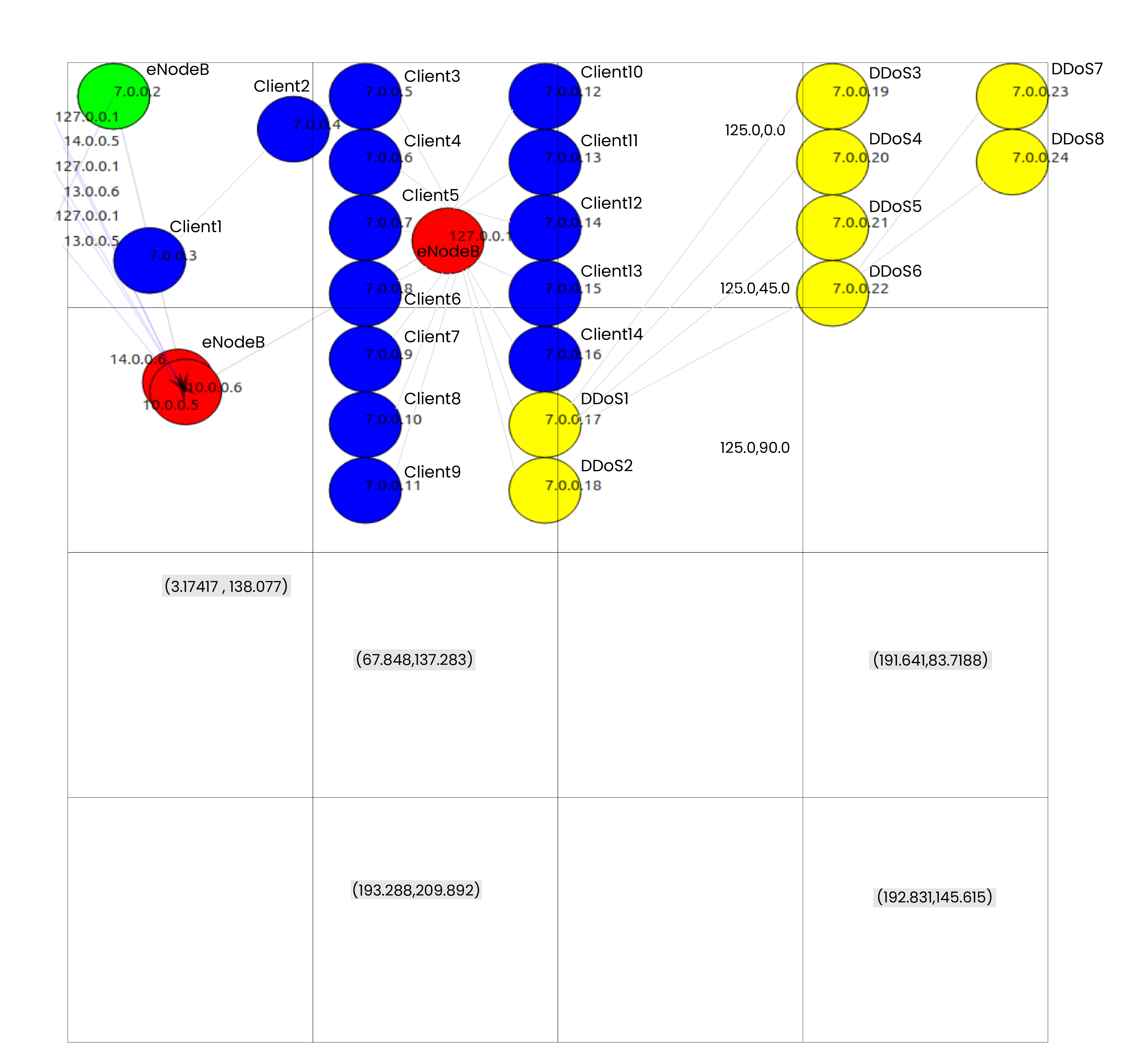}
    \caption[DDoS attack simulation over the UDP protocol]{\gls{ddos} attack simulation over the \gls{udp} protocol using ns-3 in an \gls{hiot} environment. Green nodes represent the server, red nodes represent eNodeBs, and blue nodes represent clients (Nodes 5 to 18: Client 1 to Client 14). Yellow nodes represent \gls{ddos} attackers (Nodes 19 to 23: \gls{ddos} 1 to \gls{ddos} 5). Following the baseline attack simulation setup, additional attackers (Nodes 24 to 26: \gls{ddos} 6 to \gls{ddos} 8) are dynamically introduced to reinforce the attack within the same network topology. This simulation highlights the impact of \gls{ddos} attacks on \gls{hiot} networks, emphasising patterns in normal and malicious data.}
    \label{fig:DDoSAttackNS3}
\end{figure*}

This research uses the ns-3 simulator to create a \gls{ddos} attack simulation setup in a \gls{g5}-based \gls{hiot} environment. This setup employs a hierarchical topology, organising network nodes in layers with the \gls{udp} protocol to facilitate efficient data flow and management \cite{rivas2024tefnet24}. As depicted in Figure~\ref{fig:DDoSAttackNS3}, the red nodes represent the eNodeBs (evolved Node B) and act as \gls{lte} base stations. These nodes handle wireless communication between the core network and various user equipment, such as smartphones, tablets, \gls{hiot} devices, and wearable health monitors \cite{heijligenberg2024attacks}. Multiple eNodeBs are commonly deployed to ensure broad coverage and seamless handover \cite{ross2024fixing}. A green node (single server or target UE) receives legitimate and malicious traffic in the \gls{hiot} environment. Figure~\ref{fig:DDoSAttackNS3} also depicts multiple client nodes, highlighted in blue, that simulate \gls{hiot} devices generating real-time traffic. These \gls{hiot} devices transmit health-related data, such as body temperature, oxygen levels, and heart rate, to the server (target UE). The \gls{ddos} attack-affected nodes, denoted in yellow, are configured to simulate malicious traffic directed towards the server, allowing analysis of the effects of the attacks on the network. This hierarchical structure supports high device density, low latency, and efficient traffic management, which is critical for optimal \gls{g5} performance in complex environments such as \gls{hiot}. The \gls{ddos} attack simulation setup allows observation of its impact on a centralised server within the \gls{hiot} environment.

The simulation runs for 120 seconds and processes over one million malicious and benign data records. The \gls{hiot} environment is composed of different types of nodes, including normal \gls{hiot} nodes (clients) numbered from 5 to 18 (client 1 to Client 14), which generate real-time traffic using the \gls{udp} protocol to the target UE with \gls{ip} addresses in the range 7.0.0.3 to 7.0.0.16. \gls{ddos} attack-affected nodes numbered 19 to 23 (\gls{ddos} 1 to \gls{ddos} 5), assigned \gls{ip} addresses from 7.0.0.17 to 7.0.0.21, simulate malicious traffic to overwhelm the target UE (7.0.0.2, the central server), as depicted in Figure~\ref{fig:DDoSAttackNS3}. Extra \gls{ddos} attack-affected nodes numbered from 24 to 26 (\gls{ddos} 6 to \gls{ddos} 8), with \gls{ip} addresses from 7.0.0.22 to 7.0.0.24, are introduced dynamically during the simulation to increase the intensity of the attack. These extra nodes are added at intervals of 10 seconds, starting 20 seconds into the simulation. This setup illustrates a realistic scenario where \gls{ddos} attack nodes appear as new malicious nodes at different times within the same topology. The data transmitted from these extra \gls{ddos} attack-affected nodes to the central server is then analysed to assess the impact of the new malicious nodes. This is crucial for testing the \gls{tcn} model’s ability to detect new malicious nodes that appear in the same topology at different times.

Moreover, the coordinates in Figure~\ref{fig:DDoSAttackNS3} define the nodes' relative positions in the simulated network environment. These spatial values (e.g., 125.0, 90.0) indicate the location of each node on a 2D grid in the realistic simulated behaviour \cite{bel2024co}. For instance, the distance between two \gls{iot} nodes can affect signal strength and latency. \gls{iot} nodes located farther from the central server typically experience weaker signal strength and higher transmission delays than those closer to the server. Setting these positions accurately in the simulation setup is essential for creating realistic network behaviour. In \gls{hiot} scenarios, if an \gls{iot} sensor (such as one monitoring heart rate) is far from the server, it may face challenges transmitting data reliably, especially during network disruptions caused by \gls{ddos} attacks. Understanding the impact of node placement on performance is essential for enhancing network reliability. 

\subsection{Data Acquisition with ns-3 simulator}
This research establishes a \gls{wban}-based \gls{g5} \gls{hiot} environment with \gls{udp} protocol to run simulations for data collection using the ns-3 simulator \cite{ramonet2024perspectives}. This simulation enables a detailed evaluation of each \gls{hiot} sensor's performance by analysing metadata, such as data packet size, payload length, and protocol, specific to \gls{g5} healthcare environments.

\subsection{Data Preprocessing of \gls*{udp} Dataset}
Using the ns-3 simulator, simulations run over the \gls{udp} protocol to capture a wide range of network traffic characteristics from \gls{g5}-enabled \gls{hiot} \gls{wban} sensors. The raw data includes packet sizes, transmission rates, latency, and protocol information, all recorded in a text file. The data also includes timestamps, node IDs, transmission (TX) and reception (RX) events, and IPv4 header information, such as TTL (Time To Live), which sets the maximum number of hops a packet can make before it is discarded. Another important field is the \gls{dscp}, which is used for \gls{qos} management in \gls{ip} networks. Total data packet and payload lengths are recorded, offering a comprehensive view of \gls{hiot} network behaviour.


The preprocessing of ns-3 data involves several vital steps to ensure it is clean, organised, and ready for analysis. First, essential features such as timestamps, node identifiers, protocol types, and packet-specific details are extracted by identifying patterns within the data using regular expressions (\texttt{regex}), a technique for defining search patterns in text. \texttt{Regex} allows for precise matching and extraction of specific data elements based on defined sequences of characters. Next, data cleaning processes remove irrelevant entries, address missing or invalid data, and standardise protocol and header fields for consistency. Standardising protocol fields ensures that protocol names or values follow the same format, essential for accurate analysis. Lastly, feature engineering converts categorical data into numerical form and calculates message counts. It also computes key traffic metrics such as message frequency, inter-arrival times, and monitoring frequency. These metrics enable a detailed analysis of network behaviour.

The dataset's features are described as follows.

\subsubsection{Node IDs (ids)} In ns-3 simulations for \gls{g5} \gls{hiot} environments, node IDs serve as unique identifiers for various network nodes, including base stations, servers, \gls{ue} such as smartphones, wearable health monitors, and other \gls{hiot} devices.
    
\subsubsection{Time Elapsed (time)} This feature represents the number of seconds that have passed since the first timestamp is recorded in the simulation. Here, \( time \) denotes the elapsed time. The time elapsed is calculated as the difference between the current timestamp (\( time_{\text{current}} \)) and the initial timestamp (\( time_{\text{first}} \)):
\begin{equation}
time = time_{\text{current}} - time_{\text{first}}
\end{equation}

\subsubsection{Protocol (proto)} The \gls{udp} protocol is used in the ns-3 simulation to communicate with \gls{hiot} devices. The protocol is extracted from the log line by identifying specific keywords, such as ``ns3::UdpHeader.'' Once identified, it is converted into a numerical code (e.g., \gls{udp} as 0) for further processing and analysis. In the dataset, the protocol is represented by this numerical code (e.g., 0), making it suitable for \gls{ml} models or statistical analysis.

\subsubsection{Source \gls*{ip} (src\_ip)} This feature represents the \gls{ip} address of the \gls{iot} device that originates data traffic in the healthcare system. The source \gls{ip} is extracted from the log line by locating the `ns3::Ipv4Header' section where \gls{ip} addresses are listed in the format \textless source\_ip\textgreater{}~$>$. For example, if the log line shows `7.0.0.21 $>$ 7.0.0.2', the source \gls{ip} is `7.0.0.21'. After extracting the source \gls{ip} from the logline, each unique address corresponding to an \gls{iot} device is converted into a unique numerical value (e.g., `1, 2, 3, ...') for processing and analysis. Let \( s \) denote the source \gls{ip}, and the conversion can be defined as:
\begin{equation}
numeric\_value(s_i) = index(s_i) + 1
\end{equation}
Where \( \text{Index}(s_i) \) is the zero-based index of each unique source \gls{ip}. This encoding simplifies the representation of \gls{ip} addresses, making them suitable for processing and analysis in \gls{ml} models or statistical computations. The source \gls{ip} helps identify where the traffic originates. In a \gls{ddos} attack, multiple requests might come from the same \gls{ip} (or a set of spoofed \glspl{ip}), or a botnet might use a range of \glspl{ip} \cite{khatun2019malicious}. Monitoring unusual traffic patterns from specific \gls{ip} addresses can indicate a potential \gls{ddos} attack.


\subsubsection{Destination \gls*{ip} (dst\_ip)} This feature specifies the destination \gls{ip} address in the \gls{hiot} network. The destination \gls{ip} is extracted from the log line in the ‘ns3::Ipv4Header’ section, where \gls{ip} addresses are listed in the format \verb|<source_ip> > <destination_ip>|. It is the value that appears immediately after the $>$ symbol. In the dataset, the destination \gls{ip} is always the same (e.g., ‘7.0.0.2’), representing a central server or base station that receives traffic from all \gls{hiot} devices. This encoding simplifies the representation of \gls{ip} addresses, making them suitable for \gls{ml} models or statistical computations. Monitoring the destination \gls{ip} is essential for \gls{ddos} detection, as attackers often send many data traffic requests to a single \gls{ip} address to overload the server.

\subsubsection{Payload Size (p\_size)} This feature represents the payload size, in bytes, of the data transmitted by \gls{hiot} devices in the network. The payload size is extracted from the log line using a regular expression, a pattern-matching technique that locates the size value in the ‘Payload’ section. In this dataset, the payload size remains constant at 512 bytes.

\subsubsection{Total Messages (t\_msgs)} 
It indicates the cumulative count of messages processed up to the current log line. Let \( T \) denote the total number of messages, which can be calculated as:
\begin{equation}
    t\_msgs = \sum_{i=1}^{N} M_i
\end{equation}
Where \( N \) is the total number of log lines processed and \( M_i \) represents each message count (typically 1 per log line).

\subsubsection{Total Messages Same Node (t\_msgs\_s\_node)} This feature represents the cumulative count of messages from the same node, identified by \texttt{node\_id}, processed up to the current log line. Let \( t\_msgs\_s\_node_i \) denote the total number of messages from a specific node, identified by \texttt{node\_id}. The total messages from the same node can be computed as:
\begin{equation}
    t\_msgs\_s\_node_i = \sum_{j=1}^{N} M_{\text{same\_node}_i,j}
    \label{equation:iii}
\end{equation}
Where \( N \) is the total number of log lines processed and \( M_{\text{same\_node}_i,j} \) represents the message count for the same node (specified by \texttt{node\_id}) at each log line \( j \).

\subsubsection{Frequency (f)} This feature represents the frequency (percentage) of messages from the current \texttt{node\_id} relative to the total number of messages processed up to the current log line. Let $f$ denote the frequency, which can be calculated as:
\begin{equation}
f = \left( \frac{total\_m\_same\_node}{total\_m} \right) \times 100
\end{equation}
Where \( total\_m \) is the cumulative count of all messages processed up to the current log line.

\subsubsection{Mean Frequency (m\_f)} This feature represents the running mean frequency of messages from all nodes up to the current log line. Let $m\_f$ denote the mean frequency, which can be computed as:
\begin{equation}
    m\_f = \frac{frequency\_sum}{total\_number\_of\_frequencies}
\end{equation}

Where \(\texttt{frequency\_sum}\) is the cumulative sum of all computed frequency values, and \textit{total\_number\_of\_frequencies} represents the count of all frequency values. This ensures that the \textit{mean frequency} (\texttt{m\_f}) is calculated as the average cumulative frequency values across all nodes.

Moreover, to detect potential \gls{ddos} attack-affected nodes, the \textit{monitoring frequency} (\texttt{mon\_f}) of each node is compared with the \textit{mean frequency} (\texttt{m\_f}). If a node's \texttt{mon\_f} is greater than or equal to its \texttt{m\_f}, it is assigned a value of 1, classifying it as a potential \gls{ddos} attack-affected node. Conversely, if the \texttt{mon\_f} is below the \texttt{m\_f}, the node is assigned a value of 0, indicating normal behaviour, as detailed in Table~\ref{tab:NS3DDoSdataset}.

\begin{table}[ht!] 
\caption{Dataset sample from the {\em UL-ECE-UDP-DDoS-H-IoT2025}.}
\label{tab:NS3DDoSdataset}
\setlength{\tabcolsep}{3pt}  
\renewcommand{\arraystretch}{1.1}
\begin{adjustbox}{max width=\textwidth}
\begin{tabular}{@{}|
p{0.8cm}|  
p{1.2cm}|  
c|
c|
p{1.2cm}|  
c|
p{1.4cm}|  
c|
c|
c|
c|
c|
c|
@{}}
\toprule
\textbf{ids} &
  \textbf{time} &
  \textbf{proto} &
  \textbf{src\_ip} &
  \textbf{dst\_ip} &
  \textbf{p\_size} &
  \textbf{t\_msgs} &
  \textbf{t\_msgs\_s\_node} &
  \textbf{f} &
  \textbf{m\_f} &
  \textbf{mon\_t\_msgs} &
  \textbf{MTMSN} &
  \textbf{mon\_f} \\ \midrule
\multicolumn{1}{|l|}{16} &
  \multicolumn{1}{l|}{63.1411} &
  \multicolumn{1}{l|}{0} &
  \multicolumn{1}{l|}{4} &
  \multicolumn{1}{l|}{0} &
  \multicolumn{1}{l|}{512} &
  \multicolumn{1}{l|}{49991} &
  \multicolumn{1}{l|}{1912} &
  \multicolumn{1}{l|}{3.8246} &
  \multicolumn{1}{l|}{5.5964} &
 \multicolumn{1}{l|}{4400} &
\multicolumn{1}{l|}{152} &
3.4545 \\

\multicolumn{1}{|l|}{17} &
  \multicolumn{1}{l|}{63.1411} &
  \multicolumn{1}{l|}{0} &
  \multicolumn{1}{l|}{5} &
  \multicolumn{1}{l|}{0} &
  \multicolumn{1}{l|}{512} &
  \multicolumn{1}{l|}{49992} &
  \multicolumn{1}{l|}{1912} &
  \multicolumn{1}{l|}{3.8246} &
  \multicolumn{1}{l|}{5.5964} &
 \multicolumn{1}{l|}{4400} &
\multicolumn{1}{l|}{152} &
3.4545 \\

\multicolumn{1}{|l|}{18} &
  \multicolumn{1}{l|}{63.1411} &
  \multicolumn{1}{l|}{0} &
  \multicolumn{1}{l|}{6} &
  \multicolumn{1}{l|}{0} &
  \multicolumn{1}{l|}{512} &
  \multicolumn{1}{l|}{49993} &
  \multicolumn{1}{l|}{1912} &
  \multicolumn{1}{l|}{3.8245} &
  \multicolumn{1}{l|}{5.5963} &
\multicolumn{1}{l|}{4400} &
\multicolumn{1}{l|}{152} &
3.4545 \\

\multicolumn{1}{|l|}{19} &
  \multicolumn{1}{l|}{63.1411} &
  \multicolumn{1}{l|}{0} &
  \multicolumn{1}{l|}{7} &
  \multicolumn{1}{l|}{0} &
  \multicolumn{1}{l|}{512} &
  \multicolumn{1}{l|}{49994} &
  \multicolumn{1}{l|}{3824} &
  \multicolumn{1}{l|}{7.6489} &
  \multicolumn{1}{l|}{5.5964} &
 \multicolumn{1}{l|}{4400} &
\multicolumn{1}{l|}{303} &
6.8863 \\

\multicolumn{1}{|l|}{20} &
  \multicolumn{1}{l|}{63.1411} &
  \multicolumn{1}{l|}{0} &
  \multicolumn{1}{l|}{8} &
  \multicolumn{1}{l|}{0} &
  \multicolumn{1}{l|}{512} &
  \multicolumn{1}{l|}{49995} &
  \multicolumn{1}{l|}{3824} &
  \multicolumn{1}{l|}{7.6487} &
  \multicolumn{1}{l|}{5.5964} &
 \multicolumn{1}{l|}{4400} &
\multicolumn{1}{l|}{303} &
6.8863 \\

\multicolumn{1}{|l|}{21} &
  \multicolumn{1}{l|}{63.1411} &
  \multicolumn{1}{l|}{0} &
  \multicolumn{1}{l|}{9} &
  \multicolumn{1}{l|}{0} &
  \multicolumn{1}{l|}{512} &
  \multicolumn{1}{l|}{49996} &
  \multicolumn{1}{l|}{3824} &
  \multicolumn{1}{l|}{7.6486} &
  \multicolumn{1}{l|}{5.5964} &
  \multicolumn{1}{l|}{4400} &
\multicolumn{1}{l|}{303} &
6.8863 \\

\multicolumn{1}{|l|}{22} &
  \multicolumn{1}{l|}{63.1411} &
  \multicolumn{1}{l|}{0} &
  \multicolumn{1}{l|}{10} &
  \multicolumn{1}{l|}{0} &
  \multicolumn{1}{l|}{512} &
  \multicolumn{1}{l|}{49997} &
  \multicolumn{1}{l|}{3824} &
  \multicolumn{1}{l|}{7.6484} &
  \multicolumn{1}{l|}{5.5965} &
 \multicolumn{1}{l|}{4400} &
\multicolumn{1}{l|}{303} &
6.8863 \\

\multicolumn{1}{|l|}{23} &
  \multicolumn{1}{l|}{63.1411} &
  \multicolumn{1}{l|}{0} &
  \multicolumn{1}{l|}{11} &
  \multicolumn{1}{l|}{0} &
  \multicolumn{1}{l|}{512} &
  \multicolumn{1}{l|}{49998} &
  \multicolumn{1}{l|}{3824} &
  \multicolumn{1}{l|}{7.6483} &
  \multicolumn{1}{l|}{5.5965} &
 \multicolumn{1}{l|}{4400} &
\multicolumn{1}{l|}{303} &
6.8863 \\

\multicolumn{1}{|l|}{25} &
  \multicolumn{1}{l|}{63.1478} &
  \multicolumn{1}{l|}{0} &
  \multicolumn{1}{l|}{13} &
  \multicolumn{1}{l|}{0} &
  \multicolumn{1}{l|}{512} &
  \multicolumn{1}{l|}{49999} &
  \multicolumn{1}{l|}{2008} &
  \multicolumn{1}{l|}{4.0160} &
  \multicolumn{1}{l|}{5.5965} &
 \multicolumn{1}{l|}{4400} &
\multicolumn{1}{l|}{303} &
6.8863 \\

\multicolumn{1}{|l|}{26} &
  \multicolumn{1}{l|}{63.1555} &
  \multicolumn{1}{l|}{0} &
  \multicolumn{1}{l|}{14} &
  \multicolumn{1}{l|}{0} &
  \multicolumn{1}{l|}{512} &
  \multicolumn{1}{l|}{50000} &
  \multicolumn{1}{l|}{1403} &
  \multicolumn{1}{l|}{2.806} &
  \multicolumn{1}{l|}{5.5964} &
 \multicolumn{1}{l|}{4381} &
\multicolumn{1}{l|}{303} &
6.9162 \\ \bottomrule
\end{tabular}%
\end{adjustbox}
\begin{minipage}{0.95\linewidth}
\vspace{2pt}
\scriptsize
   \textbf{Note:} \textbf{ids:} Node IDs, \textbf{time:} Time Elapsed (time), \textbf{proto:} Protocol, \textbf{src\_ip:} Source \gls{ip}, \textbf{dst\_ip:} Destination \gls{ip}, \textbf{p\_size:} Payload Size, \textbf{t\_msgs:} Total Messages,
    \textbf{t\_msgs\_s\_node:} Total Messages Same Node, \textbf{f:} Frequency, \textbf{m\_f:} Mean Frequency, \textbf{mon\_t\_msgs\_s\_node (MTMSN):}  Monitoring Total Messages Same Node, 
    \textbf{mon\_t\_msgs:} Monitoring Total Messages, \textbf{mon\_f:} Monitoring Frequency.
\end{minipage}
\end{table}

\subsubsection{Monitoring Total Messages (mon\_t\_msgs)} This equation calculates the total number of messages monitored from all \gls{hiot} nodes in a network during the simulation. It is defined as the product of the total number of messages monitored from a single node and the total number of nodes in the network.

\begin{equation}
mon\_t\_msgs = \sum_{i=1}^{n} mon\_t\_msgs\_s\_node
\end{equation}

Where:
\begin{itemize}
\item \textbf{mon\_t\_msgs}: Monitoring Total Messages for all \gls{hiot} nodes.
\item \textbf{mon\_t\_msgs\_s\_node}: The number of messages monitored from a single \gls{hiot} node.
\item $n$: The total number of nodes participating in the simulation.
\end{itemize}

\subsubsection{Monitoring Total Messages Same Node (MTMSN)} This feature calculates the rate of messages periodically monitored from an individual node during the simulation. It is based on the total number of messages the individual node sends, the total simulation time, and the monitoring interval.
\begin{equation}
MTMSN = \sum_{i=1}^{n} \left( \frac{msgs\_per\_node}{total\_simulation\_time} \right) \times \Delta t
\end{equation}
Where:
\begin{itemize}
\item \textbf{mon\_t\_msgs\_s\_node (MTMSN)}: Monitoring Total Messages from the same node, representing the rate of messages exchanged per node within the same network or simulation setup.
\item $i = 1, 2, \dots, n$: Represents the time intervals in the simulation where monitoring occurs.
\item $\Delta t = 5 \, \text{sec}$: The monitoring time interval.
\item $n$: The number of intervals during the simulation.
\item $\frac{\text{msgs\_per\_node}}{\text{total\_simulation\_time}}$: The rate of messages sent by each node over the total simulation time.
\end{itemize}

\subsubsection{Monitoring Frequency (mon\_f):} This feature represents the frequency (percentage) of messages from the same \texttt{node\_id} within the last 5 seconds relative to all messages within that period, which is shown in Table~\ref{tab:NS3DDoSdataset}. Moreover, monitoring frequency based on existing data allows for identifying any new nodes appearing within this timeframe, as the message frequency from a newly appeared node within 5 seconds may be lower than that of established nodes in the same topology.
Let \texttt{mon\_f} denote the monitoring frequency, which can be calculated as:
\begin{equation}
mon\_f = \left( \frac{mon\_t\_msgs\_s\_node}{mon\_t\_msgs} \right) \times 100
\end{equation}
Where \( \text{mon\_t\_msgs} \) is the count of all messages within the last 5 seconds, and \( \text{mon\_t\_msgs\_s\_node} \) is the count of messages from the same \texttt{node\_id} within that period.

\section{Methodology}  \label{sec:tcnmethod}
\rev{The novelty of this research lies in (i) the development of a \gls{tcn}-based mechanism for real-time \gls{ddos} detection in \gls{hiot} environments, (ii) the use of two protocol-specific datasets (\gls{mqtt} and \gls{udp}) generated from realistic simulation setups, and (iii) the integration of detection and mitigation strategies tailored for resource-constrained \gls{hiot} systems.}

Based on these contributions, this research proposes a novel approach for detecting and mitigating \gls{ddos} attacks using a \gls{tcn} classifier in \gls{mqtt} and \gls{udp}-enabled wireless body area networks within \gls{g5} \gls{hiot} environments. The proposed method is applied to two datasets, {\em UL-ECE-MQTT-DDoS-H-IoT2025} and {\em UL-ECE-UDP-DDoS-H-IoT2025}. The model predicts \gls{ddos} attacks in the first dataset based on message frequency proportions. In the second dataset, the model predicts whether \gls{hiot} nodes are malicious or non-malicious based on monitoring frequency. Following these predictions, a dynamic threshold strategy facilitates the detection and mitigation of \gls{ddos} attack nodes in \gls{g5}-based \gls{hiot} systems, as illustrated in Algorithm~\ref{algorithm:algo}. Model performance evaluation experiments are conducted on a \textit{MacBook Pro} with an \textit{M1} chip.

\subsection{\gls*{tcn} Model Architecture for \gls{hiot} Nodes Classification}
\gls{tcn} models have multiple integrated components that process input data for malicious \gls{hiot} node classification. The input data, denoted as \(X\), is reshaped to correspond to the requirements of the \gls{tcn}'s 1D convolutional layers. Therefore, \(X\) is reshaped into a 3D array, with each sample represented as a 1D vector, suitable for processing by the convolutional layers.

The \gls{tcn} model has two residual blocks, consisting of three convolutional layers with 64 filters, a kernel size of 3, a dilation rate (1, 2, and 4, respectively, for the three layers), and causal padding. With a kernel size of 3, the number of parameters (e.g., weights and biases) grows linearly. This configuration effectively captures time-series or sequential data dependencies while maintaining computational efficiency. The dilation rates of 1, 2, and 4 allow the model to capture patterns across multiple time scales. These rates progressively increase the input range the model considers without significantly increasing the computational load. Accordingly, causal padding ensures that the output at any time step \(t\) is influenced only by inputs from time step \(t\) and earlier. For example, in TensorFlow's \texttt{Conv1D} function, causal padding (\texttt{padding='causal'}) adds zeros to the left of the input sequence, ensuring that future information (i.e., inputs from time steps greater than \(t\)) is not used in the prediction, thereby preserving the temporal causality. Batch normalisation follows each convolutional layer to stabilise the learning process and improve the convergence speed. After the second residual block, a dropout layer with a rate of 0.5 is applied to reduce overfitting.

Furthermore, L2 regularisation with a factor of 0.01 is applied to each convolutional layer to further enhance regularisation. A Gaussian noise layer with a standard deviation of 0.1 is added at the beginning of the model to improve robustness. This layer enhances generalisation by reducing overfitting and helping the model handle noisy or unseen data.

Following the convolutional layers, global average pooling is applied to the feature maps, reducing them to a single vector per sample. The model then concludes with a dense layer activated by a sigmoid function used for binary classification.


\begin{figure*}[ht!]
    \centering
    \includegraphics[width=\textwidth]{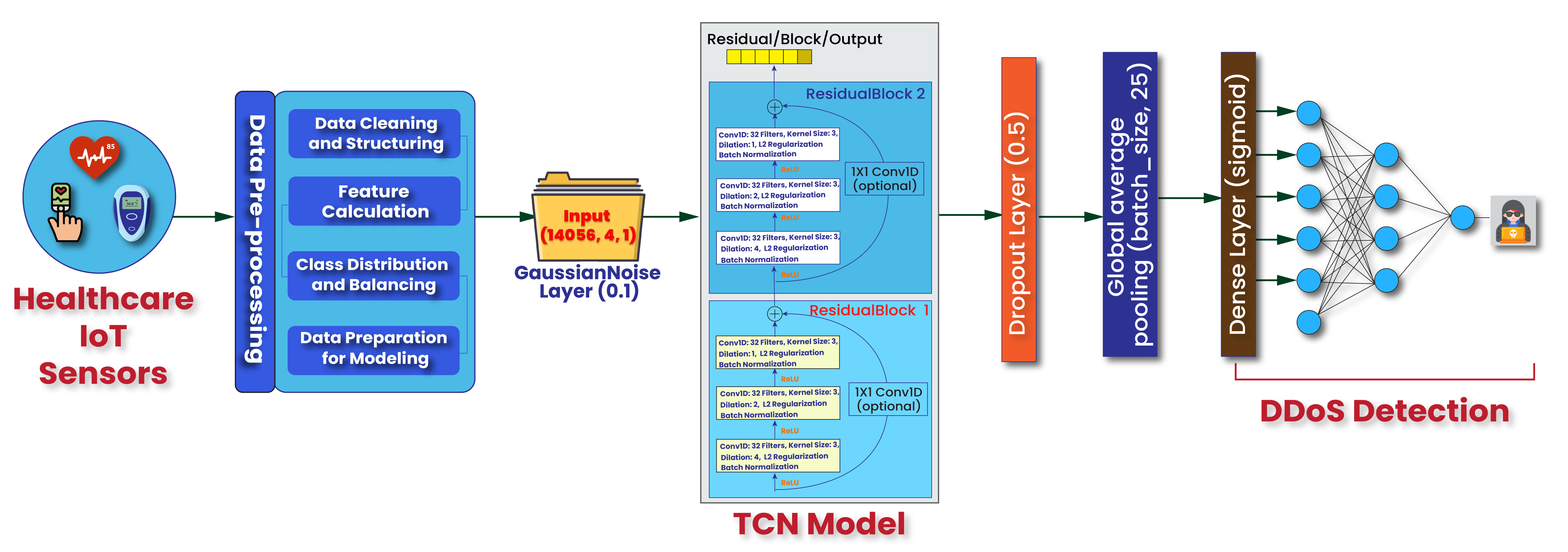}
    \caption[TCN-based architecture for DDoS detection]{\gls{tcn}-based architecture for \gls{ddos} detection in \gls{hiot} environments.}
    \label{fig:TCNarchitecture}
\end{figure*}

\begin{table}[ht!]
\caption[Key architecture details and hyperparameters]{Key architecture details and hyperparameters of the proposed \gls{tcn} model.}
\label{tab:hyperparameters}
\centering
\footnotesize
\resizebox{\textwidth}{!}{%
\begin{tabular}{@{}ll@{}}
\toprule
\textbf{Components and Hyperparameters}                                   & \textbf{Value}                                               \\ \midrule
Number of features (\gls{mqtt}/\gls{udp})                                        & 4/13                                                            \\
Number of classes                                          & 2 (Normal and Malicious)                                                            \\
Residual Blocks                   & 2 blocks, each with 3 convolutional layers                                                          \\ 
Number of filters in \gls{tcn} layers                            & 64                                                           \\
Kernel size                                                & 3                                                            \\
Dilation rate                                              & 1, 2, 4                                                      \\
Padding                            & Causal (preserves temporal order)  \\
 Batch Normalization                & Improves stability and convergence \\ 
Depth of \gls{tcn}                                               & 3                                                            \\
Optimizer                                                  & Adam                                                         \\
Learning rate                                              & 0.00005                                                      \\
Loss function                                              & Binary Crossentropy                                          \\
Batch size                                                 & 64                                                           \\
Epochs                                                     & 50                                                           \\
Regularization                                             & Dropout(0.2 - ReLU, 0.5 - Sigmoid), L2 (0.01) \\
Pooling Layer                     & Global average pooling  \\ 
Input Shape                        & $n_{samples} \times n_{features} \times n_{timesteps}$ \\
Output Layer & Dense layer with Sigmoid activation \\ 
Noise Techniques                                           & GaussianNoise(0.1)                                           \\
Threshold for filtering for the {\em UL-ECE-MQTT-DDoS-H-IoT2025} & Base \& Adjusted                                             \\
Threshold for filtering for the {\em UL-ECE-UDP-DDoS-H-IoT2025}  & Base \& Adjusted                  \\ 
Validation Split                & 20\%                                                   \\
\bottomrule
\end{tabular}}
\begin{minipage}{0.95\linewidth}
\vspace{2pt}
\scriptsize
\textbf{Note:} This table includes layer configurations, activation functions, learning rate, training parameters, and other key hyperparameters used in the proposed \gls{tcn} model for detecting \gls{ddos} attacks over the \gls{mqtt} and \gls{udp} datasets.
\end{minipage}
\end{table}

The model is trained using the Adam optimiser with a learning rate of 0.00005 (controlling the step size during training) and a batch size of 64 for 50 epochs. Early stopping and learning rate reduction on a plateau are employed as part of the training strategy. A validation split of 0.2 (20\%) is used to monitor the model's performance. The proposed \gls{tcn} architecture details and the hyperparameters on which the model performed well are illustrated in Table \ref{tab:hyperparameters}. In the ablation studies, minor adjustments are made to optimise the model based on the second dataset's different characteristics (e.g., monitoring frequency patterns). These adjustments include modifying the dropout rate, using a ReLU activation function in the hidden layer, and adjusting the learning rate, as detailed in Table \ref{tab:udp_ablation}. Despite these changes, the core architecture consistently analyses the temporal network patterns in \gls{hiot} networks, enhancing network security. Figure \ref{fig:TCNarchitecture} illustrates both datasets' detailed core \gls{tcn} architecture, highlighting the key components and layers for processing temporal patterns in \gls{hiot} environments. The class distribution and input shapes for each dataset used in the \gls{tcn} model are summarised in Tables \ref{tab:mqttclasssample} and \ref{tab:udpclasssample}. Each dimension corresponds to the number of samples, features, and the input format the model requires.

\begin{table}[ht!]
\centering
\caption{SMOTE class distribution for the \gls{mqtt} dataset.}
\label{tab:mqttclasssample}
\small
\begin{tabular}{lcc}
\toprule
\textbf{\#Samples} & \textbf{Class}                 & \textbf{Split}   \\ 
\midrule
7782                       & Malicious & \multirow{2}{*}{Train (Original)}    \\ 
6274                       & Normal    &                                      \\ 
\midrule
3348                       & Malicious & \multirow{2}{*}{Test}                \\ 
2676                       & Normal    &                                      \\ 
\midrule
7782                       & Malicious & \multirow{2}{*}{Train (After \gls{smote})} \\ 
7782                       & Normal    &                                      \\ 
\midrule
\multicolumn{3}{l}{\textbf{Train Original Data Shape: $(14056, 4, 1)$}} \\
\multicolumn{3}{l}{\textbf{Total Dataset Size: $(20080, 4, 1)$}} \\
\bottomrule
\end{tabular}
\begin{tablenotes}
    \footnotesize
    \item \textbf{Note}: Data shapes are defined as $(n_{\text{samples}}, n_{\text{features}}, n_{\text{channels}})$.
\end{tablenotes}
\end{table}

\begin{table}[ht!]
\centering
\caption{SMOTE class distribution for the \gls{udp} dataset.}


\small
\begin{tabular}{lcc}
\toprule
\textbf{\#Samples} & \textbf{Class}                 & \textbf{Split}   \\ 
\midrule
35355                       & Malicious & \multirow{2}{*}{Train (Original)}    \\ 
34565                       & Normal    &                                      \\ 
\midrule
16538                       & Malicious & \multirow{2}{*}{Test}                \\ 
13429                       & Normal    &                                      \\ 
\midrule
35355                       & Malicious & \multirow{2}{*}{Train (After \gls{smote})} \\ 
35355                       & Normal    &                                      \\ 
\midrule
\multicolumn{3}{l}{\textbf{Train Original Data Shape: $(69920, 13, 1)$}} \\
\multicolumn{3}{l}{\textbf{Total Dataset Size: $(99887, 13, 1)$}} \\
\bottomrule
\end{tabular}
\label{tab:udpclasssample}
\begin{tablenotes}
    \footnotesize
    \item \textbf{Note}: Data shapes are defined as $(n_{\text{samples}}, n_{\text{features}}, n_{\text{channels}})$.
\end{tablenotes}
\end{table}

\subsection{Dynamic Threshold-based Filtering Strategy for Mitigating the Malicious Nodes in \gls*{hiot}}
The \gls{tcn} model operates in two phases for detecting and mitigating \gls{ddos} attack-affected nodes in a \gls{g5}-based \gls{hiot} environment. First, it predicts potential malicious \gls{hiot} nodes. Second, following the prediction phase, a detection and mitigation model dynamically adjusts thresholds to improve the accuracy of detecting and blacklisting malicious nodes in \gls{hiot} environments. The proposed \gls{tcn} model focuses on selecting the appropriate features, class filter settings, kernel sizes, and depths to detect and mitigate \gls{ddos} attacks effectively. This configuration is tested against two datasets mimicking real-world \gls{ddos} scenarios in \gls{g5}-based healthcare environments.

\begin{algorithm}[ht!]
\footnotesize
\caption{Dynamic thresholding for malicious node detection and mitigation.}
 \label{algorithm:algo}
\begin{algorithmic}
\State \textbf{Input:} Features $X$, labels $y_{\text{test}}$, predictions $y_{\text{pred}}$
\State \textbf{Output:} Detection and Mitigation Accuracy, Blacklist, Threshold
\State \textbf{1. Initialize:}
\State $malicious\_nodes \gets \{\}$, $error\_count \gets 0$
\State $mitigation\_error \gets 0$
\State \textbf{2. Detection Phase:}
\For{$i = 1$ \textbf{to} the number of items in $y_{\text{test}}$}
    \State $actual \leftarrow y_{\text{test}}[i]$
    \State $predicted \leftarrow y_{\text{pred}}[i]$
    \State $node\_id \leftarrow X[i][\text{Node\_IDs}]$
    \State $error\_count \leftarrow error\_count + (predicted \neq actual)$
    \If{$predicted = 1$}
        \State $malicious\_nodes[node\_id] \leftarrow$
        \State \hspace{\algorithmicindent} $malicious\_nodes.get(node\_id, 0) + 1$
    \EndIf
\EndFor
\State \textbf{3. Calculate Detection Accuracy:}
\State $accuracy \gets (1 - error\_count / \text{len}(y_{\text{pred}})) \times 100$
\State \textbf{4. Mitigation Phase:}
\State Compute $base\_threshold$ as the average of malicious nodes' values
\State Set $adjusted\_threshold$ to $base\_threshold$ reduced by 20\%
\For{each $node\_id$, $count$ in malicious\_nodes}
    \If{$count$ exceeds $adjusted\_threshold$}
        \State Add $node\_id$ to blacklist
        \If{$node\_id$ is not in actual malicious nodes}
            \State Increment mitigation error count
        \EndIf
    \EndIf
\EndFor
\State \textbf{5. Calculate Mitigation Accuracy:}
\State $mit\_accuracy \gets 100 \times (1 - \frac{mitigation\_error}{\text{len}(blacklist)})$
\State \textbf{6. Output Results:}
\State Print ``Malicious Node Counts: '', $malicious\_nodes$
\State Print ``Blacklist Nodes: '', $blacklist$
\State Print ``Adaptive Threshold: '', $adjusted\_thresh$
\State Print ``Detection Accuracy: '', $accuracy$
\State Print ``Mitigation Accuracy: '', $mit\_accuracy$
\end{algorithmic}
\end{algorithm}
\normalsize

The dynamic threshold determines when to block nodes in the \gls{hiot} system. It is based on specific characteristics of network activity, including the number of nodes, the volume of \gls{mqtt} and \gls{udp} data transfers for both malicious and non-malicious nodes, and the ratio of malicious packets sent to the server. During the detection phase, the model compares each \gls{hiot} node's predicted behaviour with its actual class (i.e., whether the node is genuinely malicious or normal), incrementing error counts for false positives and false negatives. As a result, the mitigation technique involves iteratively checking the predicted and actual behaviours of the \gls{hiot} nodes. Malicious activity (i.e., messages) from compromised \gls{hiot} nodes is recorded in a dictionary, and detection accuracy is calculated based on the total error count, reflecting the sum of incorrect predictions, including false positives (normal nodes classified as malicious) and false negatives (malicious nodes classified as normal). Therefore, in the mitigation phase, the dynamic threshold, which serves as the base threshold for both datasets (\gls{mqtt} and \gls{udp}), is calculated by averaging the number of detected malicious messages recorded in the dictionary from the particular \gls{hiot} nodes flagged as malicious.

Moreover, in the mitigation phase for the \gls{mqtt} dataset, this research adjusts the base threshold downward by 20\% to enhance mitigation accuracy. An \gls{hiot} node with malicious activity counts exceeding the adjusted threshold is blacklisted, minimising false positives and improving network security in real-time. In contrast, the \gls{udp} dataset does not require any adjustment to the base threshold to achieve optimal mitigation accuracy. Therefore, the \gls{tcn}-based detection and mitigation technique achieves high mitigation accuracy by applying a dynamic threshold for the \gls{udp} dataset within \gls{g5} \gls{hiot} environments.

This \gls{ddos} mitigation approach can benefit real-world scenarios. \glspl{wban} in \gls{hiot} environments that use \gls{mqtt} and \gls{udp} protocols are diverse and subject to change, as protocol parameters, \gls{hiot} device configurations, and traffic patterns may vary over time. Moreover, this mitigation strategy is highly effective, handling datasets of various sizes (e.g., over one million records) with precision and accuracy, making it suitable for \gls{hiot} applications. Thus, the strategy employs risk mitigation and improved detection methods as outlined in the Algorithm~\ref{algorithm:algo}.

\subsection{Ablation Studies of the \gls*{mqtt} Dataset}  
In \gls{ml}, ablation studies are a technique used to systematically evaluate the impact of proposed model components or parameters on overall performance \cite{kasneci2024enriching}. Researchers use this technique to optimise \gls{ml} models by identifying configurations that enhance performance \cite{al2021stl}, \cite{daoud2023convolutional}. An ablation study is significant for evaluating the proposed model's performance using accuracy, precision, recall, and F1-score metrics. These standard metrics are commonly used for binary classification tasks such as fraud detection, medical diagnosis, and anomaly detection \cite{bounab2024enhancing}. Particularly in attack detection, datasets often exhibit an imbalanced distribution with more regular traffic than \gls{ddos} attack traffic or vice versa, depending on the scenario. Traditional metrics (e.g., accuracy) can be misleading, as models usually achieve high accuracy by predicting the majority class. Relying solely on a single metric, such as accuracy, does not sufficiently demonstrate the model's effectiveness in detecting critical \gls{ddos} attacks. In this research, the dataset contains slightly more malicious traffic than regular traffic. To address this slight imbalance, this research applies the \gls{smote} technique to balance the classes, which allows the model to achieve consistent performance across all evaluation metrics \cite{efendi2024dbscan}.

\onecolumn
{%
\footnotesize
\setlength{\tabcolsep}{3pt}      
\renewcommand{\arraystretch}{1.0}  
\begin{longtable}{cccccc}
\caption[Dropout and learning rate ablation study for the MQTT]{Dropout and learning rate ablation study for the \gls{mqtt} dataset.}
\label{tab:mqtt_ablation}\\
\hline
\textbf{Dropout} & \textbf{Learning Rate} & \textbf{Accuracy} & \textbf{Precision} & \textbf{Recall} & \textbf{F1-score} \\
\hline
\endfirsthead

\hline
\textbf{Dropout} & \textbf{Learning Rate} & \textbf{Accuracy} & \textbf{Precision} & \textbf{Recall} & \textbf{F1-score} \\
\hline
\endhead

\hline
\endfoot
\multicolumn{6}{c}{\textbf{ReLU Activation}}\\
\hline
0.1 & 0.00001 & 0.982570 & 1.000000 & 0.968638 & 0.984069 \\
0.1 & 0.00002 & 0.999004 & 1.000000 & 0.998208 & 0.999103 \\
0.1 & 0.00005 & 0.999668 & 1.000000 & 0.999403 & 0.999701 \\
0.2 & 0.00001 & 0.997012 & 1.000000 & 0.994624 & 0.997305 \\
0.2 & 0.00002 & 0.999004 & 1.000000 & 0.998208 & 0.999103 \\
0.2 & 0.00005 & 0.999668 & 1.000000 & 0.999403 & 0.999701 \\
0.3 & 0.00001 & 0.984728 & 1.000000 & 0.972521 & 0.986069 \\
0.3 & 0.00002 & 0.999336 & 1.000000 & 0.998805 & 0.999402 \\
0.3 & 0.00005 & 0.999668 & 1.000000 & 0.999403 & 0.999701 \\
0.4 & 0.00001 & 0.945717 & 1.000000 & 0.902330 & 0.948658 \\
0.4 & 0.00002 & 0.993360 & 1.000000 & 0.988053 & 0.993990 \\
0.4 & 0.00005 & 0.999668 & 1.000000 & 0.999403 & 0.999701 \\
0.5 & 0.00001 & 0.957503 & 1.000000 & 0.923536 & 0.960248 \\
0.5 & 0.00002 & 0.999170 & 1.000000 & 0.998507 & 0.999253 \\
0.5 & 0.00005 & 0.999668 & 1.000000 & 0.999403 & 0.999701 \\
\hline

\multicolumn{6}{c}{\textbf{Sigmoid Activation}}\\
\hline
 0.1 & 0.00001 & 0.942397 & 1.000000 & 0.896356 & 0.945346 \\
0.1 & 0.00002 & 0.943061 & 1.000000 & 0.897551 & 0.946010 \\
0.1 & 0.00005 & 0.956175 & 1.000000 & 0.921147 & 0.958955 \\
0.2 & 0.00001 & 0.942231 & 1.000000 & 0.896057 & 0.945180 \\
0.2 & 0.00002 & 0.942895 & 1.000000 & 0.897252 & 0.945844 \\
0.2 & 0.00005 & 0.955013 & 1.000000 & 0.919056 & 0.957821 \\
0.3 & 0.00001 & 0.942397 & 1.000000 & 0.896356 & 0.945346 \\
0.3 & 0.00002 & 0.942895 & 1.000000 & 0.897252 & 0.945844 \\
0.3 & 0.00005 & 0.954017 & 1.000000 & 0.917264 & 0.956847 \\
0.4 & 0.00001 & 0.942397 & 1.000000 & 0.896356 & 0.945346 \\
0.4 & 0.00002 & 0.942397 & 1.000000 & 0.896356 & 0.945346 \\
0.4 & 0.00005 & 0.949701 & 1.000000 & 0.909498 & 0.952604 \\
0.5 & 0.00001 & 0.942397 & 1.000000 & 0.896356 & 0.945346 \\
0.5 & 0.00002 & 0.943227 & 1.000000 & 0.897849 & 0.946176 \\
\textbf{0.5} & \textbf{0.00005} & \textbf{0.999883} & \textbf{0.999777} & \textbf{1.000000} & \textbf{0.999987} \\
\hline
\multicolumn{6}{c}{\textbf{Tanh Activation}}\\
\hline
0.1 & 0.00001 & 0.942397 & 1.000000 & 0.896356 & 0.945346 \\
0.1 & 0.00002 & 0.948871 & 1.000000 & 0.908005 & 0.951785 \\
0.1 & 0.00005 & 0.967131 & 1.000000 & 0.940860 & 0.969529 \\
0.2 & 0.00001 & 0.942231 & 1.000000 & 0.896057 & 0.945180 \\
0.2 & 0.00002 & 0.944721 & 1.000000 & 0.900538 & 0.947666 \\
0.2 & 0.00005 & 0.955511 & 1.000000 & 0.919952 & 0.958307 \\
0.3 & 0.00001 & 0.942729 & 1.000000 & 0.896953 & 0.945678 \\
0.3 & 0.00002 & 0.943061 & 1.000000 & 0.897551 & 0.946010 \\
0.3 & 0.00005 & 0.961819 & 1.000000 & 0.931302 & 0.964429 \\
0.4 & 0.00001 & 0.942895 & 1.000000 & 0.897252 & 0.945844 \\
0.4 & 0.00002 & 0.942895 & 1.000000 & 0.897252 & 0.945844 \\
0.4 & 0.00005 & 0.965305 & 1.000000 & 0.937575 & 0.967782 \\
0.5 & 0.00001 & 0.942397 & 1.000000 & 0.896356 & 0.945346 \\
0.5 & 0.00002 & 0.943393 & 1.000000 & 0.898148 & 0.946341 \\
0.5 & 0.00005 & 0.962483 & 1.000000 & 0.932497 & 0.965070 \\
\hline
\end{longtable}
\vspace{-10pt}
\begin{minipage}{\textwidth}
\footnotesize
\noindent\textbf{Note:} Bold values represent the proposed \gls{tcn} model prediction accuracy for the \gls{mqtt} dataset.
\end{minipage}


{
\footnotesize
\setlength{\tabcolsep}{3pt}      
\renewcommand{\arraystretch}{1.0}  
\begin{longtable}{cccccc}
\caption[Dropout and learning rate ablation study for the UDP]{Dropout and learning rate ablation study for the \gls{udp} dataset.}
\label{tab:udp_ablation} \\
\hline
\textbf{Dropout} & \textbf{Learning Rate} & \textbf{Accuracy} & \textbf{Precision} & \textbf{Recall} & \textbf{F1-score} \\
\hline
\endfirsthead

\hline
\textbf{Dropout} & \textbf{Learning Rate} & \textbf{Accuracy} & \textbf{Precision} & \textbf{Recall} & \textbf{F1-score} \\
\hline
\endhead

\hline
\endfoot
\multicolumn{6}{c}{\textbf{ReLU Activation}}\\
\hline
0.1     & 0.00001       & 0.999600 & 0.999743  & 0.999487 & 0.999615 \\
0.1     & 0.00002       & 0.999533 & 0.999808  & 0.999295 & 0.999551 \\
0.1     & 0.00005       & 0.999700 & 0.999423  & 1.000000 & 0.999712 \\
0.2     & 0.00001       & 0.999533 & 0.999167  & 0.999936 & 0.999551 \\
0.2     & 0.00002       & 0.999833 & 0.999808  & 0.999872 & 0.999840 \\
\textbf{0.2}     & \textbf{0.00005}       & \textbf{0.999933} & \textbf{0.999944}  & \textbf{ 0.999936} & \textbf{0.999940} \\
0.3     & 0.00001       & 0.999533 & 0.999615  & 0.999487 & 0.999551 \\
0.3     & 0.00002       & 0.999833 & 0.999744  & 0.999936 & 0.999840 \\
0.3     & 0.00005       & 0.999733 & 0.999551  & 0.999936 & 0.999744 \\
0.4     & 0.00001       & 0.999833 & 0.999680  & 1.000000 & 0.999840 \\
0.4     & 0.00002       & 0.999800 & 0.999808  & 0.999808 & 0.999808 \\
0.4     & 0.00005       & 0.999700 & 0.999487  & 0.999936 & 0.999712 \\
0.5     & 0.00001       & 0.999566 & 0.999167  & 1.000000 & 0.999583 \\
0.5     & 0.00002       & 0.999700 & 0.999808  & 0.999615 & 0.999711 \\
0.5     & 0.00005       & 0.999733 & 0.999551  & 0.999936 & 0.999744 \\
\hline \vspace{-1.2mm} \\ 
\multicolumn{6}{c}{\bf Sigmoid Activation} \\
\hline
0.1     & 0.00001       & 0.998732 & 0.997824  & 0.999744 & 0.998783 \\
0.1     & 0.00002       & 0.999166 & 0.998400  & 1.000000 & 0.999199 \\
0.1     & 0.00005       & 0.999533 & 0.999551  & 0.999551 & 0.999551 \\
0.2     & 0.00001       & 0.999433 & 0.999103  & 0.999808 & 0.999455 \\
0.2     & 0.00002       & 0.999566 & 0.999551  & 0.999615 & 0.999583 \\
0.2     & 0.00005       & 0.999499 & 0.999167  & 0.999872 & 0.999519 \\
0.3     & 0.00001       & 0.999633 & 0.999679  & 0.999615 & 0.999647 \\
0.3     & 0.00002       & 0.999533 & 0.999167  & 0.999936 & 0.999551 \\
0.3     & 0.00005       & 0.999399 & 0.999359  & 0.999487 & 0.999423 \\
0.4     & 0.00001       & 0.999232 & 0.999423  & 0.999102 & 0.999263 \\
0.4     & 0.00002       & 0.999366 & 0.999167  & 0.999615 & 0.999391 \\
0.4     & 0.00005       & 0.999166 & 0.998400  & 1.000000 & 0.999199 \\
0.5     & 0.00001       & 0.999499 & 0.999103  & 0.999936 & 0.999519 \\
0.5     & 0.00002       & 0.999499 & 0.999743  & 0.999295 & 0.999519 \\
0.5     & 0.00005       & 0.999533 & 0.999551  & 0.999551 & 0.999551 \\
\hline \vspace{-1.5mm} \\ 
\multicolumn{6}{c}{\bf Tanh Activation} \\
\hline
0.1     & 0.00001       & 0.999766 & 0.999936  & 0.999615 & 0.999776 \\
0.1     & 0.00002       & 0.999232 & 0.998591  & 0.999936 & 0.999263 \\
0.1     & 0.00005       & 0.999566 & 0.999487  & 0.999679 & 0.999583 \\
0.2     & 0.00001       & 0.999800 & 0.999744  & 0.999872 & 0.999808 \\
0.2     & 0.00002       & 0.999700 & 0.999808  & 0.999615 & 0.999711 \\
0.2     & 0.00005       & 0.999600 & 0.999808  & 0.999423 & 0.999615 \\
0.3     & 0.00001       & 0.999533 & 0.999103  & 1.000000 & 0.999551 \\
0.3     & 0.00002       & 0.999166 & 0.998464  & 0.999936 & 0.999199 \\
0.3     & 0.00005       & 0.999733 & 0.999487  & 1.000000 & 0.999744 \\
0.4     & 0.00001       & 0.999666 & 0.999808  & 0.999551 & 0.999679 \\
0.4     & 0.00002       & 0.999199 & 0.998527  & 0.999936 & 0.999231 \\
0.4     & 0.00005       & 0.999666 & 0.999423  & 0.999936 & 0.999680 \\
0.5     & 0.00001       & 0.999800 & 0.999808  & 0.999808 & 0.999808 \\
0.5     & 0.00002       & 0.998999 & 0.998399  & 0.999679 & 0.999039 \\
0.5     & 0.00005       & 0.999633 & 0.999679  & 0.999615 & 0.999647 \\
\hline
\end{longtable}
\vspace{-12pt}
\begin{minipage}{\textwidth}
\footnotesize
\noindent\textbf{Note:} Bold values represent the proposed \gls{tcn} model prediction accuracy for the \gls{udp} dataset.
\end{minipage}


\normalsize
Therefore, in this research, a systematic analysis with ablation studies is conducted to evaluate the stability of the \gls{tcn} model. Various configurations, including dropout rates, activation functions (ReLU, Sigmoid, and Tanh), and learning rate adjustments, are tested using metrics such as precision, recall, F1-score, and accuracy for the dataset {\em UL-ECE-MQTT-DDoS-H-IoT2025}, as outlined in Table~\ref{tab:mqtt_ablation}. Moreover, this ablation study analyses the effects of dropout and learning rates on model performance metrics. It is based on a baseline model incorporating Gaussian noise, batch normalisation, and L2 regularisation to improve generalisation and reduce overfitting \cite{hinton2012improving}. Testing a higher dropout rate (e.g.,0.6) could be an area for future exploration to assess whether it might further enhance performance or lead to overfitting. However, based on the detailed analysis and recall performance (e.g., 100\%), this research proposes the Sigmoid activation function with a dropout rate of 0.5 and a learning rate of 0.00005 as the optimal combination. A recall of 100\% indicates that the \gls{tcn} model detects all actual positive cases (i.e., all attacks), which is crucial in \gls{hiot} environments.

\subsection{Ablation Studies of the \gls*{udp} Dataset}
This research conducts an ablation study to evaluate the performance of the proposed model on the {\em UL-ECE-UDP-DDoS-H-IoT2025} dataset using various activation functions, including ReLU, Sigmoid, and Tanh. The study tests dropout and learning rate configurations with 50 epochs for each activation function, as illustrated in Table~\ref{tab:udp_ablation}. For the ReLU and Sigmoid activation functions in the hidden layer, 10 runs of 50 epochs each are performed, and the results are averaged. The proposed dropout rate of 0.2 and learning rate of 0.00005 with the ReLU activation function are positioned near the midpoint of the tested configurations. This combination demonstrates a balance with an average accuracy, precision, recall, and F1-score of 0.9999. It outperforms Sigmoid, which achieves slightly lower average accuracy and precision, with a recall of 0.9999 and an F1-score of around 0.9995. The F1-score provides a balanced metric, particularly useful when there is a trade-off between accuracy and recall \cite{baisholan2024corporate}. Therefore, based on the F1-score and overall average performance across metrics, this research selects the ReLU activation function over Sigmoid, as it provides slightly higher and more consistent results.

Furthermore, balanced output plays a key role in anomaly detection, where high recall is significant in detecting anomalies in \gls{hiot} environments. This research applies the \gls{smote} technique to balance the dataset, generating synthetic samples for the minority class, thereby improving model performance and achieving 99.99\% accuracy across all evaluation metrics. Further details on using the \gls{smote} technique are provided in Section \ref{subsubsec:smote}.

\section{Performance Evaluation and Results}
\label{sec:evaluation}
This section compares the \gls{tcn} model's performance with previous methods using the proposed datasets, as shown in Table~\ref{tab:mqttcomparative}. To evaluate its generalizability, the model is tested on previously published datasets using metrics such as accuracy, precision, recall, and F1-score, as illustrated in Table~\ref{tab:comparedataset}. Additionally, this section analyses and compares the computational efficiency of the \gls{tcn} model and \gls{bilstm} in terms of parameters, memory usage, and inference time, as outlined in Table~\ref{tab:parametercomparison}. The model's learning behaviour is analysed through accuracy and loss plots, which provide insights into its performance during training. Detection and mitigation metrics are evaluated using a dynamic threshold calculation, with results presented in Table~\ref{tab:dynamicthresholdaccuracy}. Protocol-specific impacts are analysed for the \gls{mqtt} and \gls{udp} datasets. Finally, this section discusses \glspl{tcn} in \gls{hiot} applications and concludes with a detailed discussion of mitigation strategies, demonstrating the proposed model's effectiveness in detecting and mitigating \gls{ddos} attacks.

\subsection{Comparison of \gls*{tcn} Model with Previous Methods on the Proposed Dataset}
This research compares the proposed \gls{tcn} method on the {\em UL-ECE-MQTT-DDoS-H-IoT2025} dataset with other established \gls{dl} models discussed in Section~\ref{sec:bg}. Table~\ref{tab:mqttcomparative} showcases the \gls{tcn} model's superior performance compared to other \gls{dl} and \gls{ml} models, such as \gls{dnn} \cite{chaganti2022particle}, \gls{bilstm} \cite{wagan2023fuzzy}, \gls{knn} \cite{rosay2022network}, \gls{cnnbilstmgru} \cite{kumar2023novel}, and \gls{dt} \cite{binbusayyis2022investigation}.

\begin{table}[ht!]
\caption[Comparison of baseline methods and the proposed \gls{tcn} model]{Comparison of baseline methods and the proposed \gls{tcn} model on the {\em UL-ECE-\gls{mqtt}-DDoS-H-IoT2025} dataset.}
\label{tab:mqttcomparative}
\centering
\small
\adjustbox{max width=\textwidth}{
\begin{tabular}{lllcc}
\hline
\multicolumn{1}{l}{\textbf{Ref.}} &
  \multicolumn{1}{l}{\textbf{Methods}} &
\multicolumn{1}{c}{\textbf{Accuracy (\%)}} &
  \multicolumn{1}{c}{\textbf{Time (s)}} \\ \hline
\cite{chaganti2022particle} (2022) &
  DNN &
  91.5 &
  71.5 \\
\cite{binbusayyis2022investigation} (2022) &
  \gls{dt} &
  92.5 &
  85.03 \\
\cite{wagan2023fuzzy} (2023) &
  \gls{bilstm} &
  92.7 &
  74.23 \\
\cite{rosay2022network} (2022) &
  \gls{knn} &
  93.79 &
  72.9 \\
\cite{kumar2023novel} (2023) &
  \gls{cnn}-\gls{bilstm}-\gls{gru} &
  93 &
  85.03 \\
\cite{khatun2024time} (2024) &
  \gls{cnn} &
  92.9 &
  75.03 \\ \hline
\textbf{Proposed Method} &
\textbf{\gls{tcn}} &
\textbf{99.98} &
\textbf{65.24} \\ \hline 

\end{tabular}%
}
\vspace{-8pt}
\begin{flushleft}
\scriptsize \textbf{Note:} All methods are implemented and evaluated on the {\em UL-ECE-\gls{mqtt}-DDoS-H-IoT2025} dataset under the same experimental setup. Execution time is measured on the same computing platform (Google Colab).
\end{flushleft}
\end{table}

This evaluation underscores the \gls{tcn} model's effectiveness in detecting and mitigating \gls{ddos} attacks, with prediction accuracy rates of 99.98\% and 99.99\% on the proposed {\em UL-ECE-\gls{mqtt}-DDoS-\gls{hiot}2025} and {\em UL-ECE-\gls{udp}-DDoS-\gls{hiot}2025} datasets, respectively, as shown in Table~\ref{tab:mqttcomparative}. Based on the prediction model, detection and mitigation accuracies reach 96.94\% and 95\% on the \gls{mqtt} dataset and 75\% and 80\% on the \gls{udp} dataset, respectively. The class distribution for both proposed datasets is detailed in Tables~\ref{tab:mqttclasssample} and \ref{tab:udpclasssample}, providing context for the data splits. Therefore, the \gls{tcn} model excels on both proposed datasets in the \gls{hiot} context. It offers efficient and scalable solutions well-suited to the demands of \gls{g5} networks. Its dynamic thresholding strategy in Algorithm \ref{algorithm:algo}, tested over 10 runs and 50 epochs, demonstrates stability compared to existing \gls{ml}/\gls{dl} models.

\begin{figure*}[ht!]
    \centering
    \includegraphics[width=0.9\textwidth]{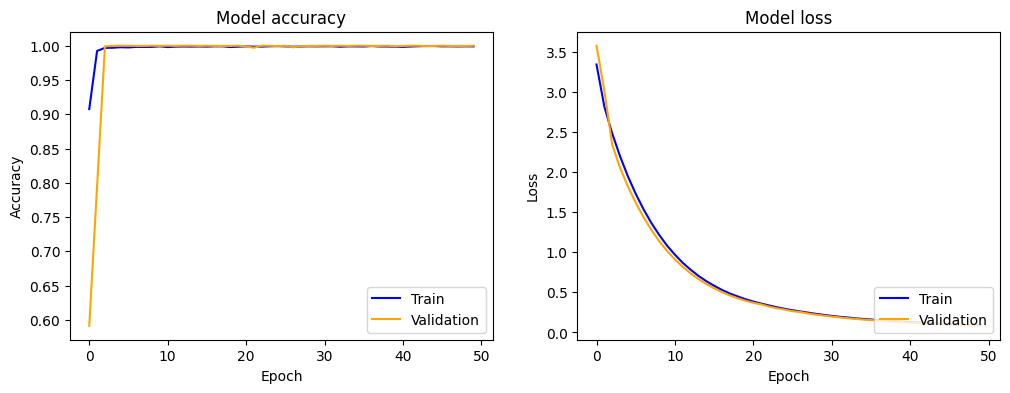}
     \caption[Training and validation accuracy and loss over the MQTT]{Training and validation accuracy and loss of the proposed method with the {\em UL-ECE-MQTT-DDoS-H-IoT2025} dataset.}
    \label{fig:mqtttrainingandvalidation}
\end{figure*}

\begin{figure*}[ht!]
    \centering
    \includegraphics[width=0.9\textwidth]{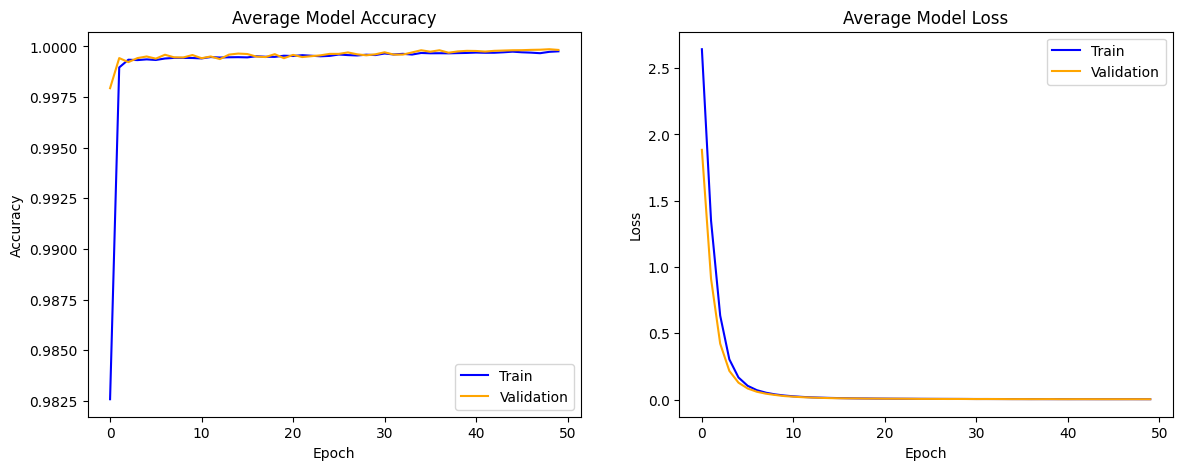}
    \caption[Training and validation accuracy and loss over the UDP]{Training and validation accuracy and loss of the proposed method with the {\em UL-ECE-UDP-DDoS-H-IoT2025} dataset.}
    \label{fig:ns3trainingandvalidation}
\end{figure*}

The proposed model's training and validation accuracy and loss are shown in Figure~\ref{fig:mqtttrainingandvalidation}. The graph illustrates the accuracy and loss trends during the model's training and validation phases. After 50 epochs, the model attains a training accuracy of approximately 99.98\%. The validation accuracy similarly reaches 99.98\%, demonstrating consistent performance between the training and validation phases.

The loss on the training dataset is recorded at approximately 0.05 after 50 epochs, while the loss on the validation dataset is slightly higher, at approximately 0.07. Due to the close correlation between the training and validation accuracy and the marginally higher validation loss than the training loss, there is no indication of overfitting.

Additionally, the training and validation accuracy and loss of the proposed model for the {\em UL-ECE-\gls{udp}-DDoS-\gls{hiot}2025} dataset are shown in Figure~\ref{fig:ns3trainingandvalidation}. The graph presents the trends over 50 epochs, averaged across 10 runs. The model achieves a training accuracy of approximately 99.99\%, with validation accuracy also reaching around 99.99\%, indicating consistent performance.

The training loss stabilises at approximately 0.02, while the validation loss remains slightly higher, at around 0.03. The close alignment between training and validation accuracy, with minimal loss, demonstrates that the model effectively performs without overfitting. 

Hence, this consistent performance across multiple executions underscores its resilience in detecting \gls{ddos} attacks in real-world \gls{hiot} systems.

\subsection{Performance Comparison with Previous Datasets Using the Proposed \gls*{tcn} Method}
This research evaluates the generalizability and scalability of the proposed \gls{tcn} method using several existing datasets, including the real-world \gls{iomt} public dataset WUSTL-\gls{ehms}-2020 \cite{jain_eeg_dataset}, as outlined in Table~\ref{tab:comparedataset}. The evaluation reflects practical \gls{iomt} scenarios using this real-world dataset, improving the reliability and applicability of the proposed \gls{tcn} method. Moreover, this research compares the proposed model across datasets from different domains, including \gls{ids}, simulated \gls{iot} environments, and real-world healthcare scenarios. Nevertheless, the proposed method achieves high accuracy across multiple datasets, including WSN\_DDoS\_Attack\_H-IoT2023 \cite{khatun2024time}, WUSTL-\gls{ehms}-2020 \cite{hady2020intrusion}, BoT-\gls{iot} 2018 \cite{koroniotis2019towards}, and CIC-IDS2017 \cite{sharafaldin2018toward}, with accuracy rates of 99.84\%, 93.34\%, 100\%, and 99.57\%, respectively. These results demonstrate the model's reliability and scalability in detecting and classifying attacks across various domains. 

Furthermore, standard data preprocessing techniques are applied to prepare each dataset, including data cleaning, standardisation, and handling of non-numeric values. For instance, these steps are essential for refining raw data, which often includes duplicates, outliers, and other quality issues in the WUSTL-\gls{ehms}-2020 and BoT-\gls{iot} 2018 datasets. Similarly, the CIC-IDS2017 dataset requires additional handling of non-numeric and infinite values before standardisation \cite{faruqui2023safetymed}. This research fine-tunes critical parameters, such as kernel size, number of filters, and dropout rate, to optimise the model’s performance across diverse datasets, including WSN\_DDoS\_Attack\_H-IoT2023~\cite{khatun2024time} and WUSTL-\gls{ehms}-2020~\cite{hady2020intrusion}. Adjusting parameters for different datasets is a standard practice in \gls{ml} and \gls{dl} to ensure optimal performance \cite{bayer2023multi}, \cite{yang2020hyperparameter}. These adjustments highlight the \gls{tcn} model's flexibility in adapting to distinct dataset features, such as data structure, input dimensions, and time-series patterns. As a result, the model achieves high accuracy across diverse datasets by resolving data quality issues and validating its effectiveness in detecting \gls{ddos} and other types of attacks.

In addition, this research evaluates the computational efficiency of the proposed model by comparing the \gls{bilstm} and \gls{tcn} architectures on the {\em UL-ECE-MQTT-DDoS-H-IoT2025} dataset. The findings reveal that the \gls{tcn} model has significantly fewer parameters than the \gls{bilstm} model. This demonstrates that the \gls{tcn} architecture is more memory-efficient, lightweight, and well-suited for resource-constrained, real-time applications in \gls{hiot} environments. Additionally, the \gls{tcn} model exhibits slightly better computational efficiency, achieving a lower average inference time per sample than \gls{bilstm}, as illustrated in Table~\ref{tab:parametercomparison}, with both models trained and evaluated over 50 epochs. Hence, the results infer improved runtime performance for \gls{tcn} compared to the \gls{bilstm} model.

\begin{table*}[ht!]
\centering
\caption[Comparison with previous datasets]{Comparison with previous datasets using the proposed and previous methods.}
\label{tab:comparedataset}
\small
\adjustbox{max width=\textwidth}{
\begin{tabular}{@{}ll llll llll@{}}
\toprule
\textbf{Ref.} & \textbf{Dataset} & 
\multicolumn{4}{c}{\textbf{Original Metrics (Prior Methods)}} & 
\multicolumn{4}{c}{\textbf{Proposed (\gls{tcn})}} \\
\cmidrule(lr){3-6} \cmidrule(lr){7-10}
& & \textbf{Acc} & \textbf{Precision} & \textbf{Recall} & \textbf{F1-score} & \textbf{Acc} & \textbf{Precision} & \textbf{Recall} & \textbf{F1-score} \\ \midrule
\cite{khatun2024time} (2024) & WSN\_DDoS\_Attack\_H-IoT2023 & 92 & 79.57 & 98.67 & 88.10 & 99.84 & 90.15 & 98.69 & 89.88 \\  
\cite{wagan2023fuzzy} (2023) & WUSTL-\gls{ehms}-2020 (real) & 89.67 & 68.42 & 68.93 & 69.02 & 93.34 & 68.66 & 87.08 & 76.78 \\  
\cite{binbusayyis2022investigation} (2022) & BoT-\gls{iot} & 100 & 100 & 100 & 100 & 100 & 100 & 100 & 100 \\  
\cite{rosay2022network} (2022) & CIC-IDS2017 & 99.31 & 99.87 & 99.89 & 99.88 & 99.57 & 96.79 & 99.82 & 98.28 \\  
\bottomrule
\end{tabular}}

\vspace{4pt}
\begin{minipage}{\linewidth}
\scriptsize \textbf{Note:} All values in the table, including accuracy, precision, recall, and F1-score, are percentages for both published methods and the proposed \gls{tcn} model across datasets. The \textbf{Original Metrics (Previous Methods)} column includes results reported using models such as \gls{cnn}, \gls{bilstm}, \gls{dt}, and \gls{knn}, as applicable. \rev{In contrast, the \textbf{Proposed (\gls{tcn})} column shows the corresponding results achieved using the proposed model on the same datasets for direct comparison.}
\end{minipage}
\end{table*}

\begin{table}[t]
\centering
\caption[Comparison of BiLSTM and TCN]{Comparison of \gls{bilstm} and \gls{tcn} in terms of parameters, memory, and inference time.}
\label{tab:parametercomparison}
\small
\adjustbox{max width=\textwidth}{
\begin{tabular}{l@{\hskip 1em}l@{\hskip 1em}l@{\hskip 1em}l@{\hskip 1em}l@{\hskip 1em}l}
\toprule
\textbf{Model}      & \textbf{Total Parameters} & \textbf{Total Memory} & \textbf{Trainable Parameters} & \textbf{Trainable Memory} & \textbf{Inference Time} \\
\midrule
Bi-\gls{lstm}             & 6,574,577                & 25.08                      & 2,191,525                    & 8.36                          & 0.000218                                   \\
\gls{tcn}                 & 189,317                  & 0.73                       & 62,849                       & 0.24                          & 0.000217                                   \\
\bottomrule
\end{tabular}
}

\vspace{5pt} 
\begin{minipage}{\linewidth}
\scriptsize \textbf{Note:} This table compares the \gls{bilstm} and \gls{tcn} models regarding total parameters, memory consumption, trainable parameters, trainable memory, and inference time. Total memory and trainable memory are in MB, and inference time represents the average time in seconds. The results highlight the \gls{tcn} model's lightweight nature compared to the \gls{bilstm}.
\end{minipage}
\end{table}

\subsection{Comparison of \gls*{tcn} Model Performance on \gls*{mqtt} and \gls*{udp} Datasets}
The {\em UL-ECE-MQTT-DDoS-H-IoT2025} and {\em UL-ECE-UDP-DDoS-H-IoT2025} evaluated using the \gls{tcn} model demonstrate high prediction performance across metrics such as accuracy, precision, recall, and F1-score. The \gls{tcn} model achieves 99.98\% accuracy and 100\% recall on the \gls{mqtt} dataset, as illustrated in Table~\ref{table:mqttperformance_metrics} and Figure~\ref{fig:mqtttrainingandvalidation}. On the \gls{udp} dataset, it achieves 99.99\% accuracy and 99.99\% recall (refer to Table~\ref{table:udpperformance_metrics} and Figure~\ref{fig:ns3trainingandvalidation}).

Although the difference in recall between the two datasets (99.99\% for \gls{udp} and 100\% for \gls{mqtt}) appears minor, even a slightly lower recall can significantly affect anomaly detection in critical applications (e.g., \gls{hiot} security). However, these minor variations in recall performance can be attributable to the differing characteristics of the \gls{mqtt} and \gls{udp} protocols and the respective network environments (Cooja-based \gls{mqtt} and ns-3-based \gls{udp} in \gls{g5}). The \gls{mqtt} dataset's higher recall (100\%) indicates that the model correctly identified all malicious samples, perhaps due to \gls{mqtt}'s structured and predictable nature \cite{al2024effective}. In contrast, \gls{udp} traffic is comparatively dynamic and less predictable, which may account for the marginally lower recall on the \gls{udp} dataset. As a connectionless protocol, \gls{udp} does not ensure order, consistency, or error correction for data packets \cite{pustovoitov2024assessment}. This leads to fluctuations from variable packet loss, reordering, or duplication in \gls{hiot} devices \cite{tiwari2024comprehensive}.

In \gls{ml}, data preprocessing techniques play a pivotal role in enhancing the performance of anomaly detection models, particularly in achieving high recall \cite{al2021review}. Therefore, the data preprocessing approach ensures clean and accurate data, significantly enhancing the model's performance across all metrics. \gls{ml}/\gls{dl} algorithms, including the \gls{tcn} model, typically excel when a model can identify consistent patterns in data. In this research, while \gls{udp} traffic is comparatively dynamic and less structured than \gls{mqtt}, this variability may challenge the model in identifying consistent patterns for predicting or classifying \gls{hiot} nodes. As a result, the model may struggle to consistently detect all instances of malicious activity, leading to a slightly lower recall rate.

The number of messages or data packets sent by non-malicious nodes in the \gls{mqtt} dataset follows a predictable, sequential pattern. For instance, node 3 sends 350 messages out of a total of 10,007 messages. Additionally, node 4 sends 344 messages out of 10,018, which is remarkably close to the transmissions of node 3, demonstrating the regular behaviour of a normal node. Notably, nodes 3 and 4 do not transmit data in sequence, yet their total packet counts remain close, highlighting the consistency of non-malicious traffic. Based on the analysis, even if a node like node 3 transmits data multiple times to the server, the messages remain consistent and close together, as is expected for a non-malicious node. Furthermore, in a dynamic topology, nodes operate in a random pattern. For instance, while node 3 may send data to the server at one moment, the next packet might be sent by a different node, such as node 18, as depicted in Table~\ref{tab:DDoSdataset}.

In contrast, nodes marked as \gls{ddos} attack nodes, such as node 18, deviate significantly from \gls{mqtt}'s structured behaviour. Node 18 transmits 1,008 messages out of a total of 10,006 messages at high transmission rates, far exceeding the message counts of normal nodes. Similarly, node 17, which sent 967 messages out of 10,011, exhibits behaviour typical of \gls{ddos} attack nodes, with message volumes close to node 18's high counts and significantly greater than those of non-malicious nodes. This structured data helps the \gls{tcn} model achieve 100\% recall.

In the \gls{udp} dataset, normal nodes like node 16 and node 18 each sent 1,912 messages out of their respective totals of 49,991 and 49,993 messages, as shown in Table~\ref{tab:NS3DDoSdataset}. Data traffic appears steady for normal nodes using the \gls{udp} protocol. In contrast, attack nodes show significant variation in message counts, indicating dynamic behaviour. For example, \gls{ddos} attack-affected nodes such as node 20 sent 3824 messages, and node 25 sent 2008. Further, while both protocols demonstrate steady traffic flow for regular and anomalous nodes in controlled \gls{hiot} simulation environments that carefully manage parameters such as packet delivery ratios and delays between malicious and non-malicious nodes, \gls{mqtt} consistently delivers data. This is due to \gls{mqtt}'s structured design, which ensures orderly message flow, and its reliable delivery mechanisms guarantee message delivery even in unstable network conditions.

Despite these differences, both datasets highlight the \gls{tcn} model's capability and versatility in detecting \gls{ddos} attacks. The slight differences in performance demonstrate the model's ability to adapt to different network protocols and attack scenarios.

\begin{table}[ht!]
\centering
\caption{\gls{tcn} model performance on the \gls{mqtt} dataset.}
\small
\adjustbox{max width=\textwidth}{
\begin{tabular}{llllllllll}
 &  &  & \textbf{Metric} &  &  & \textbf{Value} &  &  &  \\ \cline{2-9}
 &  &  & Accuracy        &  &  & 99.98\%        &  &  &  \\ \cline{2-9}
 &  &  & Precision       &  &  & 99.97\%          &  &  &  \\ \cline{2-9}
 &  &  & Recall          &  &  & 100\%        &  &  &  \\ \cline{2-9}
 &  &  & F1-score        &  &  & 99.99\%        &  &  &  \\ \cline{2-9}
\end{tabular}%
}
\label{table:mqttperformance_metrics}
\end{table}

\begin{table}[ht!]
\centering
\caption{\gls{tcn} model performance on the \gls{udp} dataset.}
\adjustbox{max width=\textwidth}{
\begin{tabular}{llllllllll}
 &  &  & \textbf{Metric} &  &  & \textbf{Value} &  &  &  \\ \cline{2-9}
 &  &  & Accuracy        &  &  & 99.99\%        &  &  &  \\ \cline{2-9}
 &  &  & Precision       &  &  & 99.99\%          &  &  &  \\ \cline{2-9}
 &  &  & Recall          &  &  & 99.99\%        &  &  &  \\ \cline{2-9}
 &  &  & F1-score        &  &  & 99.99\%        &  &  &  \\ \cline{2-9}
\end{tabular}%
}
\label{table:udpperformance_metrics}
\end{table}

\subsection{Confusion Matrix in \gls*{ddos} Attack Detection}
The confusion matrix offers a detailed analysis of the model's ability to differentiate between actual \gls{ddos} attacks and regular network traffic. This distinction is essential in analysing both missed attacks and false alarms. Furthermore, in this research, the \gls{ddos} attack detection system categorises ``Attack'' as the positive class and ``Normal'' as the negative class. The model's performance on the {\em UL-ECE-MQTT-DDoS-H-IoT2025} dataset is encapsulated in the confusion matrix, visually represented in the Equation \ref{eq:mqttconfusion}, where TN, FP, FN, and TP represent true negatives, false positives, false negatives, and true positives, respectively.


\begin{equation}
    \text{Confusion Matrix} = 
    \begin{bmatrix}
    \text{TN} & \text{FP} \\
    \text{FN} & \text{TP}
    \end{bmatrix}
    =
    \begin{bmatrix}
    2675 & 1 \\
    0 & 3348
    \end{bmatrix}
    \label{eq:mqttconfusion}
\end{equation}

On the {\em UL-ECE-UDP-DDoS-H-IoT2025} dataset, the confusion matrix follows the same structure as depicted in Equation \ref{eq:udpconfusion}. This matrix highlights the model's performance in detecting \gls{ddos} attacks, with a focus on reducing false positives and negatives.

 \begin{equation}
    \text{Confusion Matrix} = 
    \begin{bmatrix}
    \text{TN} & \text{FP} \\
    \text{FN} & \text{TP}
    \end{bmatrix}
    =
    \begin{bmatrix}
    13426 & 3 \\
    2 & 16536
    \end{bmatrix}
    \label{eq:udpconfusion}
\end{equation}


\begin{table}[ht!]
\centering
\small
\caption[Detection and mitigation accuracy]{Detection and mitigation accuracy for the proposed \gls{mqtt} and \gls{udp} datasets.}
\label{tab:dynamicthresholdaccuracy}
\Large
\resizebox{\textwidth}{!}{%
\renewcommand{\arraystretch}{1.5}
\begin{tabular}{@{}ll@{\hspace{1cm}}ll@{}}
\toprule
\multicolumn{1}{c}{\textbf{Dataset}} &
  \multicolumn{1}{c}{\textbf{Proposed Method}} &
  \multicolumn{1}{c}{\textbf{Detection Acc.}} &
  \multicolumn{1}{c}{\textbf{Mitigation Acc.}} \\ \midrule
{\em UL-ECE-MQTT-DDoS-H-IoT2025}                                  & \gls{tcn} with Dynamic Threshold & 96.94 & 95  \\ 
{\em UL-ECE-MQTT-DDoS-H-IoT2025} (20\% Reduced Dynamic Threshold) & \gls{tcn} with Dynamic Threshold & 96.94 & 100 \\ 
{\em UL-ECE-UDP-DDoS-H-IoT2025}                                   & \gls{tcn} with Dynamic Threshold & 75\%   & 80\% \\ 
{\em UL-ECE-UDP-DDoS-H-IoT2025}   (20\% Reduced Dynamic Threshold)                                & \gls{tcn} with Dynamic Threshold & 75\%   & 100\% \\ 
\bottomrule
\end{tabular}%
  }
\begin{flushleft}
    \scriptsize \textbf{Note:} This table shows the detection and mitigation accuracy of the proposed \gls{tcn} algorithm with a dynamic threshold for the datasets {\em UL-ECE-MQTT-DDoS-H-IoT2025} and {\em UL-ECE-UDP-DDoS-H-IoT2025}.
\end{flushleft}
\end{table}

\subsection{Detection and Mitigation Metrics}
Several key metrics are used to assess the performance of the proposed \gls{tcn} prediction model on both datasets, including accuracy, precision, recall, and F1-score, as illustrated in Tables~\ref{table:mqttperformance_metrics} and \ref{table:udpperformance_metrics}. Moreover, Table~\ref{tab:hyperparameters} describes the parameters used to evaluate the \gls{tcn} model. These metrics and parameters comprehensively demonstrate the model's effectiveness in detecting and mitigating \gls{ddos} attacks in \gls{hiot} environments.  Furthermore, a dynamic threshold strategy is applied to the malicious message counts across nodes. Table~\ref{tab:dynamicthresholdaccuracy} illustrates that, with this approach, the detection accuracy reaches 96.94\%, while the mitigation performance is 95\% in classifying and blacklisting malicious nodes in \gls{ddos} scenarios using the \gls{mqtt} dataset in \gls{hiot} environments.

Using a dynamic threshold, the model achieves 75\% detection accuracy and 80\% mitigation accuracy with the \gls{udp} dataset. With additional threshold filtering, the mitigation accuracy reaches 100\%, as depicted in Table~\ref{tab:dynamicthresholdaccuracy}. These results demonstrate the enhanced resilience of \gls{hiot} networks to anomalies. Further details on the distribution of both datasets are provided in Tables~\ref{tab:mqttclasssample} and \ref{tab:udpclasssample}.

Prediction model accuracy and error can be calculated using the formula below:
\begin{equation}
    Accuracy (\%) = \frac{Correct}{Total} \times 100
\end{equation}
where \(Correct\) is the number of correct predictions, and \(Total\) is the total number of predictions.
\begin{equation}
    \begin{split}
    \textit{Prediction Error} (\%) = \left( 1 - \frac{E}{T} \right) \times 100
    \end{split}
\end{equation}
where \( E \) represents the \textit{Error Count} (number of incorrect predictions), and \( T \) denotes the \textit{Total Predictions} made, which is equivalent to the length of \(y_{\text{pred binary}}\).

If the count of malicious messages for node \( n \), where \( n \) is the identifier of any node (e.g., 15, 19, 20, etc.), exceeds the threshold, node \( n \) is blacklisted. The threshold for blacklisting nodes is determined as follows:
\begin{equation}
\textit{DT} = \frac{\sum \textit{MCPN}}{\textit{N}}
\end{equation}
Where:
\begin{itemize}
    \item \(\textit{DT}\) is the Dynamic Threshold,
    \rev{\item \(\textit{MCPN}\) is the Malicious Count Per Node},
    \item \(\textit{N}\) is the Total Number of Nodes.
\end{itemize}

\subsection{Discussion on \glspl*{tcn} for \gls*{hiot} Applications}
\gls{hiot} relies on low-power devices (e.g., wearables), and \gls{tcn} has emerged as an effective architecture due to its low power consumption \cite{9283028}. Furthermore, \glspl{tcn} are recognised for their memory efficiency and lightweight design, which make them particularly suitable for time-series data analysis in \gls{hiot} applications \cite{burrello2021tcn}. In contrast to \gls{bilstm}, which processes data sequentially step-by-step and becomes slower and computationally expensive for large datasets \cite{tang2021bus}, \glspl{tcn} enable parallel processing, ensuring suitability for real-time \gls{ddos} detection in \gls{hiot} systems. Additionally, \glspl{tcn} use dilated convolutions, which process specific values in a sequence by skipping over regular intervals, such as every 1st, 2nd, or 4th position. In time-series data, such as healthcare sensor readings, each recorded measurement represents a value in the sequence. For example, a heart rate sensor might record values like 80, 82, 81, 83, and 85 every second. By processing values at regular intervals rather than every measurement, dilated convolutions enable the model to capture long-term patterns in the data efficiently.

The experiments utilise two types of data sources. Using the Cooja and ns-3 simulators, real-time simulated time series data, including malicious and non-malicious cases, is generated to evaluate the \gls{tcn} model's performance in \gls{ddos} attack scenarios for \gls{hiot} systems. However, a publicly available real-world healthcare dataset, WUSTL-\gls{ehms}-2020, the ``Electroencephalogram (EEG) Health Monitoring System Dataset" from Washington University in St. Louis \cite{jain_eeg_dataset}, is also used to evaluate the model's ability to analyse health-related time-series data. \rev{With this real-world \gls{iomt} dataset, the \gls{bilstm} model in~\cite{wagan2023fuzzy} achieves an accuracy of 89.67\%, a precision of 68.42\%, a recall of 68.93\%, and an F1-score of 69.02\%. These results provide a useful reference for performance on real-world \gls{iomt} data. Although the proposed method is evaluated on the same dataset, direct comparison with previously reported results remains limited due to differences in experimental settings (e.g., data preprocessing, train–test splits, and model configurations) and evaluation protocols.} While this dataset focuses on \gls{mitm} attack scenarios, it provides a foundation for understanding normal and abnormal patterns in physical healthcare sensor data. Nevertheless, the proposed \gls{tcn} model outperforms all metrics compared to \gls{bilstm} in real-world \gls{hiot} environments, as summarised in Table~\ref{tab:comparedataset}.  


To further demonstrate the advantages of \glspl{tcn} in \gls{hiot} applications, particularly for environmental sensors, the proposed model is evaluated using the WSN\_DDoS\_Attack\_H-IoT2023 dataset, which has been employed in previous research employing \glspl{cnn} and achieved a performance of 92\% \cite{khatun2024time}. The critical need to detect \gls{ddos} attacks in \gls{hiot} systems prompted the comparison of the model's output with previously reported metrics. As summarised in Table~\ref{tab:comparedataset}, the proposed \gls{tcn} model achieved an accuracy of 99.84\%, a precision of 90.15\%, a recall of 98.69\%, and an F1-score of 89.88\%, showing improved performance compared to the previously reported method. Moreover, the \gls{tcn} model is evaluated on two additional datasets, BoT-\gls{iot} \cite{binbusayyis2022investigation}, and CIC-IDS2017 \cite{rosay2022network}, achieving high accuracy in both cases. Therefore, this improvement highlights \gls{tcn}'s ability to effectively capture long-term dependencies in time-series data. These results demonstrate the model's ability to efficiently process resource-intensive time-series data. Additionally, this research compares the proposed model's performance with the \gls{bilstm} model across parameters, memory usage, and inference time. The findings indicate that the \gls{tcn} model is significantly more lightweight, with fewer parameters (189,317 compared to 6,574,577) and less memory consumption, as illustrated in Table~\ref{tab:parametercomparison}. This highlights the \gls{tcn} model's scalability and suitability for resource-constrained \gls{hiot} applications. \rev{Although the difference in inference time is relatively small, latency remains a critical factor in \gls{hiot} environments, where even minor delays can impact timely response and patient safety. The comparison with \gls{bilstm} is intended to highlight efficiency trade-offs between models of comparable architectural complexity. As illustrated in Table~\ref{tab:parametercomparison}, the proposed \gls{tcn} model achieves lower inference time compared to \gls{bilstm} while maintaining reduced computational complexity.}

\rev{However, the proposed \gls{tcn} model focuses specifically on \gls{ddos} detection and does not extend to multi-attack (e.g., \gls{mitm} and \gls{sf}) detection within a single model.}

\subsection{Mitigation Strategies Discussion}
The \gls{tcn} model demonstrates a detection accuracy of 96.94\% in identifying malicious activity by counting malicious messages across nodes on the \gls{mqtt} dataset in \gls{hiot} environments. As a result of the model's detection performance, it achieves a 95\% mitigation accuracy, with only one node as a false positive in the blacklist, even without additional adjustments such as a dynamic threshold-based filtering strategy. A dynamic threshold-based filtering strategy, which reduces the base threshold to improve identification for mitigating the actual \gls{ddos} attack-affected nodes without false positives, enhances the model's ability to detect and address attacks effectively. By reducing the threshold by 20\%, it enhances the security of \gls{hiot} networks by accurately blacklisting all \gls{ddos}-affected \gls{hiot} nodes, achieving 100\% final mitigation accuracy on the \gls{mqtt} dataset, as illustrated in Table~\ref{tab:dynamicthresholdaccuracy}. Moreover, this 20\% reduction in the dynamic threshold is chosen based on practical observations during the experiment. It effectively balances detection reliability and mitigation accuracy while minimising false positives and negatives.

For instance, if 3324 malicious activity messages are detected across six nodes during a specific period (16: 567, 18: 587, 10: 361, 19: 572, 17: 612, 20: 625), this total (3324) is divided by the number of nodes flagged as malicious to calculate the average. As a result, the initial dynamic threshold is approximately 554. This initial base threshold is decreased by 20\%, resulting in a refined threshold of 443.2. Once this threshold is adjusted, any node that has made malicious predictions greater than or equal to 443.2 in the specified time frame may require security actions. Afterwards, these nodes might be blacklisted. This threshold approach significantly improves network security by targeting \gls{hiot} nodes with highly suspicious behaviour, ultimately safeguarding patients' daily lives. Additionally, it provides a more nuanced response than static or arbitrary thresholds, effectively detecting and mitigating \gls{ddos} attacks in \gls{hiot} environments. 

In the proposed detection and mitigation phase, 100\% detection and mitigation accuracy is achieved on the \gls{udp} dataset. However, this result may not be reliable, as it could have occurred by chance. For instance, if the model flagged all nodes as genuinely malicious, the accuracy reached 100\%. In contrast, if the model initially flagged eight nodes, including three non-malicious nodes, it resulted in false positives. In \gls{ml}, confidence percentiles set thresholds that reduce the probability of misclassifying normal nodes as malicious \cite{king2024empirical}. Therefore, this research analyses the results using a confidence percentile threshold to adjust the initial 100\% detection accuracy to a more stable and realistic estimate \cite{xu2024code}. This threshold enables the model to focus on nodes with a high probability of malicious activity, reducing the risk of false positives and improving detection reliability. This adjustment does not modify the proposed \gls{tcn} detection and mitigation algorithm; instead, it is a practical tuning step to enhance its reliability and applicability in real-world \gls{hiot} environments.

Thus, this selective confidence percentile approach slightly lowers overall accuracy to 75\% for detection and 80\% for mitigation; however, it produces more balanced, more dependable performance by prioritising only the most consistently flagged malicious nodes. For example, the threshold might focus on nodes such as (23: 1085, 26: 1061, 24: 1041, 19: 1076, 21: 1120, 20: 1063, 22: 1050, 25: 1085, 7: 551, 6: 564), where two nodes have lower probabilities (7: 551, 6: 564). As a result, the confidence percentile excludes nodes with probabilities lower than 551 and 564. A dynamic filtering strategy is then applied, which reduces the base threshold by 20\% on the \gls{udp} dataset, achieving 100\% mitigation accuracy. Due to the high volume of data in experiments and real-world scenarios, the detection and mitigation algorithm may require additional adjustments to ensure stable, reliable results. Such refinements are critical to maintaining consistent detection and mitigation accuracy across dynamic network environments. 

\section{\rev{Threshold-Based Labelling in the \gls{mqtt} Dataset}}
\rev{In the \emph{UL-ECE-MQTT-DDoS-H-IoT2025} dataset, attack behaviour is characterised by an increased message transmission rate, represented by the frequency feature. During data preprocessing, a threshold of 5 is used to assign class labels: samples with frequency greater than or equal to 5 are labelled as attack (1), while others are labelled as normal (0). Threshold-based definitions are standard for identifying abnormal behaviour in network anomaly detection \cite{9610045}. In this research, the threshold is used to label samples as normal or attack based on data transmission behaviour.}

\rev{Moreover, the threshold is applied prior to the train-test split, ensuring that both training and test samples follow the same definition of attack behaviour. This setup avoids data leakage, as the labels are defined independently of the model training process. For example, data leakage would occur if information from the test set were used during training, such as adjusting the threshold after inspecting the test data or deriving labels from unseen samples \cite{goodfellow2016deep, davis2024framework}.}

\rev{However, to assess the impact of the predefined threshold, the \gls{tcn} model was compared with a simple rule-based classifier that assigns labels using the same threshold condition. The results show that the rule-based method achieves perfect performance on the \gls{mqtt} test set, matching the ground-truth labels exactly. The TCN model also achieves high performance under the same settings. This indicates that the predefined frequency threshold plays a dominant role in distinguishing normal and attack samples in the \gls{mqtt} dataset. Thus, the threshold alone is sufficient to distinguish between the two classes.}

\section{Conclusion} \label{sec:conclu}
This research proposes a novel method to enhance anomaly detection accuracy in \gls{g5}-based \gls{hiot} environments.
The proposed \gls{tcn} model demonstrates high prediction accuracy against \gls{ddos} attacks across both datasets, achieving 99.98\% accuracy with {\em UL-ECE-MQTT-DDoS-H-IoT2025} and 99.99\% accuracy with {\em UL-ECE-UDP-DDoS-H-IoT2025}. This precise prediction phase enables targeted mitigation strategies to enhance network security within \gls{hiot} environments. Additionally, the \gls{tcn} model achieves 96.94\% detection and 95\% mitigation efficiency over \gls{mqtt} with a dynamic threshold. In contrast, the \gls{udp} dataset achieves detection and mitigation accuracy (75\% and 80\%) using a dynamic threshold, automatically calculated as the average count of malicious activity across all nodes. One key finding of this research is that the proposed \gls{tcn} model achieves higher accuracy than previously published methods on well-known datasets (e.g., CIC-IDS2017). This highlights the \gls{tcn} model's ability to adapt to diverse behaviours, demonstrating its versatility and resilience. Furthermore, this research compares two protocols, \gls{mqtt} and \gls{udp}, using two simulators (Cooja and ns-3) within \gls{hiot} environments. The findings underscore the need for advanced security measures to prevent \gls{hiot} data breaches. 

Future research focuses on developing advanced simulated attack models that enable multi-class classification, such as detecting \gls{ddos}, \gls{sf}, and \gls{mitm} attacks within the same topology in \gls{hiot}. This approach will provide a detailed data analysis of malicious and non-malicious data from simulated \gls{hiot} devices. Additionally, it will create a new dataset specifically designed for multi-class anomaly detection in \gls{hiot} security. This dataset will train advanced \gls{dl} models well-suited to processing sequential and time-series data in \gls{hiot} environments.



\chapter{Securing \gls{iot} using Lightweight \gls{tcn} for Edge Deployment on Raspberry~Pi~4} \label{chap:edge} 

\ifpdf
    \graphicspath{{5_chapter5/figures/PNG/}{5_chapter5/figures/PDF/}{5_chapter5/figures/}}
\else
    \graphicspath{{5_chapter5/figures/EPS/}{5_chapter5/figures/}}
\fi

\noindent This research presented in this chapter has been previously published in:

\textbf{Akhi, M.}, Eising, C. and Dhirani, L.L., 2026. Securing IoT Using Lightweight TCN for Edge Deployment on Raspberry Pi 4.~\textit{IEEE Open Journal of the Communications Society},  vol. 7, pp. 442-460, 2026, doi: 10.1109/OJCOMS.2025.3649498.

The content of the paper above has been presented as published in this chapter, except for correcting grammatical errors and making slight rewording so that the narrative is more appropriate for a thesis. A brief preamble is provided to explain its relevance to the thesis narrative.

\rev{\noindent\textbf{Thesis Structure and Publication Context:}
Parts of this thesis, including Chapters~\ref{chap:dataset}, \ref{chap:tcn}, and \ref{chap:edge}, are based on related published research that share a common scope and methodological foundation. Therefore, some overlap in background, simulation setup, and methodological descriptions is included to ensure completeness.}

\clearpage

\normalsize
\noindent\textbf{\underline{Preamble}}\\[0.5em]
In this chapter, \textbf{RQ4} is addressed:

\textbf{RQ4:} What approaches enable the edge deployment of deep learning models (e.g., \gls{tcn}) on resource-constrained devices (e.g., Raspberry Pi 4) in \gls{hiot}?

\vspace{5pt}

The datasets used in this research are introduced in \textbf{RQ2}, where the {\em UL-ECE-MQTT-DDoS-H-IoT2025} and {\em UL-ECE-\gls{udp}-DDoS-H-IoT2025} datasets are constructed and analysed. The previous chapter addresses \textbf{RQ3} by developing a lightweight \gls{tcn}-based method for \gls{ddos} detection and mitigation using these datasets. This chapter reuses the same trained lightweight \gls{tcn} model for edge deployment on a Raspberry Pi 4. Specifically, \textbf{RQ4} investigates how the lightweight \gls{tcn} model can be implemented to perform real-time \gls{ddos} detection on resource-constrained hardware, such as the Raspberry Pi 4. For this edge deployment, the \gls{tcn} model is quantised and converted into \gls{tflite} format to reduce computational overhead, including model size and parameter count. The quantised model achieves a reduction in parameters (63,617~$\rightarrow$~63,136) and model size (248.5~KB~$\rightarrow$~89.1~KB), leading to improved energy efficiency and practical deployability in resource-limited environments. It also preserves detection accuracy and inference performance on the {\em UL-ECE-MQTT-DDoS-H-IoT2025} and {\em UL-ECE-UDP-DDoS-H-IoT2025} datasets. Power consumption is measured with a USB power meter, as software-based measurements on the Raspberry Pi 4 are often unreliable. The results demonstrate the feasibility of deploying the lightweight \gls{tcn} model for real-time healthcare edge applications. Accordingly, this chapter contributes to the main RQ by demonstrating the practical edge deployment of lightweight \gls{dl} models for real-time \gls{ddos} detection in \gls{hiot} environments.

\section{Abstract}
The \gls{iot} is a network of tiny sensing devices that facilitate precision monitoring, automation, and intelligent decision-making in a connected digital environment. While \gls{iot} enables high connectivity, it also presents various issues relating to compliance, operational resilience, coexistence, and cybersecurity. These factors significantly affect the robustness, reliability, and resilience of \gls{iot} devices. To mitigate emerging cybersecurity issues, an efficient and lightweight security solution is essential for resource-constrained edge devices. This research addresses security challenges by leveraging \gls{hiot} \gls{ddos} datasets to deploy a lightweight \gls{tcn} model on a standalone \gls{iot} device, the Raspberry Pi~4. It enables efficient edge deployment for detecting critical cyber threats, particularly \gls{ddos} attacks. The \gls{tcn} model is converted into \gls{tflite} format and optimised through quantisation, reducing model size and computational overhead. The deployment achieves 99.95\% accuracy with an average inference latency of 0.19\,ms on the \gls{mqtt}-based dataset, and 99.94\% accuracy with 0.27\,ms on the UDP-based dataset. The average power consumption on the Raspberry Pi~4, measured using a physical USB power meter, is 4.22\,W and 4.64\,W for the two datasets. These results demonstrate the feasibility of high-accuracy, resource-efficient \gls{ddos} attack detection on low-power edge devices for securing \gls{iot}, \gls{iiot}, and \gls{hiot} systems in real-world environments.


\section{Introduction}
\label{sec:intro}
The advancement of a secure and sustainable digital ecosystem is based on protecting the infrastructure enabled by the \gls{iiot}~\cite{dhirani2021industrial}. The \gls{iot} encompasses physical devices equipped with sensors and software that exchange data over networks~\cite{dhirani2021industrial}. These devices are often domain-specific, such as smart home appliances, wearables in \gls{hiot}, or industrial sensors in \gls{iiot}, yet all fall under the broader \gls{iot} paradigm of Internet-enabled, low-power communication. \gls{hiot} devices support real-time health monitoring and enable precision and personalised care by continuously measuring vital signs such as body temperature, oxygen saturation, heart rate, and blood pressure~\cite{bustin2023advances}. Smart sensors and real-time tracking further enhance critical operations, including medical inventory management (e.g., medications, surgical tools, and protective equipment). These capabilities help avert supply shortages and improve healthcare delivery. This shared foundation enables large-scale, low-power healthcare deployments~\cite{costa2020fasten}. At the same time, their growing connectivity increases exposure to cyber threats, underscoring the necessity of secure edge solutions to protect locally generated data. To address these challenges, edge computing bridges the gap between \gls{iot} data generation and real-time decision-making by enabling local analytics~\cite{modupe2024reviewing}. It also reduces latency and enhances data privacy, a critical requirement in healthcare and industrial ecosystems. Despite these advantages, the lack of standardised edge computing frameworks hampers interoperability across diverse \gls{iot} systems~\cite{standict2023edge,article}.

Another significant challenge is the limited computational, memory, energy, and bandwidth resources of \gls{iot} devices. These constraints, combined with the large-scale connectivity enabled by \gls{g5} networks, intensify the risk of cyber attacks. While \gls{g5} extends intelligence closer to the edge and enhances data transmission speed, it also enlarges the attack surface due to the massive number of connected devices~\cite{lohan20245g}. In particular, \gls{ddos} attacks pose a severe threat, directly compromising data availability and overwhelming \gls{iot} nodes~\cite{almuqren2025cybersecurity}. Securing these systems against \gls{ddos} attacks and achieving cyber-resilience remain pressing challenges~\cite{Meagher2024,dhirani2021industrial}. Despite significant research efforts, traditional cloud-based security solutions (e.g., \gls{ids} and centralised authentication) introduce high latency and excessive bandwidth consumption~\cite{s22103744}. Consequently, such cloud-based approaches are unsuitable for real-time \gls{ddos} detection in critical infrastructures~\cite{otorioIDS2025,article}.

Given the limitations of cloud-based solutions, edge-based security mechanisms have emerged as an alternative~\cite{he2025road}. These approaches enable low-latency processing and local detection of cyber attacks, thereby eliminating reliance on cloud resources. However, deploying \gls{dl}-based security solutions on edge devices remains challenging due to limited computational power and constrained memory resources. \gls{dl} models are resource-intensive, potentially exceeding device capacity, while frequent data transfers between memory and the CPU increase latency and power consumption. Recent studies have explored several \gls{dl} architectures, such as \gls{lstm} networks~\cite{akar2025l2d2}, \gls{ff} models~\cite{omer2025binary}, \glspl{gru}~\cite{assis2021gru}, and \glspl{cnn}~\cite{khatun2024time, aswad2023deep}, as well as traditional\gls{ml} techniques~\cite{nawaz2025lightweight}, to strengthen \gls{iot} security. Despite this progress, most existing approaches incur high computational overhead, making them unsuitable for execution on edge devices. Typically, neural network models operate with \gls{fp32} precision for weights and activations. While this representation ensures high accuracy, it also demands substantial memory and computational resources, making deployment on low-power devices impractical. Thus, reducing model size is not merely an efficiency concern; it is a prerequisite for practical deployment across a wide range of \gls{iot} devices.

\begin{figure}[!htbp] 
\centering \includegraphics[width=0.5\linewidth]{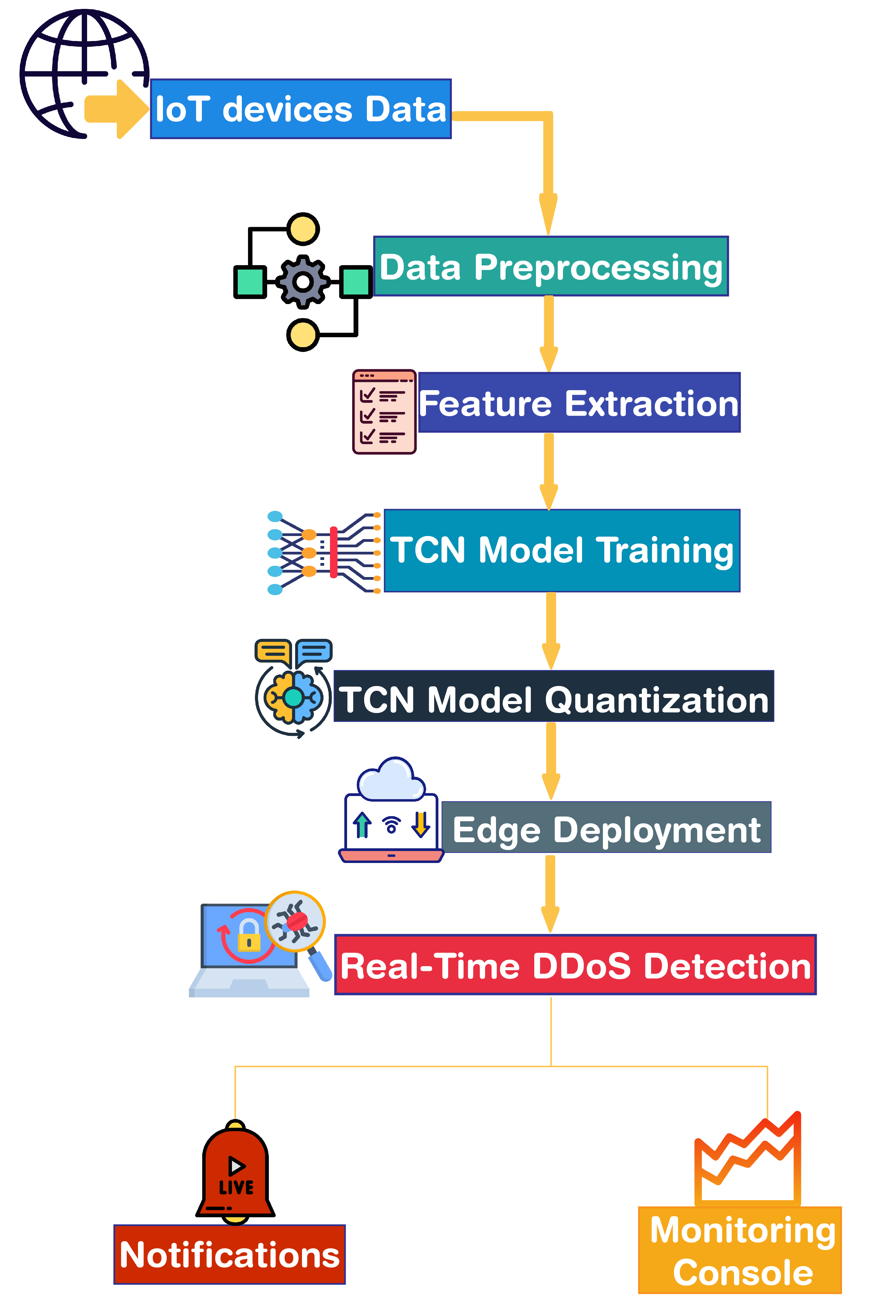} 
\caption[Real-time DDoS detection workflow within IoT]{Real-time \gls{ddos} detection workflow within \gls{iot} environments using a \gls{tcn} model deployment on edge devices.} 
\label{fig:overviewofdeployment} 
\end{figure}

To address this challenge, this research adopts the \gls{tcn} \gls{dl} model described in Chapter~\ref{chap:tcn} for edge deployment. The reuse and adaptation of previously implemented architectures align with current research trends emphasising efficient model design for resource-constrained or edge-oriented \gls{ai} systems~\cite{wang2025optimizing}. Owing to the \gls{tcn} model's lightweight nature, quantisation reduces precision from \gls{fp32} to lower-bit representations (e.g., INT8), thereby minimising resource consumption while preserving accuracy. INT8 quantisation is a widely used standard in \gls{dl} and edge deployment, particularly for \gls{tflite} and ARM-based platforms~\cite{lv2025efficient, wei2025stealthy}. This enables efficient cyberattack classification with high accuracy, low latency, and reduced power usage on edge devices. To evaluate the proposed approach on resource-constrained edge platforms, this research uses the {\em UL-ECE-MQTT-DDoS-H-IoT2025} and {\em UL-ECE-UDP-DDoS-H-IoT2025} datasets described in Chapter~\ref{chap:dataset}. The \gls{tcn} model is converted into \gls{tflite}, an optimised framework for deploying \gls{dl} models on embedded and mobile devices. During conversion, quantisation is applied, resulting in a lightweight model that enables practical \gls{ddos} detection on constrained \gls{iot} devices. Specifically, Raspberry Pi~4 is selected as the edge platform, a low-power \gls{arm}-based device widely used in \gls{iot} environments~\cite{newaz2025approach, krasteva2025implementing, technologies12060081}. Thus, the primary objective is to develop a practical deployment method for securing \gls{iot} systems against \gls{ddos} attacks across \gls{iot}, \gls{hiot}, and \gls{iiot} environments. An overview of the workflow for deploying \gls{tcn}-based \gls{ddos} detection on edge devices is depicted in Figure~\ref{fig:overviewofdeployment}. The performance and feasibility are evaluated on Raspberry Pi~4, including power consumption, inference time, parameter count, and model size. A \gls{usb} power meter is used to measure power consumption, as software-based profiling is unreliable on the Raspberry~Pi~\cite{linuxconfig2023power,bekaroo2016power}. The key contributions of this research are as follows:
\begin{itemize} 
  \item To the best of current knowledge, this is the first lightweight \gls{tcn} model deployed on Raspberry Pi~4 for real-time \gls{ddos} detection, supporting secure operation in diverse \gls{iot} domains, including healthcare and industrial systems. The deployment achieves a reduced parameter count and a smaller model size, making it feasible for \gls{iot} devices that have limited RAM while maintaining consistent performance, high accuracy, and low latency on the~{\em UL-ECE-MQTT-DDoS-H-IoT2025} and {\em UL-ECE-UDP-DDoS-H-IoT2025} datasets.

  \item The deployment is practically validated through USB power meter measurements on a Raspberry~Pi~4, demonstrating average power consumption across five runs of 4.22\,W and 4.64\,W for the~{\em UL-ECE-MQTT-DDoS-H-IoT2025} and {\em UL-ECE-UDP-DDoS-H-IoT2025} datasets, respectively.
\end{itemize}

This chapter is organised as follows. Section~\ref{sec:background} reviews related work on binary classification and edge deployment in \gls{iot}, \gls{hiot}, and \gls{iiot} environments. Section~\ref{sec:method} describes the deployment of a lightweight \gls{tcn} model on Raspberry Pi~4 for efficient \gls{ddos} attack classification across diverse \gls{iot} settings. Section~\ref{sec:Performance} presents a comparative analysis with prior studies, examines the power consumption of the edge system, and reports results on inference time, parameter count, model size, and the confusion matrix. Finally, Section~\ref{sec:conclud} summarises the contributions and discusses future research directions.

\section{Related Work}\label{sec:background}
This section reviews prior approaches, highlighting key contributions and the limitations that affect the feasibility of edge deployment. In this context, anomaly detection has been widely applied for \gls{ddos} detection and mitigation across \gls{iot}, \gls{hiot}, and \gls{iiot} environments. As highlighted in Section~\ref{sec:intro}, several \gls{dl} models are used for securing \gls{iot} environments. Beyond these, hybrid architectures such as \gls{cnn}-BiLSTM and \gls{cnn}-\gls{lstm} have also been employed in prior studies \cite{altunay2023hybrid,aswad2023deep}. These hybrid models introduce a large number of trainable parameters, which diminishes their suitability for deployment scenarios \cite{omer2025binary}. Moreover, many studies rely on older datasets (e.g., CIC-DOS2017, CIC-IDS2017, CIC-IDS2018, NSL-KDD, KDD99, UNSW-NB15) that fail to capture realistic \gls{iot} traffic and evolving threat patterns, leading to limited generalisation \cite{said2023cnn,singh2021novel,liu2020intrusion}. A summary of recent studies and their main contributions is provided in Table~\ref{tab:survey}.

\begin{table}[!htbp]
\centering
\normalsize
\setlength{\tabcolsep}{8pt} 
\renewcommand{\arraystretch}{1.6} 
\caption[Survey of existing \gls{ddos} detection methods]{Survey of existing \gls{ddos} detection methods, attack types, benefits, limitations, and edge deployment in \gls{iot}, \gls{hiot}, and \gls{iiot}.}
\label{tab:survey}
\begin{adjustbox}{max width=\textwidth}
\begin{tabular}{@{}
p{1.7cm}   
p{2.7cm}   
@{\hspace{8pt}}
p{3.2cm}   
>{\raggedright\arraybackslash}p{4.5cm}   
>{\raggedright\arraybackslash}p{4.8cm}   
c
@{}}
\toprule
\textbf{Ref.} &
\textbf{\shortstack{Technique\\Used} } &
\textbf{\shortstack{Targeted\\Attacks}} &
\textbf{\shortstack{Key\\Benefits}} &
\textbf{Limitations} &
\textbf{\shortstack{Edge\\Deployment} } \\
\midrule
\cite{akar2025l2d2} (2025) &
\gls{lstm} &
\gls{dos}, \gls{ddos}, \par and Spoofing in \gls{iomt} &
- 98\% accuracy in intrusion detection. \par - Enhances \gls{iomt} security with \gls{dl}. &
- Lacks edge deployment feasibility. \par 
- High computational cost for real-time detection. &
\ding{55} \\ 
\hline

Chapter~\ref{chap:tcn} (2025) &
\gls{tcn} &
\gls{ddos} in \gls{hiot} &
- 99.99\% accuracy in \gls{ddos} detection and mitigation. \par - Real-time attack identification by monitoring frequency. \par
- Two \gls{hiot} datasets generated with Cooja and ns-3 simulator. & 
- Focuses only on \gls{ddos} attack detection. \par 
- Edge deployment is not investigated. &
\ding{55} \\ 
\hline

\cite{omer2025binary} (2025) & Hybrid \gls{lstm}-FF & LR-\gls{dos} & - Evaluates performance on multiple datasets; binary and multi-class classification. & - High computational complexity and memory requirements. & \ding{55} \\ 
\hline
\cite{nawaz2025lightweight} (2025) & Random \par Forest, \par \acrshort{lr}, \par and \gls{nb} & \gls{ddos} in \gls{iot} & - Achieves high accuracy using traditional \gls{ml} models.\par Feature selection using Extra Trees Classifier (ETC). & - Uses older NSL-KDD dataset \cite{nslkddkaggle}; \par no deployment on \gls{iot} devices; \par no latency analysis; \par traditional \gls{ml} models add resource overhead. & \ding{55} \\ 
\hline
\cite{khatun2024time} (2024) & \gls{cnn} & \gls{ddos} in \gls{hiot} & 
- Accurate anomaly detection with high accuracy. \par
- Authors create a new dataset \cite{akhi2025ulwsn}. &
- Specific to environmental \gls{iot} sensor data. &
\ding{55} \\ 
\bottomrule
\end{tabular}%
\end{adjustbox}
\end{table}

\begin{table}[!htbp]
\ContinuedFloat
\normalsize
\centering
\setlength{\tabcolsep}{8pt} 
\renewcommand{\arraystretch}{1.6} 
\caption[]{Survey of existing \gls{ddos} detection methods, attack types, benefits, limitations, and edge deployment in \gls{iot}, \gls{hiot}, and \gls{iiot} (continued).}
\begin{adjustbox}{max width=\textwidth}
\begin{tabular}{@{}
p{1.7cm}   
p{2.7cm}   
@{\hspace{8pt}}
p{3.2cm}   
>{\raggedright\arraybackslash}p{4.5cm}   
>{\raggedright\arraybackslash}p{4.8cm}   
c
@{}}
\toprule
\textbf{Ref.} &
\textbf{\shortstack{Technique\\Used} } &
\textbf{\shortstack{Targeted\\Attacks}} &
\textbf{\shortstack{Key\\Benefits}} &
\textbf{Limitations} &
\textbf{\shortstack{Edge\\Deployment} } \\
\midrule
\cite{aswad2023deep} (2023) & \gls{cnn}-BiLSTM & \gls{ddos} in \gls{iot} & - Demonstrates high accuracy and precision; \par \gls{bilstm} is evaluated on multiple datasets (e.g., CICIDS2017, CICDDoS2019). & - Large model size; high latency; reliance on older datasets; no deployment or discussion on resource-constrained devices (e.g., Raspberry Pi). & \ding{55} \\ \hline

\cite{altunay2023hybrid} (2023) & \gls{cnn}, \gls{lstm}, \gls{cnn}+\gls{lstm} & \gls{dos}, Worms, Backdoor, Fuzzers, etc. & - Hybrid \gls{cnn}-\gls{lstm} for \gls{iiot} intrusion detection; binary and multi-class classification with up to 99.80\% accuracy. & - Datasets used (UNSW-NB15, X-IIoTID) do not explicitly cover \gls{ddos} attacks; no latency or efficiency analysis; high model complexity; not evaluated on resource-constrained \gls{iiot}/edge devices. &  \ding{55} \\ 

\bottomrule
\end{tabular}%
\end{adjustbox}
\end{table}

Akar et al. \cite{akar2025l2d2} introduce an \gls{lstm}-based intrusion detection model tailored for \gls{iomt} environments. Using the CICIoMT2024 dataset, the model achieves a classification accuracy of 98\%, demonstrating the effectiveness of \gls{dl} in strengthening \gls{iomt} security. Despite this impressive accuracy, the approach suffers from low computational efficiency, which constrains its scalability in practical \gls{iomt} deployments. The authors acknowledge that the \gls{lstm} model incurs significant computational overhead, which raises concerns about its practicality and energy efficiency in resource-constrained \gls{iomt} environments. Similarly, another study indicates that \gls{lstm}-based intrusion detection techniques are constrained by computational complexity and substantial resource requirements~\cite{dash2025optimized}.

A monitoring-frequency-based detection and dynamic-threshold mitigation method using \glspl{tcn} is proposed to secure \gls{g5} \gls{hiot} environments, as detailed in Chapter~\ref{chap:tcn}. The method analyses each node’s traffic contribution relative to overall network activity and applies adaptive thresholds for \gls{ddos} detection and mitigation. \gls{mqtt} and \gls{udp}-based datasets are generated using the Cooja and ns-3 simulators, and the \gls{tcn} model achieves high detection accuracy. However, the lack of edge deployment and resource feasibility analysis highlights a practical gap that this thesis addresses.

In \cite{omer2025binary}, a hybrid \gls{lstm}-FF neural network is proposed for detecting low-rate \gls{dos} (LR-\gls{dos}) attacks. Random Forest is used for feature selection, and the model is evaluated on CIC-DOS2017, CSE-CIC-IDS2018, and LR-HR-DDOS2024 datasets. The approach achieves high accuracy in both binary and multi-class classifications. However, the authors explicitly report that the model has limitations in terms of computational complexity and memory requirements \cite{omer2025binary}. This limitation is consistent with prior findings that \gls{lstm}-based architectures inherently require more computational resources than feedforward neural networks~\cite{mrozek2020review}.

In \cite{nawaz2025lightweight}, the NSL-KDD dataset is used to evaluate three traditional \gls{ml} methods, namely Random Forest, \gls{lr} and \gls{nb}, achieving high accuracy. The study also emphasises that these models are scalable and efficient for \gls{ddos} detection in \gls{iot} networks. However, despite achieving high accuracy, such approaches often incur significant inference latency, making them unsuitable for resource-constrained \gls{iot} devices. The lack of deployment or validation on actual \gls{iot} devices (e.g., ARM-based hardware) further raises uncertainty about their effectiveness in real-world settings. This represents a significant gap, as deployment and latency-aware evaluation are essential to ensure suitability in \gls{iot} environments.

In~\cite{khatun2024time}, \glspl{cnn} are applied to anomaly detection in time-series data for \gls{hiot} environments. The study simulates \gls{ddos} attacks in the Cooja simulator, which emulates environmental sensors, including temperature and humidity. The \gls{cnn}-based approach achieves 92\% accuracy, indicating its effectiveness in detecting anomalies within \gls{hiot} systems. To further support \gls{ddos} detection research, the authors also construct the WSN\_DDoS\_Attack\_H-IoT2023 dataset~\cite{akhi2025ulwsn}. However, the study does not examine the model’s robustness under diverse \gls{ddos} scenarios, including \gls{mqtt}-based attacks.

In~\cite{aswad2023deep}, a \gls{dl}-based \gls{ddos} detection model for \gls{iot} networks is presented. The study employs the CICIDS2017 dataset, which provides realistic traffic with both benign flows and contemporary attack patterns. A \gls{cnnbilstm} model is introduced, achieving high performance with an accuracy of 99.76\% and a precision of 98.90\%. While the results demonstrate effective detection, the centralised evaluation limits insight into the model’s scalability and feasibility in distributed \gls{iot} settings. Moreover, the study~\cite{aswad2023deep} overlooks latency and computational efficiency, which become significant constraints when the same \gls{cnn}-BiLSTM architecture is implemented on the {\em UL-ECE-MQTT-DDoS-H-IoT2025} dataset. This results in higher latency and a larger parameter count than the adopted lightweight \gls{tcn} model.

In \cite{altunay2023hybrid}, the authors present a \gls{dl}-based intrusion detection approach in \gls{iiot} environments. It utilises the UNSW-NB15 and X-IIoTID datasets, covering both binary and multi-class classifications. Three models are implemented, including \gls{cnn}, \gls{lstm}, and a hybrid \gls{cnn}+\gls{lstm}, with the hybrid architecture achieving the best performance. Specifically, the \gls{cnn}-\gls{lstm} model attains 93.21\% and 92.9\% accuracy for binary and multi-class classification on the UNSW-NB15 dataset, while it achieves 99.84\% and 99.80\% accuracy on the X-IIoTID dataset. The lack of latency and computational efficiency analysis raises concerns for \gls{iiot} and large-scale manufacturing environments, where high computational overhead or delayed detection may limit the reliability of industrial processes~\cite{tsallis2025application}. 
In comparison, the adopted lightweight \gls{tcn} model with quantisation addresses these limitations by maintaining near-original accuracy (99.98\% $\rightarrow$ 99.95\%) while reducing latency, model size, and power usage on Raspberry Pi. This optimisation makes it suitable for deployment in diverse \gls{iot} environments, including healthcare, industrial, and other resource-constrained settings.

\normalsize
\section{Methodology}
\label{sec:method}
This section presents the methodology for quantising and converting the lightweight \gls{tcn} model architecture described in Chapter~\ref{chap:tcn} into TensorFlow Lite to reduce model size and computational overhead. This adaptation enables efficient deployment on edge devices such as the Raspberry Pi~4 while maintaining high accuracy and low inference latency. The methodology consists of three key phases: data preprocessing, \gls{tcn} model training, and conversion of the trained \gls{tcn} model to the \gls{tflite} format. After conversion, the model is deployed on the edge device for inference evaluation. The training and conversion experiments are conducted in \textit{Google Colab}, a cloud-based environment for developing and testing \gls{dl} models. 

\normalsize
\subsection{Data Preprocessing}
This research uses the {\em UL-ECE-MQTT-DDoS-H-IoT2025} and {\em UL-ECE-UDP-DDoS-H-IoT2025} datasets for model training and evaluation, as described in Chapter~\ref{chap:dataset}. These datasets are derived from simulated \gls{hiot} networks (e.g., body temperature, oxygen saturation, and heart rate) using the Cooja and ns-3 simulators and include both normal traffic and \gls{ddos} attack scenarios. Related studies often reuse datasets (e.g., CIC-IDS2017) in the \gls{hiot} domain, even though those datasets are not generated from \gls{hiot}-specific devices~\cite{faruqui2023safetymed}. In this context, these previously developed datasets serve as representative benchmarks for evaluating \gls{ddos} detection in \gls{iot}, \gls{hiot}, and \gls{iiot} environments.

The preprocessing pipeline, executed before model training and subsequent \gls{tflite} generation, includes transforming features to numeric values, removing rows containing \texttt{NaN} or $\pm\infty$, label encoding of the outcome, a stratified train–test split, and z-score standardisation using a \texttt{StandardScaler} fitted on the training set and saved for reuse. To ensure numerical stability, rows with invalid entries are removed. The {\em UL-ECE-MQTT-DDoS-H-IoT2025} dataset contains no invalid entries, and its test set remains unchanged, as shown in Table~\ref{tab:mqtt_distribution}. In contrast, the {\em UL-ECE-UDP-DDoS-H-IoT2025} dataset contains slightly fewer test samples due to the removal of invalid records during preprocessing, as detailed in Chapter~\ref{chap:tcn} and summarised in Table~\ref{tab:udp_distribution}. The fitted scaler is saved to ensure consistent preprocessing across model training and Raspberry Pi inference. This approach prevents data leakage between the training and test sets.

The \gls{smote} is applied to the training set to address class imbalance, as described in Chapter~\ref{chap:tcn}. In this method, \gls{smote} generates synthetic samples for the minority class (Normal) to balance the training distribution. Finally, the preprocessed samples are reshaped into the three-dimensional format required by Conv1D layers, ensuring compatibility with the model input during training and subsequent \gls{tflite} conversion.

The conversion workflow incorporates two key refinements to ensure compatibility with \gls{tflite} deployment on Raspberry Pi. First, the feature order, scaler, and label encoder are saved as artefacts to ensure consistent preprocessing during real-time inference. Second, a flattened two-dimensional version of the training data is retained for INT8 quantisation calibration, required during conversion but not in the original \gls{tcn} model.

\begin{table}[ht!]
\centering
\footnotesize
\begin{threeparttable}
\caption[Class distribution of the {\em UL-ECE-UDP-DDoS-H-IoT2025} dataset]{Class distribution of the {\em UL-ECE-UDP-DDoS-H-IoT2025} dataset (post-preprocessing).}
\label{tab:udp_distribution}
\begin{tabular}{lccc}
\toprule
\textbf{Split} & \textbf{Class} & \textbf{\#Samples} & \textbf{Total} \\
\midrule
\multirow{2}{*}{Train (Original)} 
   & Malicious & 35355 & \multirow{2}{*}{69920} \\
   & Normal    & 34565 &  \\
\midrule
\multirow{2}{*}{Test} 
   & Malicious & 14370 & \multirow{2}{*}{29967} \\
   & Normal    & 15597 &  \\
\midrule
\multicolumn{4}{l}{\textbf{Total Dataset Size: $(99869, 13, 1)$}} \\
\bottomrule
\end{tabular}
\end{threeparttable}
\end{table}

\begin{table}[ht!]
\centering
\footnotesize
\caption[Class distribution of the {\em UL-ECE-MQTT-DDoS-H-IoT2025}]{Class distribution of the {\em UL-ECE-MQTT-DDoS-H-IoT2025} dataset.}
\label{tab:mqtt_distribution}
\begin{tabular}{lccc}
\toprule
\textbf{Split} & \textbf{Class} & \textbf{\#Samples} & \textbf{Total} \\
\midrule
\multirow{2}{*}{Train (Original)} 
   & Malicious & 7782 & \multirow{2}{*}{14056} \\
   & Normal    & 6274 &  \\
\midrule
\multirow{2}{*}{Test} 
   & Malicious & 3348 & \multirow{2}{*}{6024} \\
   & Normal    & 2676 &  \\
\midrule
\multicolumn{3}{l}{\textbf{Total Dataset Size: $(20080, 4, 1)$}} \\
\bottomrule
\end{tabular}
\end{table}

\normalsize
\subsection{\gls{tcn} Model Training}
The lightweight \gls{tcn} model, adapted from our earlier study, is trained to detect \gls{ddos} attacks across \gls{iot}, \gls{iiot}, and \gls{hiot} systems. The architecture consists of two residual blocks, each containing three Conv1D layers with 64 filters, kernel size 3, dilation rates of 1, 2, and 4, and causal padding. The depth value of 3 represents the three dilation levels (1, 2, and 4) used within each residual block, resulting in a total of six Conv1D layers. The input feature matrix \(X\) is reshaped to \((N,\;F,\;1)\) so it can be processed by 1D convolutional layers. 
Since each traffic record is treated as an independent feature vector, the input format consists of one feature dimension and one channel. The identical reshaping is applied consistently during training and before \gls{tflite} conversion for deployment on the Raspberry Pi, ensuring that the Conv1D layers process the input correctly. This design expands the receptive field across multiple time scales while maintaining computational efficiency. Causal padding ensures that predictions at time \(t\) are conditioned only on inputs up to \(t\), thereby preserving temporal causality. Batch normalisation is applied after each convolutional layer to stabilise training and improve convergence. To further reduce overfitting, $\ell_2$ regularisation with a factor of 0.01 is applied to all convolutional layers. A Gaussian noise layer ($\sigma=0.1$) is added at the input, and a Dropout layer (rate 0.5) is placed after the second residual block. As described in Chapter~\ref{chap:tcn}, the \gls{tcn} architecture employs global average pooling for temporal aggregation, followed by a sigmoid-activated dense layer for binary classification. In this chapter, the architecture remains unchanged; the focus is on its quantised conversion and deployment on low-power edge devices. For deployment, this pooling operation is implemented using TensorFlow’s built-in GlobalAveragePooling1D layer, which is supported by the model conversion pipeline and enables stable INT8 export. Model training uses the Adam optimiser (learning rate $5 \times 10^{-5}$), a batch size of 64, and up to 50 epochs, with early stopping and learning rate reduction on plateau. This architecture captures temporal dependencies across \gls{iot}, \gls{hiot}, and \gls{iiot} traffic, enabling effective \gls{ddos} classification, as detailed in Table~\ref{tab:modelparameters} and illustrated in Figure~\ref{fig:tcnarchitecture}.

\begin{figure*}[ht!]
    \centering
    \includegraphics[width=0.95\linewidth]{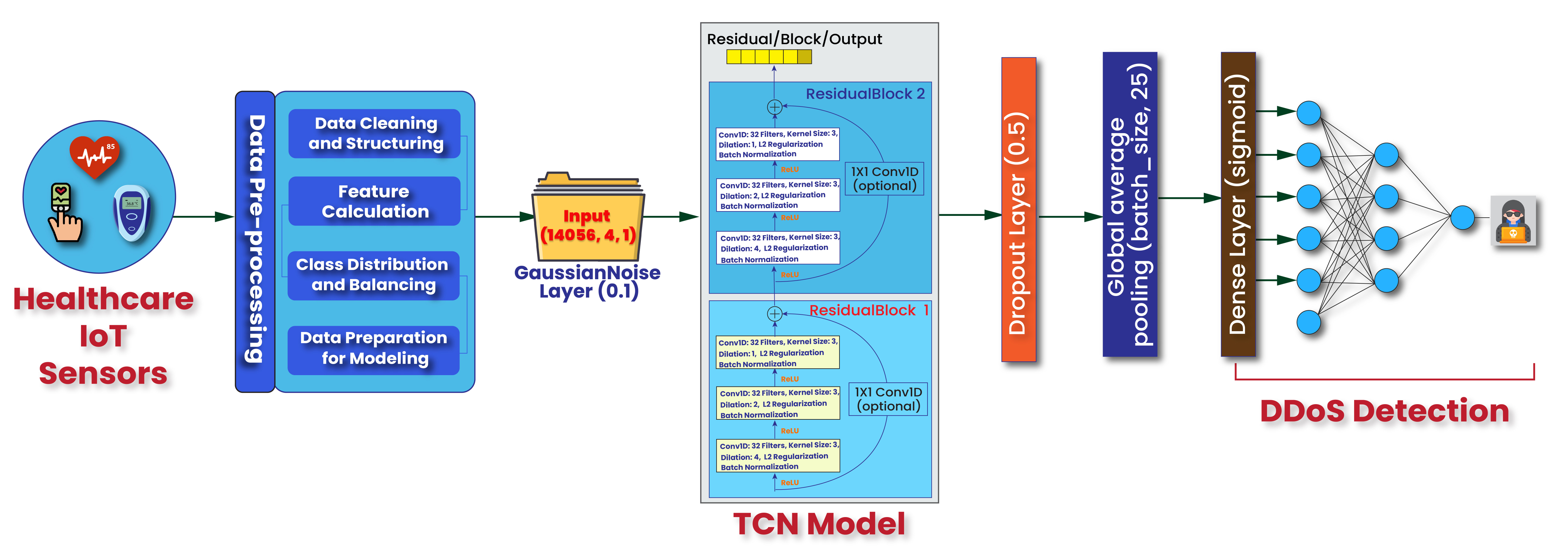}
  \caption[Architecture of the lightweight \gls{tcn} model]{Architecture of the lightweight \gls{tcn} model for \gls{ddos} detection in \gls{iot} environments, including \gls{hiot} and \gls{iiot}, based on the \gls{tcn} design described in Chapter~\ref{chap:tcn}.}
    \label{fig:tcnarchitecture}
\end{figure*}

\begin{table}[ht!]
\caption[Model configuration and training parameters]{Model configuration and training parameters of the lightweight \gls{tcn} model.}
\label{tab:modelparameters}
\centering
\footnotesize
\begin{threeparttable}
\begin{tabular}{@{}ll@{}}
\toprule
\textbf{Components and Hyperparameters}                                   & \textbf{Value}                                               \\ \midrule
Number of features (\gls{mqtt})                                        & 4                                                            \\
Number of features (\gls{udp})       & 13                                                            \\
Number of classes                                          & 2 (Normal and Malicious)                                                            \\
Number of residual blocks                   & 2 blocks, each with 3 convolutional layers                                                          \\ 
Filter count per \gls{tcn} layer                            & 64                                                           \\
Convolution kernel size                                                & 3                                                            \\
Dilation rates applied                                              & 1, 2, 4                                                      \\
Type of padding used                            & Causal (preserves temporal order)  \\
Batch normalisation effect                & Improves stability and convergence \\ 
Model depth (number of stacked \gls{tcn} layers)                                               & 3                                                            \\
Optimizer used                                                  & Adam                                                         \\
Learning rate applied                                              & 5e-5                                                       \\
Loss function                                              & Binary cross-entropy                                          \\
Batch size                                                 & 64                                                           \\
Epochs                                                     & 50                                                           \\
Regularisation techniques                                             & Dropout (0.2 - ReLU, 0.5 - Sigmoid), L2 (0.01) \\
Pooling Layer                     & Global average pooling  \\ 
Input Shape                        & $n_{samples} \times n_{features} \times n_{timesteps}$ \\

Output Layer & Dense layer with Sigmoid activation \\ 
Noise injection method                                           & GaussianNoise (0.1)                                           \\
Validation set ratio                & 20\%                                                   \\
\bottomrule
\end{tabular}
\begin{tablenotes}
\scriptsize
\item \textbf{Note:} This table presents the model configuration, activation functions, learning strategy, and other hyperparameters used in the lightweight \gls{tcn} model for \gls{ddos} detection on the \gls{mqtt} and \gls{udp}-based datasets. 
In this implementation, each feature vector corresponds to a single time step, with $n_{timesteps}=1$, 
which reduces the input shape to $(n_{samples}, n_{features}, 1)$.
\end{tablenotes}
\end{threeparttable}
\end{table}
\normalsize

\subsection{\gls{tcn} Model Conversion into \gls{tflite}}
The \gls{tcn} model, trained on the {\em UL-ECE-MQTT-DDoS-H-IoT2025} and {\em UL-ECE-UDP-DDoS-H-IoT2025} datasets, is converted into \gls{tflite} format to enable efficient deployment on resource-constrained edge devices such as the Raspberry Pi~4.

For deployment, the trained \gls{tcn} is converted using the \texttt{TFLiteConverter} with TensorFlow Lite's default optimisation strategy (\texttt{DEFAULT}), which
applies post-training optimisations such as weight quantisation and computational graph simplifications. Full-integer (INT8) quantisation is enforced by restricting the converter to integer kernels and setting both input and output types to INT8. In this configuration, all convolutional, batch normalisation, pooling, and dense layers are fully quantised with INT8 weights and activations. A representative subset of up to 300 samples from each dataset (drawn from the scaled data splits) calibrates activation ranges; a few hundred samples are sufficient to capture activation distributions while maintaining efficient conversion. For each dataset, the complete held-out test split is exported as \texttt{test.csv} to enable reproducible inference directly on the Raspberry Pi during evaluation.

This quantisation reduces model size by approximately $4\times$, since each parameter requires 1 byte in INT8 instead of 4 bytes in FP32. The converted INT8 models are saved, and their real-world file sizes are reported alongside the theoretical parameter-size reduction. In addition, the exported \gls{tflite} model uses INT8 for both input and output tensors, ensuring that runtime execution on the Raspberry Pi~4 relies on integer arithmetic rather than FP32 operations. During conversion, TensorFlow Lite serialises the model into its FlatBuffer format, a compact binary structure for storing network graphs and parameters~\cite{manor2022custom}. Any remaining non-INT8 buffers correspond to constants folded through graph optimisation (e.g., bias terms and fused batch-normalisation parameters) and therefore do not trigger floating-point computation during runtime.

Although the scale and zero-point are estimated internally by TensorFlow Lite during post-training quantisation, the process follows standard affine quantisation. For signed 8-bit quantisation,
\begin{equation*}
q = \operatorname{clip}\!\big(\operatorname{round}(f/s)+z,\,q_{\min},\,q_{\max}\big), 
\quad f \approx s\,(q - z),
\end{equation*}
with $q_{\min}=-128$ and $q_{\max}=127$. The scale and zero-point are computed as
\begin{equation*}
s=\frac{\max(F)-\min(F)}{q_{\max}-q_{\min}}, 
\qquad 
z=\operatorname{round}\!\Big(q_{\min}-\frac{\min(F)}{s}\Big),
\end{equation*}
where $F$ denotes the range of floating-point values.

In practice, the zero-point is clamped to the valid range $[q_{\min}, q_{\max}]$, 
and thus remains within $[-128,127]$ \cite{liu2025low, zhou2021octo}.

During evaluation, full INT8 quantisation of the \gls{tcn} model results in a minimal change in \gls{ddos} detection accuracy compared to the FP32 model. All evaluation metrics remain within $\pm 0.05\%$ of the original FP32 values on both the {\em UL-ECE-MQTT-DDoS-H-IoT2025} and {\em UL-ECE-UDP-DDoS-H-IoT2025} datasets. This indicates that quantisation preserves detection performance while improving deployment efficiency on the Raspberry~Pi~4.

\subsection{Real-Time Deployment of \gls{tcn} on Raspberry Pi~4}
\label{subsec:deployment}
The \gls{tcn} \gls{tflite} model is deployed on a Raspberry Pi~4 to detect \gls{ddos} traffic in \gls{iot} settings, using the {\em UL-ECE-MQTT-DDoS-H-IoT2025} and {\em UL-ECE-UDP-DDoS-H-IoT2025} datasets. The deployment consists of three steps: loading and preprocessing the dataset, loading the \gls{tflite} model, and performing inference on the edge device. The end-to-end workflow is shown in Figure~\ref{fig:pi_deployment}.

\begin{figure*}[ht!]
    \centering
    \includegraphics[width=0.9\linewidth]{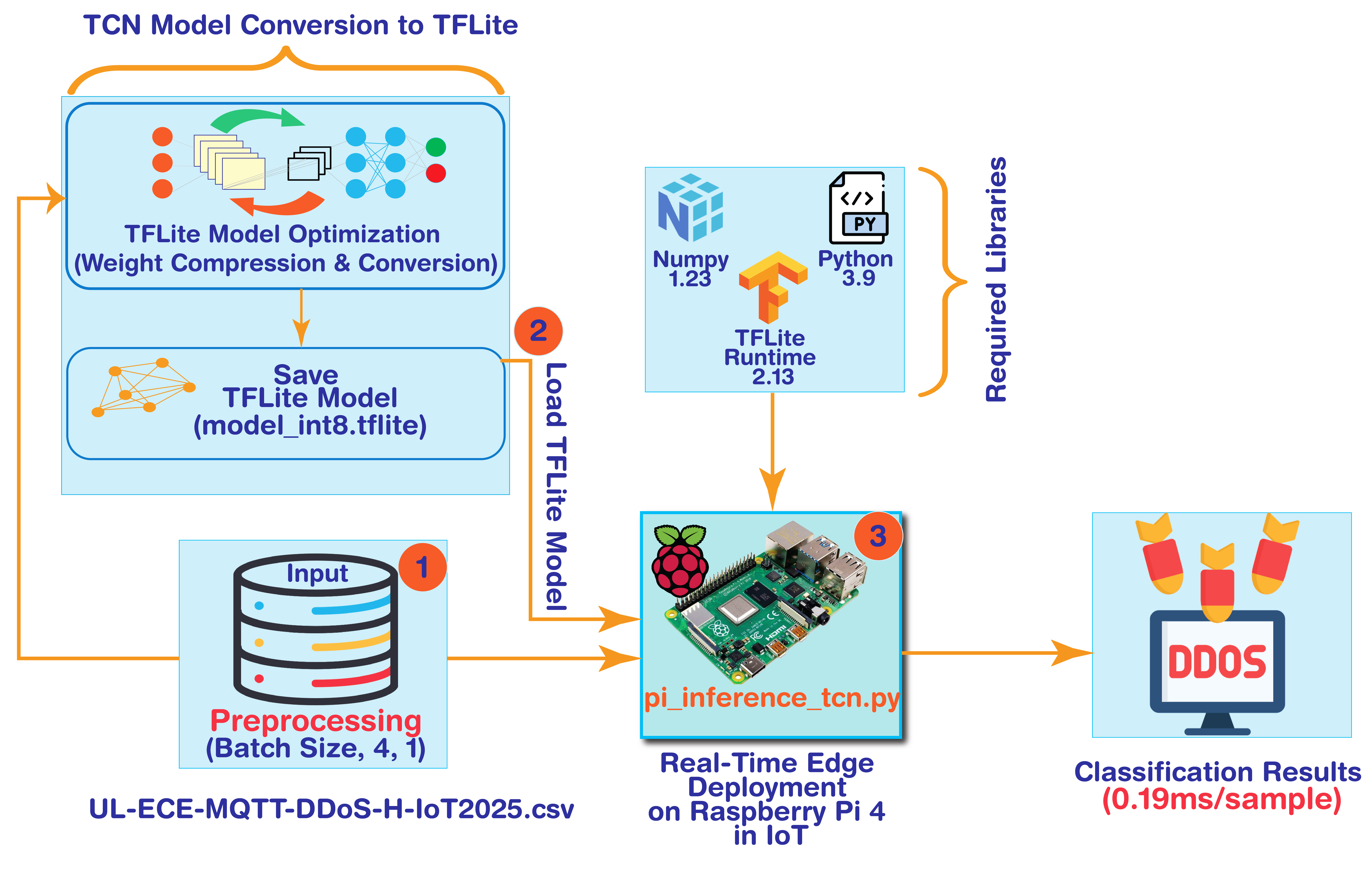}
    \caption[Real-time deployment of the lightweight \gls{tcn} on Raspberry Pi 4]{Real-time deployment of the lightweight \gls{tcn} model on Raspberry Pi 4. The process involves converting the model to \gls{tflite}, performing data preprocessing, and enabling real-time inference and classification.}
    \label{fig:pi_deployment}
\end{figure*}

\subsubsection{\textbf{Dataset Loading and Preprocessing}}
During deployment, both {\em UL-ECE-MQTT-DDoS-H-IoT2025} and {\em UL-ECE-UDP-DDoS-H-IoT2025} test datasets are loaded on the Raspberry Pi and preprocessed following the same procedure used during the original \gls{tcn} model training to ensure consistency. The feature schema, extracted from the training data and saved before \gls{tflite} conversion, is reapplied to preserve the expected feature set and order for input alignment. All feature values are coerced to numeric, $\pm\infty$ entries are converted to \texttt{NaN}, and rows containing \texttt{NaN} values are discarded. The saved \texttt{StandardScaler} and feature schema from training are applied on the Raspberry Pi before inference. \gls{smote} is not used during inference. If the outcome column is present, it is excluded from scaling and retained separately for evaluation. The processed data are reshaped to represent each sample as a sequence of input features with a single channel, as required by the Conv1D layers of the \gls{tcn} model. In this configuration, the number of features is 4 for {\em UL-ECE-MQTT-DDoS-H-IoT2025} and 13 for {\em UL-ECE-UDP-DDoS-H-IoT2025}.

\subsubsection{\textbf{Loading the \gls{tflite} Model}}
The TFLite runtime is a lightweight interpreter that executes the converted model on-device. It is configured with full INT8 input/output quantisation to reduce computational cost and enable efficient execution on edge hardware such as the Raspberry Pi~4. XNNPACK delegation is enabled when supported to reduce inference latency, and the number of inference threads is configured according to the available CPU cores (e.g., 1--4 on a Raspberry Pi~4). To prepare the model for inference, the input tensor is explicitly resized to match the expected shape $[1,F,1]$ before \texttt{allocate\_tensors()} because the interpreter requires tensor dimensions to be defined in advance to allocate memory correctly. After allocation, the interpreter metadata is used to verify that the input and output tensors are in INT8 format for each dataset. To ensure reproducibility, random seeds are fixed, and the model’s declared feature count is validated against the preprocessed dataset.

\subsubsection{\textbf{Inference and Classification}}
Inference is performed sample by sample (batch $=1$). Each normalised feature vector is quantised to INT8 using the interpreter’s input scale and zero-point parameters and then provided to the \gls{tflite} interpreter for inference. The output is subsequently dequantized to floating point using the corresponding output parameters. A threshold of $0.5$ is applied to classify each sample as \{Normal,~DDoS\}. When true class labels (ground truth) are available, the saved label encoder is used to map them, and evaluation metrics (accuracy, precision, recall, F1-score) are reported. To avoid inflated latency from one-time initialisation effects, warm-up inferences are executed and discarded before measurement. In practice, 5--10 warm-up runs are typically sufficient on \gls{tflite}; hence, ten are used for the {\em UL-ECE-MQTT-DDoS-H-IoT2025} dataset and five for the {\em UL-ECE-UDP-DDoS-H-IoT2025} dataset.

\section{Performance Evaluation and Comparison}
\label{sec:Performance}
This section presents a detailed evaluation of the adopted lightweight \gls{tcn} model, comparing it with previously published \gls{ml} and \gls{dl} methods across multiple performance metrics. It further reports the layer-wise parameter counts and per-sample inference time on Google Colab. The evaluation of the \gls{tcn} \gls{tflite} model focuses on its inference latency on the Raspberry Pi~4. In addition, power consumption is measured across the {\em UL-ECE-MQTT-DDoS-H-IoT2025} and {\em UL-ECE-UDP-DDoS-H-IoT2025} datasets, capturing temporal variations in power usage. A confusion matrix is presented to illustrate the model’s classification performance for \gls{ddos} detection on Raspberry Pi~4. It also includes cross-dataset deployment evaluation, edge security considerations, five real-world case studies, security analysis, and attack coverage across diverse \gls{iot} environments, and a discussion of \gls{tcn} model selection.

\subsection{Comparison with Previous Research}
\label{subsec:comparison}
The lightweight \gls{tcn} model is compared with existing \gls{ml} and \gls{dl} methods, including \gls{lstm} \cite{akar2025l2d2}, Hybrid \gls{lstm}-FF \cite{omer2025binary}, \gls{lr} \cite{nawaz2025lightweight}, \gls{cnn}-\gls{bilstm} \cite{aswad2023deep}, and \gls{cnn}-\gls{lstm} \cite{altunay2023hybrid}, all implemented on the {\em UL-ECE-MQTT-DDoS-H-IoT2025} dataset, as outlined in Table~\ref{tab:comparison}. The results demonstrate that the \gls{tcn} model achieves higher accuracy, precision, recall, and F1-score than previous approaches. The 95\% confidence intervals (CIs) for accuracy across both datasets indicate low variability around the mean, with changes of less than 0.1\%, as reported in Table~\ref{tab:comparison}. These narrow bounds demonstrate that the lightweight quantised \gls{tcn} model maintains stable accuracy on the Raspberry~Pi~4. It also exhibits lower inference time per sample and a reduced parameter count, obtained through layer-wise analysis, thereby enhancing its efficiency for real-time applications across \gls{iot}, \gls{iiot}, and \gls{hiot}. The average per-sample inference time on the Raspberry~Pi~4 is 0.19\,ms for the {\em UL-ECE-MQTT-DDoS-H-IoT2025} dataset and 0.27\,ms for the {\em UL-ECE-UDP-DDoS-H-IoT2025} dataset, as shown in Table~\ref{tab:comparison}. As the mean inference time remains identical across the five runs for each dataset, the standard deviation becomes zero, and the confidence interval collapses to identical lower and upper bounds, offering no additional insight. The calculation of the mean inference time is formalised as:
\begin{equation}
\label{eq:mean_inference_time}
\text{Mean Inference Time} = \frac{\sum \text{Inference Times}}{\text{Number of Runs}}
\end{equation}
where:
\begin{itemize}
    \item \(\sum \text{Inference Times}\) represents the total cumulative inference time for all processed samples.
    \item \(\text{Number of Runs}\) corresponds to the total number of inference executions.
\end{itemize}

Compared to the original \gls{tcn}, the lightweight \gls{tcn} \gls{tflite} version exhibits slightly lower accuracy after quantisation. Nevertheless, the parameter count decreases after converting the original \gls{tcn} to its \gls{tflite} version. This reduction is beneficial for edge deployment, as it lowers memory requirements and reduces power-consumption overhead on resource-constrained devices. Notably, the parameter count varies slightly between the {\em UL-ECE-MQTT-DDoS-H-IoT2025} and {\em UL-ECE-UDP-DDoS-H-IoT2025} datasets, as shown in Table~\ref{tab:comparison}. This arises from differences in the dimensions of input features after preprocessing, which affect the size of the first convolutional layer. Since the {\em UL-ECE-UDP-DDoS-H-IoT2025} dataset includes more features in the input layer, it results in slightly different weight shapes and a marginally higher parameter count (two additional parameters) compared to the {\em UL-ECE-MQTT-DDoS-H-IoT2025} dataset, as shown in Table~\ref{tab:tflite_size}. This negligible increase does not affect the overall \gls{tflite} model size. Such accuracy degradation is common in quantised models, particularly when deployed on resource-constrained hardware \cite{tensorflow2023qat}. On the other hand, inference time can vary significantly across different platforms (e.g., Google Colab vs. Raspberry Pi~4) due to differences in system architecture, operating systems, and resource availability. Specifying the computing environment is a standard practice in edge \gls{ai} benchmarking to ensure consistent results and fair comparison \cite{banbury2020benchmarking}. Therefore, the \gls{tcn} \gls{tflite} model is well-suited for low-power edge devices, balancing accuracy, efficiency, and resource consumption.

\begin{table}[!htbp]
\centering
\caption[Comparison of the lightweight \gls{tcn}]{
Comparison of the lightweight \gls{tcn} (\gls{tflite} on Pi~4) model with existing \gls{ml} and \gls{dl} methods. \rev{Baseline methods are evaluated on the {\em UL-ECE-MQTT-DDoS-H-IoT2025} dataset, while the deployed lightweight \gls{tcn} is evaluated separately on the {\em UL-ECE-MQTT-DDoS-H-IoT2025} and {\em UL-ECE-UDP-DDoS-H-IoT2025} datasets.}}
\label{tab:comparison}
\resizebox{\textwidth}{!}{%
\begin{tabular}{@{}l l c c c c c c l@{}}
\toprule
\textbf{Ref.} & \textbf{Method} &  \textbf{Acc.} & \textbf{Prec.} & \textbf{Rec.} & \textbf{F1} & \textbf{Params} & \textbf{Inf. time} & \textbf{95\% CI}\\
\midrule
\cite{akar2025l2d2} (2025)                 & \gls{lstm}                       & 81.29 & 87.71 & 77.15 & 82.09 & 167{,}377   & 0.11 &--\\

\cite{omer2025binary} (2025)            & Hybrid \gls{lstm}-FF                       & 98.49 & 99.72 & 97.54 & 98.62 & 182{,}865  & 0.24 &--\\
\cite{nawaz2025lightweight} (2025)               & \gls{lr}                       & 99.97 & 100 & 99.94 & 99.97 & 5    & 0.25 &--         \\
\cite{aswad2023deep} (2023) & \gls{cnn}-BiLSTM                      & 99.96 & 99.95 & 99.98 & 99.97 & 145{,}121        & 0.15  &--\\

\cite{altunay2023hybrid} (2023) & \gls{cnn}+\gls{lstm}                      & 99.97 & 99.97 & 99.98 & 99.97 & 589{,}953        & 0.11 &-- \\

Chapter~\ref{chap:tcn} (2025) & \gls{tcn} (Original)                  & 99.98 & 99.97 & 100 & 99.99 & 63{,}617 & 0.06 &-- \\
\midrule
\textbf{Deployed (\gls{mqtt})} & \textbf{Lightweight \gls{tcn} (\gls{tflite} on Pi~4)} & \textbf{99.95} & \textbf{99.97} & \textbf{99.94} & \textbf{99.96} & \textbf{63{,}136} & \textbf{0.19} & \textbf{[99.85, 99.98]} \\
\textbf{Deployed (\gls{udp})} & \textbf{Lightweight \gls{tcn} (\gls{tflite} on Pi~4)} & \textbf{99.94}    & \textbf{99.88}    & \textbf{100}    & \textbf{99.94}    & \textbf{63{,}138}& \textbf{0.27} & \textbf{[99.91, 99.96]} \\
\bottomrule
\end{tabular}
}
\vspace{4pt}
\begin{minipage}{\textwidth}
\scriptsize
\textbf{Note:} Results of prior methods and the Original \gls{tcn} are obtained on Google Colab. The Lightweight \gls{tcn} (\gls{tflite} on Pi~4) refers to the deployment of the Original \gls{tcn} model on the Raspberry Pi~4, with separate results reported for the {\em UL-ECE-MQTT-DDoS-H-IoT2025} and {\em UL-ECE-UDP-DDoS-H-IoT2025} datasets. \rev{For clarity, the deployed lightweight \gls{tcn} results are labelled as Deployed (\gls{mqtt}) and Deployed (\gls{udp}), indicating evaluation on the corresponding datasets.} Inference time values are reported in milliseconds per sample (ms/sample).
\end{minipage}
\end{table}

\begin{table}[!htbp]
\centering
\footnotesize
\begin{threeparttable}
\caption[Model size and parameter count]{Model size and parameter count before and after \gls{tflite} conversion of the \gls{tcn} model for both datasets.}
\label{tab:tflite_size}
\begin{tabular}{lccc}
\toprule
\textbf{\gls{tcn} Model}       & \textbf{Parameters} & \textbf{File size (KB)} \\
\midrule
Original FP32 (Keras)    & 63617              & 248.5         \\
INT8 \gls{tflite} (\gls{mqtt})    & 63136              & 89.1  \\
INT8 \gls{tflite} (\gls{udp})     & 63138              & 89.1 \\
\bottomrule
\end{tabular}
\begin{tablenotes}
\scriptsize
\item \textbf{Note:} The parameter count differs by $\pm 2$ between the {\em UL-ECE-MQTT-DDoS-H-IoT2025} and {\em UL-ECE-UDP-DDoS-H-IoT2025} datasets due to variations in input features. This does not affect model size.
\end{tablenotes}
\end{threeparttable}
\end{table}

\begin{figure}[ht!]
  \centering
  \includegraphics[width=0.9\linewidth]{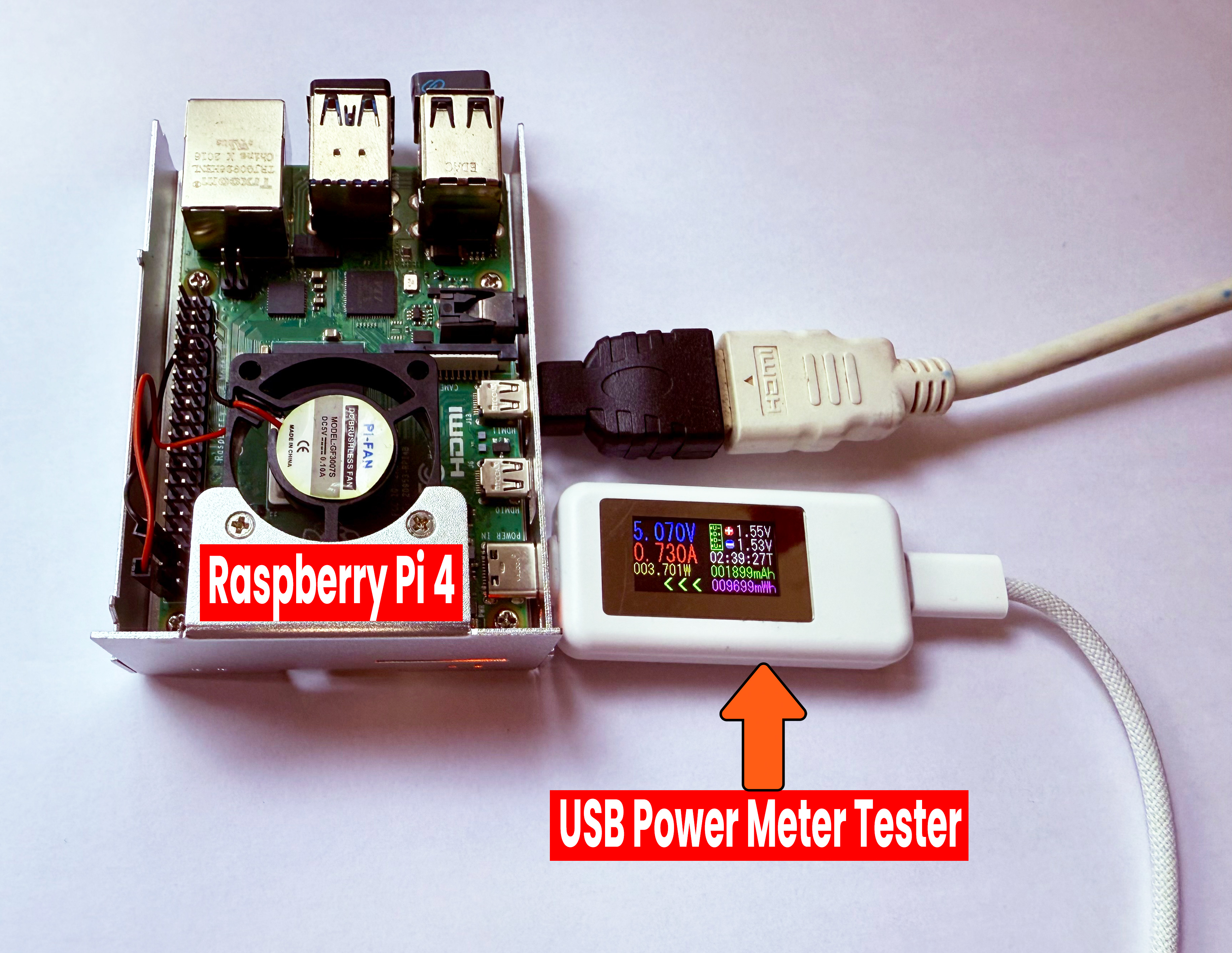}
  \caption[Raspberry Pi 4 with USB Power Meter Tester]{Raspberry Pi 4 with USB Power Meter Tester for real-time power measurement during model inference.}
  \label{fig:pi_power_meter}
\end{figure}

\subsection{Power Consumption Analysis}
The power consumption over time is measured by video recording the USB power meter connected to the Raspberry Pi~4 during execution of the \gls{tcn} \gls{tflite} model on both datasets, namely {\em UL-ECE-MQTT-DDoS-H-IoT2025} and {\em UL-ECE-UDP-DDoS-H-IoT2025}, as shown in Figure~\ref{fig:pi_power_meter}. Before the inference scripts are executed, the Raspberry Pi~4 remains in an idle state, drawing a stable baseline power ($\approx$3.7 W). This baseline reflects the power required to keep the board powered, maintain Raspberry Pi OS (64-bit) based on Debian 11 (Bullseye), and support background services, even when no inference model is executing. For both datasets, recording begins with the Raspberry Pi~4 in an idle state, and the inference script is executed at the start of the video. Recording stops when the inference completion message appears in the terminal. Consequently, the total recorded duration varies slightly between runs (typically within $\pm 1$~s) due to manual start/stop timing and minor system delays. The same inference procedure is followed across all five runs. The subsequent analysis focuses on the inference execution time for each dataset.

\begin{table*}[t!]
\centering
\caption[Power consumption measurement data points]{Power consumption measurement data points across five runs for the {\em UL-ECE-MQTT-DDoS-H-IoT2025} dataset.}
\label{tab:mqtt_powerdata}
\begin{adjustbox}{max width=\textwidth}
\begin{tabular}{cc cc cc cc cc}
\toprule
\multicolumn{2}{c}{\bf Run 1} &
\multicolumn{2}{c}{\bf Run 2} &
\multicolumn{2}{c}{\bf Run 3} &
\multicolumn{2}{c}{\bf Run 4} &
\multicolumn{2}{c}{\bf Run 5} \\
\cmidrule(lr){1-2}\cmidrule(lr){3-4}\cmidrule(lr){5-6}\cmidrule(lr){7-8}\cmidrule(lr){9-10}
Time (s) & Power (W) &
Time (s) & Power (W) &
Time (s) & Power (W) &
Time (s) & Power (W) &
Time (s) & Power (W) \\
\midrule
0.06  & 3.676   & 0.06  & 3.671 & 0.06  & 3.666 & 0.06  & 3.671 & 0.06  & 3.677 \\

1.02  & 3.676   & 1.00  & 3.671 & 0.17  & 3.650 & 1.00  & 3.671 & 1.00  & 3.677 \\

2.34  & 3.677   & 2.00  & 3.671 & 1.20  & 3.650 & 2.00  & 3.671 & 2.00  & 3.677 \\

3.41  & 3.661 & 2.87  & 3.645 & 2.30  & 3.650 & 3.11  & 3.645 & 2.80  & 3.650 \\
4.50  & 3.661 & 3.90  & 3.645 & 3.40  & 3.651 & 4.20  & 3.645 & 3.90  & 3.650 \\
5.50  & 3.661 & 5.00  & 3.645 & 4.48  & 3.661 & 5.30  & 3.645 & 5.00  & 3.650 \\
6.64  & 4.674 & 6.00  & 3.645 & 5.30  & 3.661 & 6.36  & 4.605 & 6.01  & 4.264 \\

7.71  & 4.935 & 7.04  & 5.032 & 6.62  & 4.451 & 7.37  & 5.185 & 7.07  & 5.196 \\

8.80  & 4.935 & 8.20  & 4.892 & 7.67  & 5.111 & 7.41  & 5.187 & 8.14  & 4.899 \\
9.90  & 4.935 & 9.20  & 4.892 & 8.70  & 5.111 & 8.50  & 5.186 & 9.10  & 4.899 \\
10.91 & 4.901 & 10.30 & 4.892 & 9.60  & 5.111 & 9.50  & 5.186 & 10.06 & 4.899 \\

  --    &   --    & 11.41 & 4.564 & 10.55 & 5.111 & 10.61 & 5.186 &   --    &  --     \\
   --   &    --   &    --   &    --   &    --   &     --  & 11.06 & 5.186 &  --     &   --    \\
\bottomrule
\end{tabular}
\end{adjustbox}
\vskip 3pt
\begin{minipage}{\textwidth}
\scriptsize
\noindent\textbf{Note:} Recording start and stop occur manually based on the terminal completion message. Therefore, the total measurement duration and the number of recorded samples vary slightly across runs (typically within $\pm 1$~s). Cells marked ``--'' indicate that no further power sample is available for that run because the corresponding trace finishes earlier than the others.
\end{minipage}
\end{table*}

\begin{figure*}[ht!]
  \centering
  \includegraphics[width=0.9\linewidth]{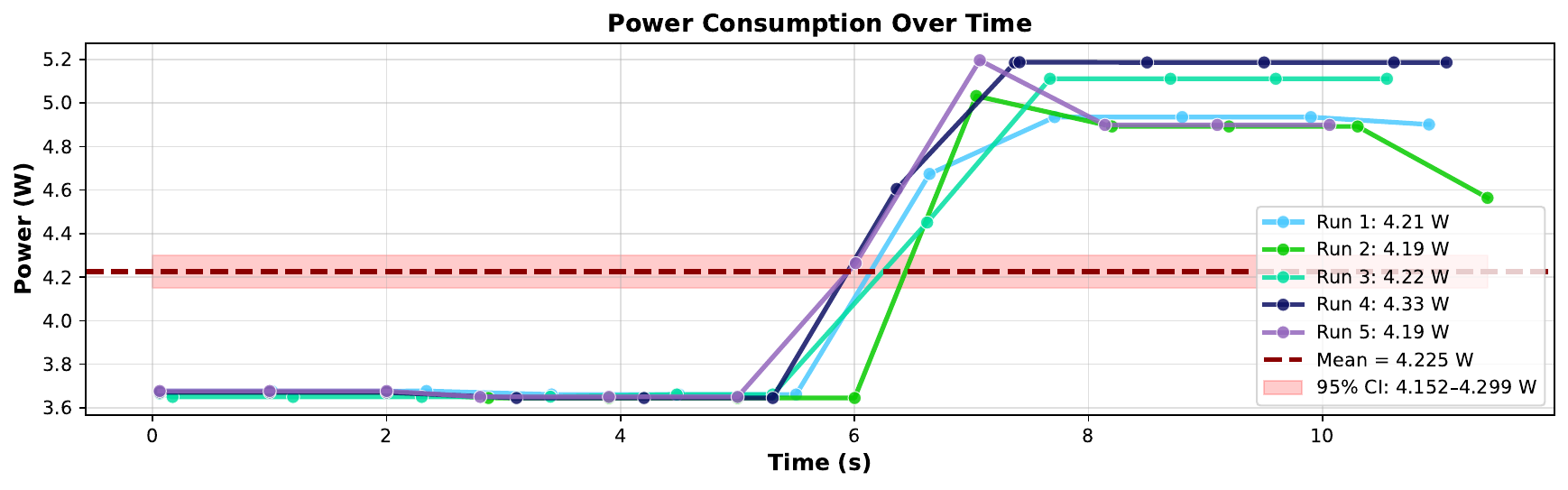}
 \caption[Instantaneous power consumption of the Raspberry~Pi~4]{Instantaneous power consumption of the Raspberry~Pi~4 during execution of the \gls{tcn} \gls{tflite} model on the {\em UL-ECE-MQTT-DDoS-H-IoT2025} dataset. The five run-level traces are shown alongside the mean power value and its 95\% confidence interval.}
   \label{fig:mqtt_power_plot}
\end{figure*}

\begin{table*}[!t]
\centering
\footnotesize
\caption[Statistical summary of Raspberry~Pi~4 power measurements]{Statistical summary of Raspberry~Pi~4 power measurements over five separate runs on the {\em UL-ECE-MQTT-DDoS-H-IoT2025} dataset.}
\label{tab:mqtt_power_stats}
\begin{tabular}{lccccc}
\toprule
\textbf{Metric} & \textbf{Mean} & \textbf{Std Dev} & \textbf{CV (\%)} & \textbf{95\% CI} & \textbf{n} \\
\midrule
Average Duration (s)         & 10.738 & 0.515 & 4.80 & [10.098, 11.378] & 5 \\
Average Total Energy (J)     & 45.379 & 2.393 & 5.27 & [42.407, 48.351] & 5 \\
Average Power (W)    & 4.225  & 0.059 & 1.40 & [4.152, 4.299]   & 5 \\
Peak Power (W)       & 5.092  & 0.110 & 2.16 & [4.956, 5.229]   & 5 \\
Minimum Power (W)    & 3.650  & 0.007 & 0.18 & [3.642, 3.658]   & 5 \\
\bottomrule
\end{tabular}
\begin{minipage}{\linewidth}
\vspace{3pt}
\scriptsize
\textbf{Note:} All statistics are computed from five experimental runs ($n=5$). The average execution duration and average total energy are first computed for each run and then aggregated across the five runs. Peak and minimum power correspond to the maximum and minimum instantaneous power values observed in each run, averaged across runs. The reported standard deviation and CV describe the dispersion of each metric, while the 95\% confidence interval reflects the uncertainty of the estimated mean values.
\end{minipage}
\end{table*}

\subsubsection{\textbf{UL-ECE-MQTT-DDoS-H-IoT2025 Dataset}}
For execution on the {\em UL-ECE-MQTT-DDoS-H-IoT2025} dataset, power readings start at 0.06~s in all runs, while the ending time differs across the five runs and reaches a maximum of 11.41~s. 
Run~1 contains 11 time–power samples, Runs~2 and~3 report 12 samples, and Run~4 consists of 13 samples, even though its ending time (11.06~s) is shorter than that of Run~2. This difference occurs because power values are recorded whenever the USB power meter display changes, whereas additional stable values are manually recorded rather than sampled at fixed time intervals. For example, in Run~4, the displayed power remains at 5.186~W from 8.50~s onwards, and these unchanged readings are included to capture the constant-power phase before termination. As a result, runs with more power fluctuations generate more recorded samples, even if the total duration is similar. This variation in sampling does not affect the energy calculation, as trapezoidal numerical integration accounts for non-uniform time spacing~\cite{chapra2011numerical}. All recorded time–power data points for each run are listed in Table~\ref{tab:mqtt_powerdata}.

The power profile of the Raspberry~Pi~4 during execution of the \gls{tflite} model on the {\em UL-ECE-MQTT-DDoS-H-IoT2025} dataset is illustrated in Figure~\ref{fig:mqtt_power_plot}, where each curve corresponds to one of five independent runs. Power remains close to the idle level (approximately 3.66~W) during the initialisation phase, then rises rapidly starting at approximately 6~s and reaching its peak around 7~s as inference begins. After this transition, power stays in a high-power plateau of roughly 4.9–5.2~W for the remainder of each run, corresponding to the active inference phase. Peak power and minimum power values are calculated individually for each run and then averaged across the five runs, as summarised in Table~\ref{tab:mqtt_power_stats}. The dashed horizontal line denotes the mean power consumption of 4.225~W, while the shaded region represents the 95\% confidence interval \([4.152, 4.299]\). The coefficient of variation (CV), defined as the standard deviation divided by the mean, is 1.40\%, indicating low relative variability across the five runs.

Moreover, a statistical summary of the Raspberry~Pi~4 power measurements on the {\em UL-ECE-MQTT-DDoS-H-IoT2025} dataset across five experimental runs is provided, including average execution duration, average total energy, average power, peak power, and minimum power, as detailed in Table~\ref{tab:mqtt_power_stats}. The average execution duration across the five runs is 10.738~s with a standard deviation of 0.515~s. The total energy consumption for each run is computed from the measured power–time samples. The mean energy across the five runs is 45.379~J, representing the time-integrated electrical demand of a single inference execution, expressed in joules (J). For each run, the average power over its execution duration is computed from the corresponding power--time samples. These five run-wise average power values are then used to calculate the aggregate statistics, including the overall mean, standard deviation, CV, and 95\% confidence interval. The standard deviation of the average power across the five runs is 0.059~W, and the CV is 1.40\%. The column $n=5$ indicates that all reported statistics are computed using five separate experimental runs. The reported 95\% confidence intervals quantify the uncertainty in the estimated means across these runs. While Figure~\ref{fig:mqtt_power_plot} presents instantaneous power trajectories over time, Table~\ref{tab:mqtt_power_stats} summarises run-level aggregate metrics. The analysis of the video-derived time data, together with the measured power values, outlines the system’s power trajectory during execution and transition toward idle. Based on this trajectory, energy consumption is calculated using the trapezoidal rule, which integrates the discrete time–power measurements \cite{chapra2011numerical, abdulhameed2021improved}.


\begin{table*}[t!]
\centering
\caption[Power consumption measurement data points]{Power consumption measurement data points across five runs for the {\em UL-ECE-UDP-DDoS-H-IoT2025} dataset.}
\label{tab:udp_data_points}
\begin{adjustbox}{max width=\linewidth}
\begin{tabular}{cccccccccccc}
\toprule
\multicolumn{2}{c}{\textbf{Run 1}} & \multicolumn{2}{c}{\textbf{Run 2}} & \multicolumn{2}{c}{\textbf{Run 3}} & \multicolumn{2}{c}{\textbf{Run 4}} & \multicolumn{2}{c}{\textbf{Run 5}} \\
\cmidrule(lr){1-2}\cmidrule(lr){3-4}\cmidrule(lr){5-6}\cmidrule(lr){7-8}\cmidrule(lr){9-10}
Time (s) & Power (W) & Time (s) & Power (W) & Time (s) & Power (W) & Time (s) & Power (W) & Time (s) & Power (W) \\
\midrule
 0.05 & 3.707    & 0.07 & 3.701   & 0.08 & 3.717   & 0.06 & 3.723   & 0.05 & 3.910 \\
 1.00 & 3.707    & 1.11 & 3.696   & 1.10 & 3.717  & 1.10 & 3.723    & 1.10 & 3.739 \\
 2.00 & 3.707    & 2.20 & 3.696   & 2.20 & 3.701  & 2.10 & 3.670    & 2.20 & 3.739 \\
 3.14 & 3.697    & 3.30 & 3.696   & 3.27 & 3.702  & 3.20 & 3.670    & 3.30 & 3.739 \\
 4.20 & 3.697    & 4.40 & 3.696   & 4.30 & 3.702  & 4.30 & 3.670    & 4.30 & 3.727 \\
 5.30 & 3.697    & 5.40 & 3.728   & 5.40 & 3.702  & 5.40 & 3.670    & 5.40 & 3.727 \\
 6.30 & 4.697    & 6.50 & 3.728   & 6.47 & 3.686  & 6.37 & 3.766    & 6.44 & 4.927 \\
 6.34 & 4.289    & 7.54 & 3.729   & 7.53 & 3.685  & 7.43 & 3.763    & 7.50 & 5.166 \\
 7.40 & 5.286    & 8.61 & 5.217   & 8.60 & 3.685  & 8.50 & 5.389    & 8.60 & 5.166 \\
 8.47 & 4.823    & 9.68 & 5.167   & 9.67 & 5.019  & 9.60 & 4.979    & 9.70 & 5.166 \\
 9.50 & 4.823    & 10.70 & 5.167  & 10.76 & 4.971  & 10.70 & 4.979   & 10.70 & 5.066 \\
10.60 & 4.823    & 11.80 & 5.167  & 11.84 & 4.936  & 11.74 & 4.945   & 10.74 & 5.024 \\
11.70 & 4.823    & 12.90 & 4.977  & 12.90 & 4.936  & 12.80 & 4.945   & 11.80 & 5.035 \\
12.73 & 4.870    & 13.96 & 4.989  & 14.00 & 4.936  & 13.87 & 5.019   & 12.90 & 5.035 \\
13.80 & 4.870    & 15.00 & 4.989  & 15.03 & 5.024  & 15.00 & 5.019   & 14.00 & 5.035 \\
14.90 & 4.870    & 16.10 & 4.989  & 16.10 & 5.024  & 16.10 & 5.019   & 15.10 & 5.035 \\
15.94 & 5.024    & 17.17 & 5.030  & 17.20 & 5.024  & 17.20 & 5.019   & 16.20 & 5.035 \\
17.00 & 5.024    & 18.20 & 5.030  & 18.30 & 5.024  & 18.13 & 5.030   & 17.30 & 5.035 \\
18.10 & 5.024    & 19.30 & 5.030  & 19.40 & 5.024  & 19.20 & 5.030   & 19.26 & 5.024 \\
19.20 & 5.024    & 20.40 & 5.030  & 20.50 & 5.024  & 20.30 & 5.030   & 20.30 & 5.024 \\
20.30 & 5.024    & 21.50 & 5.030  & 21.60 & 5.024  & 21.40 & 5.030   & 21.40 & 5.024 \\
21.40 & 5.024    & 22.60 & 5.030  & 22.70 & 5.024  & 22.50 & 5.030   & 22.50 & 5.024 \\
22.50 & 5.024    & 23.70 & 5.030  & 23.60 & 5.056  & 23.60 & 5.030   & 23.53 & 5.024 \\
23.60 & 5.024    & 24.80 & 5.030  & 24.70 & 5.056  & 24.56 & 4.701   & --   & --    \\
24.51 & 5.035    & 25.19 & 5.030  & 25.80 & 5.056  & 25.03 & 4.701   & --   & --    \\
25.60 & 5.035    & --   & --      & 26.90 & 5.056  & --   & --      & --   & --    \\
26.70 & 5.035    & --   & --      & 27.23 & 4.463  & --   & --      & --   & --    \\
27.72 & 4.435    & --   & --      & --   & --      & --   & --      & --   & --    \\
\bottomrule
\end{tabular}
\end{adjustbox}
\begin{minipage}{\linewidth}
\vspace{2pt}
\scriptsize
\textbf{Note:} Start and stop of each run are controlled manually based on the terminal completion message. Consequently, the total measurement duration and the number of recorded samples differ slightly across runs. Cells marked ``--'' indicate that no further power sample is available for that run in that row because the corresponding trace finishes earlier than the others.
\end{minipage}
\end{table*}

\subsubsection{\textbf{UL-ECE-UDP-DDoS-H-IoT2025 Dataset}}
The power measurements for the {\em UL-ECE-UDP-DDoS-H-IoT2025} dataset follow the same procedure as the \gls{mqtt} experiments. The first time-stamped sample across the five runs lies between 0.05~s and 0.08~s, as shown in Table~\ref{tab:udp_data_points}. This offset arises because the power meter and camera operate on independent refresh cycles, so the first recorded frame may not align with a meter update. All runs except Run~5 start near 3.7~W (3.70–3.72~W), while Run~5 starts slightly higher at 3.91~W before it converges to the same range; this phase corresponds to the script initialisation period. The \gls{tcn} \gls{tflite} inference script enters its active processing phase approximately between 6~s and 9~s across the five runs, marked by a distinct rise in power from the idle level to the 5.0–5.3~W range. After this transition, the traces stabilise into a steady-power plateau ranging approximately from 4.82~W to 5.06~W for the remainder of the execution window. Toward the end of each run, power decreases as the inference script completes and the Raspberry~Pi~4 returns to its idle level.

\begin{figure*}[ht!]
  \centering
  \includegraphics[width=0.9\linewidth]{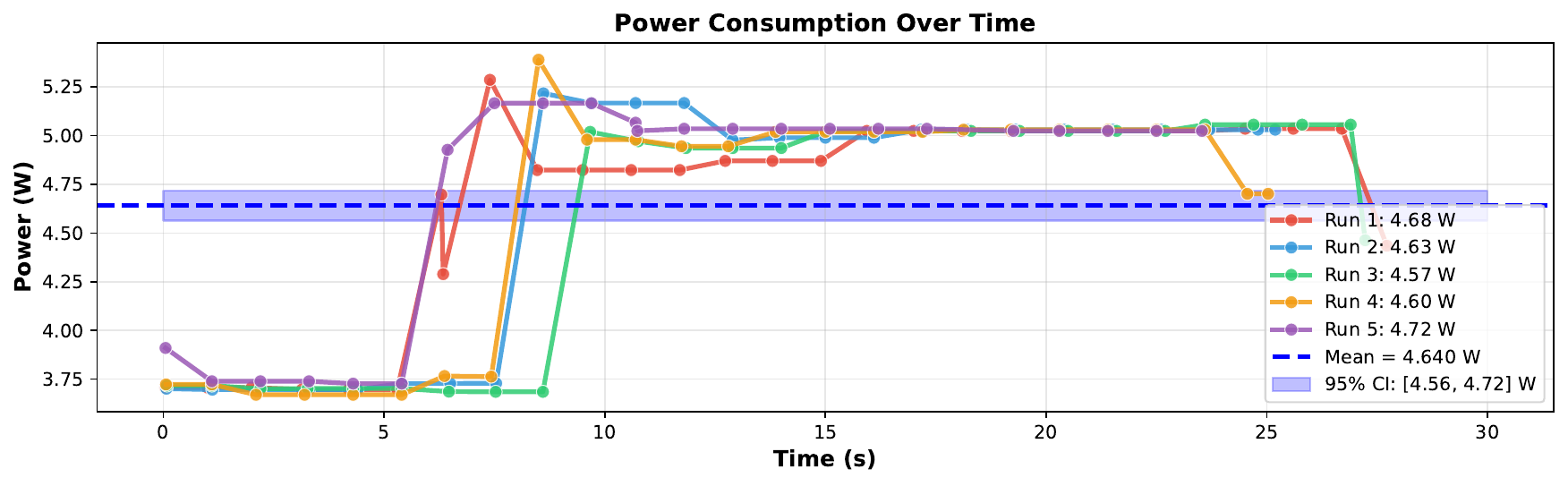}
 \caption[Power consumption of the Raspberry~Pi~4]{Power consumption of the Raspberry~Pi~4 while executing the \gls{tcn} \gls{tflite} model on the {\em UL-ECE-UDP-DDoS-H-IoT2025} dataset, with the average power highlighted.}
  \label{fig:udp_power_plot}
\end{figure*}

Moreover, the run-specific fluctuations in the idle region, transient spikes at the start of computation, and variations around the transition step are illustrated in Figure~\ref{fig:udp_power_plot}. It also includes the per-run average, the mean power across five runs of 4.640 W, and the 95\% confidence interval [4.56, 4.72], which indicates the lower and upper bounds around the mean power during the inference execution phase.

 \begin{table*}[t!]
 \footnotesize
\centering
\caption[Statistical summary of Raspberry~Pi~4 power measurements]{Statistical summary of Raspberry~Pi~4 power measurements over five runs ($n=5$) on the {\em UL-ECE-UDP-DDoS-H-IoT2025} dataset.}
\label{tab:udp_power_summary}
\begin{adjustbox}{max width=\linewidth}
\begin{tabular}{lccccc}
\toprule
\textbf{Metric} & \textbf{Mean} & \textbf{Std Dev} & \textbf{CV (\%)} & \textbf{95\% CI} & \textbf{n} \\
\midrule
Average Duration (s)      & 25.678 & 1.716 & 6.68 & [23.547, 27.809] & 5 \\
Average Total Energy (J)  & 119.124 & 7.511 & 6.31 & [109.798, 128.449] & 5 \\
Average Power (W)         & 4.640 & 0.062 & 1.34 & [4.563, 4.718] & 5 \\
Peak Power (W)            & 5.223 & 0.125 & 2.40 & [5.067, 5.378] & 5 \\
Minimum Power (W)         & 3.695 & 0.021 & 0.57 & [3.669, 3.721] & 5 \\
\bottomrule
\end{tabular}
\end{adjustbox}
\begin{minipage}{\linewidth}
\vspace{2pt}
\scriptsize
\textbf{Note:} All statistics use five experimental runs ($n=5$). Duration and total energy values are run-level means. Peak and minimum power correspond to the maximum and minimum instantaneous power observed in each run, averaged across runs. The reported standard deviation and CV describe the dispersion of each metric, while the 95\% confidence intervals indicate the uncertainty in the estimated mean values.
\end{minipage}
\end{table*}

A statistical summary of the Raspberry~Pi~4 power consumptions over five runs on the {\em UL-ECE-UDP-DDoS-H-IoT2025} dataset is listed in Table~\ref{tab:udp_power_summary}, reporting the average execution duration, total energy, mean power, peak power, and minimum power. The Raspberry~Pi~4 maintains a consistent power profile during active execution, with a low coefficient of variation of 1.34\%, indicating that the mean power stays stable across runs. The total energy varies between approximately 109.80 and 128.45 J, primarily due to differences in execution duration. The instantaneous power values also remain clustered, with peak levels between 5.067 and 5.378 W and minimum levels between 3.669 and 3.721 W.

A comparison of the two datasets shows a clear difference in sustained power demand during \gls{tcn} inference. The \gls{mqtt}-based dataset produces an average active power of 4.225~W, whereas the \gls{udp}-based dataset produces 4.640~W, as depicted in Figures~\ref{fig:mqtt_power_plot} and~\ref{fig:udp_power_plot}. Subtracting the shared idle region near 3.7~W yields an effective computational overhead of approximately 0.53~W for \gls{mqtt} and 0.94~W for \gls{udp}. This higher requirement in the \gls{udp}-based dataset aligns with the heavier packet processing observed in its transition and plateau regions. Despite this difference, both workloads maintain a tightly grouped plateau and exhibit stable behaviour across runs, supporting the suitability of the lightweight \gls{tcn} model for deployment in resource-constrained \gls{iot} environments.

To ensure reproducibility, an example USB power meter recording and the corresponding time–power data files for both datasets are publicly available in our GitHub repository \cite{akhi2025uleceiotddos}. The repository also illustrates the whole measurement procedure to support method transparency, while the chapter reports the extracted data from five independent runs.

\subsection{Confusion Matrix Analysis}
The confusion matrix is employed to evaluate the binary classification performance of the lightweight \gls{tcn} model on the Raspberry Pi~4 within \gls{iot} environments, including \gls{hiot} and \gls{iiot}. It demonstrates the model’s ability to distinguish Normal from \gls{ddos} traffic with high accuracy and minimal misclassification.

As shown in~\eqref{eq:mqtt_confusion}, the confusion matrix for the {\em UL-ECE-MQTT-DDoS-H-IoT2025} dataset reports 2675 true negatives (TN), 3346 true positives (TP), 1 false positive (FP), and 2 false negatives (FN). This corresponds to only three misclassifications out of 6024 samples, indicating high classification accuracy.

For the {\em UL-ECE-UDP-DDoS-H-IoT2025} dataset, the model achieves 14352 TN and 15597 TP, with 18 FP and no FN, as reported in~\eqref{eq:udp_confusion_matrix}. This outcome indicates consistent performance, with few false positives and no false negatives, ensuring that all \gls{ddos}-affected nodes are correctly identified.
\begin{equation}
    \label{eq:mqtt_confusion}
     \text{Confusion Matrix} = 
    \begin{bmatrix}
    \text{TN} & \text{FP} \\
    \text{FN} & \text{TP}
    \end{bmatrix}
    =
    \begin{bmatrix}
    2675 & 1 \\
    2 & 3346
    \end{bmatrix}
\end{equation}

\begin{equation}
\label{eq:udp_confusion_matrix}
    \text{Confusion Matrix} = 
    \begin{bmatrix}
    \text{TN} & \text{FP} \\
    \text{FN} & \text{TP}
    \end{bmatrix}
    =
    \begin{bmatrix}
    14352 & 18 \\
    0 & 15597
    \end{bmatrix}
\end{equation}

\subsection{Deployment for Cross-Dataset Evaluation} \label{subsec:cross-dataset}
This section presents the deployment of a lightweight \gls{tcn} on the Raspberry Pi~4 using the public CICIoT2023 and CICIoT2025 datasets, which represent generic \gls{iot} and \gls{iiot} traffic, respectively. This cross-datasets evaluation assesses model generalisation beyond the \gls{hiot} \gls{ddos} datasets used in this study.

\subsubsection{\textbf{CICIoT2023 Dataset}}
A binary classification pipeline is designed for cross-dataset evaluation using CICIoT2023 traffic data. The process starts by constructing a \gls{ddos} \gls{udp} flood traffic dataset. Initially, four \gls{udp} flood traffic files are randomly selected from the CICIoT2023 dataset: DDoS-UDP\_Flood\{1, 5, 10, 15\}.pcap.csv. These files are merged to produce a consolidated dataset, merged\_CICIoT2023\_DDoSUDP.csv, which captures varying \gls{udp} flood intensities. Two benign traffic samples are subsequently collected from BenignTraffic.pcap.csv and BenignTraffic2.pcap.csv and concatenated into a single benign dataset. A binary outcome column is introduced to distinguish between normal and attack traffic. Feature column consistency is enforced, and the benign dataset is aligned with the same feature structure as the \gls{ddos} \gls{udp} dataset. The benign subset is then merged with the consolidated \gls{ddos} \gls{udp} traffic to construct the final binary dataset named CICIoT2023\_DDoS\_vs\_Normal\_binary.csv. The combined dataset is randomly shuffled to avoid ordering bias. After merging, duplicate samples are inspected, and all exact duplicate rows are removed to reduce redundancy and prevent sampling bias. The final deduplicated dataset is saved as CICIoT2023\_clean\_binary.csv. A summary of the dataset composition and final feature dimension after preprocessing and leakage removal is presented in Table~\ref{tab:ciciot2023_summary}.

\begin{table}[t!]
\centering
\footnotesize
\caption{Summary of the CICIoT2023 and CICIIoT2025 datasets.}
\label{tab:ciciot2023_summary}
\begin{tabular}{l l}
\toprule
\multicolumn{2}{c}{\textbf{CICIoT2023 Dataset}} \\
\hline
\textbf{Metric} & \textbf{Count} \\
\midrule
Total samples             & 1180686 \\
\gls{ddos} attack samples       & 510000 \\
Benign samples            & 670686 \\
Final feature dimension   & 38 \\
Training samples (70\%)   & 826480 \\
Testing samples (30\%)    & 354206 \\
\midrule
\multicolumn{2}{c}{\textbf{CICIIoT2025 Dataset}} \\
\hline
Total samples             & 238828 \\
Attack samples            & 129236 \\
Benign samples            & 109592 \\
Final feature dimension   & 71 \\
Training samples (70\%)   & 167179 \\
Testing samples (30\%)    & 71649 \\
\bottomrule
\end{tabular}
\end{table}

During data preprocessing, a leakage analysis is applied to the cleaned dataset. All features are converted to numeric values, and rows containing NaN or infinite values are removed. Label and identifier features such as label, class, attack\_type, and is\_attack are explicitly checked and removed if present. Timestamp columns are also removed to prevent indirect temporal leakage. Pearson correlation is then computed between each remaining feature and the target label. The ``Number'' column shows a high correlation with the target (Pearson correlation $\approx 0.999$), indicating direct label leakage. This column is removed from both the training and test sets to prevent artificially inflated performance. After these removals, no remaining feature exhibits near-perfect correlation with the target label, confirming the absence of detectable feature-level leakage.

A stratified 70--30 train--test split is applied to preserve class distribution. Feature normalisation is performed using a StandardScaler fitted only on the training data. The same scaler and feature order are reused during deployment to maintain preprocessing consistency.

The lightweight \gls{tcn} model is fully quantised to INT8 using TensorFlow Lite and deployed on the Raspberry Pi~4. The deployment procedure follows the same pipeline used for the \gls{mqtt} and \gls{udp}-based datasets, as described in Section~\ref{subsec:deployment}. Evaluation on the CICIoT2023 dataset demonstrates a deployment accuracy of 98.43\%, with a precision of 97.13\% and a recall of 99.30\%. All deployment performance metrics are reported in Table~\ref{tab:crossdatasets_results}.

\begin{table}[ht!]
\centering
\small
\caption[Deployment performance of the lightweight TCN]{Deployment performance of the lightweight \gls{tcn} (\gls{tflite} on Raspberry Pi~4) on the CICIoT2023 and CICIIoT2025 datasets.}
\label{tab:crossdatasets_results}
\begin{tabular}{l c c c c @{\hspace{4pt}} c c}
\toprule
\textbf{Dataset} & \textbf{Acc.} & \textbf{Prec.} & \textbf{Rec.} & \textbf{F1} & \textbf{ROC-AUC} & \textbf{Inf. time (ms/sample)} \\
\midrule
CICIoT2023 & 98.43 & 97.13 & 99.30 & 98.20 & 99.55 & 0.70 \\
\hline
CICIIoT2025 & 65.80 & 62.99 & 89.26 & 73.86 & 75.35 & 1.29 \\
\bottomrule
\end{tabular}
\end{table}

The confusion matrix shows 1068 false negatives and 151995 true positives, indicating that the model rarely misclassifies attack traffic as normal, which is critical for security-sensitive \gls{iot} environments. A total of 4494 false positives is observed, reflecting a bias toward attack detection. This behaviour is acceptable in intrusion-detection scenarios, where failing to detect an attack poses a higher operational risk than triggering occasional false alarms. The confusion matrix is given as follows:


\begin{equation*}
    \label{eq:ciciot_confusion_matrix}
     \text{Confusion Matrix} = 
    \begin{bmatrix}
    \text{TN} & \text{FP} \\
    \text{FN} & \text{TP}
    \end{bmatrix}
    =
    \begin{bmatrix}
    196649 & 4494 \\
    1068 & 151995
    \end{bmatrix}
\end{equation*}

The average inference latency on the Raspberry Pi~4 is 0.70~ms per sample, as computed using (\ref{eq:mean_inference_time}) and reported in Table~\ref{tab:crossdatasets_results}. This corresponds to a maximum theoretical throughput of approximately 1429 samples per second on a single device. Such latency enables near real-time packet-level inference at the network edge, which is essential for timely \gls{ddos} detection in constrained environments. This cross-dataset evaluation demonstrates the model’s generalisation capability under realistic hardware and quantisation constraints in a low-power edge environment.

\subsubsection{\textbf{CICIIoT2025 Dataset}}
The CICIIoT2025 dataset is a recent \gls{iiot} traffic dataset that contains benign flows and multiple attack behaviours. The dataset provides traffic segments in separate CSV files, each containing flows captured over different time windows, such as 2, 3, 5, 8, and 10 seconds. The attack category reflects network-level malicious activity, including packet bursts and traffic patterns consistent with \gls{ddos} behaviour. In this work, the dataset is prepared in a binary format to support intrusion detection.

The process starts by combining all attack and benign CSV files into a single dataset. A set of fields that behave as explicit labels or metadata identifiers is removed to avoid leakage. This removal includes fields such as Number, label, attack\_type, and related categorical indicators. The target variable is the outcome, and the feature matrix is constructed by removing the target field. Timestamp-related columns, including timestamp, are also removed because they act as implicit identifiers and degrade model generalisation.

Numeric fields are retained, infinite values are replaced with the median, missing values are filled with column-wise medians, and constant features are removed. After preprocessing, the dataset contains 238828 samples, with 129236 attack samples and 109592 benign samples. The final feature dimension is 71. Stratified sampling is applied to create training and test splits of 70\% and 30\%, respectively. The summary of the CICIIoT2025 dataset, including total samples, attack and benign samples, is summarised in Table~\ref{tab:ciciot2023_summary}.

All features are scaled using StandardScaler, fitted on the training data, and applied to the test data. These steps establish a consistent, reproducible configuration for training the \gls{tcn} model and converting it to \gls{tflite}.

Following model conversion, the INT8 \gls{tflite} model is deployed on a Raspberry Pi~4 using the proposed edge pipeline. The \gls{tcn} model is evaluated on the CICIIoT2025 dataset under identical settings. The deployment metrics, including accuracy and precision, recall, F1-score, ROC-AUC, and average inference time, are reported in Table~\ref{tab:crossdatasets_results}. On the CICIIoT2025 dataset, the model produces lower accuracy (e.g., 65\%) due to greater feature overlap between benign and malicious flows. However, the recall of 89.26\% and ROC-AUC remain consistent with the overall detection objective. The inference time on the Raspberry Pi~4 increases to 1.29 ms/sample. The latency difference is independent of the number of samples in the dataset. The Raspberry Pi processes one sample at a time, and computation time depends on the size of the feature vector fed into the model. CICIIoT2025 provides 71 input features per sample, while CICIoT2023 provides 38 features per sample. A larger feature vector increases the number of operations in the first convolutional layers, thereby increasing per-sample processing time. This effect explains why CICIIoT2025 exhibits a slightly higher inference time of 1.29 ms than CICIoT2023 (0.70 ms). The difference remains small and stays within real-time requirements for \gls{iiot} settings.

Moreover, in the confusion matrix, the first row corresponds to benign samples, and the second row to attack samples. The model identifies a high proportion of attack samples, resulting in high recall. However, a substantial number of benign samples appear in the attack column, and this pattern reduces accuracy and precision in Table~\ref{tab:crossdatasets_results}. This behaviour indicates the close statistical similarity between benign and attack traffic within CICIIoT2025. As a result, the classification boundary becomes harder to maintain on edge hardware.


\begin{equation*}
    \label{eq:ciciiot_confusion_matrix}
     \text{Confusion Matrix} = 
    \begin{bmatrix}
    \text{TN} & \text{FP} \\
    \text{FN} & \text{TP}
    \end{bmatrix}
    =
    \begin{bmatrix}
    12542 & 20336 \\
    4165 & 34606
    \end{bmatrix}
\end{equation*}

The \gls{tcn} model is evaluated from two perspectives: using a large dataset with fewer features (CICIoT2023) and a smaller dataset with a larger feature set (CICIoT2025), both on the Raspberry Pi~4. The results demonstrate that per-sample latency is related to the dimensionality of the input features rather than to the dataset size \cite{thapa2025generalizable}. This trend also appears in our \gls{mqtt} and \gls{udp}-based \gls{ddos} datasets, where the {\em UL-ECE-UDP-DDoS-H-IoT2025} dataset, with 13 features, exhibits a higher latency of 0.27 ms per sample.

\subsection{Security Considerations for Edge \gls{tflite} Deployment}
Security considerations focus on preserving the integrity, confidentiality, and authenticity of data, deployment code, and \gls{tflite}-based \gls{ml}/\gls{dl} models in security-critical edge deployments~\cite{padmavathi2025security}. This work focuses on application-layer \gls{ddos} detection, in which the Raspberry Pi~4 operates as a local edge node in diverse \gls{iot} environments. In this setting, traffic-derived features are analysed locally on the Raspberry Pi, and the \gls{ddos} detection outputs are displayed via a local monitoring interface. Deploying a \gls{tflite} model on a resource-constrained edge device introduces risks such as misconfiguration, model tampering, privilege misuse, and resource abuse, which require explicit analysis at the deployment level~\cite{jeyalakshmi2025security, huang2025systematic}. The potential deployment of cyber threats is as follows:

\begin{itemize}
       \item \textbf{Model and deployment tampering:} 
    A malicious actor modifies the \gls{tflite} model file or its deployment scripts to alter detection logic or disable the \gls{ddos} detection service.

    \item \textbf{Unauthorized device access:} 
    An adversary gains unauthorised physical or remote access to the edge device (e.g., a Raspberry Pi~4), enabling the detection service to be stopped, bypassed, or reconfigured.

    \item \textbf{Inference service denial-of-service:} 
    An adversary floods the inference service with excessive or malformed traffic-derived input samples, causing CPU exhaustion and increased \gls{ddos} detection latency or service interruption.
    
    \item \textbf{Model extraction and reverse engineering:} 
    An unauthorised entity copies the \gls{tflite} model file stored on the Raspberry Pi to analyse its structure or reuse it without authorisation, exposing its internal architectural and quantisation design details.
\end{itemize}

Given the identified deployment-level threats, a proactive mitigation strategy is necessary to preserve the integrity and reliability of the edge inference process. This strategy focuses on protecting the deployed \gls{tflite} model, the host edge device (e.g., Raspberry Pi~4), and the inference pipeline against practical attack surfaces in resource-constrained edge environments~\cite{padmavathi2025security}. The corresponding mitigation mechanisms are as follows:

\begin{itemize}
    \item \textbf{Model integrity protection:} The deployment design includes integrity verification of the \gls{tflite} model and associated unauthorised cryptographic file hashing, enabling detection of unauthorised modification before runtime loading.

    \item \textbf{Privilege minimisation:} The inference service is designed to operate under restricted system privileges or a dedicated user account to limit the impact of device compromise.

    \item \textbf{Resource protection:} Rate-control mechanisms can be applied at the inference interface to prevent abuse by excessive request injection or CPU resource exhaustion.
    
  \item \textbf{Process isolation:}
    The inference process can be deployed in a lightweight container or sandbox environment to isolate it from the host system and limit lateral impact in the event of a compromised process.

   \item \textbf{Data locality and exposure reduction:}
    All inference is executed locally on the edge device, and extracted traffic features are not transmitted outside the node, reducing exposure to remote adversaries.
\end{itemize}

The above measures address the primary practical security risks associated with deploying a \gls{tflite}-based \gls{ddos} detection model on a low-power edge platform.

\subsection{Case Studies}
To highlight the practical relevance of deploying the lightweight \gls{tcn} model in \gls{tflite} form on low-power edge devices, this section outlines five real-world case studies that demonstrate its integration across \gls{iot}, \gls{hiot}, and \gls{iiot} environments.

\begin{enumerate}
\item \textbf{Home Healthcare Gateways:} Home \gls{iot} gateway devices that support precision healthcare and aggregate data from wearable sensors such as heart-rate monitors, oximeters, and motion trackers could use the model to detect abnormal spikes in network traffic \cite{ianculescu2025enhancing}. These spikes (e.g., sudden bursts of \gls{mqtt} publish requests) may indicate attempts by compromised devices to overwhelm the gateway. The low-energy overhead enables continuous, on-device monitoring without affecting battery-powered nodes.

 \item \textbf{Smart Hospital Ward Monitoring:} A Raspberry~Pi can operate as a local edge gateway in hospital wards, where it connects to \gls{hiot} devices such as heart-rate sensors, temperature monitors, and wearable patient trackers \cite{serepas2025lightweight}. In this setting, the \gls{tcn} \gls{tflite} model could run directly on the edge device to detect abnormal network patterns generated by compromised devices. The measured mean inference latency (0.19--0.27\,ms) enables real-time identification of \gls{ddos}-induced congestion that may disrupt patient-critical data streams.

\item \textbf{Remote Health Wearables and Telemedicine Kits:} Portable diagnostic and telemedicine kits (e.g., glucose meters and ECG patches) used in rural clinics, emergency response teams, or mobile medical vans often depend on weak local WiFi or temporary portable routers to transmit patient readings \cite{kumar2025evaluation}. These unreliable connections can become congested by unsolicited network traffic, which may introduce delays during remote consultations or reduce the responsiveness of diagnostic data transfers. Deploying the \gls{tcn} \gls{tflite} model on a Raspberry~Pi that controls the kit provides a lightweight mechanism for tracking connection quality and maintaining timely data delivery in resource-limited medical settings.

\item \textbf{Industrial Sensor Clusters:} Edge controllers in \gls{iiot} environments, such as manufacturing floors and process plants, are capable of embedding the model to protect PLCs and high-frequency sensor clusters from \gls{ddos} bursts \cite{latif2025hardware}. The model’s ability to operate reliably across different input dimensions and communication patterns makes it suitable for \gls{iiot} deployments where sensors often transmit heterogeneous, protocol-specific traffic. For example, the cross-dataset evaluation on the edge device, discussed in Section~\ref{subsec:cross-dataset}, demonstrates that the model remains stable even when operating on a higher-dimensional feature space (38 features).

\item \textbf{Smart City Environmental Nodes:} Air-quality sensors, roadside traffic detectors, and weather-monitoring units typically forward periodic measurements through lightweight edge gateways with limited compute resources \cite{pozzebon2025real}. These gateways are susceptible to abnormal volumes of unsolicited network traffic originating from outside the sensor network, which can saturate their communication links. Deploying the \gls{tcn} \gls{tflite} model on the gateway enables early detection of traffic anomalies associated with \gls{ddos} activity, supporting uninterrupted delivery of sensor data to municipal servers.
\end{enumerate}

\subsection{Security Analysis and Attack Coverage across Diverse \gls{iot} Environments}
In this chapter, the lightweight \gls{tcn} \gls{tflite} model is deployed to detect traffic anomalies on the Raspberry~Pi~4. Its primary focus is identifying \gls{ddos} attacks, which typically manifest as rapid increases in packet rates or other abnormal communication patterns across \gls{hiot}, \gls{iot}, and \gls{iiot} devices. Beyond \gls{ddos}, \gls{mitm} variants such as packet injection, duplication, or relaying can also introduce traffic deviations in these environments \cite{aliyu2025enhancing}. Packet injection, duplication, or relay at higher-than-normal rates alters message frequency and traffic volume, and these changes are reflected in the \gls{mqtt} and \gls{udp}-based features in our datasets. Under these conditions, the \gls{tcn} \gls{tflite} model can flag the anomaly because it has learned these frequency-based patterns during training.

General \gls{iot} and \gls{iiot} threats also arise from irregular communication patterns rather than increased traffic volume \cite{sheng2025network}. Replay attempts may introduce repeated message patterns, brute-force bursts may produce short, periodic authentication clusters, and malware scanning may generate probe sequences across ports or devices \cite{flood2025enhancing}. These pattern-level deviations differ from normal device behaviour and can be detected by the \gls{tcn} \gls{tflite} model when they appear in the input features. The cross-dataset results in Table~\ref{tab:crossdatasets_results} demonstrate that the model continues to identify these traffic-pattern anomalies on the CICIIoT2025 dataset, despite the diverse device behaviours and noisier \gls{iiot} traffic.

Other attack classes, such as firmware compromise, eavesdropping, side-channel leakage, or ransomware activation, operate at layers that do not produce distinctive changes in traffic patterns \cite{jeyalakshmi2025security}. These classes are typically mitigated through cryptographic, firmware-level, or device-level protections~\cite{huang2025systematic}. In this setting, the model could contribute to a broader multilayer defence by covering the network-traffic dimension of the threat surface.

\subsection{Discussion of \gls{tcn} Selection and Edge Deployment}
This discussion focuses on the rationale for selecting the \gls{tcn} model and its deployment on the Raspberry Pi~4. The \gls{tcn} model is used in this research due to its lightweight architecture and computational efficiency, as demonstrated in Chapter~\ref{chap:tcn}, where it is shown to require fewer parameters than \glspl{cnn}, \glspl{lstm}, and traditional models such as \glspl{dt} and \gls{knn}. This advantage is further reinforced in the present research through comparison with prior models, as discussed in Section~\ref{subsec:comparison}. In addition to its efficiency, the \gls{tcn} model demonstrates preliminary evidence of generalisation and scalability across the evaluated datasets and experimental settings. It achieves high accuracy, as reported in Chapter~\ref{chap:tcn}, on the {\em UL-ECE-MQTT-DDoS-H-IoT2025} and {\em UL-ECE-UDP-DDoS-H-IoT2025} datasets and demonstrates competitive performance on selected existing datasets, including the real-world public \gls{iomt} dataset WUSTL-\gls{ehms}-2020~\cite{jain_eeg_dataset}. These results indicate the potential applicability of \gls{tcn} for handling diverse \gls{iot} and \gls{iomt} scenarios within the evaluated experimental setup. Therefore, the \gls{tcn} model is converted to \gls{tflite} and deployed on a resource-constrained edge device, such as the Raspberry Pi~4.

Although the reduction in parameter count is relatively small (63,617 $\rightarrow$ 63,136), this outcome is expected, as \gls{tflite} conversion focuses on optimising model storage and computation rather than on changing the underlying architecture. The practical benefit is reflected in the substantial decrease in model size (248.5 KB $\rightarrow$ 89.1 KB), which directly improves feasibility for deployment on edge devices with limited storage and memory resources.

The power evaluation further indicates that the additional energy overhead of deploying the \gls{tcn} \gls{tflite} model is minimal on the Raspberry~Pi~4, even when handling different feature dimensions (4 in the \gls{mqtt}-based dataset versus 13 in the \gls{udp}-based dataset). The model maintains efficiency across \gls{mqtt} and \gls{udp}-based \gls{hiot} traffic, supporting its feasibility for deployment in \gls{iot}, \gls{hiot}, and \gls{iiot} environments.

\rev{However, a direct comparison of energy consumption with existing methods is not included because power measurements for those methods are not reported under the same Raspberry Pi 4 setup~\cite{akar2025l2d2, omer2025binary, nawaz2025lightweight, aswad2023deep}. As a result, comparison under identical experimental conditions is not feasible. Therefore, this study evaluates the energy consumption of the lightweight \gls{tcn} using both \gls{mqtt} and \gls{udp}-based datasets under a consistent Raspberry Pi 4 deployment setting. Input characteristics, such as feature complexity and data processing requirements, can influence power consumption. Additionally, compared to conventional deep learning models such as \gls{cnnbilstm} and \gls{lstm}, the lightweight \gls{tcn} uses fewer parameters and reduced computational complexity, supporting its suitability for energy-efficient deployment.}

\section{Conclusion}
\label{sec:conclud}
This chapter adopts a previously proposed \gls{tcn} model and investigates its deployment on low-power edge devices for \gls{ddos} detection in \gls{iot}, \gls{hiot}, and \gls{iiot} environments. The model, trained and evaluated on the {\em UL-ECE-MQTT-DDoS-H-IoT2025} and {\em UL-ECE-UDP-DDoS-H-IoT2025} datasets, achieves high accuracy in identifying \gls{ddos} attacks. Furthermore, conversion to the \gls{tflite} format enables lightweight deployment on resource-constrained edge devices (e.g., Raspberry Pi~4). Through quantisation, the model maintains high detection accuracy while reducing both latency and model size. The deployment of the \gls{tcn} \gls{tflite} model on the Raspberry~Pi~4 demonstrates low, stable power consumption during inference across both datasets, highlighting its feasibility for resource-constrained edge environments within the scope of the study. The deployment evaluation also encompasses public \gls{iot} and \gls{iiot} datasets, enabling a broader, more representative assessment across multiple deployment domains. Future work involves deploying the $\sim$89.1 KB INT8 \gls{tcn} on resource-constrained \gls{iot} devices (e.g., Raspberry Pi Pico and ESP32) using TensorFlow Lite for Microcontrollers. Additionally, future work will evaluate the model on multiclass datasets and explore alternative, lightweight architectures to enhance performance and resilience against emerging cyber attacks. This research is limited to a single edge platform (Raspberry Pi~4), which may limit the direct generalizability of the deployment results to other hardware platforms.



\chapter{Residual Temporal \glspl*{cnn} for Emerging Cyber Threat Detection in \gls*{hiot}} \label{chap:restcn} 

\ifpdf
    \graphicspath{{6_chapter6/figures/PNG/}{6_chapter6/figures/PDF/}{6_chapter6/figures/}}
\else
    \graphicspath{{6_chapter6/figures/EPS/}{6_chapter6/figures/}}
\fi

\noindent The research presented in this chapter has been previously published in:


\textbf{Akhi, M.}, Eising, C. and Dhirani, L.L., 2026. Residual Temporal \glspl*{cnn} for Emerging Cyber Threat Detection in Healthcare \gls{iot}.~\textit{Discover Internet of Things}.

The content of the paper above has been presented as published in this chapter, except for correcting grammatical errors and making slight rewording so that the narrative is more appropriate for a thesis. A brief preamble is provided to explain its relevance to the thesis narrative.

\clearpage

\noindent\textbf{\underline{Preamble}}\\[0.5em]
In this chapter, \textbf{RQ5} is addressed:

\textbf{RQ5:} How can deep learning models such as Residual-\gls{tcn} be utilised for multiclass classification to distinguish among multiple attack categories, including \gls{sf}, \gls{mitm}, and \gls{ddos}, in \gls{hiot} environments?

\vspace{5pt}

Building on the edge deployment of the \gls{tcn} model for binary \gls{ddos} attack detection on a resource-constrained Raspberry Pi 4 in the previous chapter, which addressed \textbf{RQ4}, this research extends the investigation to multiclass cyber threat classification among heterogeneous attack types in \gls{hiot} environments.

This chapter contributes to the overarching research question by leveraging distinct data transmission behaviours of normal nodes and the malicious activities associated with multiple cyber threats (e.g., \gls{sf}, \gls{mitm}, and \gls{ddos}), thereby enabling the generation of a realistic dataset.
This dataset supports the evaluation of lightweight \gls{dl} models for multiclass cyber threat detection in \gls{hiot} environments.

The primary focus of this chapter is to utilise a \gls{restcn} model for multiclass classification across diverse cyber threats in \gls{hiot} environments. With respect to \textbf{RQ5}, this research investigates how the \gls{restcn} model can distinguish among multiple emerging cyber threats, including \gls{sf}, \gls{mitm}, and \gls{ddos}, within a unified \gls{hiot} topology. A realistic setup is developed in the Cooja simulator, comprising 24 \gls{hiot} nodes and one server node, where 14 nodes act as normal and 10 nodes perform different attack behaviours (\gls{sf}: 3, \gls{mitm}: 4, \gls{ddos}: 3). During the simulation, normal nodes transmit data at one-packet-per-60-second intervals, while \gls{ddos} nodes transmit at one-packet-per-20-second intervals. \gls{sf} nodes randomly drop approximately 50\% of packets every 60 seconds, and \gls{mitm} nodes modify payloads at 60-second intervals. Accordingly, as the simulation relies solely on synthetic network traffic derived from physiological and network parameters, \gls{gdpr} does not apply to this research.

Although several existing studies have applied \gls{ml} and \gls{dl} models with promising results, they often overlook realistic simulation conditions and protocol-specific variations that affect model reliability. This research addresses this critical gap by developing a realistic simulation setup tailored for precision healthcare, ensuring that the generated traffic accurately reflects real-world communication behaviours in \gls{hiot} systems. The dataset is generated through a statistical feature extraction process that integrates heterogeneous physiological and network-layer parameters, enabling a detailed representation of attack patterns in \gls{hiot} environments. The resulting dataset, {\em UL-ECE-MultiAttack-H-IoT2025}, is then used to train and evaluate the Res-\gls{tcn} model for multiclass attack classification in \gls{hiot} environments.

The \gls{smote} is applied to balance class distribution and minimise the risk of overfitting during model development. The model is also validated using publicly available datasets, including CICIoMT2024, CICIoT2023, and \gls{iomt}-TrafficData for external evaluation. The Res-\gls{tcn} demonstrates improved performance compared to rule-based DT, \gls{cnn}, and 1D-CLSTM models. For instance, the rule-based DT correctly classifies Normal (class 0), \gls{mitm} (class 2), and \gls{ddos} (class 3) samples. In contrast, it fails to detect Selective Forwarding (class 1) attacks, misclassifying all of them as Normal.

\section{Abstract}
The rapid advancement of \gls{iot} technologies has accelerated the emergence of \gls{hiot} systems. These systems rely on wearable devices to monitor patient vitals and enable timely alerts in precision healthcare settings. Despite these benefits, a single \gls{hiot} network topology might be exposed to multiple simultaneous threats, particularly those attacks designed to manipulate medical sensor data at the application layer. This poses significant challenges for real-time detection and classification of diverse attack behaviours. To address this, a realistic application-layer attack model is developed using the Cooja simulator, modelling \gls{hiot} nodes that track body temperature, oxygen level, and heart rate under concurrent \gls{sf}, \gls{mitm}, and \gls{ddos} attacks. Based on this setup, a dataset is generated to train the proposed \gls{dl} model. This research proposes a \gls{dl} model, a \gls{restcn}, designed to classify multiclass attacks while maintaining low latency per sample in \gls{hiot} environments. It also uses the \gls{smote} during training to mitigate class imbalance and reduce overfitting. The proposed model achieves a high classification accuracy of 99.32\% and outperforms traditional \gls{ml} and \gls{dl} methods. This demonstrates its effectiveness in real-time decision-making for securing \gls{hiot} systems. Based on these findings, the \gls{restcn} model is potentially well-suited for deployment in resource-constrained \gls{hiot} environments.

\section{Introduction}
\label{sec:ch6-introduction}
The advancement of a smart and sustainable healthcare ecosystem relies on the security of infrastructures powered by the \gls{iiot}~\cite{dhirani2021industrial}. The \gls{iot} comprises interconnected physical devices that communicate through embedded sensors, software, and wireless communication technologies \cite{dhirani2021industrial, Dhirani2024SecuringI5}. In recent years, implementing \gls{iot} has enabled the vision to achieve personalised and precision healthcare~\cite{priyadarshi2026comprehensive}. The growing use of wearable \gls{hiot} devices has made real-time monitoring of critical physiological parameters, such as body temperature, heart rate, and oxygen saturation, practically achievable within digital healthcare systems. Among the key components of digital healthcare systems, \glspl{wban} are central in enabling continuous monitoring for remote healthcare applications~\cite{chawla2025smart, alturki2024iomt}. Typically, wearable \gls{hiot} devices generate physiological data at fine temporal granularity (e.g., millisecond-level intervals)~\cite{akhi2023securingHIoT}, which is transmitted to the application layer for timely access and review by physicians or patients \cite{gopichand2024use}. Moreover, these devices communicate using various protocols, including \gls{mqtt}, \gls{udp}, \gls{tcp}, \gls{icmp}, and others, depending on system requirements and application-specific needs \cite{anandhavalli2025monitoring}. 

\begin{figure}[ht!]
    \centering
    \includegraphics[width=\textwidth]{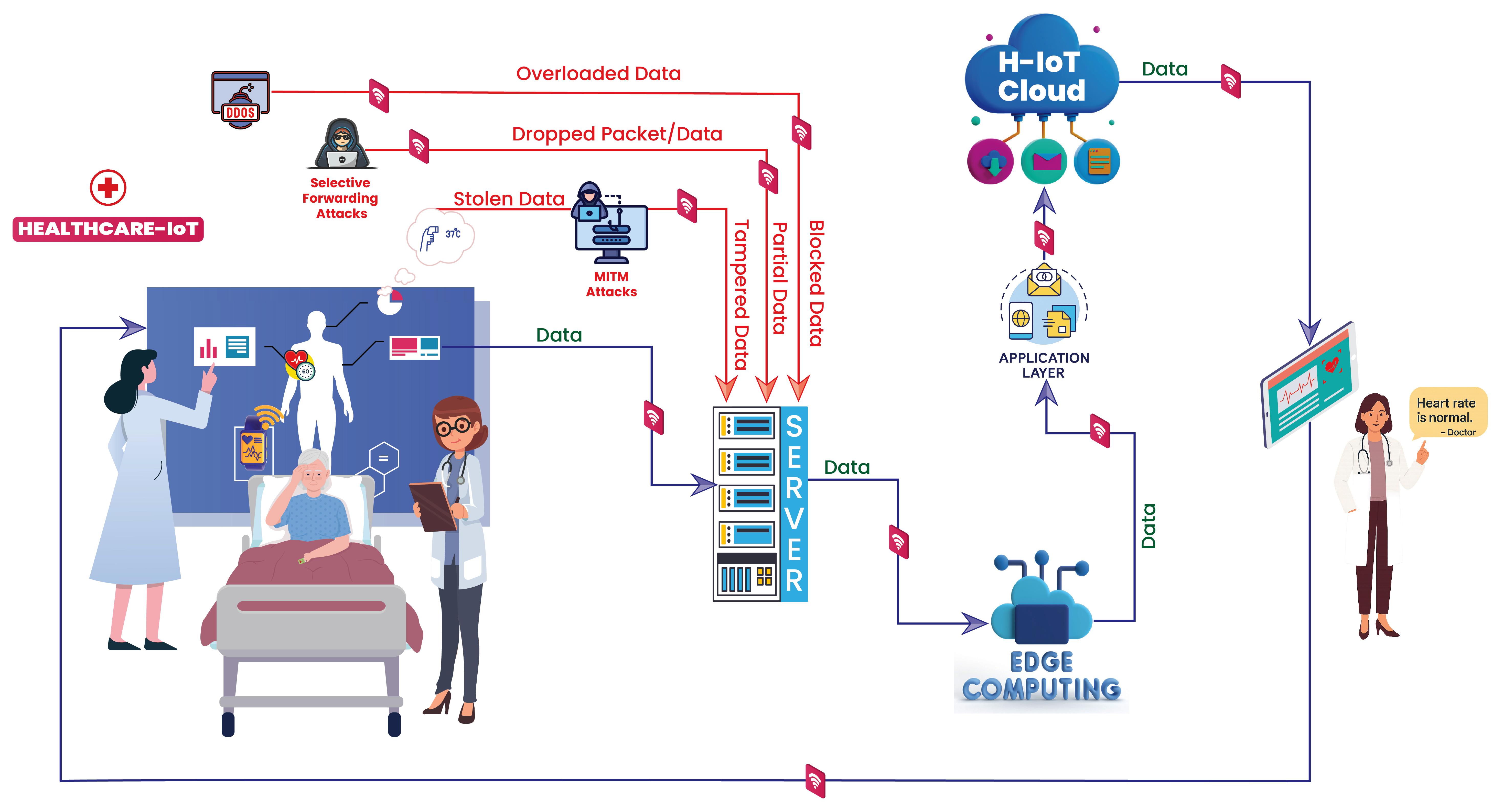}
   \caption[The diagram illustrates potential attacks and data flow]{The diagram illustrates potential attacks and data flow disruptions in \gls{g5}-based \gls{hiot} environments. It highlights \gls{ddos} attacks causing ``Overloaded Data,'' Selective Forwarding Attacks leading to ``Dropped Packet/Data,'' and \gls{mitm} attacks resulting in ``Stolen Data.'' The impacts of these attacks, including Blocked Data, Partial Data, and Tampered Data, affect the server, edge computing, application layer, and \gls{hiot} cloud levels.}
    \label{fig:flagship}
\end{figure}

However, despite the benefits of continuous health monitoring enabled by the enhanced speed and low latency of \gls{g5} networks, the application layer remains highly vulnerable in \gls{hiot} environments. It handles and visualises sensitive health data, making it a frequent target of cyber attacks~\cite{sadhwani20255g}. Such attacks often take the form of \gls{sf}, \gls{mitm}, and \gls{ddos}, particularly in healthcare settings~\cite{banerjee2025analysis, fereidouni2025iot}. In \gls{sf} attacks, critical health packets are intentionally dropped, disrupting the data transmission flow~\cite{rajasekar2025evaluating}. In \gls{mitm} attacks, an attacker silently intercepts and manipulates communications to access or tamper with health data exchanged between \gls{hiot} devices, such as body temperature readings \cite{fereidouni2025iot, cekerevac2025security}. One common technique is to send forged \gls{arp} spoofing messages that associate the attacker’s MAC address with the \gls{ip} address of a legitimate device, effectively enabling \gls{ip} spoofing \cite{fereidouni2025iot}. This allows the attacker to intercept, alter, or block communications between the \gls{hiot} nodes~\cite{al2016address}. Continuing the attack chain within the same topology, a \gls{ddos} attack may overwhelm the network and disrupt the operation of the \gls{hiot} system, as illustrated in Figure~\ref{fig:flagship}. Thus, building cyber resilience and defence capabilities remains a significant challenge amid the evolving threat landscape in \gls{hiot} systems \cite{Meagher2024}.

Various standards for securing healthcare have been developed by standardisation bodies (i.e., ISO/IEC, ETSI). A comprehensive list of the standards has been provided in \cite{Dhirani2024SecuringI5} that provide security assurance (ISO/IEC 15443), information security for access controls (ISO/IEC 15816), security requirements for cryptographic modules (ISO/IEC 19790), anonymous digital signatures (ISO/IEC 20008), privacy (ISO/IEC 20889), vulnerability disclosure and management (ISO/IEC 29147 and 30111), Identity management and discovery for \gls{iot} (ETSI DTS/CYBER 0014 (TS 103 486)), redaction of authenticated data (ISO/IEC DIS 23264-1), privacy and verification (ETSI DTS/CYBER-0013 (TS 103 485)), etc.~\cite{Dhirani2024SecuringI5, cybersec4europe2020standards}. Although the standards provide a strong baseline for assessing security measures, they alone would not be sufficient to counter the emerging threats, anomalies, and supply-chain risks arising from \gls{ai} in the \gls{hiot} ecosystem \cite{dhirani2024best}. Hence, to protect, a similar \gls{ai}-based approach would be required as a defence. The \gls{enisa} has also enacted regulations (i.e., EU \gls{ai} Act, General Data Protection Regulation (GDPR), Data Act, Data Governance Act, \gls{nis2d}) and frameworks (\gls{ehds}) for setting clear guidelines for the usage of \gls{ai} in the healthcare environment, building trust and cyber resilience in the infrastructure \cite{ecAIHealthcare2025, Meagher2024}.

Besides this, the usage of stronger controls (i.e., encrypting \gls{iot} data) has been widely implemented in healthcare sectors, fostering data confidentiality and ensuring data is secure in all three states (in transit, in storage, and while in use) \cite {dhirani2021industrial}. However, encryption alone cannot mitigate attacks (e.g., \gls{ddos} attacks). It is also important to note that \gls{iot} devices travel via the server, where the data is encrypted, and the server checks (decrypts) every message sent in, requiring additional memory and computing power \cite{sadhwani20255g}. In a \gls{ddos} scenario, where the server is already flooded with messages, this adds extra load to the environment. This is why it is essential to have other mitigation techniques (i.e., throttling, auto-scaling, anomaly detection, and prevention) aligned with it. Besides this, \gls{ddos} attacks rely on traffic frequency and volume, which can potentially exploit weak endpoints on the edge by sending recursive messages and flooding APIs or the \gls{mqtt} broker with requests. Hence, a unique \gls{ai}-based approach is required to detect and prevent malicious attacks and anomalies in \gls{hiot} \cite{eshmawi2026machine}.

To address the above security issues, this research conducts a simulation-based experiment that emulates \gls{hiot} nodes, such as body temperature, oxygen level, and heart rate sensors, communicating over the \gls{udp} protocol using the Cooja simulator within an \gls{hiot} environment. In this simulation, an attack model is integrated at the application layer, where three attacks (\gls{sf}, \gls{mitm}, and \gls{ddos}) occur concurrently in the same \gls{hiot} topology. The included sophisticated attack types are relevant to \gls{hiot} environments, as these target common vulnerabilities in health data transmission, interception, and forwarding processes. During the simulation, malicious and normal nodes transmit data simultaneously, reflecting a realistic attack scenario in real-time healthcare systems. This simulation approach also complies with privacy regulations (e.g., \gls{gdpr} and \gls{ehds}) by generating simulated or synthetic healthcare traffic rather than sensitive personal data \cite{ecAIHealthcare2025, Meagher2024}. Subsequently, the simulated traffic is collected and analysed to extract relevant features for training a \gls{dl} model for multiclass classification.

Several \gls{dl} and \gls{ml} models have been explored in previous studies for anomaly detection using both binary and multiclass classification in general \gls{iot} and \gls{hiot} environments. Among these, \glspl{cnn} and \gls{bilstm} Networks~\cite{zahid2025enhancing, 10628718}, \gls{lstm} Networks~\cite{ali2025intrusion, akar2025l2d2}, and \gls{dt}~\cite{binbusayyis2022investigation} have gained increasing attention due to their ability to capture temporal and spatial features in various \gls{iot} datasets. In this thesis, a \gls{tcn}-based architecture is developed for \gls{ddos} detection
and mitigation, with methodological details presented in Chapter~\ref{chap:tcn}. However, multiclass classification in \gls{hiot} often involves highly imbalanced datasets, where certain attack types have significantly fewer samples than benign classes. This is frequently expected, as one or two specific attack types may naturally generate fewer transmissions than \gls{hiot} devices' regular activity. As a result, predictive \gls{ml} or \gls{dl} models may favour the majority classes, potentially reducing overall detection performance. In such cases, the \gls{smote} is crucial for synthetically generating samples for minority classes, thereby improving detection performance and reducing overfitting during training.

However, practical solutions should also address regulatory obligations outlined by ENISA and the \gls{nis2d}, which emphasise the continuous operation of healthcare services under cyber attacks \cite{giogiou2025regulatory, negreiro2021nis2}. In response to these mandates, this study proposes a \gls{dl} model, Res-\gls{tcn}, to classify multiclass attacks in \gls{hiot} environments. From a GDPR perspective, ENISA highlights challenges such as data localisation, cross-border data transfers, consent mechanisms, and the right to erasure, all of which become more complex in cloud-based healthcare systems \cite{corrales2025multidisciplinary, isibor2024regulation}. In the proposed \gls{ids} workflow, regulatory requirements are practically integrated by processing simulated healthcare network traffic entirely within a local environment, without storing, transmitting, or accessing any real patient data. As the proposed approach uses simulated network traffic without any personal or patient-identifiable data, the GDPR does not apply to this research. This approach inherently embeds privacy preservation and data protection into the \gls{ids} design. Beyond compliance, the model ensures low latency and efficient resource use, supporting the \gls{enisa} and \gls{nis2d} mandates for resilient and uninterrupted healthcare operations during cyber attacks. A realistic dataset is developed from simulated \gls{hiot} scenarios encompassing both regular and malicious traffic. The Res-\gls{tcn} model is trained and evaluated on this dataset, which captures realistic traffic patterns while complying with \gls{gdpr} and \gls{ehds}. Specifically, this chapter emphasises achieving consistent detection performance while minimising overfitting using \gls{smote} and demonstrating low per-sample latency in an \gls{hiot} environment. Maintaining this balance is crucial for supporting real-time decision-making in continuous health monitoring scenarios. The key contributions of this research are as follows:
\begin{itemize}
   \item This research presents a realistic and practical experiment by emulating \gls{hiot} nodes such as body temperature, oxygen level, and heart rate sensors under three sophisticated concurrent attacks (\gls{sf}, \gls{mitm}, and \gls{ddos}) within a unified precision healthcare topology using the Cooja simulator.  
 \rev{\item To the best of current knowledge, this work presents an open-access multiclass \gls{hiot} dataset, {\em UL-ECE-MultiAttack-H-IoT2025}, containing both regular and malicious traffic generated by simulated attacks. Unlike existing multiclass IoT datasets that primarily focus on network-level traffic, the proposed dataset integrates heterogeneous physiological and network-layer parameters through a rolling-window statistical feature-extraction approach. This enables detailed profiling of attack patterns in \gls{hiot} environments. The dataset is publicly available in the Zenodo repository~\cite{akhi2025multiattack}.}
 
   \item This research introduces a novel \gls{dl} model, Res–\gls{tcn}, for multiclass attack detection in \gls{hiot} environments. The model integrates residual connections and temporal convolutions with systematically selected kernel sizes, residual depths, dropout rates, and dense layer sizes, designed explicitly for resource-constrained settings. It achieves state-of-the-art performance on the proposed dataset, {\em UL-ECE-MultiAttack-H-IoT2025}, for multiclass classification (99.32\% accuracy), while maintaining a low per-sample inference time among existing methods. The model also demonstrates reduced complexity, with the lowest parameter count among the compared \gls{dl} architectures. Its low complexity makes the model potentially suitable for edge deployment on \gls{hiot} devices in real-world settings.
\end{itemize}

This chapter is organised as follows. Section~\ref{sec:relatedwork} reviews related work on multiclass attack classification in \gls{hiot} environments. Section~\ref{sec:simulationdata} describes the simulation setup, development of three attacks in the same \gls{hiot} topology, and details on data collection, preprocessing, and analysis to create the proposed dataset.
Section~\ref{sec:methodology} outlines the proposed Res-\gls{tcn} model for efficient multiclass attack classification in \gls{hiot} systems, includes an ablation study to select the suitable model, and presents the design rationale for its configuration. Section~\ref{sec:performance} presents a comparison with existing work and shows results on inference time, memory usage, confusion matrix, and evaluation of the proposed Res-\gls{tcn} on public datasets. Finally, Section~\ref{sec:conclusion} summarises the work and discusses future research directions.

\section{Related Work}
\label{sec:relatedwork}
This section discusses existing anomaly and intrusion detection methods in \gls{hiot} and general \gls{iot} settings, targeting various attack types (e.g., \gls{dos}/\gls{ddos}, \gls{sf}, and \gls{mitm}). It also compares benchmark methods in \gls{iomt} environments, focusing on landmark studies \cite{akar2025l2d2, binbusayyis2022investigation, balhareth2024optimized} to highlight differences in scope, dataset development, and real-time feasibility (such as computational and resource demands). An often-overlooked aspect in previous studies is the evaluation of inference time, which is critical for real-time \gls{hiot} applications. Several \gls{ml} and \gls{dl} models report high accuracy, yet the applicability of these models for time-sensitive \gls{hiot} scenarios remains unclear without latency analysis. Another key issue is adversarial robustness, defined as a model’s capacity to process intentionally manipulated inputs correctly \cite{ravikumar2024securing}. \rev{Based on existing literature, adversarial robustness remains largely unaddressed in \gls{iomt} intrusion detection studies \cite{kumari2025secure, akar2025l2d2, shombot2025maximizing, binbusayyis2022investigation}. This issue is further compounded by the continued reliance on public datasets (e.g., CIC-IDS2017, BoT-\gls{iot}), which restricts the realism of \gls{hiot} research.} This research addresses these gaps by focusing on latency evaluation and creating realistic datasets with three attacks (\gls{sf}, \gls{mitm}, and \gls{ddos}), as detailed in Sections~\ref {sec:simulationdata} and~\ref{subsec:inference_time}. Adversarial robustness is identified as an open direction for future work, as this chapter focuses on \gls{hiot} realism, dataset development, and latency evaluation rather than assessing such defence techniques. 
Table \ref{tab:surveyofrelatedwork} synthesises recent studies by comparing their methods, attack coverage, and datasets, and highlights reported limitations.
It also provides analytical insights from the critical observations on aspects such as dataset imbalance, confusion-matrix analysis, latency evaluation, adversarial robustness, and other key architectural elements.

Kumari et al. \cite{kumari2025secure} propose a \gls{dl} approach based on \glspl{vae} to detect cyber threats in \gls{hiot} environments. The approach, evaluated using real-time data from the WUSTL-EHMS system, achieves a detection accuracy of 91.61\%, which is relatively lower than several recent \gls{hiot} \gls{ids} frameworks employing temporal convolution or hybrid \gls{dl} that surpass 95\% accuracy. The study also includes comparisons with traditional \gls{ml} models, such as \glspl{svm}, XGBoost, and Random Forest. However, the evaluation omits recent \gls{dl} models, such as \gls{tcn}-based architectures, which are effective in capturing time-dependent traffic variations and inter-sensor behaviour in \gls{hiot} systems. While the effectiveness of the \gls{vae}-based method in identifying \gls{mitm} attacks is highlighted, a key concern is the need for rapid detection to prevent patient risk in \gls{hiot} systems. Nevertheless, the study does not evaluate inference latency, limiting insight into its responsiveness in time-sensitive environments.

\begin{table}[htbp!]
\centering
\setlength{\tabcolsep}{4pt}
\caption[Survey of existing anomaly and intrusion-detection approaches]{Survey of existing anomaly and intrusion-detection approaches in \gls{iot}, \gls{iomt}, and \gls{hiot} environments. It highlights key methods, attacks, benefits, limitations, multiclass classification, and analytical insights from each prior study.}
\label{tab:surveyofrelatedwork}
\renewcommand{\arraystretch}{0.75} 
\resizebox{\textwidth}{!}{
\begin{tabular}{@{}%
p{2.2cm}  
p{1.5cm}  
p{2.3cm}  
>{\raggedright\arraybackslash}p{3.5cm}
>{\raggedright\arraybackslash}p{3.5cm}
p{2.6cm}  
p{5.8cm}  
c         
@{}}
\toprule
\textbf{Ref.} & 
\textbf{Method} & 
\textbf{Attacks} & 
\textbf{Benefits} & 
\textbf{Limitations} & 
\textbf{Dataset} & 
\textbf{Analytical Insight} &
\textbf{Multiclass} \\ 
\midrule
\cite{kumari2025secure} (2025) & \glspl{vae} & \gls{mitm} & \gls{vae}-based \par detection with high accuracy on real-time data. & Inference time not evaluated. \par Evaluation does not address adversarial robustness. & WUSTL-EHMS-2020 & Demonstrates that the authors report the dataset is skewed (87.5\% normal vs.\ 12.5\% attack) and use a balanced subset (8159 each) for evaluation without explaining the balancing process. No class-imbalance method (e.g., SMOTE) is reported. A Streamlit demo classifies ``Normal'' vs.\ ``Spoofing'' from eight vital signs; however, no live URL is provided, and the figure is too dark to clearly show the interface. 
 & \ding{55} \\
\hline
\cite{akar2025l2d2} (2025) &
\gls{lstm} &
\gls{dos}, \gls{ddos}, \par Spoofing &
- Achieves 98\% accuracy in detecting intrusions. \par - Improves \gls{iomt} security through \gls{dl} techniques. &
- High computation; not suitable for real-time \gls{iomt}. &
CICIoMT2024 &  Although \gls{lstm} configurations with 128 units achieve only 75\% accuracy, the 64-unit configuration attains 98\% for intrusion detection. However, the study omits the total parameter count, latency, and adversarial-robustness evaluation. It also states that the \gls{lstm} model is unsuitable for detecting binary attacks on resource-constrained \gls{iot} devices due to its computational intensity. &
\ding{51} \\ 
\hline

Chapter~\ref{chap:tcn} (2025) &
\gls{tcn} &
\gls{ddos} &
- 99.99\% \gls{ddos} detection and mitigation in \gls{hiot}.  \par - Real-time detection via monitoring frequency. \par - Two \gls{hiot} \par datasets created using simulators. &
- Limited to \gls{ddos} attack \par detection. &
Generated \gls{hiot} \gls{ddos} datasets & The authors state that the \gls{udp}-based \gls{hiot} simulation adds extra \gls{ddos} nodes every 10\,s from 20\,s onward within the same topology. 
This setup simulates the appearance of new malicious \gls{hiot} nodes over time. 
Nevertheless, the evaluation does not examine adversarial robustness beyond this setup. &
\ding{55} \\ 
\hline
\cite{kumar2025hybrid} (2025) & 
1D-CLSTM &
Multiple cyber attacks &
- High \gls{dl}-based detection accuracy in \gls{iomt}. \par - Uses undersampling and boosting to address overfitting. &
- Slower inference per sample. \par - Larger weighted model than Res-\gls{tcn}.
&
WUSTL-EHMS-2020, ECU-IoHT & 1D-CLSTM attains 98.55\% multiclass accuracy on ECU-IoHT. Minority classes remain weaker (\gls{dos} 88\% and Nmap Port Scan 81\% vs. $\approx$100\% for \gls{arp} Spoofing/Smurf) even with RUSBoost and class weighting. Without weighting, the \gls{dos} drops to 44.59\%, indicating high sensitivity to imbalance handling.  
 &
\ding{51} \\ 
\hline
\cite{10628718} (2024) & 
\gls{cnn} & 
\gls{ddos} & 
- High accuracy anomaly detection. \par - New dataset generated by authors. &
- Environmental \gls{hiot} sensors. &
Generated \gls{hiot} \gls{ddos} datasets \cite{akhi2025ulwsn} & \gls{cnn} attains high accuracy and trains faster, 1 m 40 s for 10 epochs compared with 2 m 17 s for \gls{svm}, which indicates a lower computational cost. &
\ding{55} \\ 
\bottomrule
\end{tabular}
}
\end{table}

\begin{table}[htbp!]
\centering
\ContinuedFloat
\setlength{\tabcolsep}{5pt}
\caption[]{Survey of existing anomaly and intrusion-detection approaches in \gls{iot}, \gls{iomt}, and \gls{hiot} environments. It highlights key methods, attacks, benefits, limitations, multiclass classification, and analytical insights from each prior study (continued).}
\renewcommand{\arraystretch}{0.75} 
\resizebox{\textwidth}{!}{
\begin{tabular}{@{}%
p{2.2cm}  
p{1.5cm}  
p{2.3cm}  
>{\raggedright\arraybackslash}p{3.5cm}
>{\raggedright\arraybackslash}p{3.5cm}
p{2.6cm}  
p{5.8cm}  
c         
@{}}
\toprule
\textbf{Ref.} & 
\textbf{Method} & 
\textbf{Attacks} & 
\textbf{Benefits} & 
\textbf{Limitations} & 
\textbf{Dataset} & 
\textbf{Analytical Insight} &
\textbf{Multiclass} \\ 
\midrule
\cite{binbusayyis2022investigation} (2022) & 
\gls{dt} & 
\gls{dos}, \gls{ddos}, Scanning & 
Achieves 100\% accuracy as an \gls{ids} in \gls{iomt} environments. &
- Potential risk of overfitting. \par 
- Not evaluated on \gls{hiot} data. \par 
- Inference time not analysed. &
BoT-\gls{iot} & The \gls{dt} model achieves 100\% accuracy and correctly detects all attacks on both datasets. At the same time, its false alarm rate rises to 12\% on the imbalanced set, showing that high accuracy can mask many false positives.
 &
\ding{51} \\
\hline
\cite{s22093367} (2022) & 
\gls{dt} & \gls{dos}, \gls{ddos} & - Demonstrates 100\% accuracy in \gls{iot}\par environments. & - Potential overfitting with the \gls{dt} model. \par 
- Extensive inference time per sample.  &
BoT-\gls{iot} & \glspl{dt} outperform \gls{dl}, relying on shallow (7–10) entropy splits, where feature correlations may cause overfitting and limit generalisation, despite high flow rates, while the authors report no majority-class bias. &\ding{51} \\ 

\hline
\cite{10.1007/978-3-031-50583-6_11} (2024) & XGBoost & \gls{iot} Network Attacks. & - High accuracy detection. & - Unspecified regularisation controls. \par - Uneven per-class accuracy. & UNSW-NB15 & The model relies on an older dataset, and even after synthetic data augmentation, attack-specific accuracies differ markedly. XGBoost achieves a peak of 99.63\% for Worms, yet then drops to 87.67\% for Analysis and 83.06\% for Reconnaissance, indicating uneven generalizability across classes despite high overall metrics. & \ding{51} \\ 
\hline
\cite{shombot2025maximizing} (2025) & \gls{ml} pipeline with multiple models & \gls{dos}, \gls{arp} Spoofing, Nmap port scan, and Smurf attack. & - High multiclass accuracy 90\% & 
- Lacks statistical features for advanced detection. & ECU-IoHT & Confusion matrices reveal persistent misclassification of Nmap Port Scan as Normal traffic; although Gradient Boosting attains 96.80\% precision with 91.57\% recall, Decision Trees raise recall to 92.61\%, indicating non-negligible false negatives. & \ding{51} \\ 
\hline
\cite{balhareth2024optimized} (2024) & Tree-based \gls{ml} with feature selection & Attack category merged (benign vs. malicious) & \gls{ids} with a low false-alarm rate (0.007 FAR); improved efficiency by reducing features. & - Outdated and imbalanced dataset.\par - Possible generalisation issues.\par - No hyperparameter tuning. & CIC-IDS2017 & Although the feature set is reduced from 30 to 15 using Mutual Information–XGBoost, the study still relies on an outdated traffic corpus. The absence of adversarial-robustness evaluation limits generalizability to multi-class and contemporary \gls{iomt} attacks, reducing relevance to current \gls{iomt} traffic. & \ding{55} \\ 
\bottomrule
\end{tabular}
}
\renewcommand{\arraystretch}{1}  
\end{table}

Akar et al. \cite{akar2025l2d2} present an advanced \gls{lstm}-based intrusion detection model specifically designed for \gls{iomt} environments. The model is evaluated on the CICIoMT2024 dataset and achieves a high classification accuracy of 98\%, demonstrating the potential of \gls{dl} for enhancing \gls{iomt} security. However, despite its promising results, the model incurs high computational overhead, making it less suitable even for binary detection tasks in resource-constrained \gls{iomt} environments. Moreover, the study lacks a simulation-based attack model and an in-depth evaluation of \gls{iomt}-specific threat scenarios, which limits its applicability in real-world \gls{hiot} contexts. Simulated attacks are crucial for gaining a comprehensive understanding of \gls{hiot} systems, as they enable researchers to analyse system behaviour and identify vulnerabilities of medical sensors under threat conditions~\cite{KALARIA2024104037}. Such insights are vital for developing lightweight, responsive, and practically deployable \gls{ml} and \gls{dl}-based intrusion detection mechanisms in \gls{hiot} environments. Furthermore, the study does not examine adversarial robustness, a key aspect for assessing the model’s resilience under worst-case \gls{iomt} attack conditions.

A monitoring frequency-based detection and dynamic threshold mitigation approach using \glspl{tcn} to secure \gls{g5} \gls{hiot} environments is developed in this thesis, as detailed in Chapter~\ref{chap:tcn}. The method evaluates each \gls{hiot} node’s message count as a percentage of the total traffic and dynamically adjusts detection and mitigation thresholds. Two datasets, UL-ECE-MQTT-DDoS-H-IoT2025 and UL-ECE-UDP-DDoS-H-IoT2025, are generated using \gls{mqtt} and \gls{udp} protocols via Cooja and ns-3 simulators in a \gls{g5} \gls{hiot} setup, as described in Chapter~\ref{chap:dataset}. The approach achieves 99.99\% accuracy with \gls{tcn}, demonstrating its effectiveness in securing \gls{hiot} systems. However, it primarily focuses on binary classification and a single \gls{ddos} attack. The study does not examine robustness against adversarial evasion attempts, which are increasingly critical for maintaining reliable intrusion detection under evolving \gls{hiot} threat conditions.

Kumar et al. \cite{kumar2025hybrid} propose a \gls{dl}-based model for detecting cyber attacks and authenticating \gls{iomt} data; however, the approach largely focuses on data authenticity rather than comprehensive attack characterisation in healthcare environments. The model employs an embedded \gls{el} method for feature selection, aiming to minimise redundancy and overfitting; yet it fails to ensure sufficient feature diversity to represent complex \gls{hiot} attack behaviours. The authors develop a \gls{1dclstm} model for cyber attack classification. However, the architecture lacks explainability and fails to capture the temporal irregularities often present in heterogeneous \gls{hiot} data. The developed \gls{1dclstm} model is evaluated using two datasets, ECU-IoHT and WUSTL-EHMS-2020, for binary and multiclass classification tasks, although both datasets exhibit limited variability in traffic behaviour and attack diversity. While the ECU-IoHT dataset includes labelled \gls{arp} spoofing instances, it remains confined to basic packet-level attributes such as protocol type, \gls{ip} addresses, and \gls{tcp} flags, which limits its capability to reveal subtle behavioural deviations essential for accurate \gls{mitm} detection \cite{ahmed2021ecu}. Such limited features are inadequate for reliably detecting \gls{arp} spoofing, a variant of the \gls{mitm} attack previously introduced in Section~\ref{sec:ch6-introduction}, as they fail to capture the statistical deviations and temporal dynamics indicative of malicious activity. Reliable detection of these critical threats requires enriched statistical features such as standard deviation, minimum, and maximum to capture subtle behavioural irregularities beyond basic network metrics in \gls{hiot} environments. The inclusion of derived features enhances the capability of \gls{ml} and \gls{dl} algorithms to capture abnormal \gls{hiot} device behaviour often overlooked by models relying solely on basic network metrics. In contrast to earlier studies relying on basic network attributes, the proposed research extracts enriched statistical features and integrates them into a comprehensive dataset designed for detecting \gls{mitm}, \gls{sf}, and \gls{ddos} attacks. The resulting {\em UL-ECE-MultiAttack-H-IoT2025} dataset~\cite{akhi2025multiattack} supports multiclass classification of \gls{mitm}, \gls{sf}, and \gls{ddos} attacks using statistical features in \gls{hiot} environments.

In \cite{10628718}, \glspl{cnn} are investigated for anomaly detection in time-series data within \gls{hiot} environments. However, the use of a simple \gls{cnn} architecture restricts its scalability and adaptability to large and diverse \gls{hiot} datasets. The authors simulate \gls{ddos} attacks using the Cooja simulator, which emulates environmental sensors such as temperature and humidity. The \gls{cnn}-based method achieves 92\% accuracy; however, the evaluation is limited to a single-attack scenario without multi-protocol or cross-domain validation. Additionally, the authors develop the WSN-DDoS-Attack-H-IoT2023 dataset to support \gls{ddos} attack detection in \gls{hiot} systems \cite{akhi2025ulwsn}. However, the dataset focuses on environmental \gls{iot} data that does not fully capture the complexity of healthcare-oriented network traffic, which may limit its generalizability to broader \gls{hiot} sensor contexts.

Binbusayyis et al. \cite{binbusayyis2022investigation} explore key \gls{iomt} security challenges, focusing on \gls{dos} and \gls{ddos} attacks. The authors evaluate several conventional \gls{ml} algorithms, including K-Nearest Neighbours, Naïve Bayes, and SVMs, and propose a decision tree model trained on the BoT-\gls{iot} dataset~\cite{koroniotis2019towards}.
However, since this dataset primarily represents generic \gls{iot} traffic and broad attack categories such as \gls{dos}, \gls{ddos}, scanning, and data theft, the study lacks emphasis on healthcare-specific communication dynamics or medical sensor vulnerabilities critical to \gls{hiot} intrusion detection. The model reportedly achieves 100\% accuracy on this general-purpose \gls{iot} dataset. Despite the high accuracy, the study does not assess inference efficiency or latency, thus providing limited insight into its practicality for real-time \gls{hiot} settings. The study also overlooks the model’s resistance to adversarial attacks, leaving the traditional \gls{ml} techniques employed vulnerable to minor disruptions that can adversely impact detection outcomes in \gls{iomt} environments. Even without explicit adversarial attack patterns in the dataset, a similar lack of robustness is observed during the multiclass evaluation, where the baseline rule-based \gls{dt} model misclassifies the \gls{sf} attack as normal, as discussed in Section~\ref{subsec:modelcomparison}. In contrast with this study, the proposed approach further integrates multiple strategies to improve reliability and efficiency. It focuses on a realistic \gls{hiot} environment by emulating medical sensor nodes and modelling concurrent sophisticated attacks. To mitigate overfitting, regularisation techniques such as dropout and early stopping are employed, while SMOTE is applied to address class imbalance. Multiple configurations of the \gls{dl} model are evaluated, and inference time analysis is conducted to ensure suitability for resource-constrained \gls{hiot} devices. These considerations make the proposed method both practically viable and reliable for precision healthcare systems.

Another notable study leveraging the Bot-\gls{iot} dataset employs traditional \gls{ml}, particularly decision trees, to detect transport and application layer \gls{ddos} attacks in \gls{iot} networks \cite{s22093367}. Using the entropy criterion, the authors specify this model's hyperparameters, including a maximum depth of 10 and 11. While providing these settings is a positive step toward transparency, the potential for overfitting remains a concern given the high reported accuracy of 100\%. Without further analysis, such as an ablation study or cross-validation across varying depths, the model's generalizability in real-world \gls{iot} or \gls{hiot} scenarios may still be questionable \cite{breiman1984classification, james2021introduction}.

Expanding the focus beyond \gls{ddos} attacks and the \gls{hiot} environment, the authors in~\cite{10.1007/978-3-031-50583-6_11} apply multiple \gls{ml} models to detect various \gls{iot} network attack types using the UNSW-NB15 dataset, reporting 96.71\% accuracy with XGBoost. Although the study presents promising results in a general \gls{iot} context, its applicability to \gls{hiot} scenarios remains limited due to the model’s computational demands. Specifically, XGBoost has a per-sample inference complexity of $O(MN)$, where $M$ is the number of trees and $N$ is the number of nodes evaluated per tree~\cite{9504587}, making it less suitable for real-time healthcare applications.

Shombot et al. \cite{shombot2025maximizing} evaluate an \gls{ml} pipeline comprising Gradient Boosting, \glspl{dt}, Random Forest, and Multilayer Perceptron on the ECU-IoHT dataset for multiclass attack classification in \gls{ioht} environments. While the study reports classification accuracy exceeding 90\% for \gls{arp} spoofing, \gls{dos}, Nmap port scan, and Smurf attacks, the reliance on basic protocol-level features may limit the model’s ability to detect more sophisticated attack patterns in \gls{hiot} networks. Effective threat detection in such networks requires statistical features beyond protocol-level data to enhance the stability and reliability of \gls{ml}/\gls{dl} models, as discussed earlier in this section. This limitation further reinforces the earlier observation concerning the lack of statistical feature depth in the ECU-IoHT dataset \cite{ahmed2021ecu}.

Balhareth et al. \cite{balhareth2024optimized} propose an \gls{ids} for \gls{iomt} networks using tree-based \gls{ml} classifiers such as \gls{dt}, Random Forest, XGBoost, and CatBoost. The approach applies a filter-based feature selection strategy using Mutual Information and XGBoost, intersecting the selected features to enhance detection accuracy and efficiency. While the method reduces feature dimensionality, it is limited to binary classification (benign vs. malicious) using the CICIDS2017 dataset. The model achieves 98.79\% accuracy with a 0.007 false-alarm rate; however, its reliance on an outdated and imbalanced dataset~\cite{ccetin2022effective} limits generalizability to current \gls{iomt} traffic. The study further omits essential implementation details, such as hyperparameter configurations (e.g., depth, number of trees, learning rate) and the tuning procedure, which affects reproducibility and interpretability. It also overlooks adversarial-robustness evaluation, leaving model resilience under evolving \gls{iomt} threats unexplored.

Collectively, these landmark \gls{iomt} and \gls{hiot} studies~\cite{akar2025l2d2, binbusayyis2022investigation, balhareth2024optimized} reveal uneven progress in balancing accuracy, realism, and scalability. Yet, a consistent method for assessing model reliability under adversarial or evolving attack conditions remains absent. This comparison highlights the need to evaluate \gls{ml} and \gls{dl} models that jointly ensure adversarial resilience, methodological rigour, and deployability across \gls{iomt} and \gls{hiot} environments.

\section{Simulation Setup and Dataset Creation}\label{sec:simulationdata}
This section describes the simulation setup and attack model implemented using the Cooja simulator to generate a dataset for multiclass classification in \gls{hiot} environments. The simulation includes three attack scenarios: \gls{sf}, \gls{mitm}, and \gls{ddos}, conducted on the same topology. This topology consists of twenty-four nodes operating under both normal and malicious conditions. Consequently, raw data are collected for further processing. This section then outlines the data preprocessing pipeline, which includes feature extraction, handling missing data, addressing class imbalance, data structuring, and reshaping. It further includes exploratory data analysis (EDA), including feature correlation heatmaps and time-series-based data splitting, to prepare the dataset for the proposed Res-\gls{tcn} model.

\subsection{Multiattack Simulation over a Single \gls*{hiot} Topology} \label{subsec:simulation}
This research employs the Cooja simulator with Sky mote sensors to design a network topology for the precision \gls{hiot} environment. In Cooja, motes are called nodes; both terms are used interchangeably. The simulation setup enables modelling multiple attack scenarios. This research then emulates \gls{wban} \gls{hiot} sensors, including body temperature, oxygen level, and heart rate monitors, over the \gls{udp} protocol within the same \gls{hiot} topology to create a realistic multiattack model. Specific attacks, such as \gls{sf}, \gls{mitm}, and \gls{ddos}, are simulated in this topology. All other simulation configuration parameters are detailed in Table~\ref{tab:simulation-details}.

\begin{table}[ht!]
\centering
\caption[Simulation details for the baseline topology]{Simulation details for the baseline topology with data transmission rates of normal and attack nodes (\gls{sf}, \gls{mitm}, and \gls{ddos}) in \gls{hiot}.}
\label{tab:simulation-details}
\small 
\begin{tabular}{p{5.2cm}|p{9.5cm}}
\hline
\multicolumn{1}{c|}{\textbf{Category}} & \multicolumn{1}{c}{\textbf{Details}}         \\ 
\hline
Platform                               & Cooja Contiki 3.0                            \\
Total Number of Nodes                        & 24 nodes, 1 server                           \\
Normal nodes                        & 14 nodes (IDs 2-15, orange)                          \\
Selective Forwarding Attack & 3 nodes (IDs 16-18, purple)                                           \\
\gls{mitm} Attack & 4 nodes (IDs 19-22, yellow)                                           \\
\gls{ddos} Attack & 3 nodes (IDs 23-25, cyan) \\
Topology                               & Star                                         \\
Area                                   & 300 m x 300 m                                \\
Normal Node Transmission Rate      & 1 packet/60\,s interval \\

\gls{sf} Transmission Behaviour & randomly drops $\approx$50\% of packets (at 60\,s interval) \\
\gls{mitm} Transmission Behaviour & modifies payload (at 60\,s interval) \\

\gls{ddos} Transmission Rate   & 1 packet/30\,s interval \\
Simulation run time                    & 5 hours                                       \\
Application Layer & 	Simulated Health Data (Temperature, O2 Level, BP)  \\
Radio Medium Model                     & Unit Disk Graph (UDGM) for Distance Loss \\
Mote Type                              & Sky mote                                     \\
Duty Cycle                             & ContikiMAC                                   \\
Range of Nodes                         & Rx and Tx: 50 m, Interference: 100 m         \\
Physical Layer                         & IEEE 802.15.4                                \\
MAC Layer                              & ContikiMAC                                   \\
MAC Driver                             & CSMA/CA                                      \\
Channel Selection                      & By default, 26                               \\
Network Layer                          & ContikiRPL                                   \\
Transport Layer                        & \gls{udp}                                          \\
Trafﬁc model                           & Constant Bit Rate                            \\ \hline
\end{tabular}
\end{table}

\begin{figure}[ht!]
    \centering
    \includegraphics[width=0.6\textwidth]{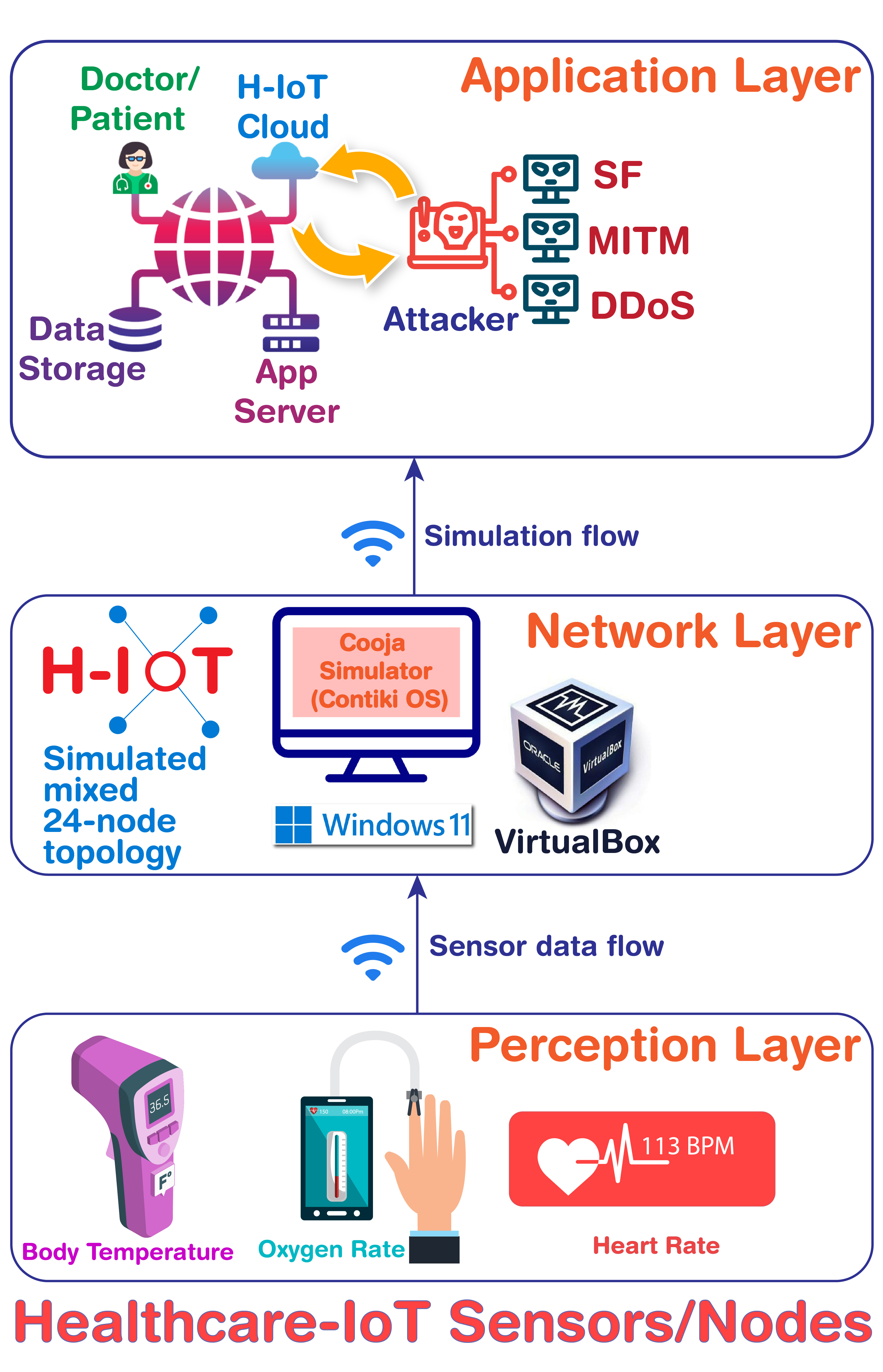}
    \caption[Simulation setup for data collection in the \gls{hiot} environment]{Simulation setup for data collection in the \gls{hiot} environment. The layered architecture represents healthcare sensor data generation at the perception layer, transmission through a simulated mixed 24-node topology in the network layer using Cooja, and attacks (\gls{sf}, \gls{mitm}, \gls{ddos}) carried out at the application layer, where data is processed and visualised.}
    
    \label{fig:ch6datacapture}
\end{figure}

In this simulated \gls{hiot} network topology, the perception layer comprises the \gls{hiot} sensors. The network layer includes a simulated twenty-four-node topology in Cooja, along with the node distributions and application-layer behaviours associated with \gls{sf}, \gls{mitm}, and \gls{ddos} attacks, as detailed in Table~\ref{tab:simulation-details}. The selection of twenty-four nodes reflects real-world \gls{hiot} deployment scales. For instance, a typical precision healthcare or home-care monitoring environment typically does not involve hundreds of devices within a confined area. Instead, deployments commonly involve 20–30 interconnected devices, such as patient wearables, medical sensors, and environmental monitors, all operating under a local gateway (e.g., a hospital access point, home \gls{iot} hub, or 6LoWPAN border router) \cite{miranda2018review}. Additionally, prior studies demonstrate that realistic home-care and \gls{hiot} deployments often rely on a limited number of sensors (e.g., approximately 24–28 motion sensors) to enable accurate monitoring of daily living activities \cite{ramos2021daily, cook2012casas}. As a result, this research adopts a realistic scale with a limited number of nodes to emulate \gls{hiot} settings. It also aligns with the practical limits of the Cooja simulator, where larger topologies with more sensors become computationally expensive. Such setups introduce event synchronisation delays and logging bottlenecks that reduce reproducibility, and these effects may vary across systems.

A \gls{g5} Wi-Fi bridge connects the simulated Cooja network to the host system, enabling real-time packet capture and forwarding to the application layer for further processing. This setup reflects the layered \gls{hiot} architecture: data are generated at the perception layer, passed through the network layer, and then targeted by application-layer attacks. The overall data collection process across these layers is illustrated in Figure~\ref{fig:ch6datacapture}. At the application layer, data are visualised for healthcare professionals and patients, enabling real-time monitoring and analysis. This architecture allows for the generation of multiclass attack scenarios for precision healthcare analysis. Two scenarios are defined within the same \gls{hiot} simulation setup: a baseline scenario with only normal nodes and another scenario where \gls{sf}, \gls{mitm}, and \gls{ddos} attacks occur alongside normal nodes.

\subsubsection{Scenario 1 – Baseline Topology}
This scenario represents the baseline \gls{hiot} simulation setup, where data are transmitted using the \gls{udp} protocol over IPv6, compressed using \gls{6lowpan}. This configuration reflects standard behaviour in Contiki-based \gls{iot} environments. The baseline topology contains a total of fifteen nodes, including fourteen \gls{wban} \gls{hiot} sensor nodes and one server node, deployed in a simulation setup without malicious activity. These \gls{hiot} nodes include sensors for body temperature, heart rate, and oxygen level, randomly distributed to emulate patient monitoring. The setup consists of one server node (ID 1, represented in green) and fourteen normal sensor nodes (IDs 2–15, denoted in orange), as illustrated in Figure~\ref{fig:normalnodes}. 

\begin{figure}[ht!]
    \centering
    \includegraphics[width=0.9\textwidth]{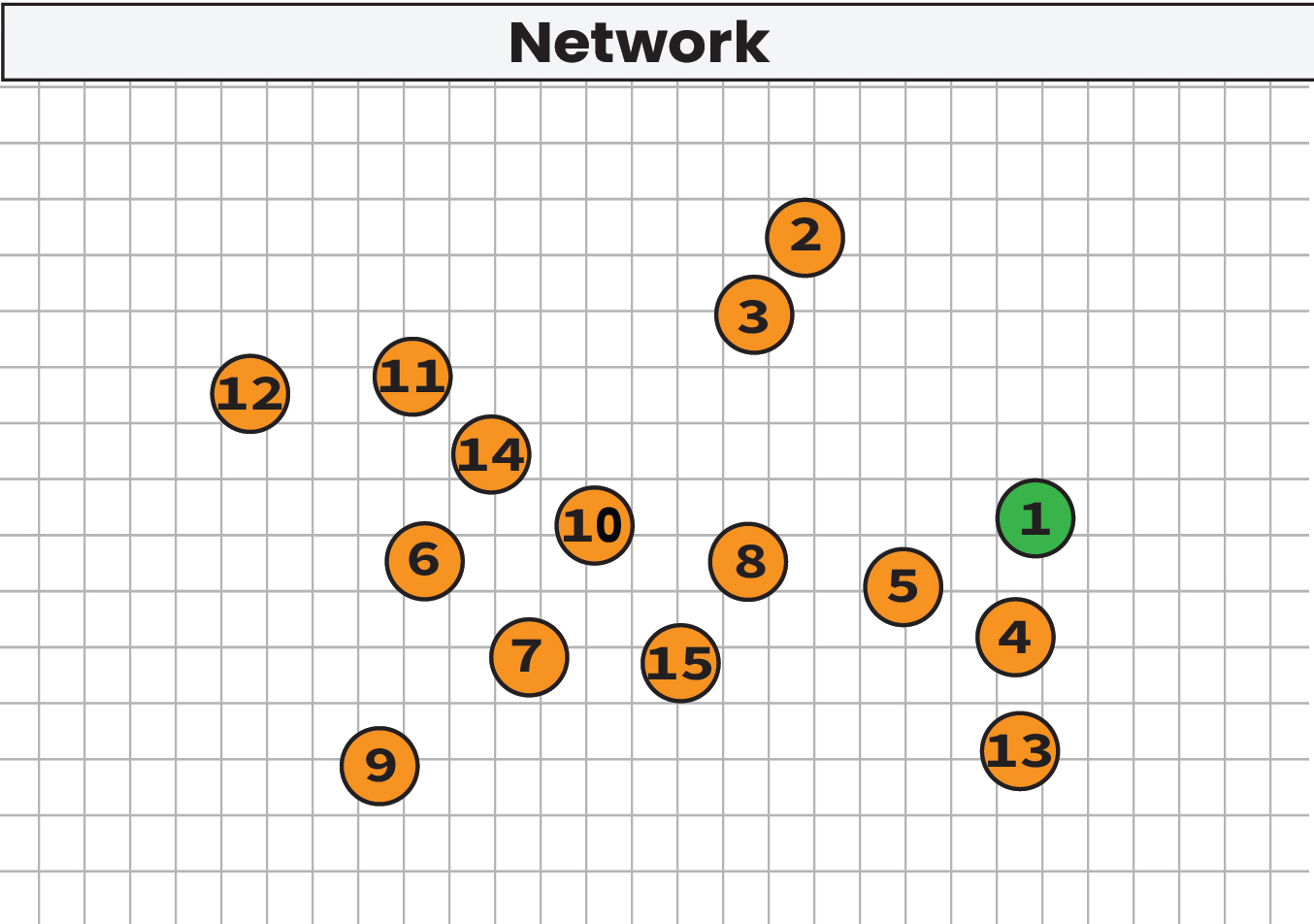}
    \caption[This diagram illustrates the baseline \gls{hiot} topology]{This diagram illustrates the baseline \gls{hiot} topology in the simulation setup. The network includes 15 nodes, comprising one server node (in green) and 14 normal nodes (in orange). This configuration represents Scenario 1, where all nodes operate without malicious behaviour.}
    \label{fig:normalnodes}
\end{figure}

\subsubsection{Scenario 2 – Malicious Topology}
The main difference between normal nodes (IDs 2-15) and attack nodes is their data transmission behaviour. In real-world \gls{hiot} settings, data transmission intervals vary across devices; for example, body temperature sensors or wearable health monitors such as smartwatches and fitness bands may transmit data every 60 s \cite{setia2020smarttemp}. These devices are typically considered low-frequency devices compared to heart rate and ECG sensors, which transmit at much shorter intervals. Reflecting this heterogeneity, the simulation configures all normal nodes with a 60 s transmission interval \cite{paul2025load, said2020machine}, aligning with low-frequency \gls{hiot} sensors. Notably, Said et al. \cite{said2020machine} report that each sensor mote transmits one packet per minute to the sink node, indicating that a 60-second transmission interval closely reflects realistic communication behaviour in smart hospital \gls{iot} environments. Similarly, ultra-low-frequency reporting has also been observed in other \gls{hiot} studies \cite{tekeste2018ultra}, further supporting the selected transmission rate. In contrast, malicious nodes are configured with shorter intervals (e.g., 20 s) to emulate the increased traffic intensity typically observed during attacks. \gls{ddos} attacks, in particular, use short packet intervals to simulate high traffic intensity~\cite{aslam2022adaptive, yang2013study}. This difference in transmission frequency represents attack intensity in the simulation, where malicious nodes generate rapid request traffic resembling \gls{ddos} behavior~\cite{9798301}. Therefore, this setup emulates both typical low-frequency medical sensors and the high-intensity communication bursts observed during malicious activity in \gls{hiot} networks, with the corresponding attack intensity and attempted transmission rates summarised in Table~\ref{tab:per_node_intensity}. This variation in transmission frequency also provides distinguishable temporal patterns crucial for \gls{ml}/\gls{dl} model training and performance evaluation. The corresponding simulation scripts and configurations illustrating these intensity settings are publicly available in the referenced GitHub repository \cite{akhi2025_multiattacksimulation}. The multiple malicious nodes are categorised as follows:

\begin{itemize} 
   \item \textbf{\gls{sf}-affected nodes} (IDs 16–18) intentionally drop selected packets during forwarding to cause data loss,
    \item \textbf{\gls{mitm}-affected nodes} (IDs 19–22) transmit modified data at the same interval as normal nodes, and
    \item \textbf{\gls{ddos}-affected nodes} (IDs 23–25) transmit data every 30 seconds, emulating the rapid request bursts typical of real-world flooding attacks.
\end{itemize}

\begin{table}[ht!]
\centering
\setlength{\tabcolsep}{6pt}
\renewcommand{\arraystretch}{1.3}
\caption[Per-node data transmission intervals]{Per-node data transmission intervals, attempted rates, intensity, observed delivery, and total packets under normal and attack roles in an \gls{hiot} setting.}
\label{tab:per_node_intensity}
\resizebox{\textwidth}{!}{
\begin{tabular}{@{}%
p{2.4cm}  
p{1.8cm}  
p{2.4cm}  
p{3.9cm}  
p{3.8cm}  
p{3.4cm}  
p{2.9cm}  
@{}}
\toprule
    \textbf{Node type} & \textbf{IDs} & \textbf{Interval (s)} & \textbf{Attempted~rate (pkt/min)} & \textbf{Intensity} & \textbf{Delivered (pkt/min)} & \textbf{Total pkts/node}  \\ 
    \midrule
 Normal (14) & 2--15 & 60 & 1 & $1\times$ & One packet per minute & $\approx$718 \\
    \gls{sf} (3) &  16--18  & 60 & 1 & Attempt $1\times$ & Delivered $\approx 0.5$ (50\%) &  $\approx$359 \\
    \gls{mitm} (4) & 19--22 & 60 & $1$ & Attempt $1\times$ & Payload Tampered & $\approx$718 (altered)  \\
    \gls{ddos} (3) & 23--25 & 30 & 2 & Attempt $2\times$ & $\approx2\times$ packets/min & $\approx$1436 \\
    \bottomrule
\end{tabular}
}
\vskip 3pt
\begin{minipage}{\linewidth}
\scriptsize 
  \textbf{Note:} \gls{sf} and \gls{mitm} nodes follow the same transmission interval as normal nodes, whereas \gls{ddos} nodes transmit more frequently (every 30 s) to emulate approximately $2\times$ higher intensity under high-density network conditions. The attempted rates represent per-node transmission frequencies, while the intensity column indicates the relative rate scaling across node types. The term \textit{Attempt} is used only for attack nodes to denote deliberately increased or modified transmission behaviour. The \textit{Payload tampered} indicates modification of transmitted data (biased toward high or low health readings) during \gls{mitm} attacks. The example total packets per node assumes each normal node transmits approximately 718 packets. In comparison, a \gls{ddos} node generates roughly 1,436 packets, resulting in a total traffic volume exceeding 40,000 packets in a single simulation run. It is also observed that when the \gls{ddos} send interval is set to 20~s, a normal node (60~s) produces about 718 packets, and a \gls{ddos} node produces about 2,154 packets.
  \end{minipage}
\end{table}

\begin{figure}[ht!]
    \centering
    \includegraphics[width=0.7\textwidth]{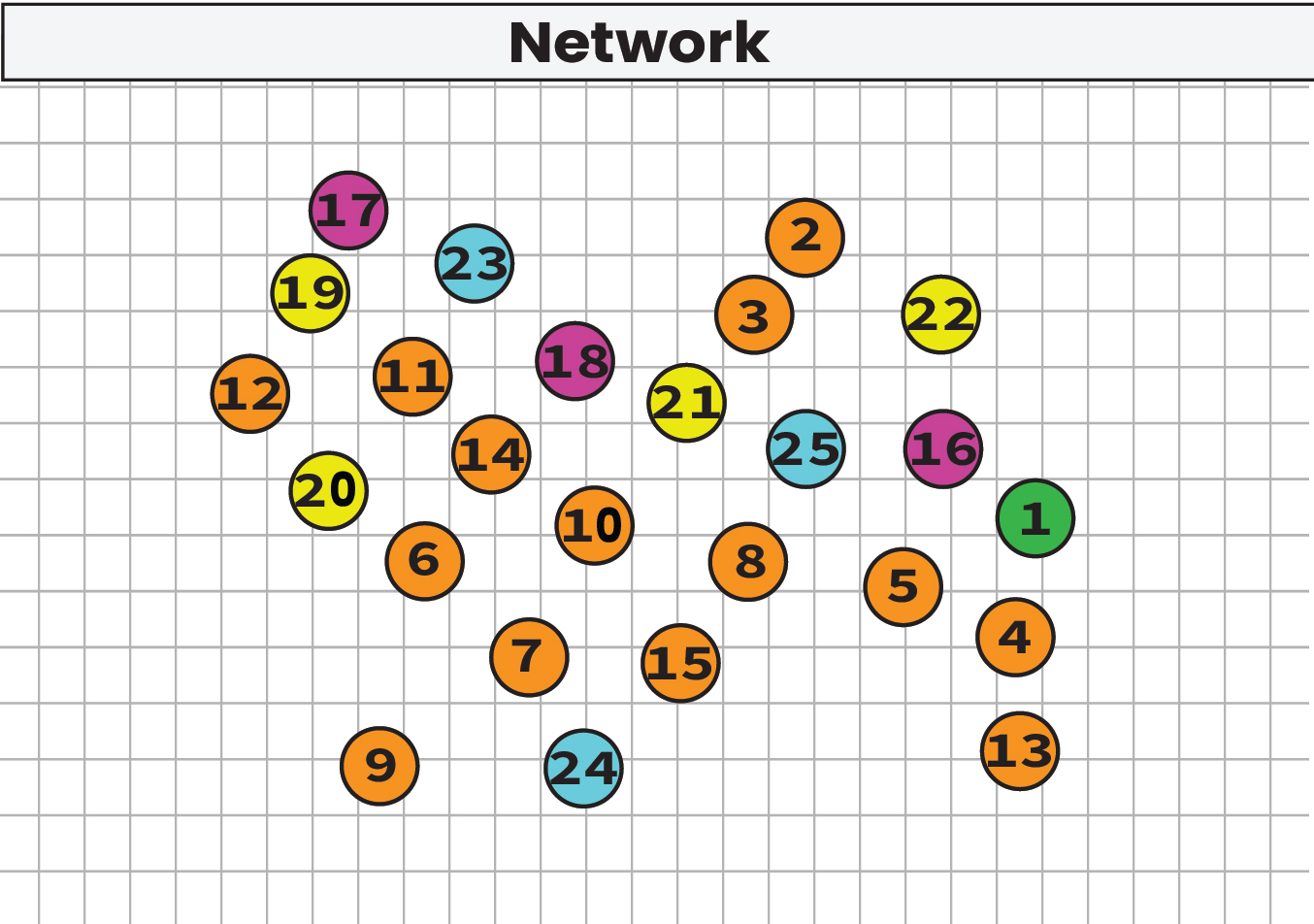}
    \caption[The common topology includes both malicious and non-malicious]{The common topology includes both malicious and non-malicious nodes. Normal nodes are highlighted in orange (IDs 2–15), and the server node is green (ID 1). \gls{sf} attack nodes are denoted in purple (IDs 16–18), \gls{mitm} attack nodes in yellow (IDs 19–22), and \gls{ddos} nodes in cyan (IDs 23–25).}
    \label{fig:DDoS-Selective-MITM}
\end{figure}

These malicious nodes are integrated into the topology used in the baseline scenario to construct the attack model. As a result, a unified simulation setup is created with malicious and non-malicious nodes, where three sophisticated attacks, \gls{sf}, \gls{mitm}, and \gls{ddos}, are executed within the same \gls{hiot} topology. This setup enables the collection of both normal and malicious traffic data within the \gls{hiot} environment, as depicted in Figure~\ref{fig:DDoS-Selective-MITM}. In addition, the corresponding data flow and attack behaviours within the same topology are illustrated in Figure~\ref{fig:DataFlow-HIoT}.

\begin{figure}[ht!]
  \centering
  \includegraphics[width=\textwidth]{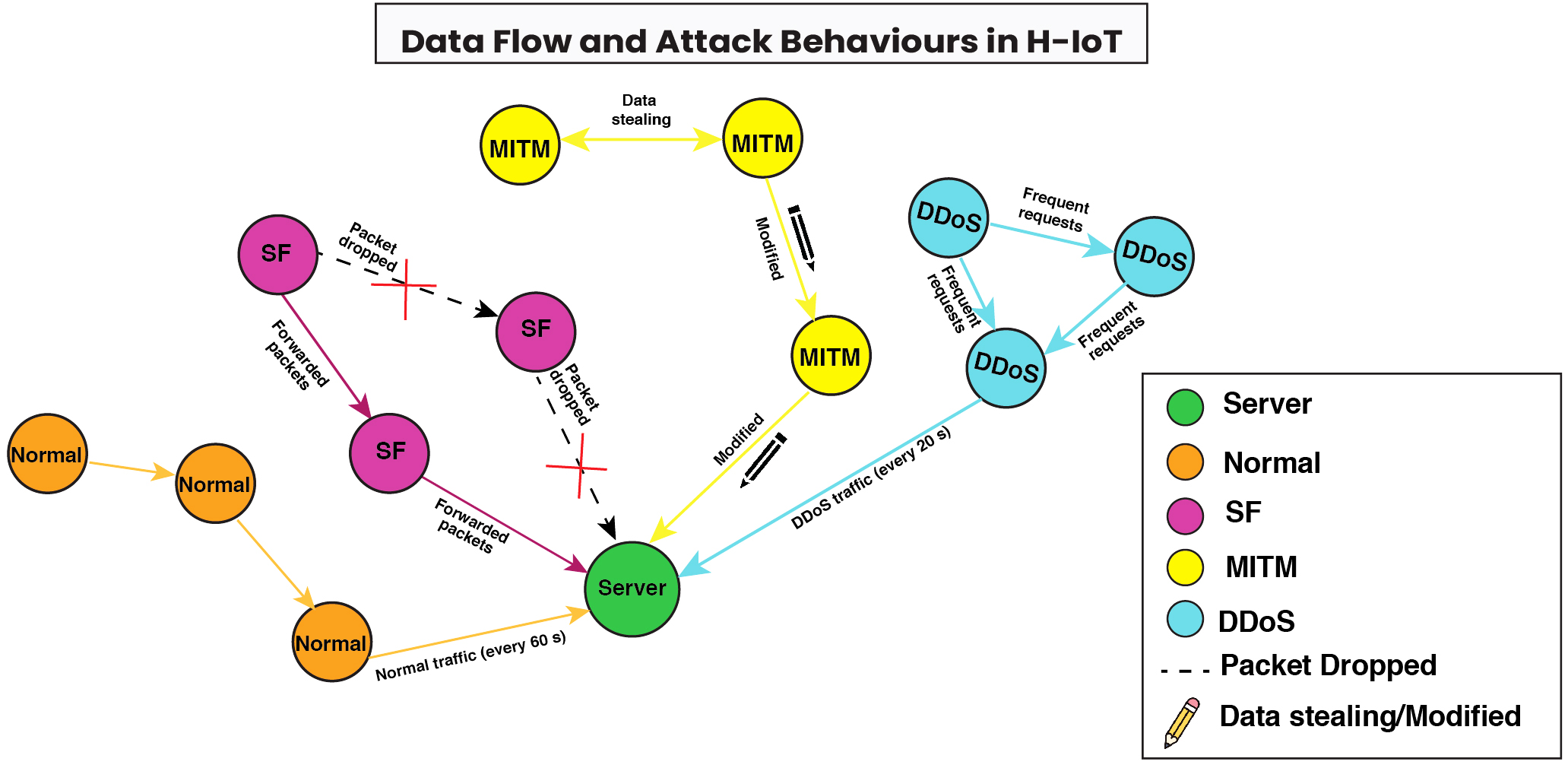}
  \caption[Data flow and attack behaviours in the simulated \gls{hiot}]{Data flow and attack behaviours in the simulated \gls{hiot} environment.
Orange nodes (Normal) send traffic every 60 s.
Purple nodes (\gls{sf}) selectively forward packets or drop them (Packet dropped).
Yellow nodes (\gls{mitm}) modify data and exfiltrate it (data stealing).
Cyan nodes (\gls{ddos}) send frequent requests every 30 s (\gls{ddos} traffic).
Solid arrows from \gls{sf} nodes indicate forwarded packets, while dashed arrows with \ding{55} indicate dropped packets.}
  \label{fig:DataFlow-HIoT}
\end{figure}

\textbf{Selective Forwarding Attack:}
In this attack, three nodes (IDs 16–18, purple), as depicted in Figure~\ref{fig:DDoS-Selective-MITM}, behave as compromised nodes. Whenever an \gls{sf}-affected node attempts to transmit data, it generates a random number. If the number is even, the packet is dropped; if it is odd, it is forwarded to the server (ID 1). This behaviour results in approximately 50\% packet loss, as packets are randomly dropped before transmission. Suppose a normal node (ID 15) transmits a sensor reading (e.g., 1122) to the server. However, the next-hop node (ID 16), which is compromised by a selective forwarding attack, may forward a different value (e.g., 591). In some cases, it may drop the packet entirely. This demonstrates how selective forwarding can lead to data loss or inconsistencies in \gls{hiot} environments. Furthermore, since these nodes transmit fewer packets, continuous patient monitoring may be disrupted. As a result, it can delay timely medical responses or trigger false alarms in \gls{hiot} systems.

\textbf{Man-in-the-Middle Attack:}
This attack comprises four yellow-colored nodes (IDs 19-22), as illustrated in Figure~\ref{fig:DDoS-Selective-MITM}, which act as compromised \gls{mitm}-affected nodes. In the simulation, these nodes generate falsified sensor readings, including temperature, oxygen level, and heart rate. The data are transmitted to the server by mimicking the behaviour of legitimate nodes. Each transmission contains abnormally high or low values based on a random selection. In the high-value scenario, the node sends tampered readings indicating critical health risks (e.g., 105.00°F temperature, 110 oxygen level, and 150 heart rate). Conversely, in the low-value case, the node transmits dangerously low values (e.g., 95.00°F, 70 oxygen level, and 40 heart rate). Thus, these fake readings may falsely indicate severe health conditions, such as high fever, low oxygen levels, or irregular heart rate. As a result, such falsified data can lead to incorrect diagnoses and prolong necessary medical treatments in the \gls{hiot} environment.

\textbf{Distributed Denial of Service:}
This attack involves three blue-colored nodes (IDs 23–25), as presented in Figure~\ref{fig:DDoS-Selective-MITM}, which behave as \gls{ddos}-affected nodes. These nodes transmit data at double the rate of legitimate nodes, resulting in excessive traffic that overloads the server~\cite{10628718}. As a result, the server may experience delays in processing legitimate health data. This increases the risk of missed alerts or late medical decisions in \gls{hiot} environments where timely communication is critical.

\subsection{Data Collection}
The first step is to create a baseline profile for generating datasets from \gls{hiot} sensors' raw data packets. The baseline profile consists of three features, including \textbf{Time}, \textbf{Mote}, and \textbf{Message}. The raw text file contains sensor data readings with timestamps from the emulated \gls{hiot} nodes, formatted as communication messages. These include temperature, oxygen level, heart rate, node IDs of sending and receiving packets, and attack indicators (e.g., \gls{mitm} high/low reading or packet drops).

\subsection{Data Preprocessing}
After acquiring the raw data, a series of preprocessing steps is applied to make it suitable for analysis. These include: label encoding of categorical attributes, normalisation using MinMax scaling, class balancing via SMOTE, and reshaping the data for time-series modelling with Res-\gls{tcn}.

\textbf{Data Filtering:} The first step in the filtering process excludes certain records based on the node ID value. Text strings within the ``Event Description'' are then identified using regular expressions and string matching techniques \cite{singh2025datavinci}.  
System initialisation logs and configuration details are filtered out to focus on the most relevant information. Among these entries are ``Rime started with address'' and ``Contiki-2.6-2450-geaa8760 started,'' both considered standard system or network initialisations. Several configuration specifics are also excluded, such as ``null-sec CSMA ContikiMAC, channel check rate 8 Hz, radio channel 26, CCA threshold -45.'' This filtering approach ensures that only essential data points are retained for analysis, minimising noise and irrelevant information.

\textbf{Feature Extraction:}
Key features, including node IDs, total, average, and per-node message counts, are extracted from packet traces and log files using pattern matching~\cite{khatun2019malicious}. Specific features (e.g., Total Messages), detailed in Chapter~\ref{chap:tcn},
are incorporated into the {\em UL-ECE-MultiAttack-H-IoT2025} dataset. Twenty-two parameters are extracted to analyse three attack types: \gls{sf}, \gls{mitm}, and \gls{ddos}. Each parameter corresponds to a dataset column, with representative names (e.g., Body\_Temperature\_Mean\_Same\_Node) for \gls{mitm}. A detailed list of parameters is presented in Table~\ref{tab:dataset_combined}.

Several statistical features are computed using a rolling-window technique to capture short-term temporal patterns~\cite{dimoudis2023utilizing}. Rolling-window statistics are derived for diverse \gls{hiot} features across both physiological (body temperature, oxygen level, BPM) and communication (message count, frequency, packet drop, and last sender) domains. The resulting parameters are analysed jointly to characterise \gls{sf}, \gls{mitm}, and \gls{ddos} attack patterns in a unified dataset, {\em UL-ECE-MultiAttack-H-IoT2025}. Its preprocessing scripts are publicly available on Zenodo~\cite{akhi2025multiattack} to support reproducibility. Furthermore, the structure and features of the proposed dataset facilitate reuse in \gls{iot} security research, particularly in analyses of communication-layer parameters.

\begin{table}[ht!]
\centering
\caption[Dataset sample from the \emph{UL-ECE-MultiAttack-H-IoT2025}]{Dataset sample from the \emph{\textbf{UL-ECE-MultiAttack-H-IoT2025}}, showing sensor readings (e.g., temperature, oxygen level, BPM) and communication metrics (e.g., message frequency, packet drop, and last sender) for each node.}
\label{tab:dataset_combined}
\renewcommand{\arraystretch}{1.2} 
\resizebox{\textwidth}{!}{%
\normalsize
\begin{tabular}{llllllllllll}

\hline
\textbf{IDs} 
& \textbf{Temp} 
& \textbf{Temp\_Mean} 
& \textbf{Temp\_Std} 
& \textbf{Temp\_Min} 
& \textbf{Temp\_Max} 
& \textbf{O$_2$\_l} 
& \textbf{O$_2$\_l\_Mean} 
& \textbf{O$_2$\_l\_Std} 
& \textbf{O$_2$\_l\_Min} 
& \textbf{O$_2$\_l\_Max}
& \textbf{BPM} \\
\hline
21 & 95.0  & 100.3333 & 5.1639 & 95.0  & 105.0 & 70 & 91.3333  & 20.6559 & 70.0  & 110.0 & 40 \\

25 & 98.24 & 97.586 & 2.0107 & 96.0  & 101.0 & 94 & 89.9333  & 12.3431 & 80.0  & 110.0 & 64 \\

9  & 98.06 & 97.9213 & 1.6787 & 96.0  & 101.0  & 96 & 91.4666   & 10.1900  & 80.0  & 110.0  & 66 \\

8  & 101.0 & 98.3699 & 2.0155 & 96.0  & 101.0 & 110 & 93.6666  & 12.0218 & 80.0  & 110.0 & 120 \\

24 & 98.75 & 98.1486 & 1.6103 & 96.0  & 101.0  & 95 & 91.5333  & 9.4481  & 80.0  & 110.0  & 95 \\

25 & 101.0 & 97.9193 & 2.1394 & 96.0  & 101.0  & 110 & 91.9333  & 13.0300  & 80.0 & 110.0 & 120 \\
8  & 98.34 & 98.5260 & 1.9066 & 96.0  & 101.0  & 94 & 94.6  & 11.4130  & 80.0 & 110.0 & 74 \\

22 & 95.0 & 97.6666 & 4.5773 & 95.0  & 105.0 & 70 & 80.6666  & 18.3095 & 70.0  & 110.0 & 40 \\

13 & 96.0 & 97.6613 & 2.0410 & 96.0  & 101.0 & 80 & 90.1333  & 12.3916 & 80.0  & 110.0 & 50 \\

23 & 96.0 & 97.9600 & 2.3464 & 96.0  & 101.0 & 80 & 92.0 & 14.1975  & 80.0 & 110.0 & 50 \\
\hline
\end{tabular}
}

\vspace{0.2cm} 
\renewcommand{\arraystretch}{1.2} 
\resizebox{\textwidth}{!}{%
\footnotesize
\begin{tabular}{llllllllllll}
\hline
\textbf{IDs} 
& \textbf{BPM\_Mean} 
& \textbf{BPM\_Std} 
& \textbf{BPM\_Min} 
& \textbf{BPM\_Max} 
& \textbf{T\_Msgs} 
& \textbf{T\_Msgs\_S\_N} 
& \textbf{Mean\_T\_Msgs} 
& \textbf{Msgs Freq} 
& \textbf{Packet Drop (\%)} 
& \textbf{Last Sender} \\
\hline
21 & 98.6666 & 56.8037 & 40.0 & 150.0 & 9991 & 405 & 416 & 4.0536 & 2.9961 & 24 \\
25 & 73.2666 & 29.2927 & 50.0 & 120.0 & 9992 & 810 & 416 & 8.1064 & 0.0000 & 21 \\
9 & 72.8 & 23.7372  & 50.0 & 120.0  & 9993 & 405 & 416 & 4.0528 & 3.0412 & 25 \\
8  & 79.0 & 28.4981 & 50.0 & 120.0 & 9994 & 405 & 416 & 4.0524 & 3.0412 & 9  \\
24  & 77.5333 & 23.9549 & 50.0 & 120.0  & 9995 & 811 & 416 & 8.1140 & 0.0000 & 8  \\
25 & 77.9333 & 30.8555 & 50.0 & 120.0 & 9996 & 811 & 416 & 8.1132 & 0.0000 & 24 \\
8 & 80.6 & 27.4064 & 50.0 & 120.0 & 9997 & 406 & 406 & 4.0612 & 3.0412 & 25 \\
22  & 69.3333 & 50.3511 & 40.0 & 150.0 & 9998 & 406 & 416 & 4.0608 & 2.9961 & 8  \\
13 &  71.4666 & 28.6826 & 50.0 & 120.0 & 9999 & 406 & 416 & 4.0604 & 3.0412 & 22 \\
23 & 76.0 & 32.9089 & 50.0 & 120.0 & 10000 & 811 & 416 & 8.1100 & 0.0000 &  13 \\
\hline
\end{tabular}
}
\vskip 3pt
\begin{minipage}{\linewidth}
\scriptsize \textbf{Note:} \textbf{IDs}=Node IDs (repeated in both parts of the table to facilitate sequential computations of the following features), \textbf{Temp} = Body\_Temperature\_Same\_Node, \textbf{Temp\_Mean} = Temperature\_Mean\_Same\_Node, \textbf{Temp\_Std} = Temperature\_Std\_Same\_Node, \textbf{Temp\_Min} = Temperature\_Min\_Same\_Node, \textbf{Temp\_Max} = Temperature\_Max\_Same\_Node, \textbf{O$_2$\_l} = Oxygen\_Level\_Same\_Node, \textbf{O$_2$\_l\_Mean} = Oxygen\_Level\_Mean\_Same\_Node, \textbf{O$_2$\_l\_Std} = Oxygen\_Level\_Std\_Same\_Node, \textbf{O$_2$\_l\_Min} = Oxygen\_Level\_Min\_Same\_Node, \textbf{O$_2$\_l\_Max} = Oxygen\_Level\_Max\_Same\_Node, \textbf{BPM} = Beats Per Minute (Heart\_Rate\_Same\_Node), \textbf{BPM\_Mean} = BPM\_Mean\_Same\_Node, \textbf{BPM\_Std} = BPM\_Std\_Same\_Node, \textbf{BPM\_Min} = BPM\_Min\_Same\_Node, \textbf{BPM\_Max} = BPM\_Max\_Same\_Node, \textbf{T\_Msgs} = Total Messages, \textbf{T\_Msgs\_S\_N} = Total Messages (Same Node), \textbf{Mean\_T\_Msgs} = Mean Total Messages, \textbf{Msgs Freq} = Messages Frequency (Same Node), \textbf{Packet Drop} = Packet Drop Rate (Same Node), \textbf{Last Sender} = Previous Sender Node.
\end{minipage}
\end{table}

 This form of temporal and cross-domain feature integration remains relatively unexplored in \gls{hiot} attack studies~\cite{shombot2025maximizing, ahmed2021ecu}. Specifically, some prior studies employ single-domain feature fusion methods for anomaly-based intrusion detection in \gls{iomt} environments. For example, Zachos et al. \cite{10909110} use low-level system metrics (e.g., CPU ticks, interrupts, memory counters) from edge devices, while the BlueTack dataset~\cite{zubair2022secure} focuses solely on Bluetooth communication attributes (e.g., protocol fields and packet parameters). Similarly, the CICIoMT2024 dataset~\cite{dadkhah2024ciciomt2024} provides multi-protocol network traffic (Wi-Fi, \gls{mqtt}, and Bluetooth). These datasets neither integrate physiological data nor capture temporal behaviour. In contrast, the feature-engineering method introduces rolling-window statistics and combines physiological and communication domains in \gls{hiot}.

\textbf{Rolling Window-Based Features:}
This method considers a fixed number of sequential data points (\(w\)), called the window size, to compute mean, standard deviation, minimum, and maximum features \cite{thedatabeast2025window}. For this research, the window size is set to \(w = 5\), meaning each node's calculation uses the most recent five data points. This choice balances the need to capture short-term variations while ensuring enough data within the window to compute meaningful statistical features. Moreover, a window size of 5 is a commonly used heuristic for many rolling-window applications in time-series analysis. Accordingly, the rolling window processes the data values step by step by sequentially sliding through the dataset. This sliding process refers to how the window moves one data point at a time, including the most recent \(w\) data points, while removing the oldest and adding the next data point at each step. This ensures that the calculated features are dynamically updated as the window progresses through the dataset.

Moreover, this method captures localised trends and variability for each node by analysing sequential subsets of data within the rolling window. Localised trends refer to patterns or changes within these smaller subsets, such as a gradual increase or sudden spike in values over a short period. Variability measures how much the data fluctuates within the same subgroup. It is often quantified using standard deviation or minimum metrics.

This analysis of localised trends and variability highlights the benefits of analysing smaller, sequential subsets of data. Focusing on these subsets instead of the entire dataset makes detecting short-term patterns, anomalies, or abrupt changes easier. For instance, sudden shifts in temperature, oxygen level, or heart rate can be identified through the evolution of rolling window features, which might otherwise be overlooked when analysing the dataset.

\subsubsection{Node IDs}
The Cooja simulations use IDs to identify nodes and servers, facilitating tracking and managing client-server interactions.

\subsubsection{Body Temperature (Same Node)}
The recorded temperature of a node, typically in degrees Fahrenheit (\(^\circ\)F) or Celsius (\(^\circ\)C), represents its human body temperature status. For example, if a node or \gls{hiot} device records a temperature of 98.6\(^\circ\)F, the ``body temperature'' value for that node is \(98.6\).

\subsubsection{Body Temperature Mean (Same Node)}The ``Body\_Temperature\_Mean\_Same\_Node'' feature is calculated as the rolling mean of a node's body temperature using the most recent \(w = 5\) data points within a fixed window. For example, if the temperatures in the window are \([98.0, 98.6, 99.0, 97.8, 98.4]\), the mean temperature \(\bar{T}\) is \(98.36\). As shown in (1), the equation computes the average of the body temperature readings (\(T_i\)), where \(i\) denotes the time step, ensuring the calculation captures short-term fluctuations in temperature data for each node.
\begin{equation}
\text{Body\_Temperature\_Mean\_Same\_Node} = \frac{1}{w} \sum_{i=1}^{w} T_i
\label{eq:temperature_mean}
\end{equation}
where,
\begin{itemize}
    \item \(T_i\): Temperature value at time \(i\).
    \item \(i\): Denotes the index or time step within the window.
    \item \(w\): Window size (set to 5).
\end{itemize}

\subsubsection{Body Temperature Standard Deviation (Same Node)}
This feature is derived by computing the variation in temperature readings for each node over a fixed number of recent data points using a rolling window. It calculates how much the temperature values deviate from their mean within the specified window size \(w = 5\). For example, if the rolling window contains the temperatures \([98.0, 98.6, 99.0, 97.8, 98.4]\), the mean temperature \(\bar{T}\) is \(98.36\), and the ``Body\_Temperature\_Std\_Same\_Node'' (standard deviation) is approximately \(0.43\), indicating the degree of fluctuation in the readings. The calculation is performed as follows:
\begin{equation} 
\text{Body\_Temperature\_Std\_Same\_Node} = \sqrt{\frac{1}{w - 1} \sum_{i=1}^{w} (T_i - \bar{T})^2}
\label{eq:temperature_std} \vspace{-3mm}
\end{equation}
where,
\begin{itemize}
    \item \(T_i\): Temperature value at time \(i\).
    \item \(\bar{T}\): Mean temperature in the rolling window (e.g., 98.36).
    \item \(w\): Window size (set to 5).
\end{itemize}


\subsubsection{Body Temperature Minimum (Same Node)}The ``Body\_Temperature\_Min\_Same\_Node'' feature is obtained by selecting the minimum temperature value within a rolling window, which considers a fixed number of consecutive data points. For example, if the rolling window contains the temperatures \([98.0, 98.6, 99.0, 97.8, 98.4]\), the minimum temperature is \(97.8\). The calculation is performed as defined in (6.3).
\begin{equation}
\text{Body\_Temperature\_Min\_Same\_Node} = \min(T_1, T_2, \ldots, T_w)
\label{eq:temperature_min}
\end{equation}
where,
\begin{itemize}
  \item \(T_i\) is the temperature value at time \(i\) in the rolling window.
\end{itemize}

\subsubsection{Body Temperature Maximum (Same Node)} The ``Body\_Temperature\_Max\_Same\_Node'' feature is determined using the maximum value within a rolling window of consecutive temperature readings. For example, if the rolling window contains the temperatures \([98.0, 98.6, 99.0, 97.8, 98.4]\), the ``Body\_Temperature\_Max\_Same\_Node'' is \(99.0\). The computation is shown in (6.4).
\begin{equation}
\text{Body\_Temperature\_Max\_Same\_Node} = \max(T_1, T_2, \ldots, T_w)
\label{eq:temperature_max}
\end{equation}
where,
\begin{itemize}
  \item \(T_i\) is the temperature value at time or index \(i\) in the rolling window.
\end{itemize}

\subsubsection{Oxygen Level (Same Node)}
This feature contains the recorded oxygen level reported by each node in the \gls{hiot} network.
The recorded oxygen level is used to assess the health of the person being monitored, as oxygen levels above or below the normal range (\(80\%-110\%\)) may indicate abnormalities or potential health issues.
For example, if a node reports an oxygen level of \(92\%\), it falls within the normal range, suggesting that the system functions correctly. However, an oxygen level of \(75\%\) would be flagged as abnormal, indicating potential issues that require further investigation.

\subsubsection{Oxygen Level Mean (Same Node)}The ``Oxygen\_Level\_Mean\_Same\_Node'' feature is calculated as the average oxygen level for a specific node in the \gls{hiot} network, using the most recent fixed-size window of 5 data points. The rolling average is used to observe short-term fluctuations in the oxygen level of that node. For example, if the window contains oxygen levels \([92, 95, 96, 94, 93]\), the mean oxygen level \(\bar{O}\) is \(94.0\). The calculation of the rolling mean is expressed as:
\begin{equation}
\text{Oxygen\_Level\_Mean\_Same\_Node} = \frac{1}{w} \sum_{i=1}^{w} O_i
\label{eq:oxygen_level_mean}
\end{equation}
where,
\begin{itemize}
    \item \(O_i\): Oxygen level value at index \(i\) in the rolling window.
    \item \(w\): Window size (set to 5).
\end{itemize}
As shown in (6.5), the values in this feature column are updated dynamically as the window slides through the dataset, capturing short-term trends in oxygen levels for individual nodes.

\subsubsection{Oxygen Level Standard (Same Node)}
The ``Oxygen\_Level\_Std\_Same\_Node'' feature is calculated as the standard deviation of oxygen saturation levels over a fixed number of consecutive data points using a rolling window. This value quantifies the variability in oxygen levels within the specified window size.
For instance, if the window contains oxygen levels \([92, 95, 96, 94, 93]\), the mean oxygen level \(\bar{O}\) is \(94.0\), and the standard deviation of oxygen levels is approximately \(1.58\). The calculation is expressed as follows:
\begin{equation} 
\text{Oxygen\_Level\_Std\_Same\_Node} = \sqrt{\frac{1}{w - 1} \sum_{i=1}^{w} (O_i - \bar{O})^2}
\label{eq:oxygen_level_std}
\end{equation}
where,
\begin{itemize}
    \item \(O_i\): Oxygen level value at index \(i\) in the rolling window.
    \item \(\bar{O}\): Mean oxygen level in the rolling window (e.g., \(94.0\)).
    \item \(w\): Window size (set to 5).
\end{itemize}

\subsubsection{Oxygen Level Minimum (Same Node)}
This feature is obtained by selecting the minimum oxygen saturation level within a rolling window, which considers a fixed number of consecutive data points. Suppose the window contains oxygen levels \([92, 95, 96, 94, 93]\), the minimum oxygen level is \(92\). The calculation is outlined in (6.7).
\begin{equation}
\text{Oxygen\_Level\_Min\_Same\_Node} = \min(O_1, O_2, \ldots, O_w)
\label{eq:oxygen_level_min}
\end{equation}
where,
\begin{itemize}
    \item \(O_i\): Oxygen level value at index \(i\) in the rolling window.
\end{itemize}

\subsubsection{Oxygen Level Maximum (Same Node)}
The ``Oxygen\_Level\_Max\_Same\_Node'' feature is determined by selecting the maximum oxygen saturation level within a rolling window, considering a fixed number of consecutive data points. For instance, if the window contains oxygen levels \([92, 95, 96, 94, 93]\), the maximum oxygen level is \(96\). The computation is defined in (6.8).
\begin{equation}
\text{Oxygen\_Level\_Max\_Same\_Node} = \max(O_1, O_2, \ldots, O_w)
\label{eq:oxygen_level_max}
\end{equation}
where,
\begin{itemize}
    \item \(O_i\): Oxygen level value at index \(i\) in the rolling window.
\end{itemize}

\subsubsection{Heart Rate (Same Node)} The ``Heart\_Rate\_Same\_Node'' feature is defined as the number of heartbeats per minute (BPM) recorded for each node in the \gls{hiot} network. Monitoring heart rate is essential for assessing the health status of the system, as abnormal values may indicate potential health issues. The normal range for heart rate is defined as \(50\)-\(120\) BPM. Suppose a node records a heart rate of \(90\) BPM, which is considered normal, indicating a healthy status. However, a heart rate of \(130\) BPM would fall outside the normal range, suggesting an abnormal condition that requires further attention.

\subsubsection{Heart Rate Mean (Same Node)} The rolling mean of heart rate values for each node in the \gls{hiot} network computes the ``Heart\_Rate\_Mean\_Same\_Node'' feature. It is calculated using a fixed-sized window \(w\), where the average heart rate over the most recent \(w\) time steps is computed. This rolling mean provides localised insights into short-term trends in heart rate for individual nodes. For example, if a node reports heart rate values of \(80\), \(85\), \(90\), \(88\), and \(92\) BPM over the last \(5\) time steps (\(w = 5\)), the ``Heart\_Rate\_Mean\_Same\_Node'' is calculated as \(87\). The calculation of the rolling mean is expressed as follows:
\begin{equation}
\text{Heart\_Rate\_Mean\_Same\_Node} = \frac{1}{w} \sum_{i=1}^{w} H_i
\label{eq:heart_rate_mean}
\end{equation}
where,
\begin{itemize}
    \item \(H_i\): Heart rate value at time \(i\).
    \item \(w\): Window size (set to 5).
\end{itemize}
This feature helps identify patterns or sudden changes in heart rate, which could indicate potential issues in the monitored system.

\subsubsection{Heart Rate Standard (Same Node)}The ``Heart\_Rate\_Std\_Same\_Node'' feature is calculated as the standard deviation of heart rate values using a rolling window over a fixed number of consecutive data points. This metric reflects the variability in heart rate values within the specified window size. For example, if the rolling window size \(w = 5\) and the heart rates in the window are \(80\), \(85\), \(90\), \(88\), and \(92\), the mean heart rate \(\bar{H}\) is \(87.0\), and the ``Heart\_Rate\_Std\_Same\_Node'' is approximately \(4.69\). The formula is outlined in (6.10).
\begin{equation}
\text{Heart\_Rate\_Std\_Same\_Node} = \sqrt{\frac{1}{w - 1} \sum_{i=1}^{w} (H_i - \bar{H})^2}
\label{eq:heart_rate_std}
\end{equation}
where,
\begin{itemize}
    \item \(H_i\): Heart rate value at index \(i\) in the rolling window.
    \item \(\bar{H}\): Mean heart rate in the rolling window (e.g., \(87.0\)).
    \item \(w\): Window size (set to 5).
\end{itemize}

\subsubsection{Heart Rate Minimum (Same Node)}
The ``Heart\_Rate\_Min\_Same\_Node'' feature is determined by the minimum heart rate value within a rolling window, considering a fixed number of consecutive data points. For example, if the rolling window contains heart rates \([98.0, 98.6, 99.0, 97.8, 98.4]\), the ``Heart\_Rate\_Min\_Same\_Node'' is \(97.8\). The calculation is shown in (6.11).
\begin{equation}
\text{Heart\_Rate\_Min\_Same\_Node} = \min(H_1, H_2, \ldots, H_w)
\label{eq:heart_rate_min}
\end{equation}
where,
\begin{itemize}
    \item \(H_i\): Heart rate value at index \(i\) in the rolling window.
    \item \(w\): Window size (set to 5).
\end{itemize}

\subsubsection{Heart Rate Maximum (Same Node)}
The ``Heart\_Rate\_Max\_Same\_Node'' feature is identified by the maximum heart rate value within a rolling window, which considers a fixed number of consecutive data points. For example, if the rolling window contains heart rates \([98.0, 98.6, 99.0, 97.8, 98.4]\), the ``Heart\_Rate\_Max\_Same\_Node'' is \(99\). The computation is shown in (6.12).
\begin{equation}
\text{Heart\_Rate\_Max\_Same\_Node} = \max(H_1, H_2, \ldots, H_w)
\label{eq:heart_rate_max}
\end{equation}
where,
\begin{itemize}
    \item \(H_i\): Heart rate value at index \(i\) in the rolling window.
    \item \(w\): Window size (set to 5).
\end{itemize}

\subsubsection{Total Messages}  
The ``Total Messages'' metric measures the number of messages transmitted between nodes (e.g., 24 nodes) in the \gls{hiot} network. This count increases when a new message is transmitted, allowing the effectiveness of network communication to be evaluated. 
In this case, the total number of messages \( T \) up to the \( N \)-th observation is defined in (6.13):
    \begin{equation}
        T = \sum_{i=1}^{N} M_i
        \label{equation:ch6tm}
    \end{equation}
where,
    \begin{itemize}
        \item Let \( M_i \) denote the presence of a message at the \( i \)-th interval, with a value of 1 if a message is present and 0 otherwise.
        \item Let \( N \) denote the total number of data records.
    \end{itemize}
The equation sums each message indicator \( M_i \) together to determine the total number of messages.
    
\subsubsection{Total Messages (Same Node)} 
The ``Total Messages Same Node'' feature contains the total count of messages transmitted by an individual node in the \gls{hiot} network. This value serves as an indicator of the activity level of each node by summing up all the messages it has sent. The total number of messages sent by the node \(i\) is computed as follows:
    \begin{equation}
        \text{TotalMessagesSameNode}_i = \sum_{j=1}^{N} M_{\text{node}_i,j}
    \end{equation}
where, 
\begin{itemize}
  \item \( M_{\text{node}_i,j} \) counts the messages sent by node \( i \) at the \( j \)-th time period.
  \item The sum of these counts from \( j = 1 \) to \( N \) provides the total number of messages sent by node \( i \).
\end{itemize}

\subsubsection{Mean Total Messages}
The ``Mean\_Total\_Msgs'' feature is the average number of messages sent across all nodes in the \gls{hiot} network. It serves as a baseline for comparing the messaging behaviour of individual nodes and identifying deviations or anomalies. The calculation of the mean total messages is shown in (6.15).
\begin{equation}
\text{Mean\_Total\_Msgs} = \frac{\text{Total Messages}}{\text{Number of Nodes}}
\label{eq:mean_total_messages}
\end{equation}

\subsubsection{Messages Frequency (Same Node)}
This feature indicates the percentage of total network messages each node sends over a specific period. This value reflects each node’s contribution to the overall message volume during that time. Let \( M_{\text{node}_i} \) be the total messages from each node, and \( M_{\text{total}} \) be the total number of messages. The frequency percentage can be calculated using (6.16), where \(\text{Frequency (\%)}\) represents the percentage of total messages sent by a specific node:
    \begin{equation}
        \text{frequency (\%)} = \left( \frac{M_{\text{node}_i}}{M_{\text{total}}} \right) \times 100
    \end{equation}

\subsubsection{Packet Drop Rate (Same Node)}
The ``Packet\_Drop\_Rate'' feature contains the percentage of dropped packets for each node, obtained by computing the difference between the expected average messages (``Mean\_T\_Msgs'') and the actual messages sent by the node (``Total\_Messages\_Same\_Node''). This value helps identify nodes showing abnormal messaging behaviour, such as selective forwarding or network issues. The packet drop rate for each node is calculated as shown in (6.17).
{\small
\begin{equation}
\text{Packet\_Drop\_Rate} = \max\left(0, \left(1 - \frac{\text{T\_Msg\_S\_Node}}{\text{Mean\_T\_Msgs}}\right) \times 100 \right)
\label{eq:packet_drop_rate}
\end{equation}
}

\subsubsection{Last Sender} 
The last sensor node transmits data before the current sensor node, for instance, the previous sensor node ID:8 and the current node ID:2.

\textbf{Handling Missing Data:}
During the feature calculation phase, especially when applying rolling window operations (e.g., mean, standard deviation), some entries may produce missing values at the sequence boundaries. To ensure a complete and model-ready dataset, all missing values are filled with zeros, a commonly used imputation technique in time-series preprocessing to handle edge-case missing values \cite{zheng2025missing}. This approach is appropriate given the deterministic nature of the {\em UL-ECE-MultiAttack-H-IoT2025} dataset and helps maintain consistent input shapes for the Res-\gls{tcn} model.

\textbf{Handling Class Imbalance:}
The proposed dataset is inherently imbalanced, with unequal distributions across the four classes: Normal, \gls{sf}, \gls{mitm}, and \gls{ddos}. To address this issue, \gls{smote} is used, which generates synthetic samples for the minority classes. This ensures a balanced distribution across all classes before model training. The application of SMOTE, along with the class distributions before and after oversampling, is presented in Section~\ref{subsec:Res-TCN}, where the model training and validation settings are also described. 

This research uses \gls{smote} instead of more complex generative approaches, such as \glspl{dnngan}, \glspl{vae}, or conditional autoencoders, to maintain a reproducible and straightforward rebalancing process without adding model complexity or risking data leakage \cite{shirvan2025deep}. Moreover, prior studies show that SMOTE-based techniques can enhance classification performance with less computational overhead compared to deep generative methods \cite{joloudari2023effective}.

\textbf{Data Structuring/Normalisation:} Using the Pandas library, all extracted and derived features, including \path{node_ids}, \path{body_temperature_same_node}, \path{heart_rate_same_node}, \path{oxygen_level_same_node}, and other relevant features, are merged into a structured dataset. Rolling statistics such as mean, standard deviation, minimum, and maximum values are computed for selected features. Min-Max scaling is applied before model training, ensuring consistent data ranges.

\textbf{Data Reshaping:} The data are reshaped into a format compatible with \gls{restcn} layers, which are based on 1D convolution. Specifically, the dataset is converted into a 3D array structure, where the dimensions represent samples, timesteps, and features, respectively.

\subsection{Exploratory Data Analysis}
As part of the exploratory data analysis, the {\em UL-ECE-MultiAttack-H-IoT2025} dataset is used to examine feature correlations and temporal dependencies before model training. This analysis includes: generating a correlation heatmap to identify feature dependencies, time-series-aware splitting to avoid data leakage, and inspecting the raw class distribution before balancing. 

\subsubsection{Feature Correlation Heatmap}
The Pearson correlation technique is applied before normalisation to analyse feature correlations within the {\em UL-ECE-MultiAttack-H-IoT2025} dataset \cite{lopez2025pixel}. The resulting heatmap helps identify linear dependencies and potential redundancy among features. However, the dataset exhibits consistent correlations among semantically similar features, as illustrated in Figure~\ref{fig:heatmap}. In particular, the temperature, oxygen level, and heart rate features from the same \gls{hiot} node (e.g., mean, max, min, and standard deviation variants) display closely aligned patterns, indicating consistency across physiological measurements. Leveraging their statistical characteristics, these signals are key in distinguishing between regular activity and anomalous patterns associated with various attack scenarios in \gls{hiot} environments. This structured correlation pattern reflects statistical alignment, which supports robust learning and generalisation in \gls{dl} models.
Additionally, features such as ``message\_frequency'' and ``total\_message'' demonstrate moderate correlations with vital signs, reflecting relevant behavioural patterns in \gls{hiot} communication under benign and malicious conditions.

\begin{figure}[ht!]
    \centering
    \includegraphics[width=\textwidth]{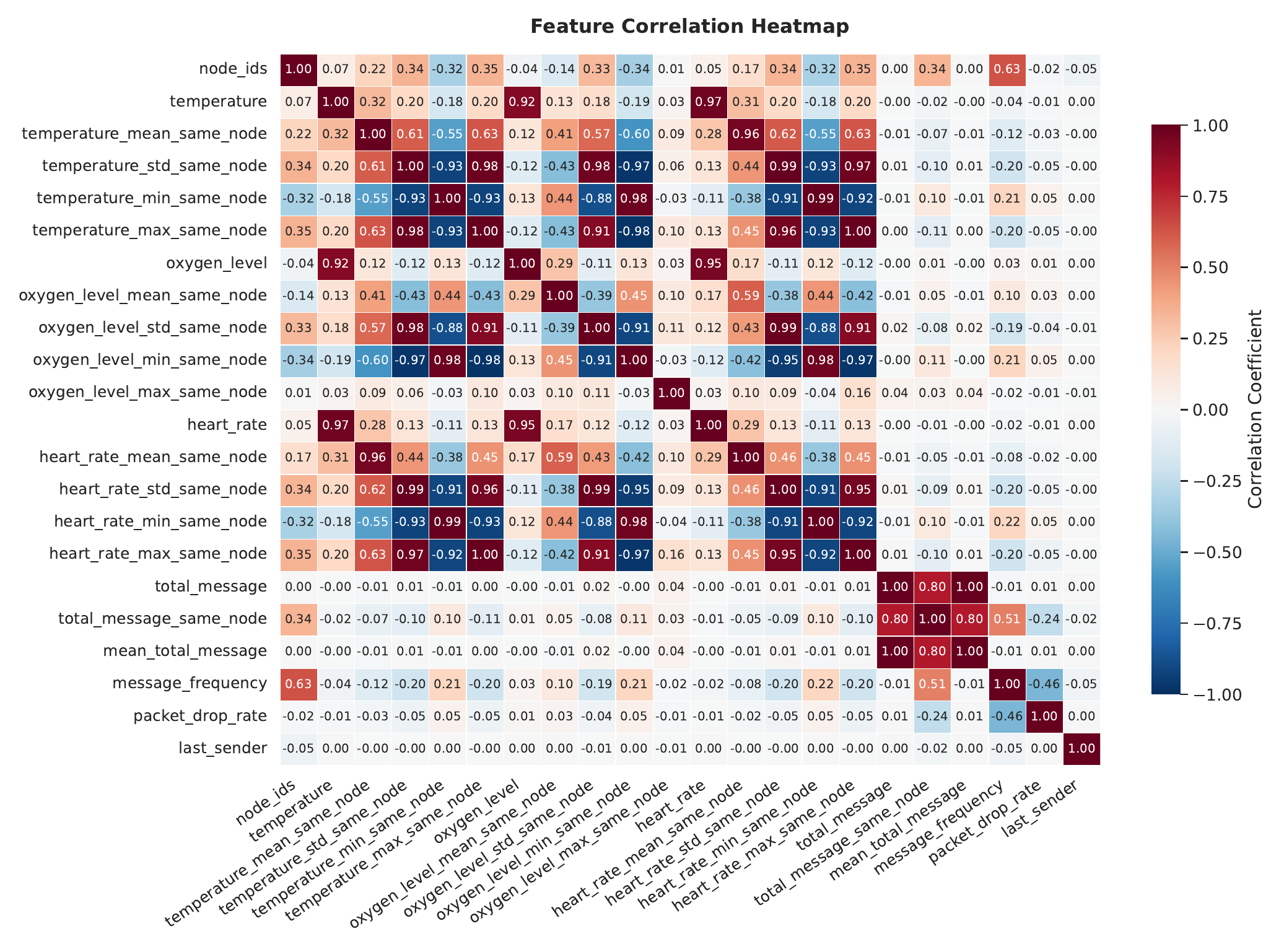}
    \caption[Feature correlation heatmap]{Feature correlation heatmap of the {\em UL-ECE-MultiAttack-H-IoT2025} dataset. The heatmap illustrates consistent correlations among semantically related physiological features, such as temperature, oxygen level, and heart rate from the same node, indicating a coherent structure suitable for learning.}
    \label{fig:heatmap}
\end{figure}

\subsubsection{Time-series Data Splitting} When handling time-series data, paying attention to the temporal structure is crucial. A time-series dataset differs from a standard dataset in that data points are arranged sequentially rather than independently. This sequential arrangement is essential for identifying underlying patterns and relationships in \gls{hiot} environments. A specialised method is used when splitting the dataset to preserve these temporal dependencies during model training and evaluation. This approach divides the dataset based on sequential order rather than random selection. Specifically, the first portion of the dataset (e.g., 70\%) is used for training, while the remaining portion is reserved for testing. As a result, any potential data leakage is eliminated since the training dataset always precedes the test dataset in sequence. This allows predictive models to learn from past events and make informed predictions on future data, thereby maintaining the natural order of the sequence, which is crucial for anomaly detection in \gls{hiot} data \cite{altindal2024anomaly}.

\section{Methodology}
\label{sec:methodology}
This research proposes a novel \gls{dl}-based method for multiclass classification of \gls{hiot} anomalies. This methodology is specifically tailored to \gls{hiot}-emulated sensor data. The proposed model is implemented and analysed using Python 3.9, integrated into a framework designed for neural networks. To ensure robust and efficient \gls{dl} processing, all experiments are conducted on \textit{Google Colab}, a cloud-based environment for \gls{ml} and \gls{dl} model development and testing \cite{googlecolab}.

\subsection{Proposed Res-\gls{tcn} Model Architecture}
\label{subsec:Res-TCN}
The proposed Res-\gls{tcn} model is selected after experimenting with eight different configurations, including variations in the number of Res-\gls{tcn} blocks, dense layer units, and dropout rates, as well as the use of SMOTE versus non-SMOTE data, as part of the ablation study summarised in  Table~\ref{tab:ablationstudy}.
This model is designed to perform multiclass classification of cyber attacks using the proposed dataset, {\em UL-ECE-MultiAttack-H-IoT2025}, in an \gls{hiot} environment. The architecture begins with an input layer reshaped to a three-dimensional format $(64, 22, 1)$, where $22$ represents the number of features, and $1$ denotes a single time step. This reshaping is required to make the input compatible with the temporal convolutional layers and to capture sequential correlations in the data.

A Gaussian noise layer with a standard deviation of 0.3 is applied after the input layer for regularisation. This layer adds minor random variations to the input data during training, which helps prevent overfitting by encouraging the model to learn robust and generalizable patterns. The value 0.3 is empirically selected based on its effectiveness in improving validation performance during model tuning. Following the noise layer, an initial 1D convolutional layer with 64 filters and a kernel size of 3 is applied for feature mapping. This layer extracts local temporal patterns from the input sequence by sliding a kernel (a small window of size 3) over adjacent time steps. This enables the model to capture short-range dependencies across the sequence. Using 64 filters balances model capacity and computational efficiency. The kernel size of 3 is selected to detect fine-grained temporal changes without excessively smoothing the signal.

\begin{figure}[ht!]
    \centering
    \includegraphics[width=\textwidth]{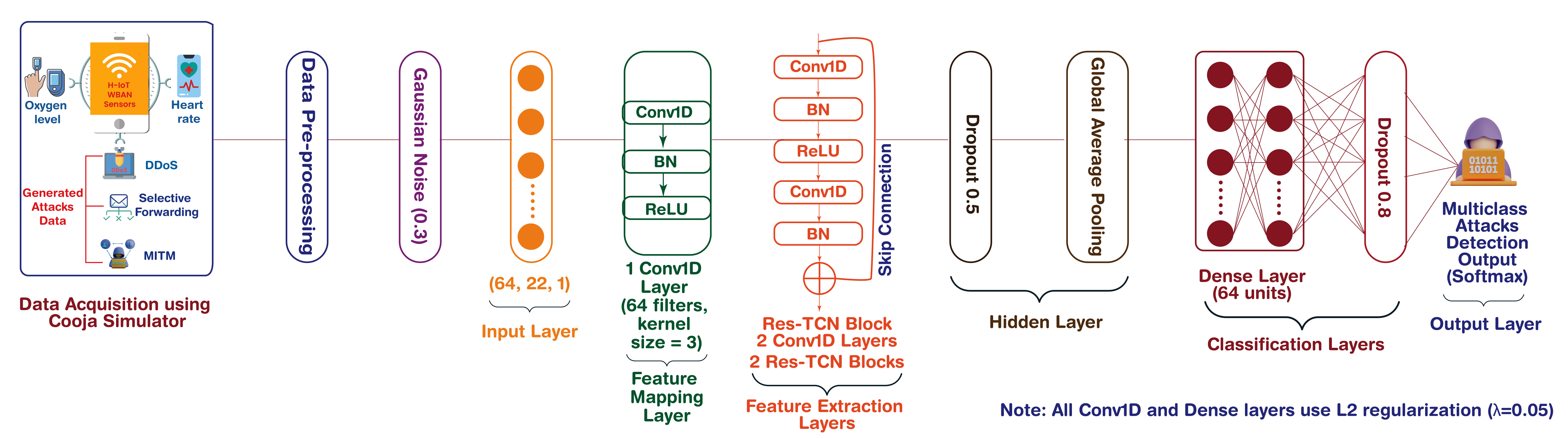}
    \caption[Res-TCN architecture for multiclass attack detection in \gls{hiot}]{Res-\gls{tcn} architecture for multiclass attack detection in \gls{hiot}. The diagram illustrates data acquisition using the Cooja simulator, preprocessing with Gaussian noise for data augmentation, and processing through Conv1D layers with Batch Normalisation (BN) and ReLU activation. Feature extraction is efficiently performed using Res-\gls{tcn} blocks. The resulting feature representations are refined by a Dense layer (64 units), followed by a Dropout layer (0.8) to prevent overfitting. The output layer employs softmax activation for multiclass attack classification.}
    \label{fig:restcnarchitecture}
\end{figure}

\begin{figure}[ht!]
    \centering
    \includegraphics[width=\textwidth]{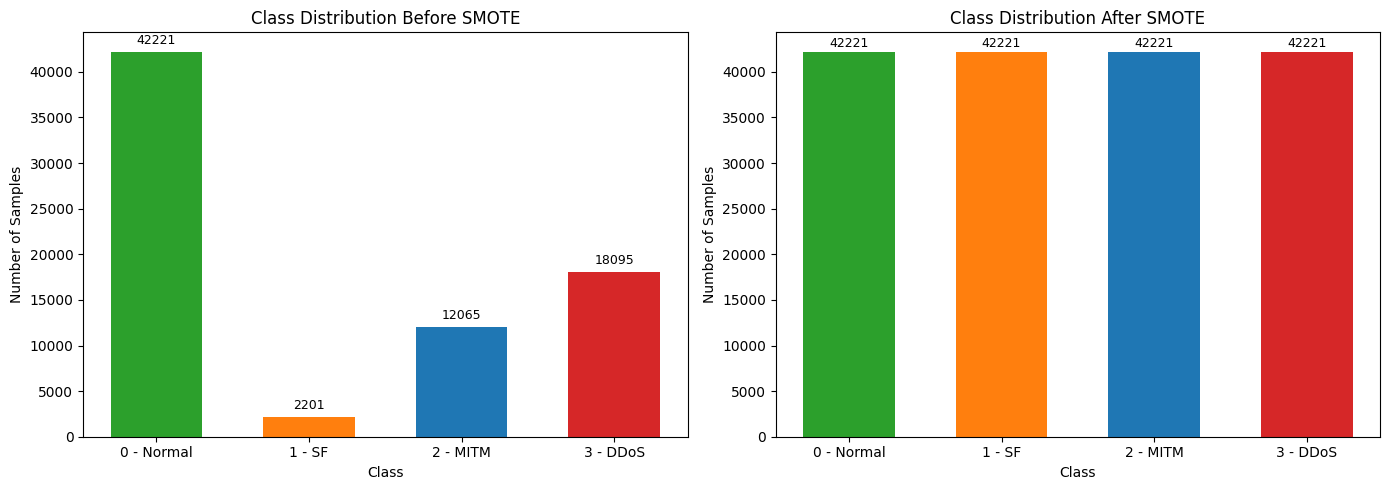}
    \caption[Class distribution before and after applying SMOTE]{Class distribution before and after applying SMOTE to balance the training dataset. The minority classes (\gls{sf}, \gls{mitm}, and \gls{ddos}) are oversampled to match the majority class (Normal), ensuring an equal number of samples (42221) across all four classes.}
    \label{fig:smoteclass}
\end{figure}


\begin{table}[ht!]
\centering
\caption[Dataset distribution for training and testing phases]{Dataset distribution for training and testing phases (without SMOTE).}
\label{tab:dataset-distribution-without-smote}
\begin{tabular}{lllc}
\toprule
\rowcolor[HTML]{EFEFEF} \textbf{No.} & \textbf{\#Samples} & \textbf{Class} & \textbf{Split} \\ \midrule
0  & 42221 & Normal                & Train \\
\rowcolor[HTML]{EFEFEF} 
1  & 2201  & Selective Forwarding  & Train \\
2  & 12065 & \gls{mitm}                  & Train \\
\rowcolor[HTML]{EFEFEF} 
3  & 18095 & \gls{ddos}                  & Train \\ 
\midrule
4  & 18045 & Normal                & Test  \\
\rowcolor[HTML]{EFEFEF} 
5  & 1029  & Selective Forwarding  & Test \\
6  & 5157  & \gls{mitm}                  & Test \\
\rowcolor[HTML]{EFEFEF} 
7  & 7733  & \gls{ddos}                  & Test \\ \bottomrule
\end{tabular}
\end{table}

\begin{table}[!htbp]
\centering
\caption{Hyperparameters of proposed Res-\gls{tcn} Model.}
\label{tab:hyperparams}
\small
\begin{tabular}{ll}
\toprule
\textbf{Hyperparameter} & \textbf{Value} \\ \midrule
\textbf{Dataset Attributes} &  \\  
Number of Features & 22 \\  
Number of Classes & 4 (Normal, \gls{sf}, \gls{mitm}, \gls{ddos})  \\  

Input Shape & (22, 1) \\
\midrule 
\textbf{Architecture} &  \\  
Initial Conv Layer & 1 \\
Kernel Size (Conv1D) & 3 \\ 
Number of Filters (Conv1D) & 64 \\
Number of Res-\gls{tcn} blocks & 2 \\ 
Conv1D layers per Res-\gls{tcn} block & 2 \\  
Total Conv1D layers & $1\ (\text{Initial}) + (2 \times 2) = 5$ \\
Number of Dense Units & 64 \\  
Activation Function (Hidden Layers) & ReLU \\  
Activation Function (Output Layer) & Softmax \\  
Skip Connections & Applied (within Res-\gls{tcn} blocks) \\ 

Normalization Technique & Batch Normalization (after each Conv1D) \\

Global Average Pooling & Applied \\  
\midrule
\textbf{Training Configuration} &  \\ 
Early Stopping & Patience = 15, restore best weights = true \\  
Optimizer & Adam \\  
Loss Function & Sparse Categorical Cross-Entropy \\  
Batch Size & 64 \\  
Epochs & 50 \\  
Learning Rate & 0.001 \\  
Training-testing split & 70\%-30\% \\
SMOTE Sampling Strategy              & 42221 samples per class \\ 
Noise Augmentation & Applied 0.3 (train), 0.1 (test) \\
\midrule 
\textbf{Regularization} &  \\  
Dropout Rate (Residual Blocks) & 0.5 \\  
Dropout Rate (Final Layer) & 0.8 \\ 
L2 Regularization (Conv1D \& Dense) & $\lambda = 0.05$ \\
\bottomrule
\end{tabular}
\normalsize
\end{table}

Afterwards, two Res-\gls{tcn} blocks are stacked, each comprising two 1D convolutional layers with batch normalisation and ReLU activation. Batch normalisation improves training efficiency and stabilises learning by normalising the activations within each layer. ReLU introduces non-linearity, allowing the model to learn more complex patterns related to different types of cyber attacks. As a core component of the Res-\gls{tcn} architecture, skip connections are integrated within each residual block. The output of each Res-\gls{tcn} block is computed as follows:

{\small
\begin{equation*}
\mathbf{H}^{(l)} = \text{ReLU}\left( \text{BN} \left( \mathbf{W}_2^{(l)} * \text{ReLU} \left( \text{BN} \left( \mathbf{W}_1^{(l)} * \mathbf{X}^{(l)} \right) \right) \right) + \mathbf{X}^{(l)} \right)
\label{eq:restcnblock}
\end{equation*}
}
In this formulation, $\text{BN}$ denotes Batch Normalization, $\mathbf{W}_1^{(l)}$ and $\mathbf{W}_2^{(l)}$ are the learnable convolutional filters in the $l$-th Res-\gls{tcn} block, and $\mathbf{x}^{(l)}$ is the input feature tensor at that layer.

This structure highlights how skip connections preserve temporal dependencies, which is particularly beneficial in \gls{hiot} environments for multiclass classification, where minor differences between attack types must be retained. These connections also help preserve learned features across layers and support the efficient flow of gradients during training. In multiclass settings, skip connections enhance the model’s ability to retain subtle variations in feature patterns, including differences in message frequency, inter-packet intervals, and packet drop behaviour. Such variations in the temporal features are critical for distinguishing between closely related attack types, such as \gls{sf} and \gls{ddos}, which often exhibit overlapping characteristics. A dropout layer with a rate of 0.5 is added after the residual blocks to reduce overfitting by randomly deactivating neurons during training, encouraging the learning of more robust representations. Following feature extraction, a global average pooling layer aggregates the temporal information. The pooled output is then passed through a dense layer with 64 units, followed by a dropout layer with a rate of 0.8. Finally, a softmax activation function is used to perform multiclass classification. All Conv1D and Dense layers are regularised using L2 with $\lambda = 0.05$.  
The core architecture of the proposed Res-\gls{tcn} model for multiclass classification in an \gls{hiot} environment is illustrated in Figure~\ref{fig:restcnarchitecture}.

The proposed Res-\gls{tcn} model is trained using the Adam optimiser with a learning rate of 0.001 and a batch size of 64. This learning rate is widely adopted as a default setting in many \gls{dl} frameworks and is effective for stable convergence in this model. Training is conducted for up to 50 epochs, with early stopping enabled to prevent overfitting, using a patience value of 15 and restoring the best weights upon termination. The model is trained using the sparse categorical cross-entropy loss function, with the dataset split into 70\% for training and 30\% for testing, as shown in Table~\ref{tab:dataset-distribution-without-smote}. To address the class imbalance in the training set, the SMOTE is applied, which generates synthetic samples for the minority classes to balance the distribution, as illustrated in Figure~\ref{fig:smoteclass}. Hyperparameters and architectural details of the proposed model are summarised in Table~\ref{tab:hyperparams}.

\begin{table}[ht!]
\centering
\caption[Ablation study exploring variations in Res-\gls{tcn} architecture]{Ablation study exploring variations in Res-\gls{tcn} architecture: effects of blocks, dense layers, SMOTE, dropout rates, and model complexity on detection performance and resource utilisation.}
\label{tab:ablationstudy}
\renewcommand{\arraystretch}{1.2} 
\resizebox{0.9\textwidth}{!}{%
\begin{tabular}{lcccccc} 
\hline
\textbf{Model} 
& \textbf{Blocks} 
& \textbf{Dropout} 
& \textbf{Dense}
& \textbf{SMOTE} 
& \textbf{Inf. Mem. (MB)} 
& \textbf{Total Params}
\\
\hline

Baseline & \multicolumn{1}{l}{3} & \multicolumn{1}{l}{0.3/0.5} & \multicolumn{1}{l}{128} & \multicolumn{1}{l}{No} & \multicolumn{1}{l}{1.21} & \multicolumn{1}{l}{316804}\\

Baseline & \multicolumn{1}{l}{3} & \multicolumn{1}{l}{0.3/0.5} & \multicolumn{1}{l}{128} & \multicolumn{1}{l}{Yes} & \multicolumn{1}{l}{1.21} & \multicolumn{1}{l}{316804}\\

Reduced Res-\gls{tcn} & \multicolumn{1}{l}{1} & \multicolumn{1}{l}{0.3/0.8} & \multicolumn{1}{l}{128} & \multicolumn{1}{l}{Yes}  & \multicolumn{1}{l}{0.70} & \multicolumn{1}{l}{183428} \\

Augmented Res-\gls{tcn} & \multicolumn{1}{l}{2} & \multicolumn{1}{l}{0.2/0.8} & \multicolumn{1}{l}{128} & \multicolumn{1}{l}{Yes}  & \multicolumn{1}{l}{0.83} & \multicolumn{1}{l}{216708}  \\

\textbf{Res-\gls{tcn} (Proposed)} & \multicolumn{1}{l}{\textbf{2}} & \multicolumn{1}{l}{\textbf{0.5/0.8}}  & \multicolumn{1}{l}{\textbf{64}}  & \multicolumn{1}{l}{\textbf{Yes}} & \multicolumn{1}{l}{\textbf{0.21}} & \multicolumn{1}{l}{\textbf{55364}}\\

Deeper Res-\gls{tcn} & \multicolumn{1}{l}{7}  & \multicolumn{1}{l}{0.9/0.8}  & \multicolumn{1}{l}{256} &  \multicolumn{1}{l}{No} & \multicolumn{1}{l}{2.79} & \multicolumn{1}{l}{732164} \\

Deeper Res-\gls{tcn} & \multicolumn{1}{l}{7}   & \multicolumn{1}{l}{0.9/0.8} & \multicolumn{1}{l}{256}  & \multicolumn{1}{l}{Yes}  & \multicolumn{1}{l}{2.79} & \multicolumn{1}{l}{732164} \\

Max-Deeper Res-\gls{tcn} & \multicolumn{1}{l}{10} & \multicolumn{1}{l}{0.5/0.8} & \multicolumn{1}{l}{128}   & \multicolumn{1}{l}{Yes} & \multicolumn{1}{l}{3.87} & \multicolumn{1}{l}{1013892}  \\
\hline
\end{tabular}
}

\vspace{0.2cm} 
\renewcommand{\arraystretch}{1.2} 
\resizebox{0.9\textwidth}{!}{%
\begin{tabular}{lcccccc} 
\hline
\textbf{Model} 
& \textbf{Blocks} 
& \textbf{Accuracy} 
& \textbf{Precision} 
& \textbf{Recall} 
& \textbf{F1-score} 
& \textbf{Inf. Time (ms)} 
\\
\hline

Baseline & \multicolumn{1}{l}{3} & \multicolumn{1}{l}{77.79} & \multicolumn{1}{l}{66} & \multicolumn{1}{l}{78} & \multicolumn{1}{l}{71} & \multicolumn{1}{l}{0.08} \\

Baseline & \multicolumn{1}{l}{3} & \multicolumn{1}{l}{99.77} & \multicolumn{1}{l}{100} & \multicolumn{1}{l}{100} & \multicolumn{1}{l}{100} & \multicolumn{1}{l}{0.07} \\

Reduced Res-\gls{tcn} & \multicolumn{1}{l}{1} & \multicolumn{1}{l}{99.84} & \multicolumn{1}{l}{100} & \multicolumn{1}{l}{100} & \multicolumn{1}{l}{100} & \multicolumn{1}{l}{0.06} \\

Augmented Res-\gls{tcn} & \multicolumn{1}{l}{2} & \multicolumn{1}{l}{99.88} & \multicolumn{1}{l}{100} & \multicolumn{1}{l}{100} & \multicolumn{1}{l}{100} & \multicolumn{1}{l}{0.08}  \\

\textbf{Res-\gls{tcn} (Proposed)} & \multicolumn{1}{l}{\textbf{2}} & \multicolumn{1}{l}{\textbf{99.32}} & \multicolumn{1}{l}{\textbf{99}} & \multicolumn{1}{l}{\textbf{99}} & \multicolumn{1}{l}{\textbf{99}} & \multicolumn{1}{l}{\textbf{0.06}}  \\

Deeper Res-\gls{tcn} & \multicolumn{1}{l}{7} & \multicolumn{1}{l}{85.94} & \multicolumn{1}{l}{89} & \multicolumn{1}{l}{86} & \multicolumn{1}{l}{87} & \multicolumn{1}{l}{0.11}   \\

Deeper Res-\gls{tcn} & \multicolumn{1}{l}{7} & \multicolumn{1}{l}{97.84} & \multicolumn{1}{l}{98} & \multicolumn{1}{l}{98} & \multicolumn{1}{l}{98} & \multicolumn{1}{l}{0.10} \\

Max-Deeper Res-\gls{tcn} & \multicolumn{1}{l}{10} & \multicolumn{1}{l}{99.95} & \multicolumn{1}{l}{100} & \multicolumn{1}{l}{100} & \multicolumn{1}{l}{100} & \multicolumn{1}{l}{0.16}  \\

 \hline
\end{tabular}
}
\begin{minipage}{\textwidth}
\vspace{3pt}
 \scriptsize
\textbf{Note:} This two-part table includes two common columns, \textbf{Model} and \textbf{Blocks}, to ensure consistent cross-reference between architecture and performance metrics. The column \textbf{Blocks} refers to the number of residual blocks used in each model. \textbf{Inf. Mem.} denotes Inference Memory (MB); \textbf{Inf. Time} refers to Inference Time (ms); \textbf{Params} indicates the total number of trainable parameters. This ablation study highlights a progression from the baseline Res-\gls{tcn} architecture to a more complex variant with the highest number of trainable parameters among all evaluated models. The highlighted bold row, \textbf{Res-\gls{tcn} (Proposed)}, refers to the final selected architecture based on high detection accuracy, absence of overfitting, and efficient resource usage, including low inference time and memory consumption. It also maintains stable detection performance and, particularly, the lowest number of parameters among all model variants. For each architecture, identical model names indicate that the model is evaluated under both settings (with and without SMOTE). 
\end{minipage}
\end{table}

\subsection{Ablation Study of Res-\gls{tcn} Architecture and Configuration} \label{subsec:ablation}
Ablation studies systematically evaluate the impact of individual design choices in a model by altering one or more components. This ablation study explores both architectural variations and training-related configurations of the Res-\gls{tcn} model. The architectural variations include changes in the number of residual blocks and the dense layer(s) design. The training-related variants involve adjustments to the dropout rate and the use of SMOTE. The study begins with a baseline architecture and progresses toward a model with the highest number of trainable parameters. The goal is to identify a suitable model that offers an optimal trade-off between multiattack detection performance, including \gls{sf}, \gls{mitm}, and \gls{ddos}, and resource efficiency in \gls{hiot} environments. To extend this investigation, the study also evaluates the performance of each model, both with and without SMOTE, to assess its impact on detection outcomes. 

Overall, the ablation results show that increasing residual depth or dense-layer width leads to only marginal accuracy gains while consistently increasing the number of parameters, inference memory, and latency, as summarised in Table~\ref{tab:ablationstudy}. These patterns underscore how each configuration balances detection accuracy against its parameter count, memory footprint, and latency.

The labels of all models, such as \textbf{Reduced Res-\gls{tcn}}, \textbf{Deeper Res-\gls{tcn}}, and \textbf{Res-\gls{tcn} (Proposed)}, are assigned post-experimentation based on architectural configurations and observed performance outcomes. All models are executed multiple times to ensure the stability and reproducibility of the results.

\subsection{Design Rationale for Proposed Res–\gls{tcn} Settings}
\label{subsec:designrational}
This section outlines the rationale for selecting the proposed Res–\gls{tcn} configuration based on the ablation study results. Its design is further guided by low-power resource constraints and traffic patterns of \gls{hiot} environments. The following design considerations are presented to justify the proposed model settings:

\textbf{Kernel size.} With five stacked convolutional layers, a kernel width of 3 yields an effective receptive field (ERF) of 11 time steps, computed as $1+(k-1)\times L$ with $k=3$ and $L=5$. This receptive field defines the temporal span over which past inputs influence each output activation. It is sufficient to capture the short-range temporal dependencies and localised variations in packet inter-arrival times characteristic of H–IoT traffic. Larger kernels further increase computational cost while providing only marginal ERF expansion~\cite{zhang2025scaling}. In comparison, smaller convolution kernels achieve equivalent accuracy with considerably lower computational overhead~\cite{mohammadpour2022survey}. Therefore, a kernel size of 3 ensures both computational efficiency and adequate temporal coverage.

\textbf{Residual depth.} Employing two residual blocks provides sufficient temporal abstraction while maintaining architectural simplicity. Theoretically, the required model depth depends on the receptive field, i.e., the number of past time steps the model can observe, relative to the total input sequence length~\cite{lopes2023network}. In this intrusion detection workflow, two residual blocks achieve adequate receptive field coverage, capturing both short-term network patterns and longer temporal dependencies.
As shown in Section~\ref{subsec:ablation}, deeper variants can achieve higher accuracy; however, they require substantially more parameters, inference memory, and latency. For example, the baseline model with three residual blocks consumes over 316\,000 parameters and 1.21\,MB of memory while yielding lower accuracy (77.79\%), making it unsuitable for constrained H–IoT settings. This aligns with the principle of optimal model capacity \cite{belkin2019reconciling, vapnik2013nature, vapnik1999overview}, where model complexity should match data complexity to avoid overfitting. Moreover, Figure~\ref{fig:pareto_tradeoffs} shows that the proposed design lies on the Pareto frontier, achieving Pareto-optimal trade-offs for resource-constrained environments~\cite{badar2024fairtrade}.

\textbf{Dropout rates.} A mid-layer dropout rate of 0.5 helps prevent overfitting by reducing dependency among filters, which leads to better generalisation \cite{srivastava2014dropout}. This rate is theoretically optimal as it maximises ensemble diversity through the largest number of sub-network combinations~\cite{baldi2013understanding}. Based on the ablation results (e.g., the baseline model), a higher dropout rate (0.8) is applied before the final layer, serving as a confidence calibration mechanism to mitigate overconfident predictions in \gls{hiot} environments. This interpretation follows the Bayesian view of dropout as an approximation of variational inference, where sampling dropout masks during prediction introduces predictive uncertainty and mitigates overconfident decisions~\cite{gal2016dropout}.  In the ablation study, $p{=}0.8$ yielded the lowest \gls{ece}, confirming improved uncertainty calibration without degrading accuracy. These selections align with established practices in temporal \gls{cnn} design for small-to-medium datasets.

\textbf{Dense layer size.} The final dense layer uses 64 units, which provides sufficient representational capacity for multiclass classification while avoiding over-parameterisation. In the ablation study (Table~\ref{tab:ablationstudy}), models with larger dense layers (e.g., Max-Deeper Res-\gls{tcn} with 128 units) achieve 99.95\% accuracy, accompanied by 100\% precision, recall, and F1-score, indicating possible overfitting. These configurations also increase parameter count and inference memory, reinforcing 64 units as the balanced choice.

\begin{figure}[ht!]
    \centering
    \includegraphics[width=\textwidth]{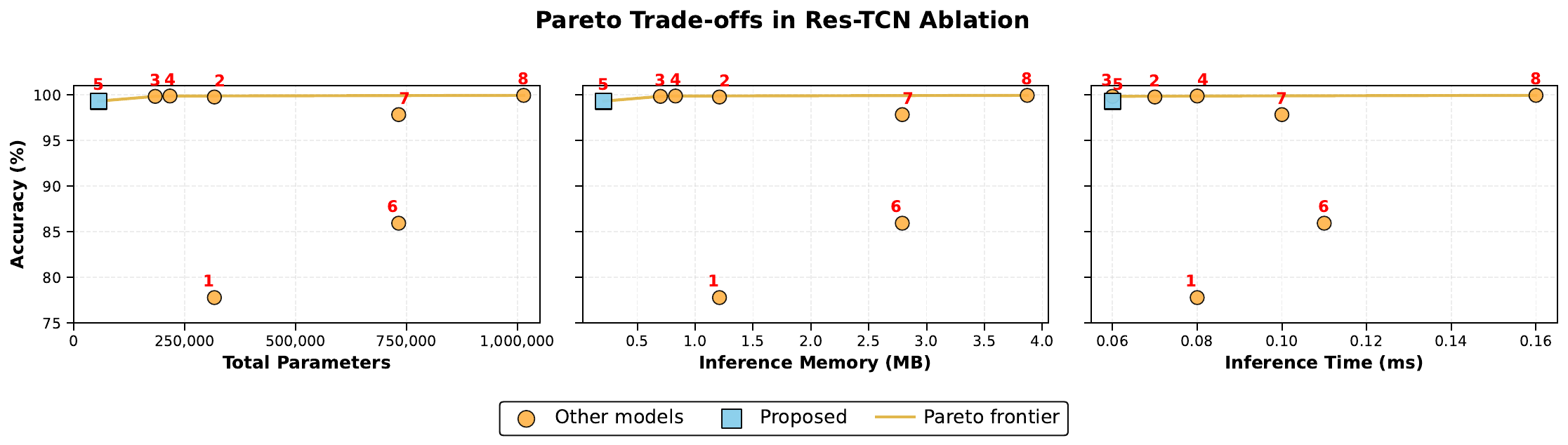}
 \caption[Pareto trade-offs between accuracy and total parameters]{Pareto trade-offs between accuracy and total parameters, inference memory (MB), and inference time for Res-\gls{tcn} variants. Numbers correspond to models in Table~\ref{tab:ablationstudy}. The golden line denotes the Pareto frontier; the proposed Res-\gls{tcn} (\#5) lies on this frontier, balancing accuracy and efficiency better than deeper (\#8) or baseline (\#1) models. 
The inference-time value of the proposed model (0.06) coincides with that of the Reduced Res-\gls{tcn} (\#3) in the plot.}
    \label{fig:pareto_tradeoffs}
\end{figure}

\textbf{Resource-aware selection.} Among the eight Res–\gls{tcn} variants evaluated, the proposed model (\#5) achieves Pareto-optimal trade-offs across accuracy, inference memory, parameter count, and latency, as shown in Figure~\ref{fig:pareto_tradeoffs}, making it a practical choice for securing H–IoT~\cite{badar2024fairtrade}. 
In the plots, the golden line represents the Pareto frontier, which connects models that are not dominated in terms of accuracy and resource cost. A model is on this Pareto frontier when no other variant achieves higher accuracy with fewer resources.  
As illustrated, the proposed Res-\gls{tcn} (\#5) lies on the Pareto frontier, combining high accuracy with markedly lower parameter count, memory, and latency. In contrast, deeper variants such as Max-Deeper (\#8) achieve comparable accuracy at substantially higher computational cost, while the baseline model (\#1) delivers much lower accuracy despite using more resources than the proposed model.

\section{Performance Evaluation}
\label{sec:performance}
This section presents the experimental results and evaluates the effectiveness of the proposed approach in comparison with existing \gls{ml} and \gls{dl} methods. The evaluation covers multiple aspects, including accuracy, precision, recall, F1-score, number of parameters, and inference memory. The proposed model's learning behaviour is also examined using training and validation plots, along with corresponding error curves. In addition, this section includes an analysis of the confusion matrix to evaluate the performance of multiclass classification. It then compares the inference times of different methods designed for \gls{hiot} applications. Finally, the generalizability of the proposed Res-\gls{tcn} model is validated through both multiclass and binary classification tasks. This evaluation utilises two publicly available \gls{hiot} datasets: CICIoMT2024 and CICIoT2023.

\subsection{Comparison with Previous Research}
\label{subsec:modelcomparison}
The proposed \gls{restcn} model is compared with existing \gls{dl} methods, including \gls{lstm} \cite{akar2025l2d2}, \gls{tcn}, \gls{1dclstm} \cite{kumar2025hybrid}, \gls{cnn} \cite{10628718}, lightweight \gls{ml} baselines such as \gls{dt} \cite{binbusayyis2022investigation}, and XGBoost \cite{shombot2025maximizing} as summarized in Table~\ref{tab:compare}. It outperforms these approaches in terms of accuracy, precision, recall, and F1-score. In addition to evaluation metrics, this research also compares the number of parameters, the inference time and the memory footprint across models, as outlined in Table~\ref{tab:compare}. The proposed model has a significantly lower parameter count compared to other \gls{dl} baselines, except for the \gls{cnn} model~\cite{10628718}, which adopts a simpler architecture. \gls{dt} and XGBoost, being lightweight \gls{ml} models, exhibit tiny inference-time memory footprints in the evaluation environment. These models lack large trainable weight matrices, making their parameter counts incomparable to those of \gls{dl} models. The classification accuracies of these lightweight shallow models are significantly lower than those of the proposed Res-\gls{tcn} model. This result suggests that the proposed model combines consistently high predictive performance with manageable memory usage, indicating its potential suitability for deployment on resource-constrained \gls{hiot} devices. 

\begin{table}[!htbp]
\centering
\caption[Comparison of the proposed and existing methods]{Comparison of the proposed and existing methods evaluated on the proposed dataset.}
\label{tab:compare}
\small
\begin{tabular}{@{}l l l l l l l @{}}  
\toprule
\textbf{Ref.} & \textbf{Acc.} & \textbf{Prec.} & \textbf{Rec.} & \textbf{F1} & \textbf{Params}  & \textbf{Inf. Mem. (MB)}\\ 
\midrule
\cite{akar2025l2d2} (2025) & 98.32 & 99.00 & 99.00 & 99.00 & 83460 & 0.31 \\  

Chapter~\ref{chap:tcn} (2025) & 73.80 & 87.00 & 87.92 & 83.43 & 62849 & 0.23 \\  

\cite{kumar2025hybrid} (2025) & 83.45 & 67.56 & 74.91 & 70.86 & 108228 & 0.41 \\  

\cite{10628718} (2024) & 83.22 & 89.76 & 92.57 & 89.20 & 6308 & 0.02 \\  

\cite{binbusayyis2022investigation}  (2025) & 96.84 & 73.67 & 75.00 & 74.32 & - & -  \\ 

\cite{shombot2025maximizing} (2025) & 99.00 & 100.00 & 99.00 & 100.00 & - & - \\ 

\textbf{Proposed Method} & \textbf{99.32} & \textbf{99.00} & \textbf{99.00} & \textbf{99.00} & \textbf{55364} & \textbf{0.21} \\
\bottomrule
\end{tabular}
\begin{minipage}{\linewidth}
\vspace{3pt}
\scriptsize \textbf{Note:} The methods listed correspond to \gls{lstm}, \gls{tcn}, \gls{1dclstm}, \gls{cnn}, Decision Tree, and XGBoost. Abbreviations in the table are as follows: Ref.: Reference, Acc.: Accuracy, Prec.: Precision, Rec.: Recall, F1: F1-score, Params: Total Parameters, Inf. Mem. (MB): Inference Memory (in megabytes). Inference memory for DT and XGBoost is minimal; parameter counts are not provided. Bold values correspond to the performance of the proposed method for ease of comparison with existing approaches. All methods are evaluated using the proposed dataset \emph{UL-ECE-MultiAttack-H-IoT2025} to ensure a consistent comparison. Evaluation metrics, including accuracy, precision, recall, and F1-score, are expressed as percentages (\%).
\end{minipage}
\end{table}

Each \gls{ml} and \gls{dl} model differs in type, depth, and architectural complexity. The effect of model capacity is evaluated using a \gls{dt} \cite{binbusayyis2022investigation} model on the proposed {\em UL-ECE-MultiAttack-H-IoT2025} dataset. The simplest configuration (max\_depth = 1) represents a shallow tree with limited capacity and correspondingly reduced accuracy, reflecting underfitting. Increasing the depth to two (max\_depth = 2) introduces moderate capacity and markedly improves generalisation and accuracy. At a depth of three (max\_depth = 3), the DT achieves 100\% accuracy on the test set, indicating that it has fully captured the dataset’s decision boundaries and may risk overfitting to unseen data. These findings illustrate how model type, depth, and complexity shape predictive behaviour. This analysis supports the selection of balanced depth and capacity for the proposed Res-\gls{tcn} model. Table~\ref{tab:dt_depth} summarises the effect of DT depth on performance.

To further illustrate interpretability, a rule-based DT is applied with controlled parameters to balance model capacity and generalisation. At \texttt{max\_depth} = 3, the DT achieves 100\% accuracy, indicating overfitting. When the depth is reduced to \texttt{max\_depth} = 2 with \texttt{min\_samples\_leaf} = 3, the accuracy decreases to 96.97\%, with macro-precision, recall, and F1-scores of 73.73\%, 75.00\%, and 74.35\%, respectively. The DT produces human-readable \textit{if–then} rules that explicitly describe its decision logic, providing a transparent understanding of how predictions are made. The resulting rules are based on \texttt{packet\_drop\_rate} and \texttt{temperature\_max\_same\_node}, which the DT identifies as the most discriminative features based on information gain, defined as the reduction in class impurity achieved at each split. For instance, if \texttt{packet\_drop\_rate} $\leq$ 1.487 then class = 3 (\gls{ddos}) and if \texttt{packet\_drop\_rate} $>$ 1.487 and \texttt{temperature\_max\_same\_node} $\leq$ 103 then class = 0 (Normal). These rules illustrate interpretable decision boundaries that map feature conditions to class outcomes. The rule-based DT correctly identifies Normal (class 0), \gls{mitm} (class 2), and \gls{ddos} (class 3) instances, whereas all \acrlong{sf} (class 1) samples are misclassified as Normal. This misclassification likely arises from overlapping feature patterns and similar transmission characteristics between Normal and Selective Forwarding nodes. These results indicate that \texttt{packet\_drop\_rate} and \texttt{temperature\_max\_same\_node} provide effective discrimination for \gls{mitm} and \gls{ddos}, while offering limited separability for \gls{sf} attacks, which exhibit near-normal packet characteristics. Hence, the rule-based \gls{dt} model demonstrates limited effectiveness for multiclass scenarios, reinforcing the need for a higher-capacity model such as Res-\gls{tcn} that can learn non-linear decision boundaries and temporal patterns. The DT’s inability to distinguish between critical threats and normal \gls{hiot} behaviour highlights the limitations of simple rule-based approaches. It emphasises the importance of balanced model complexity for handling diverse \gls{hiot} attack behaviours.

In the evaluation, the 1D-CLSTM model is implemented according to the design described in \cite{kumar2025hybrid}. The architecture combines an initial causal one-dimensional convolutional layer with 32 filters (kernel size 3) and a 128-unit \gls{lstm} layer, followed by dense layers and dropout. This deeper recurrent–convolutional design increases model capacity compared to lightweight \gls{ml} models such as Decision Trees, enabling it to capture temporal dependencies and complex feature interactions in the multiclass classification task. 
Despite its higher capacity, the 1D-CLSTM achieves 83.45\% accuracy on the proposed dataset, which is lower than the 96.84\% obtained by the moderately deep \acrlong{dt} (DT-2) model, as detailed in Table~\ref{tab:dt_depth}. 
The \gls{1dclstm} model also has a larger parameter set, resulting in higher inference memory compared with shallower convolutional or tree-based models, as shown in Table~\ref{tab:compare}.

\begin{table}[!htbp]
\centering
\caption[Effect of model depth on Decision Tree performance]{Effect of model depth on Decision Tree performance using the proposed dataset.}
\begin{tabular}{llllll}
\toprule
\multicolumn{1}{c}{Model} & \multicolumn{1}{c}{max\_depth} & \multicolumn{1}{c}{Accuracy} & \multicolumn{1}{c}{Precision} & \multicolumn{1}{c}{Recall} & \multicolumn{1}{c}{F1-score} \\
\midrule
DT-1 (Shallow) & 1 & 80.48 & 43.56 & 50.00 & 46.30 \\
DT-2 (Moderate) & 2 & 96.84 & 73.67 & 75.00 & 74.32 \\
DT-3 (Deep) & 3 & 100 & 100 & 100 & 100 \\
\bottomrule
\end{tabular}
\begin{minipage}{\linewidth}
\vspace{2pt}
   \scriptsize \textbf{Note:} DT-1, DT-2, and DT-3 correspond to shallow, moderately deep, and deep Decision Tree models, respectively, based on different \texttt{max\_depth} configurations.
\end{minipage}
\label{tab:dt_depth}
\end{table}

Standard \gls{tcn} employs a shallower stack of dilated (e.g., 1, 2, 4) convolutions, which limits its receptive field compared with the balanced residual blocks in the proposed Res-\gls{tcn}, as shown in Table~\ref{tab:ablationstudy}. 
The integration of residual connections with temporal convolutional layers in Res-\gls{tcn} enlarges the effective receptive field yet keeps the number of parameters low \cite{luo2016understanding}, as detailed in Section~\ref{subsec:designrational}. 
This design shows that residual depth and temporal kernel width directly influence performance, leading to high accuracy with low inference memory usage compared to the standard \gls{tcn}, as detailed in Table~\ref{tab:compare}.

\begin{figure}[ht!]
    \centering
    \includegraphics[width=0.9\columnwidth]{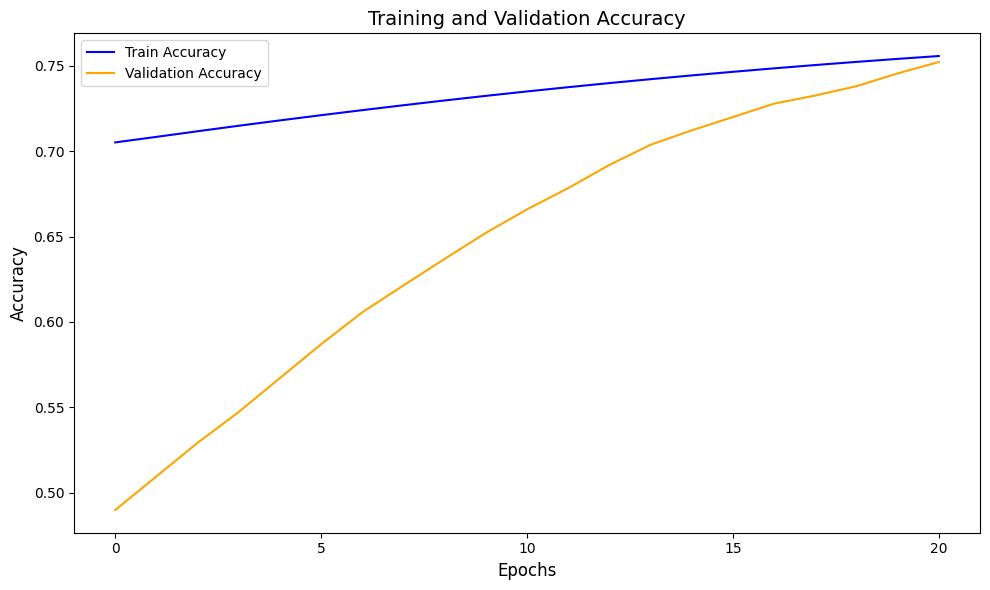}
    \caption[Training and validation accuracy]{Training and validation accuracy of the proposed Res-\gls{tcn} model.}
    \label{fig:acc_plot}
\end{figure}

\begin{figure}[ht!]
    \centering
    \includegraphics[width=0.9\columnwidth]{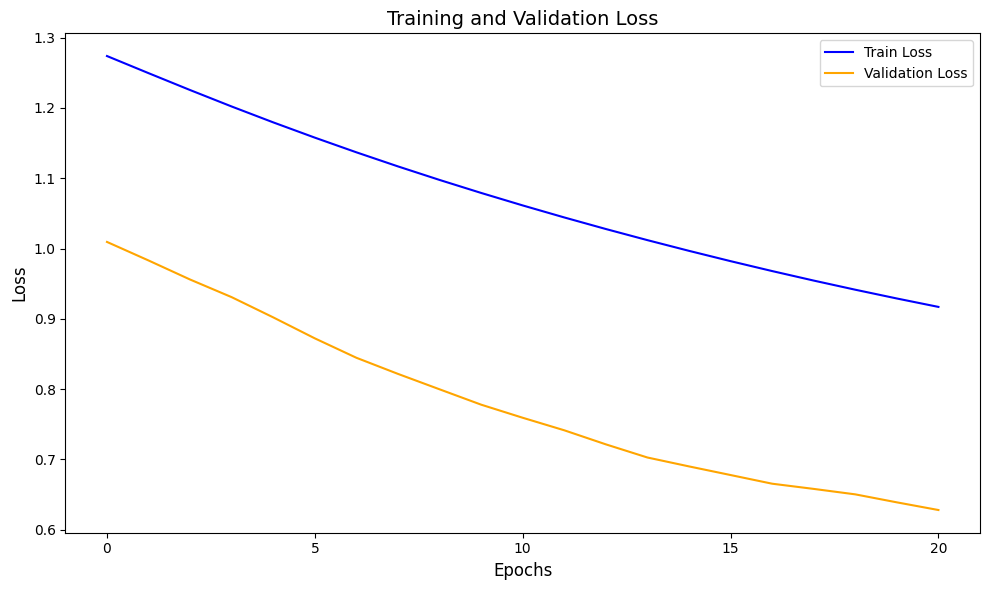}
    \caption[Training and validation loss]{Training and validation loss of the proposed Res-\gls{tcn} model.}
    \label{fig:loss_plot}
\end{figure}

The training and validation accuracy curves of the proposed Res-\gls{tcn} model over 50 epochs remain consistently high, as shown in Figure~\ref{fig:acc_plot}. A smoothing approach, combining exponential smoothing and moving average, is applied to the validation curve to reduce short-term fluctuations and highlight general epoch-wise trends \cite{jayanth2024developing, suprijanto2024new}. The resulting smoothed curve suggests consistent generalisation performance throughout training. Initially, the training accuracy starts around 70\%, while the validation accuracy begins near 50\%, indicating the model’s early adjustment to unseen data. As training progresses, the validation accuracy steadily increases and gradually converges with the training curve. It reflects improved generalisation to unseen samples. The narrowing gap between the two curves over epochs, without signs of divergence, further confirms stable and balanced learning behaviour. Thus, this smoothing technique highlights the general trend across epochs without altering the actual behaviour of the model. It also makes the learning dynamics easier to interpret and reinforces the consistent performance reflected in the stable validation accuracy.

\rev{The model was trained for a maximum of 50 epochs with early stopping. Training terminated at 20 epochs based on validation loss, and the curves in Figure~\ref{fig:loss_plot} reflect the executed epochs. The training and validation loss curves show a steady decline across the training epochs, indicating consistent learning behaviour, although the lower validation loss than the training loss reflects the effect of regularisation. This behaviour has been reported in the literature, where training loss can exceed validation loss when regularisation is applied during training but not during validation \cite{salehin2023review, geron2019hands}.} The validation loss closely follows the training loss without significant divergence, indicating stable learning and no signs of overfitting. This behaviour suggests that the model generalises without memorising the training data. 

\begin{figure}[ht!]
    \centering
\includegraphics[width=0.9\columnwidth]{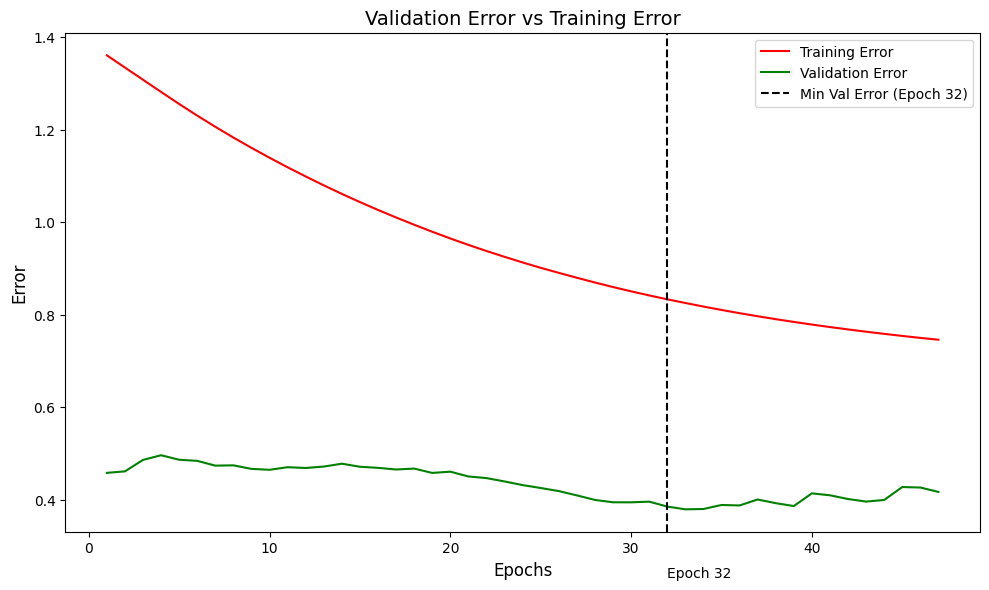}
    \caption[Training vs. validation error]{Training vs. validation error of the proposed Res-\gls{tcn} model.}
    \label{fig:error_plot}
\end{figure}

This is further supported by the error comparison in Figure~\ref{fig:error_plot}, where the validation error remains low and stable throughout training. Meanwhile, the training error continues to decline over epochs. The dashed vertical line at Epoch 32 indicates the point at which the validation loss reaches its minimum value. This is automatically detected by the EarlyStopping callback during training, and the corresponding model weights are restored.

The model achieves a final accuracy of 99.32\%, which remains consistent across multiple executions, demonstrating reliable generalisation performance for multiclass classification in \gls{hiot} environments.

Moreover, the total inference memory of the proposed model is estimated by summing two components: the memory required to store model parameters (e.g., weights and biases) and the memory consumed by the outputs of intermediate layers. The total number of parameters is obtained from the model summary. The model uses 32-bit floating-point precision (float32), where each parameter occupies 4 bytes of memory. Activation memory is estimated using the output shape of each layer with a batch size of 1, reflecting the memory footprint of per-sample inference. All values are converted from kilobytes to megabytes to ensure consistency with previously reported models. The inference memory footprint, expressed in megabytes (MB), is computed using (6.18), following the standard method described in \cite{denneman2022memory}.

{\footnotesize
\begin{equation}
\text{Total Inf. Mem.} = \frac{\text{Total Params} \times 4}{1024^2} + \sum_{\text{layers}} \frac{\text{Activation Size} \times 4}{1024^2}
\label{eq:inference_memory}
\end{equation}
}

\begin{figure}[ht!]
    \centering
    \includegraphics[width=0.9\columnwidth]{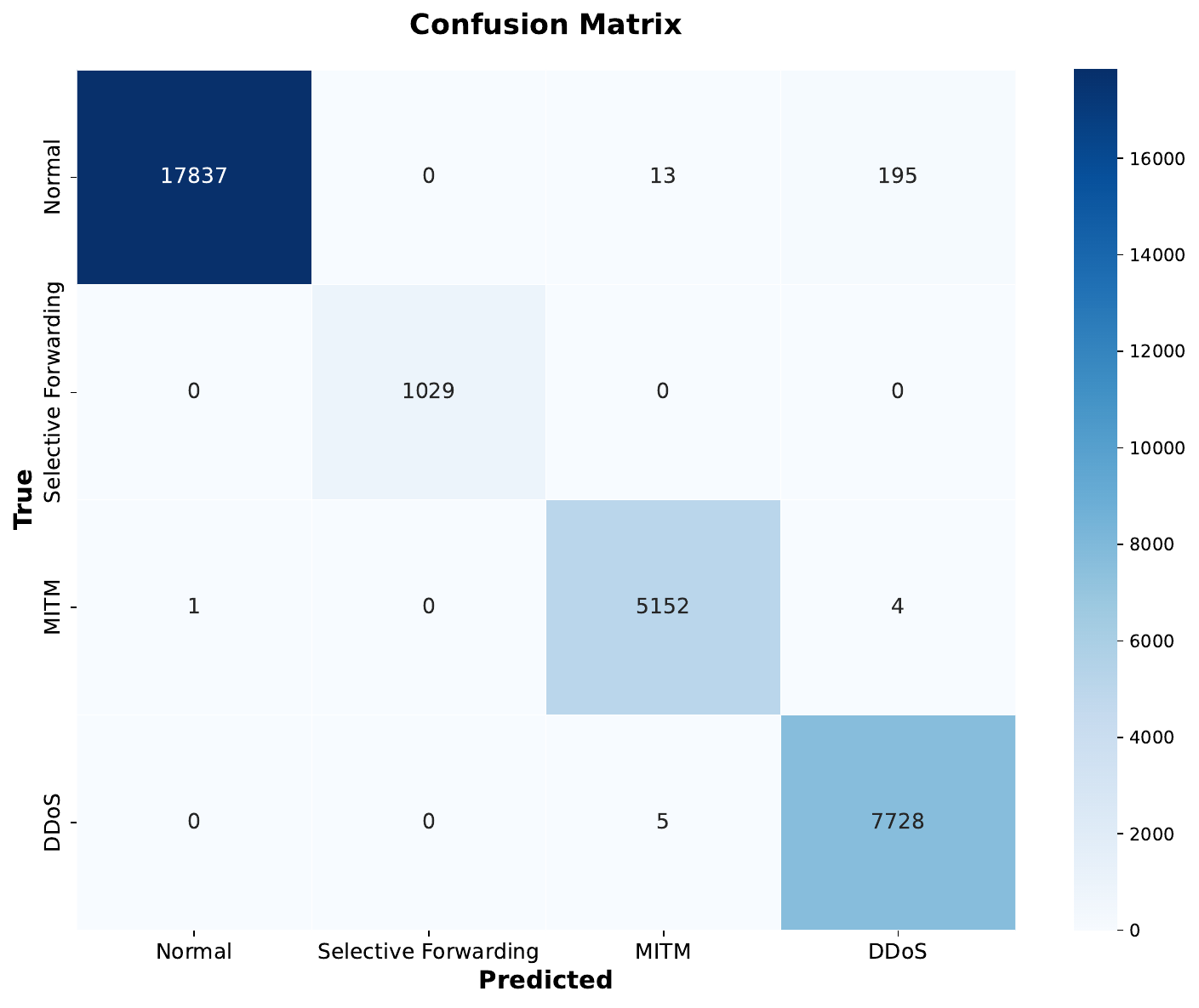}
   \caption[Confusion matrix of the proposed Res-TCN model]{Confusion matrix of the proposed Res-\gls{tcn} model for multiclass classification of \gls{hiot} traffic on the proposed dataset \emph{UL-ECE-MultiAttack-H-IoT2025}, showing class-wise prediction performance across Normal, \acrlong{sf}, \gls{mitm}, and \gls{ddos} categories.}
    \label{fig:cm_plot}
\end{figure}

\subsection{Confusion Matrix in Multiclass Classification}
The confusion matrix offers class-wise insight into the model’s prediction behaviour. It helps complement aggregate evaluation metrics by highlighting inter-class distinctions. The confusion matrix of the proposed Res-\gls{tcn} model illustrates the classification performance across four classes: Normal, Selective Forwarding, \gls{mitm}, and \gls{ddos} on the proposed dataset \emph{UL-ECE-MultiAttack-H-IoT2025}, as depicted in Figure~\ref{fig:cm_plot}.

The model effectively distinguishes between these classes, with the majority of samples correctly classified in each case and minimal confusion across classes. Notably, the \acrlong{sf} class shows no overlap with other categories, indicating high precision. Both \gls{mitm} and \gls{ddos} are also identified with consistent accuracy, while only a small number of Normal samples are misclassified. The patterns observed in the confusion matrix align with the model’s high class-wise precision and recall. This outcome demonstrates the model’s effectiveness in detecting multiple types of attacks in \gls{hiot} environments.

\begin{figure}[ht!]
    \centering   
    \includegraphics[width=0.9\columnwidth]{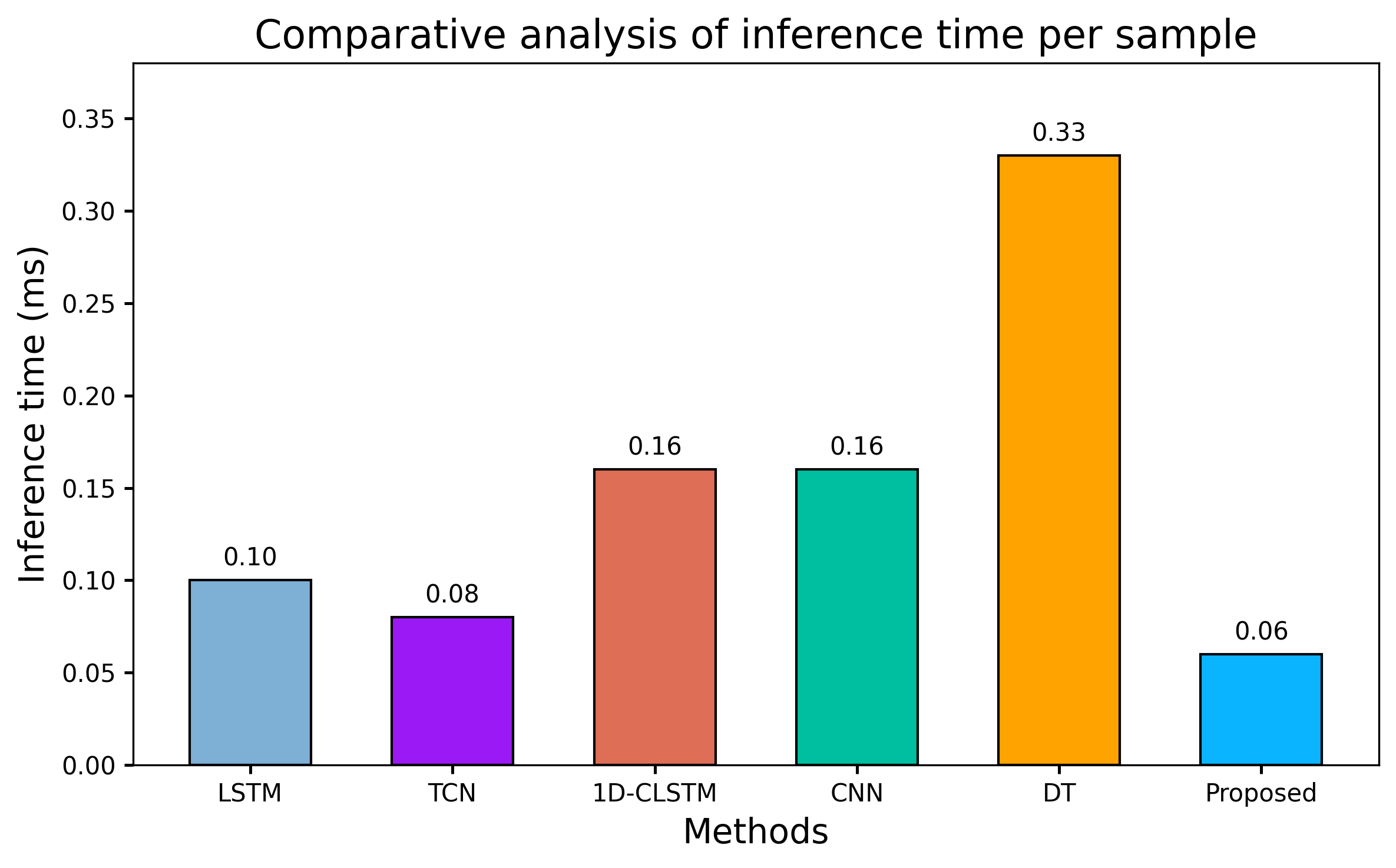}
    \caption[Comparative analysis of inference time per sample]{Comparative analysis of inference time per sample (in milliseconds) across different methods, highlighting the efficiency of the proposed model. The XGBoost model is excluded from this comparison due to its higher inference time (0.35 ms).}
    \label{fig:inferencetime}
\end{figure}

\subsection{Inference Time Comparison for \gls{hiot} Applications}
\label{subsec:inference_time}
This research presents an analysis of inference time per sample across several previously published models, including \gls{lstm}~\cite{akar2025l2d2}, \gls{tcn}, \gls{1dclstm}~\cite{kumar2025hybrid}, \gls{cnn}~\cite{10628718}, and \gls{dt}~\cite{binbusayyis2022investigation}. All models are trained using the proposed dataset, {\em UL-ECE-MultiAttack-H-IoT2025}, to ensure a consistent evaluation. The proposed Res-\gls{tcn} model achieves the lowest inference time per sample, outperforming all baseline methods. This highlights its potential suitability for real-time applications in low-power \gls{hiot} devices, as illustrated in Figure~\ref{fig:inferencetime}. Among the compared models, XGBoost~\cite{10.1007/978-3-031-50583-6_11} exhibits the highest inference time (0.35 ms); it is excluded from Figure \ref{fig:inferencetime} to prevent scale distortion and maintain a meaningful comparison. This comparison is particularly effective for multiclass attack classification, where both speed and efficiency are critical. All inference times are measured using the Google Colab platform to ensure a consistent and fair execution environment across all models.

The inference time per sample is computed by measuring the total prediction time for the entire test set and dividing it by the number of test samples, as shown below:
\begin{equation}
\text{Per-sample Inf. Time (ms)} = \frac{(T_{\text{end}} - T_{\text{start}}) \times 1000}{N}
\end{equation}
where \( T_{\text{start}} \) and \( T_{\text{end}} \) represent the start and end timestamps (in seconds) during prediction, and \( N \) is the total number of test samples. The factor of 1000 is used to convert the inference time from seconds to milliseconds, since \(1\,\text{second} = 1000\,\text{milliseconds}\). Milliseconds provide a finer temporal resolution, which is more appropriate for evaluating low-latency models. This calculation follows a standard approach commonly used in runtime evaluation of \gls{ml} models \cite{markaicode_inference}.

\subsection{Res-\gls{tcn} Performance on Public \gls{hiot} Datasets}
The proposed Res-\gls{tcn} model is evaluated on publicly available datasets to assess its performance and generalizability in \gls{hiot} environments. Three recent datasets, CICIoMT2024, IoMT-TrafficData, and CICIDS2023, are employed for multiclass classification. A separate binary classification dataset is also derived from CICIoMT2024 by extracting \gls{icmp}-based \gls{ddos} attacks and benign traffic. The multiclass and binary versions of the CICIoMT2024 dataset are referred to as CICIoMT2024-Multiclass (MC) and CICIoMT2024-Binary (BI), respectively. These datasets provide multiple testbeds for evaluating the model’s performance and generalizability in both multiclass and binary settings involving protocol-centric threats.

The CICIoMT2024-MC dataset comprises traffic samples categorised into BENIGN, \gls{ddos}-\gls{tcp}, \gls{ddos}-\gls{icmp}, and \gls{arp} spoofing classes, selected from the well-known CICIoMT2024 dataset~\cite{dadkhah2024ciciomt2024}. This dataset is chosen for its relevance to protocol-centric multiclass classification in \gls{hiot} environments.

\subsubsection{CICIoMT2024-MC Dataset}
\label{subsubsec:CICIoMT2024MC} The CICIoMT2024-MC dataset, derived from the CICIoMT2024 dataset~\cite{dadkhah2024ciciomt2024}, is chosen for its relevance to protocol-centric multiclass classification. This dataset comprises traffic generated by \gls{tcp}, \gls{icmp}, and \gls{arp} communication, specifically selected to assess the proposed Res-\gls{tcn} model’s performance in distinguishing protocol-specific attacks in \gls{hiot} environments~\cite{cic_iomt2024}. The dataset includes traffic representing BENIGN, \gls{ddos}-\gls{tcp}, \gls{ddos}-\gls{icmp}, and \gls{arp} spoofing scenarios. Each traffic type is assigned a custom label to form a balanced multiclass dataset suitable for \gls{hiot} environments. Among these, the volumetric (\gls{ddos}) and \gls{arp} spoofing (\gls{mitm}) classes are particularly aligned with the attack types considered in the proposed \gls{udp}-based dataset. Thus, this evaluation incorporates additional protocol-specific traffic, providing a broader context for assessing the Res-\gls{tcn} model’s performance in \gls{hiot} environments.

\begin{table}[!htbp]
\centering
\caption{Feature categories in the CICIoMT2024 dataset.}
\label{tab:ciciomt_features}
\small 
\begin{tabular}{llp{9cm}}
\toprule
\textbf{Category} & \textbf{Count} & \textbf{Description} \\
\midrule
Protocol Features     & 7  & Indicates the protocol type used for each packet, including \gls{tcp}, \gls{udp}, \gls{dhcp}, \gls{arp}, \gls{icmp}, IGMP, and LLC. \\
Flag Features         & 11 & Contains binary indicators for \gls{tcp} control flags such as SYN, ACK, FIN, and related counts. \\
Header Features       & 2  & Captures structural information from packet headers, e.g., header length and protocol type. \\
Statistical Features  & 10 & Aggregated statistics such as total sum, average, standard deviation, radius, and covariance. \\
Time-based Features   & 5  & Time-related metrics like flow duration, source rate (Srate), destination rate (Drate), and inter-arrival time (IAT). \\
Service Features      & 7  & Indicates use of application-layer protocols such as HTTP, DNS, Telnet, SMTP, SSH, IRC, and HTTPS. \\
Other Features        & 3  & Miscellaneous attributes, including \gls{ip} version, total packet size, and magnitude-based metrics. \\
\textbf{Total Features} & \textbf{45} & \textbf{Sum of all features across categories} \\
\bottomrule
\end{tabular}
\end{table}

Several preprocessing steps are applied to prepare the protocol-centric multiclass dataset for Res-\gls{tcn} classification. Label encoding is initially performed using a custom mapping to convert categorical labels into numerical form. The dataset is then separated into two parts: the information used for making predictions (e.g., protocol type, packet size, and flow duration), and the actual category of each traffic record (e.g., benign or \gls{ddos}). This separation allows the model to learn patterns in the input data that correspond to each attack type. A detailed summary of the feature categories used for prediction in the CICIoMT2024-MC dataset is provided in Table~\ref{tab:ciciomt_features}. Feature values are normalised using MinMax scaling to ensure consistency across input variables. To address class imbalance, SMOTE is applied to upsample the minority classes in the training set to match the size of the majority class. Additionally, Gaussian noise is added to both training and test sets to promote generalisation and reduce the risk of overfitting. Finally, the feature data is reshaped to match the input format required by the Res-\gls{tcn} architecture. 

\begin{table*}[ht!]
\centering
\caption[Classification performance of the Res-TCN model]{Classification performance of the Res-\gls{tcn} model on \gls{hiot}-relevant public and proposed datasets.}
\label{tab:publicdatasetclassification}
\renewcommand{\arraystretch}{1.2} 
\resizebox{0.9\textwidth}{!}{%
\begin{tabular}{@{}l@{\hskip 10pt}l@{\hskip 10pt}l@{\hskip 10pt}l@{\hskip 10pt}l@{\hskip 10pt}l@{\hskip 10pt}l@{}}
\toprule
\textbf{Dataset} & \textbf{Acc.} & \textbf{Prec.} & \textbf{Rec.} & \textbf{F1} & \textbf{Inf. Time} & \textbf{Class Type} \\
\midrule
CICIoMT2024-MC & 98.82 & 99.00 & 99.00 & 99.00 & 0.06 & Multiclass \\

CICIoMT2024-BI & 99.90 & 100 & 100 & 100 & 0.06 & Binary \\

IoMT\_Flows\_Multiclass & 96.72 & 97.46 & 96.72 & 96.79 & 0.07 & Multiclass \\

CICIoT2023-MC & 74.91 & 75.16 & 74.91 & 70.96 & 0.07 & Multiclass \\

{\em UL-ECE-MultiAttack-H-IoT2025} & \textbf{99.32} & \textbf{99.00} & \textbf{99.0} & \textbf{99.0} & \textbf{0.06} & \textbf{Multiclass} \\
\bottomrule
\end{tabular}%
}
\begin{minipage}{0.9\linewidth}
\vspace{3pt}
\scriptsize 
\textbf{Note:} All metrics are expressed as percentages. \textit{Inf. Time} refers to the average inference time per sample in milliseconds. The proposed Res-\gls{tcn} model achieves 99.32\% accuracy on the internally generated multiclass \gls{hiot} dataset, {\em \textbf{UL-ECE-MultiAttack-H-IoT2025}}. Bold values indicate the performance of the Res-\gls{tcn} model evaluated on the proposed dataset.
\end{minipage}
\end{table*}

The proposed Res-\gls{tcn} model achieves 98.82\% accuracy on the CICIoMT2024-MC dataset, as shown in Table~\ref{tab:publicdatasetclassification}. Moreover, it outperforms the previously published \gls{lstm}-based model~\cite{akar2025l2d2}, which reports 98\% accuracy on the same dataset.

\rev{Due to differences in datasets, including protocol types (e.g., \gls{udp}), class distributions (binary vs. multiclass), and evaluation settings, direct one-to-one comparison with baseline models is limited; however, the results still provide useful insights into model performance across different \gls{hiot} scenarios.} \rev{The results demonstrate that the proposed \gls{restcn} model performs consistently well, particularly in multiclass scenarios, indicating its effectiveness in handling complex \gls{hiot} traffic.} \rev{Additionally, baseline models evaluated on the proposed dataset (e.g., \gls{dt}) show reduced performance, further supporting the advantage of the proposed approach.}

\subsubsection{CICIoMT2024-BI Dataset}
The binary classification evaluation utilises the CICIoMT2024-BI dataset, which consists of \gls{icmp}-based \gls{ddos} attacks and benign traffic. Features with constant values, such as a column containing only zeros or a single repeated value across all instances, are removed, and the remaining inputs are normalised using MinMax scaling. The dataset is then split into training and testing sets. To address class imbalance, SMOTE is applied to the training set. Gaussian noise is added to both sets to improve generalisation and reduce overfitting. The data is reshaped to fit the 3D input format of the Res-\gls{tcn} model. The model is trained using a sigmoid activation function in the output layer and binary cross-entropy loss.

The proposed model achieves 99.90\% accuracy on the CICIoMT2024-BI dataset, as illustrated in Table~\ref{tab:publicdatasetclassification}. This high accuracy reflects the model's ability to capture the distinct patterns associated with \gls{icmp}-based \gls{ddos} traffic.

\subsubsection{IoMT-TrafficData (IoMT\_Flows\_Multiclass)}
This research utilises the publicly available IoMT-TrafficData multiclass dataset from Zenodo~\cite{areia2023iomttrafficdata} to evaluate the proposed model’s generalizability. Multiple single-attack CSV files are merged into a unified dataset (IoMT\_Flows\_Multiclass). This flow-based dataset represents aggregated communication behaviours between \gls{iomt} devices and includes diverse attack types alongside normal traffic, providing a comprehensive benchmark for model evaluation. The combined dataset contains 718{,}060 flow samples across nine categories, covering both benign and malicious traffic. The majority classes (e.g., CAM, NETSCAN, NORMAL, and RUDY) each contain approximately 120{,}000 samples, while the minority classes, such as \gls{arp} and SLOWREAD, include approximately 11{,}000 and 9{,}000 samples, respectively. This moderate imbalance is later mitigated during training using SMOTE to enhance class representation consistency.

The ``IoMT\_Flows\_Multiclass'' dataset is preprocessed to ensure consistency and suitability for model training. The preprocessing retains numerical features while converting categorical target labels into integer indices. All non-finite values are sanitized by replacing \(\{\pm\infty, \text{NaN}\}\) with 0. A stratified split is applied to divide the data into train and test sets (80--20), followed by drawing a validation set from the training portion (10\% of train), yielding an effective \(\approx 72\%\) train, \(8\%\) validation, and \(20\%\) test partition. This three-way split facilitates hyperparameter tuning and validation monitoring while keeping the test set completely unseen for unbiased evaluation. A MinMaxScaler is fitted on the training features only and applied to the validation/test sets to avoid leakage. To address class imbalance, SMOTE is applied to the training split (triggered when the max/min class ratio exceeds \(3\times\)), with \(k_{\text{neighbours}} = \min(5, n_{\min}-1)\). Features are reshaped to \((F,1)\) for temporal convolution, and the preprocessed data are then used to train the proposed Res-\gls{tcn} model.

To mitigate overfitting, the model corresponding to the epoch with the lowest validation loss is retained (early stopping), and a minimal train--validation accuracy gap is confirmed. To guard against data leakage, the absence of exact or near-duplicate overlaps across splits is ensured using row hashing rounded to six decimal places. Single-feature proxy labels are eliminated through depth-$1$ decision stumps, mutual-information scores are verified to remain within a reasonable range, and all preprocessing steps are fitted exclusively on the training data. Low train--test separability is further verified via adversarial validation, using logistic-regression AUC $< 0.80$ as a practical upper bound to ensure that splits are not easily distinguishable while avoiding overly strict thresholds that may flag natural variance as leakage. Under this pipeline, the held-out test set achieves 96.72\% accuracy with weighted precision/recall/F1 of 97.46/96.72/96.79, respectively, as outlined in Table~\ref{tab:publicdatasetclassification}.

\subsubsection{CICIoT2023-MC Dataset}
The proposed Res-\gls{tcn} model is evaluated on the severely imbalanced CICIoT2023 dataset in a multiclass classification setup. The data is collected from 105 general \gls{iot} devices, including home appliances (e.g., indoor \gls{ip} cameras) \cite{CICIoT2023}. This dataset contains 712311 samples and 39 features. All available features, including protocol flags and packet-level statistics (e.g., rate, total size, variance), are retained. Session-level indicators (e.g., average value, standard deviation, number of packets) are also included to assess the model’s generalisation ability across a diverse feature space. It also contains 34 distinct classes, including benign, \gls{ddos}, \gls{dos}, scanning-related, \gls{mitm}, and web-level attacks. This research selects 15 classes that align more closely with \gls{hiot} attack patterns, such as flooding-based \gls{ddos}/\gls{dos} and \gls{mitm}-\gls{arp} spoofing. The selected classes and the number of samples for each are listed in Table~\ref{tab:ciciot_class}.  This selected subset of attacks is more representative of \gls{hiot} environments. 

\begin{table}[ht!]
\centering
\caption[Class distribution of the CICIoT2023]{Class distribution of the CICIoT2023 multiclass \gls{hiot}-relevant dataset (after train-test split, before SMOTE).}
\label{tab:ciciot_class}
\small
\begin{adjustbox}{max width=\linewidth}
\begin{tabular}{@{}l l l@{}}
\toprule
\textbf{Class ID} & \textbf{Class Name} & \textbf{Samples} \\
\midrule
0  & BENIGN                        & 11602  \\
1  & DDOS-ACK\_FRAGMENTATION       & 3015   \\
\textbf{2}  & \textbf{DDOS-ICMP\_FLOOD}              & \textbf{76063}  \\
3  & DDOS-ICMP\_FRAGMENTATION      & 4748   \\
4  & DDOS-PSHACK\_FLOOD            & 43519  \\
5  & DDOS-RSTFINFLOOD              & 43156  \\
6  & DDOS-SYNONYMOUSIP\_FLOOD      & 38324  \\
7  & DDOS-SYN\_FLOOD               & 43020  \\
8  & DDOS-TCP\_FLOOD               & 47801  \\
9  & DDOS-UDP\_FLOOD               & 57408  \\
10 & DDOS-UDP\_FRAGMENTATION       & 2985   \\
11 & DOS-SYN\_FLOOD                & 21434  \\
12 & DOS-TCP\_FLOOD                & 28274  \\
13 & DOS-UDP\_FLOOD                & 35259  \\
14 & MITM-ARPSPOOFING              & 3213   \\
\bottomrule
\end{tabular}%
\end{adjustbox}
\begin{minipage}{0.97\linewidth}
\vspace{3pt}
\scriptsize
\textbf{Note:} The SMOTE oversampling strategy uses the majority class (\textbf{DDOS-ICMP\_FLOOD, 76063 samples}), shown in bold, as the target sample count to balance all minority classes.
\end{minipage}
\end{table}

To prepare the data for training, several preprocessing steps are applied. First, non-numeric and invalid values (e.g., NaN, Inf) are removed. Label encoding is performed on the class labels. This transformation converts categorical labels, such as BENIGN, DDOS-SYN\_FLOOD, etc., into numerical values (e.g., 0, 1, 2, \dots), where each unique class is assigned a distinct integer. This allows the model to process labels efficiently while preserving their semantic meaning. The dataset is then split into training and testing subsets using stratified sampling to maintain the class distribution. A variance threshold filter is applied to remove features that show very little variation across all samples, such as flags that remain nearly constant (e.g., cwr\_flag\_number or ece\_flag\_number), which often take the value 0 in most records. These low-variance features are unlikely to help the model distinguish between different attack types and may introduce unnecessary noise. After this step, min-max normalisation is applied to scale all feature values to the range \([0,1]\), ensuring that no feature dominates due to its scale and enhancing the stability of the training process. To address class imbalance, SMOTE is used to oversample all minority classes up to the maximum training class size (76063 samples).

Moreover, this research evaluates the proposed Res-\gls{tcn} architecture on multiclass \gls{hiot}-relevant datasets (e.g., CICIoMT-MC and CICIoT2023-MC) without applying hyperparameter tuning, such as adjusting dropout rate, learning rate, or regularisation. This decision preserves the integrity of the original model design. Nevertheless, the model achieves a test accuracy of 74.04\%, providing a reliable baseline for comparison. Other evaluation metrics, including precision, recall, and F1-score, are presented in Table~\ref{tab:publicdatasetclassification}. Further fine-tuning of hyperparameters may lead to improved classification outcomes in future work. Therefore, using this filtered subset of classes with the full feature set results in a slightly enhanced accuracy of 74.91\%, compared to 74.04\% with the reduced feature set, as outlined in Table~\ref{tab:publicdatasetclassification}.

\subsubsection{External Validation on IoMT-TrafficData}
The IoMT-TrafficData test subset provides binary labels, differing from the multiclass CICIoT2023 setup and allowing cross-domain (e.g., \gls{iot} and \gls{iomt}) generalisation assessment. To evaluate the cross-dataset generalizability of the proposed Res-\gls{tcn} model, an external validation is conducted using the IoMT-TrafficData ``\gls{ip}-Based Flows Pre-Processed'' subset. In contrast to the multiclass CICIoT2023 benchmark, containing 15 attack categories and benign traffic, the IoMT-TrafficData dataset contains only two classes: BENIGN and ATTACK. It captures both packet and flow characteristics from healthcare-oriented \gls{iot} devices. The CICIoT2023 dataset includes diverse traffic-level metrics, such as packet rates, total sizes, protocol flags, and session-level aggregates (e.g., averages and variances), which are primarily derived from simulated \gls{iot} communication. In contrast, IoMT-TrafficData focuses on healthcare-specific network flows that emphasise directional traffic behaviour, including packet counts, byte ratios, payload volume, and flow asymmetry metrics. The ``\gls{ip}-Based Flows Pre-Processed Test'' subset, used for external validation, contains 146{,}436 samples and 32 features.

The Res-\gls{tcn} model is first trained on CICIoT2023-MC to distinguish among multiple \gls{iot} attack types within the same distribution. It is then directly evaluated on the ``\gls{ip}-Based Flows Pre-Processed'' dataset without retraining to assess its ability to generalise across datasets that differ in protocols, traffic behaviours, and device contexts. The same trained model parameters and preprocessing pipeline (feature selector, scaler, and input shape) are applied to the \gls{iomt} dataset. All samples are reshaped to match the model input dimension, and predictions are obtained through the softmax probabilities of the trained multiclass model. As IoMT-TrafficData provides binary labels, the multiclass predictions are mapped into BENIGN and ATTACK categories. To mitigate confidence bias caused by domain shift, a confidence-based mapping is applied during external validation. This mapping evaluates the predicted probability of the BENIGN class, and samples with BENIGN confidence below a defined threshold (e.g., 0.7) are classified as ATTACK. Such threshold-based calibration has been widely adopted in cross-domain intrusion detection to enhance robustness when models are applied to unseen environments~\cite{zhang2025heterogeneous, ma2023adversarial}.

The external validation yields an accuracy of approximately 51.78\%, with a precision of 51.78\%, a recall of 100\%, and an F1-score of 65.58\%. The perfect recall indicates that all attack samples are correctly detected. However, the moderate precision reflects a higher false-positive rate under domain shift between general \gls{iot} and \gls{iomt} datasets rather than a model limitation \cite{zhang2025heterogeneous, ma2023adversarial}. The noticeable drop in cross-dataset accuracy empirically demonstrates that \gls{iomt} traffic exhibits distinct distributional and behavioural patterns, such as packet counts, byte ratios, and payload volume, compared to the broader \gls{iot} traffic characteristics observed in CICIoT2023 (e.g., packet rates, total sizes, and protocol flag distributions). While this result highlights the existing domain gap, it can be addressed through feature alignment or fine-tuning strategies to adapt the model more effectively to IoMT-specific data distributions, as discussed in Section~\ref{subsec:discussion}. The reduced accuracy under this external validation is expected due to the domain shift between general \gls{iot} and healthcare-specific \gls{iomt} traffic, illustrating real-world generalisation challenges.

\begin{table}[!htbp]
\centering
\caption[Per-class performance of the Res-TCN model]{Per-class performance of the Res-\gls{tcn} model on the \textbf{CICIoMT2024} dataset.}
\label{tab:ciciomt_metrics}
\small
\begin{tabular}{llllll}
\toprule
\multicolumn{1}{l}{\textbf{Class}} & 
\multicolumn{1}{l}{\textbf{Accuracy}} & 
\multicolumn{1}{l}{\textbf{Precision}} & 
\multicolumn{1}{l}{\textbf{Recall}} & 
\multicolumn{1}{l}{\textbf{F1-Score}} & 
\multicolumn{1}{l}{\textbf{Support}} \\
\midrule
BENIGN         & 80.00 & 98.00 & 80.00 & 88.00 & 11282 \\
\gls{ddos}-\gls{icmp}      & 100.00 & 100.00 & 100.00 & 100.00 & 58482 \\
\gls{ddos}-\gls{tcp}       & 100.00 & 100.00 & 100.00 & 100.00 & 54780 \\
\gls{arp} spoofing   & 75.00 & 16.00 & 75.00 & 27.00 & 523 \\
\midrule
\textbf{Overall Multiclass} & 98.82 & 99.00 & 99.00 & 99.00 & -- \\

\bottomrule
\end{tabular}

\begin{minipage}{\textwidth}
\vspace{3pt}
\scriptsize
\textbf{Note:} This table presents the per-class performance of the proposed Res-\gls{tcn} model on the CICIoMT2024 dataset. The final row provides the overall multiclass classification results. All evaluation metrics are expressed as percentages (\%)
\end{minipage}
\end{table}

\subsection{Protocol Impact on Res-\gls{tcn} Performance in \gls{hiot}}
This research evaluates protocol-specific classification performance using two multiclass \gls{hiot} datasets: the publicly available CICIoMT2024 and a \gls{udp}-based proposed dataset, {\em UL-ECE-MultiAttack-H-IoT2025}. The proposed Res-\gls{tcn} model achieves an overall accuracy of 98.82\% on CICIoMT2024 and 99.32\% on the proposed {\em UL-ECE-MultiAttack-H-IoT2025} multiclass dataset, as outlined in Table~\ref{tab:publicdatasetclassification}. However, this research further examines class-wise metrics in both datasets to identify potential causes of performance variation, such as protocol characteristics, data distribution, or feature relevance. Detailed information on the class labels and features used in the CICIoMT2024 dataset is provided in Section~\ref{subsubsec:CICIoMT2024MC}. The model achieves 100\% accuracy, precision, and recall for \gls{ddos}-\gls{tcp} and \gls{ddos}-\gls{icmp} classes on the CICIoMT2024 dataset, indicating that attack patterns significantly influence detection performance, as shown in Table~\ref{tab:ciciomt_metrics}. Similarly, the proposed {\em UL-ECE-MultiAttack-H-IoT2025} dataset demonstrates high detection accuracy for \gls{udp}-based \gls{ddos} classes, as shown in Table~\ref{tab:udp_metrics}. While the nature of the attack (e.g., \gls{ddos} flooding) often shapes traffic behaviour, such as high packet rates, protocol-specific communication patterns can also impact detection performance. A key distinction is whether a protocol is connection-oriented (e.g., \gls{tcp} and \gls{mqtt}~\cite{permana2024osi}) or connectionless (e.g., \gls{udp} and \gls{icmp}~\cite{li2025stateshield}). These structural differences determine whether the protocol keeps a session active and how it manages starting and ending communication between endpoints. Despite these protocol differences, \gls{ddos} flooding attacks across \gls{tcp}, \gls{udp}, and \gls{icmp} demonstrate similar behavioural patterns, including repetitive messaging, high traffic volumes, and resource exhaustion~\cite{suprijanto2024new, yu_ddos_types}. Therefore, it is not the protocol alone; instead, how it is exploited during a \gls{ddos} flooding attack influences its traffic behaviour. The temporal convolutional architecture effectively learns these patterns, enabling more accurate detection.

\begin{table}[!htbp]
\centering
\caption[Per-class performance of the Res-\gls{tcn} model on the proposed]{Per-class performance of the Res-\gls{tcn} model on the proposed {\em \textbf{UL-ECE-MultiAttack-H-IoT2025}} dataset.}
\label{tab:udp_metrics}
\begin{tabular}{llllll}
\toprule
\multicolumn{1}{l}{\textbf{Class}} & 
\multicolumn{1}{l}{\textbf{Accuracy}} & 
\multicolumn{1}{l}{\textbf{Precision}} & 
\multicolumn{1}{l}{\textbf{Recall}} & 
\multicolumn{1}{l}{\textbf{F1-Score}} & 
\multicolumn{1}{l}{\textbf{Support}} \\
\midrule
Normal               & 99.50  & 1.000 & 0.970 & 0.985 & 18045 \\
\gls{sf} & 100.00 & 1.000 & 1.000 & 1.000 & 1029 \\
\gls{mitm}                 & 99.90  & 0.999 & 0.996 & 0.998 & 5157 \\
\gls{ddos}                 & 99.96  & 0.932 & 1.000 & 0.965 & 7733 \\
\midrule
\textbf{Overall Multiclass} & 99.32 & 99.00 & 99.00 & 99.00 & --\\

\bottomrule
\end{tabular}
\begin{minipage} {\textwidth}
\vspace{3pt}
\scriptsize
\textbf{Note:} All metric values (accuracy, precision, recall, F1-score) are shown as percentages. This entry highlights the classwise performance of the proposed Res-\gls{tcn} model on the internally generated {\em \textbf{UL-ECE-MultiAttack-H-IoT2025}} dataset. The last row reports the overall multiclass classification metrics. All evaluation metrics are expressed as percentages (\%).
\end{minipage}
\end{table}

In contrast, the model achieves only 75\% accuracy on the \gls{arp} spoofing class, with low precision (0.16) and F1-score (0.27), as shown in Table~\ref{tab:ciciomt_metrics}. Unlike flooding-based attacks, which generate easily distinguishable traffic patterns, \gls{arp} spoofing attacks operate at the data link layer (Layer 2 of the OSI model)~\cite{bruschi2024ensuring}, which is responsible for node-to-node communication within a local network. These attacks involve sending falsified \gls{arp} responses to associate a malicious MAC address with the \gls{ip} address of a legitimate device, enabling unauthorised interception or manipulation of network traffic. Since the \gls{arp} protocol lacks built-in authentication, spoofed packets often appear as legitimate messages. This similarity may lead to higher false positives, making \gls{arp} spoofing challenging to distinguish from benign traffic using \gls{ml} and \gls{dl} models. Previous studies have also highlighted the challenges in detecting \gls{arp}-based attacks due to the similarity between malicious and benign \gls{arp} traffic~\cite{majumder2025arp}. Despite these challenges, the proposed model achieves 99.90\% accuracy on the \gls{mitm} class over the UDP-based dataset in \gls{hiot} environments, as outlined in Table~\ref{tab:udp_metrics}. This suggests that with effective data preprocessing and relevant feature extraction, \gls{ml}/\gls{dl} models can still perform well even on subtle attack types such as \gls{arp}-based \gls{mitm}.

Moreover, the variation in classification performance across these protocols indicates that accuracy is not solely determined by model architecture or training method. Instead, it depends on several interrelated aspects, such as protocol-aware features~\cite{tafreshian2024defensive}, input data quality~\cite{ogbu2024conceptual}, and traffic-level distinguishability~\cite{wu2024bot}.

\subsection{Discussion on Applicability and Usability in \gls{hiot}} \label{subsec:discussion}
This section discusses the applicability and usability of the proposed Res-\gls{tcn} method in \gls{hiot}. It highlights how the method and the underlying simulation setup address security challenges and operational constraints specific to real-world precision-healthcare environments. As described in Section~\ref{subsec:simulation}, simulated application-layer attack scenarios emulate \gls{wban} \gls{hiot} sensor communication and encompass three critical attacks (\gls{sf}, \gls{mitm}, and \gls{ddos}). The simulation reproduces typical \gls{hiot} traffic using the configured transmission intervals (normal vs. malicious node frequencies), along with the exchange of health metrics between simulated nodes. Accordingly, the {\em UL-ECE-MultiAttack-H-IoT2025} dataset is used to train and evaluate the proposed model under these simulated \gls{hiot} settings. Wearable devices in such environments face two key constraints: low latency and minimal memory consumption \cite{garg2025energy}. The Res-\gls{tcn} model addresses these constraints, enabling efficient processing of sensor data while maintaining reliable performance for latency-sensitive clinical monitoring.

In addition, the simulation includes attacks that maintain regular transmission intervals, particularly \gls{sf} and \gls{mitm}, which drop or alter packets; the proposed Res-\gls{tcn} model classifies these behaviours. This suggests that the proposed model has the capacity to generalise across different attack patterns and transmission behaviours in \gls{hiot} settings. In this research, the external validation demonstrates that the domain shift between general \gls{iot} and \gls{iomt} traffic leads to degraded overall performance, despite perfect recall. However, this chapter presents pre-deployment research; the proposed model can be post-deployment fine-tuned on additional emerging attack patterns beyond \gls{sf}, \gls{mitm}, and \gls{ddos} by updating the final classification layer while retaining previously learned knowledge (i.e., applying a transfer learning approach) \cite{wu2024deep}. This approach minimises the requirement for retraining the entire model on edge devices, thereby mitigating the computational overhead inherent in \gls{hiot} deployments. Nevertheless, deploying the proposed Res-\gls{tcn} model across diverse \gls{hiot} infrastructures requires rigorous evaluation of device heterogeneity, supported protocols, secure model updates (e.g., authenticated Over-the-Air (OTA) updates), and regulatory requirements (e.g., \gls{gdpr}) to ensure compliance.

\section{Conclusion}
\label{sec:conclusion}
This research proposes a novel \gls{dl} method, \gls{restcn}, for classifying multiple attacks on the proposed dataset {\em UL-ECE-MultiAttack-H-IoT2025}, in \gls{hiot} environments. The model achieves consistent accuracy (99.32\%) in detecting three concurrent attacks: \gls{sf}, \gls{mitm}, and \gls{ddos}, which transmit data simultaneously within the same \gls{hiot} topology. It also demonstrates low inference latency, which is essential for real-time healthcare applications.

However, the proposed dataset focuses on \gls{hiot} scenarios, which limits the scope and may affect how well the model generalises to other \gls{iot} environments. Therefore, additional testing across varied \gls{iot} settings, such as different devices, protocols, and network conditions, including latency, jitter, and packet loss, is required to validate the model’s robustness and generalisation beyond \gls{hiot}.

Adversarial robustness is not addressed in this research; future work involves evaluating the model on evasive or modified attack traffic (e.g., malicious nodes imitating normal behaviour or novel attack variants in the Cooja simulator) to assess the model's resilience to intentionally manipulated data. This also includes deploying the proposed \gls{restcn} model on diverse edge hardware platforms (e.g., Raspberry Pi variants and microcontrollers) and systematically measuring its performance (e.g., latency and energy usage) to compare these metrics across edge devices. Such real-time attack detection on resource-constrained platforms could help safeguard sensitive patient data and critical medical services in \gls{hiot} environments.



\chapter{Conclusions} 


\ifpdf
    \graphicspath{{7_conclusions/figures/PNG/}{7_conclusions/figures/PDF/}{7_conclusions/figures/}}
\else
    \graphicspath{{7_conclusions/figures/EPS/}{7_conclusions/figures/}}
\fi


This thesis addresses one overarching research question supported by five interconnected RQs focused on securing \gls{hiot} systems.

\textbf{Main RQ:} How can differences in data transmission behaviour between normal and malicious nodes support the development of realistic datasets and lightweight deep learning models for detecting and mitigating emerging cyber threats in \gls{hiot} environments?

Firstly, it investigates \gls{hiot} traffic transmission behaviours, including protocol-specific data patterns, which are essential for creating realistic datasets.
Secondly, it examines how these datasets enhance detection performance in \gls{dl}-based models by developing a \gls{tcn} for binary classification in \gls{hiot}.
It then evaluates the lightweight \gls{tcn} model on Raspberry~Pi~4 to demonstrate its feasibility for edge deployment. Finally, it proposes a \gls{restcn}-based solution for robust multiclass detection, covering \gls{mitm}, \gls{sf}, and \gls{ddos} attacks in \gls{hiot} environments. The key outcomes of this research are summarised with respect to the overarching RQ and its supporting research questions:

\textbf{RQ1: \rev{What key challenges and research gaps are identified in existing literature on \gls{ml}/\gls{dl}-based cyber attack detection in \gls{hiot} systems, and how can a conceptual deep learning framework address these limitations?}}

Chapter~\ref{chap:review} of this thesis provides a comprehensive review of existing approaches for securing \gls{hiot} systems, identifying several key research gaps, including the lack of realistic simulations of inter-device communication activities, the scarcity of domain-specific datasets for training and validation, and the need for resilient anomaly detection and mitigation mechanisms. With respect to \textbf{ RQ1}, the review also introduces a \gls{cnn}-oriented approach to strengthen \gls{dl} applications for secure and efficient \gls{hiot} systems. A simplified \gls{cnn} implementation based on this concept uses environmental sensor data (e.g., humidity) to demonstrate its practical value in \gls{hiot}~\cite{khatun2024time}.

\textbf{RQ2: How can realistic synthetic datasets be generated to represent \gls{ddos} attack behaviours in \gls{hiot} environments?} 

Chapter~\ref{chap:dataset} of this thesis addresses this research question by developing two realistic \gls{ddos} datasets that reflect the communication patterns of \gls{hiot} devices. The first dataset is generated using the Cooja simulator over the \gls{mqtt} protocol, while the second is created using the ns-3 simulator over the \gls{udp} protocol. Both datasets include normal and attack traffic, capturing realistic sensor communication behaviour observed in \gls{hiot} environments. The research also defines a systematic data preprocessing pipeline and feature selection strategy to ensure integrity, consistency, and usability for training \gls{ml}/\gls{dl}-based models. These datasets collectively provide a reproducible foundation for evaluating detection and mitigation mechanisms in protocol-specific \gls{hiot} scenarios.

\textbf{RQ3: Are deep learning techniques (e.g., \gls{tcn}) capable of detecting and mitigating critical cyber threats (e.g., \gls{ddos}) across \gls{hiot} systems?}

Chapter~\ref{chap:tcn} of this thesis addresses this question by employing a \gls{tcn} model to detect and mitigate \gls{ddos} attacks in \gls{g5} \gls{hiot} environments. The model is trained and evaluated using the \gls{mqtt} and \gls{udp}-based datasets developed earlier, integrating a frequency monitoring mechanism to identify \gls{ddos}-affected \gls{hiot} nodes. It further incorporates a dynamic threshold mechanism for real-time mitigation, enabling automated blacklisting of detected nodes. The results demonstrate that the \gls{tcn} achieves high detection and mitigation accuracy across both protocol-specific datasets, outperforming conventional \gls{dl} models such as \gls{cnn} and \gls{bilstm}.

\textbf{RQ4: What approaches enable the edge deployment of deep learning models (e.g., \gls{tcn}) on resource-constrained devices (e.g., Raspberry Pi 4) in \gls{hiot}?}

Chapter~\ref{chap:edge} of this thesis explores the feasibility of deploying a lightweight \gls{tcn} model for \gls{ddos} detection on an edge platform, Raspberry Pi 4. The model is quantised and converted into \gls{tflite} format to reduce computational complexity and model size, while maintaining detection accuracy and inference time. Power consumption is also measured using a physical power meter to evaluate energy consumption during real-time operation. The results demonstrate that the lightweight \gls{tcn} achieves low latency and energy efficiency, demonstrating the feasibility of deploying \gls{dl} models on resource-constrained \gls{hiot} devices.

\textbf{RQ5: How can deep learning models such as \gls{restcn} be utilised for multiclass classification to distinguish among multiple attack categories, including \gls{sf}, \gls{mitm}, and \gls{ddos}, in \gls{hiot} environments?} 

Chapter~\ref{chap:restcn} of this thesis highlights multiclass cyber threat detection using a \gls{restcn} model in \gls{hiot} environments. A realistic simulation of multiple cyber threats is developed on a unified \gls{hiot} topology using the Cooja simulator. It generates network traffic comprising normal and malicious activities across three attack categories: \gls{sf}, \gls{mitm}, and \gls{ddos}. The \gls{restcn} integrates residual and temporal convolutional layers to enhance feature learning and ensure stable classification performance. Class imbalance is mitigated using \gls{smote} during training. The results demonstrate that the model achieves high multiclass detection accuracy and generalisation, highlighting its effectiveness in classifying concurrent cyber threats in \gls{hiot} environments.

\section{Overarching Discussion on Securing \gls{hiot}}
This section presents an overarching discussion that unifies the key outcomes of this research across dataset generation, \gls{dl}–based detection, mitigation, and edge deployment in \gls{hiot} environments. It highlights how analysing differences in data transmission behaviour between normal and malicious nodes led to the development of realistic datasets and lightweight models capable of detecting, mitigating, and classifying emerging cyber threats under resource constraints. The discussion further connects these findings to the broader goal of enhancing cybersecurity resilience in \gls{hiot} systems through practical, deployable \gls{dl} solutions.

The data transmission interval varies between 20 seconds and 30 seconds for malicious \gls{hiot} nodes across different simulations using the Cooja simulator. It represents various levels of transmission aggressiveness, mimicking how attack nodes may send data more frequently to overload the network, allowing the dataset to capture both moderate and high-intensity malicious behaviours. As a result, the \gls{ml}/\gls{dl} models are exposed to diverse traffic patterns and inter-arrival times, improving their adaptability to real-world \gls{iot} attack dynamics. In the unified simulated \gls{hiot} topology, normal and \gls{sf} nodes transmit at 60-second intervals, whereas \gls{ddos}-affected nodes transmit at 30-second intervals. This configuration ensures that malicious nodes generate roughly twice the traffic of normal nodes, allowing balanced total message collection and sufficient attack samples for dataset creation. It also helps the model learn stable temporal patterns, as statistical features such as mean, standard deviation, minimum, and maximum values become consistent and less affected by noise. This noise refers to irregular fluctuations in packet timing and volume within the network topology. By reducing such variability, the model can accurately identify changes in network behaviour and distinguish between different attack intensities, whether low or high.

With respect to \textbf{RQ1} in Chapter~\ref{chap:review}, the review-based investigation identifies critical challenges, including limited device-to-device interactions, a lack of realistic datasets, and barriers to privacy compliance. As part of that analysis, a conceptual \gls{cnn}-based anomaly-detection framework is introduced to demonstrate how \gls{dl} can be applied to secure and privacy-aware operations within \gls{hiot}. To support \textbf{RQ1}, a simplified \gls{cnn} implementation is used to demonstrate the feasibility of the proposed conceptual approach for anomaly detection in \gls{hiot} systems. Hence, this research builds on the conceptual insights from the review, translating the introduced \gls{cnn}-based anomaly detection method into a practical validation. The insights derived from this work lead to the creation of realistic \gls{hiot} datasets addressing different research questions and support the design of lightweight temporal \gls{dl} models for cyber threat detection and mitigation.

In addressing \textbf{RQ2} in Chapter~\ref{chap:dataset}, the analysis focuses on generating realistic synthetic datasets for \gls{hiot} environments in Chapter~\ref{chap:dataset}. In the \gls{mqtt}-based dataset~\emph{UL-ECE-MQTT-DDoS-\gls{hiot}2025}, four relevant features, such as node IDs, total messages, total messages per node, and frequency, are used, and the \gls{tcn} model achieves high performance. This finding suggests that a compact and relevant feature set can effectively represent traffic behaviour in \gls{hiot} systems. The feature selection depends on the communication protocol and the network-level parameters (e.g., inter-arrival time) accessible in the simulated \gls{hiot} topology. In \gls{mqtt}-based \gls{ddos} scenarios, a higher transmission rate, reflected by reduced inter-arrival time and increased packet counts, is a key indicator of malicious behaviour. These parameters enable \gls{ml}/\gls{dl} models to capture temporal dependencies between events over time and distinguish \gls{ddos}-related anomalies in \gls{hiot} network activity.

Regarding \textbf{RQ3} in Chapter~\ref{chap:tcn}, this research focuses on the \gls{dl}-based detection and mitigation of malicious nodes in \gls{hiot} systems. While most existing studies focus solely on anomaly detection, this research extends the investigation to immediate mitigation after detection, ensuring continuous network stability. In this context, mitigation involves isolating detected malicious nodes to prevent further disruption while maintaining normal device communication.

The proposed \gls{tcn} model detects \gls{ddos}-affected nodes by monitoring frequency, where each node’s message activity is calculated over a five-second observation period as a proportion of total messages. Once detected, the malicious nodes are automatically added to a blacklist for real-time mitigation in \gls{hiot} settings. Several mitigation strategies exist, including rule-based blocking, rate limiting, and threshold-based adaptation. This research employs the dynamic threshold approach to mitigate identified cyber threats, given its adaptive behaviour and computational efficiency. It is characterised by minimal processing time and power consumption, making it suitable for resource-constrained \gls{hiot} environments. The key advantage of this approach lies in its ability to adjust the mitigation threshold dynamically based on node-level activity, such as the number of detected malicious messages and overall network traffic behaviour. For instance, in the \gls{mqtt}-based dataset~\emph{UL-ECE-MQTT-DDoS-\gls{hiot}2025}, the base threshold is reduced by 20\% to enhance mitigation accuracy while minimising false positives, ensuring that only actual \gls{ddos}-affected nodes are blacklisted. This fine-tuning process demonstrates flexibility across different simulation scenarios using the Cooja and NS-3 simulators. Furthermore, Algorithm~\ref{algorithm:algo} illustrates how the threshold is iteratively adjusted within a feedback loop to maintain accuracy and responsiveness for real-time mitigation in \gls{hiot} settings.

Concerning \textbf{RQ4} in Chapter~\ref{chap:edge}, this research deploys the \gls{tcn} model on the edge platform, Raspberry~Pi~4, in \gls{hiot} environments. On the \gls{mqtt}-based dataset \emph{UL-ECE-MQTT-DDoS-\gls{hiot}2025}, the original \gls{tcn} model achieves 99.98\% accuracy, whereas the converted \gls{tflite} model on the Raspberry~Pi~4 maintains 99.95\% accuracy. The latency is 0.19~ms on the edge device and 0.06~ms on Google~Colab before conversion to \gls{tflite}. This minor trade-off is offset by the reduction in parameters (63,617~$\rightarrow$~63,136) and memory footprint (248.5~KB $\rightarrow$ 89.1 KB), improving energy efficiency and deployment feasibility under constrained resources. Specifically, the \gls{tcn} model exhibits low average power consumption, measured at 4.22\,W for the \emph{UL-ECE-MQTT-DDoS-\gls{hiot}2025} dataset and 4.64\,W for the \emph{UL-ECE-UDP-DDoS-\gls{hiot}2025} dataset, confirming its suitability for sustained real-time operation on edge devices.

As for \textbf{RQ5} discussed in Chapter~\ref{chap:restcn}, this research develops the multiclass dataset~\emph{UL-ECE-MultiAttack-\gls{hiot}2025}, which includes normal and malicious data from \gls{sf}, \gls{mitm}, and \gls{ddos} attacks. The dataset integrates diverse physiological indicators (e.g., minimum and maximum body temperature) with network-layer metrics (e.g., total message count), providing a comprehensive basis for analysing different cyberattack behaviours in \gls{hiot} environments. Although physiological metrics are incorporated, the dataset is entirely synthetic, ensuring that it remains outside the scope of \gls{gdpr} constraints. The proposed \gls{restcn} model is trained on this dataset to detect emerging cyber threats while maintaining privacy-preserving and resource-efficient characteristics. These aspects align with the \gls{enisa} and \gls{nis2d} objectives for resilient and uninterrupted healthcare operations during cyber attacks.

Another point regarding \textbf{RQ5} is that the \gls{restcn} model is finalised through a systematic ablation study across eight configurations, varying the number of residual blocks, the number of dense layer units, and the dropout rates. The study also compares performance between \gls{smote}-balanced and original datasets, as the data are inherently imbalanced due to the inclusion of diverse cyber threats within a unified \gls{hiot} network topology. The results indicate that applying \gls{smote} balancing ensures consistent performance across all attack classes while balancing accuracy, stability, and computational efficiency. Therefore, the \gls{restcn} model demonstrates high efficacy in detecting emerging cyber threats within \gls{hiot} systems.

Therefore, this thesis addresses the overarching research question by integrating dataset generation, \gls{dl}–based detection and mitigation, and edge deployment for \gls{hiot} environments. It analyses differences in message transmission patterns between normal and malicious nodes and evaluates the proposed models on resource-constrained edge devices. These contributions address the overarching research question by linking dataset design, model development, and deployment feasibility in \gls{hiot}.

\section{Key Contributions of the Thesis}
The key contributions of this research are as follows:
\begin{itemize}

\item \textbf{Creation of realistic \gls{hiot} cybersecurity datasets:} development of two protocol-specific synthetic datasets, \emph{UL-ECE-MQTT-DDoS-\gls{hiot}2025} and \emph{UL-ECE-UDP-DDoS-\gls{hiot}2025}, for evaluating \gls{ml}/\gls{dl}-based \gls{ddos} detection and mitigation methods in \gls{hiot} environments. A multiclass dataset, \emph{UL-ECE-MultiAttack-\gls{hiot}2025}, is also developed to support broader cyber threat classification.

    \item \textbf{\gls{dl} model for \gls{ddos} detection and mitigation:} design and implementation of a \gls{tcn} model for detecting and mitigating \gls{ddos} attacks through frequency monitoring and dynamic threshold-based adaptation in \gls{hiot} environments.

\item \textbf{Lightweight edge deployment of the \gls{tcn} model:} implementation of the quantised \gls{tcn} on Raspberry~Pi~4 using \gls{tflite}, validated through real-time latency and power consumption measurements to ensure efficiency under constrained resources.

\item \textbf{Emerging cyber threat detection using \gls{restcn} in \gls{hiot}:} implementation of the \gls{restcn} model using the dataset \emph{UL-ECE-MultiAttack-\gls{hiot}2025} for classifying \gls{ddos}, \gls{sf}, and \gls{mitm} attacks. The model utilises \gls{smote} to address data imbalance, achieving consistent and reliable classification performance across all attack types.

\item \textbf{Privacy-compliant and resource-efficient \gls{ids} workflow for \gls{hiot}:} establishment of a \gls{dl}-based \gls{ids} aligned with cyber resilience objectives.  
It ensures that \gls{gdpr} constraints do not apply, as the research is based on synthetic data, thereby supporting the development of secure and sustainable \gls{hiot} systems.

\end{itemize}

\section{Future Work}
This research demonstrates significant progress in addressing cybersecurity challenges within \gls{hiot} systems. Despite the promising outcomes, several areas remain open for further investigation. The following points outline potential directions for future research:

\begin{enumerate}
 \item \textbf{Expanding Threat Coverage and Adversarial Robustness:}  
Extend the current detection approach to identify a broader range of emerging cyber threats in \gls{hiot}, including \gls{g5} exploits (e.g., device or protocol-level vulnerabilities) and adversarial attacks that attempt to mislead \gls{ml}/\gls{dl}-based \gls{ddos} detection methods. Detecting such adversarial behaviours requires effective feature extraction, and this research already incorporates key features, such as the 5-second monitoring frequency and mean frequency, for identifying newly appearing malicious \gls{hiot} nodes, as discussed in~Chapter~\ref{chap:dataset}. Nevertheless, detecting adversarial attacks may require introducing additional features, such as packet inter-arrival variation, sudden changes in traffic burst intensity, entropy (a measure of packet randomness) in packet size and flow distribution, and correlations in traffic behaviours across multiple nodes to identify coordinated or disguised attack patterns. Future work may focus on developing adaptive detection mechanisms that enhance robustness against adversarial and \gls{g5}-related threats.

\item \textbf{Integration of Proposed \gls{dl} Methods with Software Defined Networking:}  
\gls{sdn} is a programmable networking paradigm that separates the control plane from the data plane, enabling centralised and flexible traffic management, as discussed in Chapter~\ref{chap:review}. Integrating the proposed \gls{dl}-based detection models (e.g., \gls{tcn}) with SDN can facilitate automated network reconfiguration, such as rerouting legitimate traffic, isolating compromised nodes, and dynamically adjusting flow rules based on real-time detection outcomes. For example, in this research, the lightweight \gls{tcn} model, deployed on edge devices such as a Raspberry~Pi~4, can share \gls{ddos} detection results with \gls{sdn} controllers. This integration enables the \gls{hiot} network to automatically apply mitigation policies in response to detected malicious nodes. It also enhances coordination between edge intelligence and network control, improving responsiveness and overall system resilience. Future work may utilise lightweight \gls{sdn} simulation environments, such as Mininet with the Ryu controller~\cite{sheikh2024qualitative}, to evaluate scalability and performance within \gls{hiot} networks.

   \item \textbf{Real-World Deployment and Energy Optimisation :} As discussed in Chapter~\ref{chap:edge}, this thesis demonstrates edge deployment on Raspberry Pi 4; future work can extend this setup using more realistic \gls{hiot} traffic simulations with sensor nodes transmitting to a central server. The deployed \gls{tcn} model then operates within this testbed to detect abnormal or suspicious node behaviours in real time. This practical setup can be implemented using Mininet to emulate \gls{hiot} nodes, a Mosquitto broker for application traffic, a Ryu \gls{sdn} controller for mitigation, and a Raspberry Pi 4 running the quantised \gls{tcn}. This setup would enable evaluation of sustained detection performance, stability, and power efficiency under near-operational conditions in resource-constrained \gls{hiot} environments. 
    
\item \textbf{Explainable and Trustworthy \gls{ai} in \gls{hiot}:}  
Beyond the scope of this thesis, future research can explore explainability techniques, such as \gls{shap} and \gls{lime}, to clarify which network features or temporal patterns influence the detection outcomes of models such as \gls{tcn} and \gls{restcn} in \gls{hiot} environments~\cite{salih2025perspective}. This approach extends beyond correlation by examining how variations in key network features (e.g., packet rate) directly lead to malicious behaviours. For example, a sudden spike in packet frequency may cause bandwidth saturation or trigger abnormal communication patterns. The objective is to understand how specific network behaviours influence the model’s classification results, enabling more straightforward interpretation of why particular traffic is identified as normal or malicious. Future work can also incorporate human-in-the-loop validation, where healthcare operators review model explanations and provide feedback to help cybersecurity experts refine model behaviour in line with real-world \gls{hiot} network operations. This collaboration can further strengthen trust, accountability, and the safe adoption of \gls{ai}-driven security mechanisms in practical \gls{hiot} systems.

\end{enumerate}

\rev{Overall, this thesis demonstrates the development and evaluation of \gls{dl}–based approaches for detecting and mitigating cyber threats under realistic network and traffic conditions in \gls{hiot} systems.} The scope of the thesis encompasses multiple attack scenarios aligned with practical cyber threat landscapes in \gls{hiot} settings. It further underscores the relevance of modelling diverse attack scenarios to continued research in this domain. In this context, the use of emulated wearable devices and edge-based deployment on Raspberry Pi 4 enables evaluation under practical constraints related to computation, memory, and inference latency. As a result, the proposed approach addresses the core components of the cybersecurity workflow in a coherent, practical design within \gls{hiot} settings.

The findings of this thesis provide a foundation for scalable and resilient cybersecurity mechanisms in \gls{hiot} deployments. These support future academic research and practical implementations, thereby contributing to the development of secure, deployable \gls{hiot} infrastructures.






\appendix

\chapter{Conference and Poster Presentations} \label{appendix:presentations}

The author presented parts of this doctoral research at the following conferences:

\begin{enumerate}[leftmargin=1.2cm,label={\arabic*.}]
  \item \textbf{Poster Presentation at the IEEE International Conference, USA:} \textit{Time Series Anomaly Detection with \gls{cnn} for Environmental Sensors in Healthcare-IoT}, presented at the \textbf{IEEE 12th International Conference on Healthcare Informatics (ICHI 2024)}, University of Florida, Orlando, USA. Available at: \href{https://www.researchgate.net/publication/383333927_Time_Series_Anomaly_Detection_with_CNN_for_Environmental_Sensors_in_Healthcare-IoT}{ResearchGate Publication Page}.
  

  \item \textbf{National Conference Presentation, Ireland:} \textit{Securing Healthcare IoT using \gls{ai}/\gls{ml}}, presented at the \textbf{HEAnet Conference 2023}, organised by Ireland’s National Education and Research Network (HEAnet). Available at: \href{https://www.heanet.ie/talk/securing-healthcare-to-iot-using-ai-ml?conference=2023}{HEAnet Conference Lightning Talks 2023}.

\end{enumerate}

\chapter{Research Code Repositories} \label{appendixB:code}

The author has made the simulation and experimental code publicly available to support reproducibility and further research. All repositories are accessible via the author's GitHub profile:  
\href{https://github.com/mirzaakhi}{mirzaakhi} 



\begin{enumerate}[leftmargin=1.2cm,label={\arabic*.}]
  \item \textbf{UL-ECE-H-IoT-Multiattack-Simulation:} C-based simulation scripts developed in Cooja to generate multiclass cyberattack datasets in \gls{hiot} environments. Used in Chapter~\ref{chap:restcn}.
  \textit{Repository link:} \href{https://github.com/mirzaakhi/UL-ECE-H-IoT-Multiattack-Simulation}{Author’s GitHub repository}

  \item \textbf{UL-ECE-IoT-DDoS-Detection-Power-Consumption-RaspberryPi4:} Power meter data and TensorFlow Lite inference evaluation scripts for \gls{tcn}-based \gls{ddos} detection on Raspberry~Pi~4. Used in Chapter~\ref{chap:edge}.
  \textit{Repository link:} 
  \href{https://github.com/mirzaakhi/UL-ECE-IoT-DDoS-Detection-Power-Consumption-RaspberryPi4}{Author’s GitHub repository}
\end{enumerate}

\chapter{Technical Articles and Online Resources}
\label{app:article}

The author has published the following technical articles on Medium, directly related to this doctoral research.
The articles are available via the author’s Medium profile: \url{https://medium.com/@mirzaakhi}

\begin{enumerate}
    \item \textbf{How to install ContikiOS and run Cooja Simulator on Windows 11 with Oracle VM VirtualBox}.  
   Describes Contiki-OS installation and Cooja simulator setup on Windows 11 for IoT simulation experiments.
    Available at: \href{https://medium.com/@mirzaakhi/how-to-install-contikios-and-run-cooja-simulator-on-windows-11-with-oracle-vm-virtualbox-2691fce267af}{Article 1 link} (2023).

    \item \textbf{How to Create Cooja Simulator Executable File on ContikiOS 3.0}.  
   Describes the process of building a Cooja simulator executable for Contiki-OS 3.0. 
    Available at: \href{https://medium.com/@mirzaakhi/how-to-create-cooja-simulator-executable-file-on-contikios-3-0-ec44f1053fa5}{Article 2 link} (2023).    
      \item \textbf{MQTT-SN Configuration on Contiki: A Step-by-Step Guide}.  
   Describes MQTT-SN configuration on Contiki using a step-by-step approach. Practical configuration steps from this article are used in Chapters~\ref{chap:dataset} and~\ref{chap:tcn} of this thesis. 
    Available at:
    \href{https://medium.com/@mirzaakhi/mqtt-sn-configuration-on-contiki-a-step-by-step-guide-679f42ae31fc}{Article 3 link} (2023).
    
   \item \textbf{How to Install NS-3 on Ubuntu with Oracle VM VirtualBox}.  
   Describes the installation of NS-3 on Ubuntu using Oracle VM VirtualBox. Used in Chapters~\ref{chap:dataset} and~\ref{chap:tcn} of this thesis.  
    Available at: 
     \href{https://medium.com/@mirzaakhi/how-to-install-ns-3-on-ubuntu-with-oracle-vm-virtualbox-f779496ae545}{Article 4 link} (2024).
      \item \textbf{A Practical Guide to CoAP Protocol Setup in IoT Using the Cooja Simulator}.  
 This article describes a practical CoAP protocol setup in IoT using the Cooja simulator. It relates to the research scope but is not directly applied in the thesis. 
    Available at: 
     \href{https://medium.com/@mirzaakhi/a-practical-guide-to-coap-protocol-setup-in-iot-using-the-cooja-simulator-b5ecef2d1956}{Article 5 link} (2025).
\end{enumerate}

\printbibliography[heading=bibintoc,title={References}]







\end{document}